%% file: main.tex
\documentclass[mnsc,nonblindrev]{informs3_homepage}

\usepackage{natbib}
\usepackage{adjustbox,float,comment,booktabs,diagbox,caption,subcaption,graphicx,bm, makecell,textcomp,mathtools,multirow,fancyvrb,upgreek,verbatim,longtable
}
\usepackage[table]{xcolor}
\usepackage[hidelinks,breaklinks=true]{hyperref}

\TheoremsNumberedBySection

\usepackage{tocloft}
\cftsetindents{section}{0em}{3em}
\cftsetindents{subsection}{3em}{4em}
\cftsetindents{subsubsection}{7em}{4.5em}

\newcommand{\nocontentsline}[3]{}
\newcommand{\tocless}[2]{\bgroup\let\addcontentsline=\nocontentsline#1{#2}\egroup}

\usepackage[ruled,norelsize]{algorithm2e}
\def\*#1{\mathbf{#1}}
\def\+#1{\mathbb{#1}}

\newcommand{\I}{{-1}}
\newcommand{\T}{\top}

\newcommand{\Lp}{\left(}
\newcommand{\Rp}{\right)}
\newcommand{\Ls}{\left[}
\newcommand{\Rs}{\right]}
\newcommand{\tlo}{\mathcal{O}}

\newcommand{\var}{\mathrm{Var}}
\newcommand{\Cov}{\textnormal{Cov}}

\newcommand{\rnd}{\textnormal{rnd}}
\newcommand{\OPT}{\textnormal{opt}}

\newcommand{\BA}{\mathrm{ba}}
\newcommand{\FF}{\mathrm{ff}}
\newcommand{\FFBA}{\mathrm{ffba}}

\newcommand{\norm}[1]{\left\lVert#1\right\rVert}
\newcommand{\RNum}[1]{\uppercase\expandafter{\romannumeral #1\relax}}
\newcommand{\diag}{\mathrm{diag}}
\newcommand{\tr}{\mathrm{tr}}

\newcommand{\plim}{\mathrm{plim}}

\newcommand{\fcs}{\mathrm{ntu}}

\newcommand{\ad}{\mathrm{atu}}

\newcommand{\iid}{\mathrm{iid}}

\newcommand{\all}{\mathrm{all}}
\newcommand{\within}{\textsf{within }}

\newcommand{\Prec}{\mathrm{Prec}}
\newcommand{\funfrac}{g_\tau}
\newcommand{\estsigmasq}{\widehat{\sigma^2}}
\newcommand{\samplesigmasq}{\widetilde{\sigma^2}}
\newcommand{\estxisq}{\widehat{\xi^2}}
\newcommand{\estxidaggersq}{\widehat{\xi^{\dagger 2}}}

\makeatletter
\usepackage{tikz}
\usetikzlibrary{calc}
\newcommand*\circled[1]{\tikz[baseline=(char.base)]{
    \node[shape=circle, draw, inner sep=1pt, 
        minimum height={\f@size*1.6},] (char) {\vphantom{WAH1g}#1};}}
\makeatother

\let\oldsection\section
\renewcommand{\section}{
\setcounter{equation}{0}
  \renewcommand{\theequation}{\thesection.\arabic{equation}}
  \oldsection}

\usepackage{tikz}
\usetikzlibrary{decorations.pathreplacing,calc}

\tikzset{brace/.style={decorate, decoration={brace}},
	brace mirrored/.style={decorate, decoration={brace,mirror}},
}

\newcounter{brace}
\newcounter{arrow}
\def\cmtfinal#1{{\textcolor{black}{#1}}}

\def\cmtmb#1{{\textcolor{black}{#1}}}

\setcitestyle{open={(},close={)}}

\bibpunct[, ]{(}{)}{,}{a}{}{,}
\def\bibfont{\small}

\definecolor{Red}{rgb}{1,0,0}
\definecolor{Blue}{rgb}{0,0,1}
\definecolor{Olive}{rgb}{0.41,0.55,0.13}
\definecolor{Green}{rgb}{0,1,0}
\definecolor{MGreen}{rgb}{0,0.8,0}
\definecolor{DGreen}{rgb}{0,0.55,0}
\definecolor{Yellow}{rgb}{1,1,0}
\definecolor{Cyan}{rgb}{0,1,1}
\definecolor{Magenta}{rgb}{1,0,1}
\definecolor{Orange}{rgb}{1,.5,0}
\definecolor{Violet}{rgb}{.5,0,.5}
\definecolor{Purple}{rgb}{.75,0,.25}
\definecolor{Brown}{rgb}{.75,.5,.25}
\definecolor{Grey}{rgb}{.5,.5,.5}
\definecolor{Pink}{rgb}{1,0,1}
\definecolor{DBrown}{rgb}{.5,.34,.16}

\definecolor{Black}{rgb}{0,0,0}

\def\blue{\color{Black}}

\def\cmtrx#1{{\textcolor{black}{#1}}}

\begin{document}

	\RUNAUTHOR{}
	\RUNTITLE{Optimal Experimental Design for Staggered Rollouts}
	\TITLE{Optimal Experimental Design for Staggered Rollouts{\Large \footnote{We thank seminar participants at Boston University, Columbia, Cornell, Cornell Tech, Emory, University of Florida, London Business School, National University of Singapore, Stanford, Toronto Rotman, University of Washington, UT Austin, Yale, Lyft Rideshare Labs, and participants at several conferences. We thank the editor, associate editor, and two anonymous referees for their insightful and helpful comments. Alphabetical author order other than the first author. Athey 
		and Imbens were supported by the Office of Naval Research under grant N00014-19-1-2468. Bayati was supported by the National Science Foundation grant CMMI: 1554140.
		}}}
	
	\ARTICLEAUTHORS{
		\AUTHOR{Ruoxuan Xiong}
		\AFF{Department of Quantitative Theory and Methods, Emory University, \EMAIL{ruoxuan.xiong@emory.edu}}
		\AUTHOR{Susan Athey}
		\AFF{Graduate School of Business, Stanford University, \EMAIL{athey@stanford.edu}}
		\AUTHOR{Mohsen Bayati}
		\AFF{Graduate School of Business, Stanford University, \EMAIL{bayati@stanford.edu}}
		\AUTHOR{Guido Imbens}
		\AFF{Graduate School of Business, Stanford University, \EMAIL{imbens@stanford.edu}}
	} 

	\ABSTRACT{
		In this paper, we study the design and analysis of experiments conducted on a set of units over multiple time periods where the starting time of the treatment may vary by unit. The design problem involves selecting an initial treatment time for each unit in order to most precisely estimate both the instantaneous and cumulative effects of the treatment. We first consider non-adaptive experiments, where all treatment assignment decisions  are made prior to the start of the experiment. For this case, we show that the optimization problem is generally NP-hard and we propose a near-optimal solution. Under this solution the fraction entering treatment each period is initially low, then high, and finally low again. Next, we study an adaptive experimental design problem, where both the decision to continue the experiment and treatment assignment decisions are updated after each period's data is collected. For the adaptive case we propose a new algorithm, the Precision-Guided Adaptive Experiment (PGAE) algorithm, that addresses the challenges at both the design stage and at the stage of estimating treatment effects, ensuring valid post-experiment inference accounting for the adaptive nature of the design. Using realistic settings, we demonstrate that our proposed solutions can reduce the opportunity cost of the experiments by over 50\%, compared to static design benchmarks.

	}
	

	\maketitle
 
\input{section_1}

    \input{section_2}

    \input{section_3}

\input{section_4}

	\input{section_5}

	\input{section_6}

	\bibliographystyle{informs2014}
	\bibliography{reference}

    \ECSwitch
    \ECHead{E-Companion}

	\input{appendix_A}

    \include{appendix_B}

	\include{appendix_C}

    \include{appendix_D}

	\include{appendix_E}
	
	\include{appendix_F}

	\include{appendix_G}



\end{document}

%% file: section_1.tex
\section{Introduction}\label{sec:intro}
	Large technology companies run tens of thousands of experiments (also known as A/B tests) per year to evaluate the impact of various decisions, new features, or products \citep{gupta2019top}. 
	In many cases, outcomes are observed 
  for multiple periods, where the units can change treatment status over time.  We refer to these experiments as \cmtfinal{panel experiments}; variants of these are sometimes referred to as experiments with staggered rollout or staggered adoption \citep{athey2018matrix}, or stepped wedge designs \citep{hemming2015stepped}. 

	\begin{example}[Driver Experience]\label{ex:driver-safety}
	Consider a ride-hailing platform that plans to test the impact of a new app feature that improves driver experience. They wish to run an experiment to estimate the effect of providing the app feature to all drivers. To avoid biases from interference between drivers it is useful to randomize at the city level where the starting time for the intervention may vary by city. 
	\end{example}
	\begin{example}[Public Health Intervention]\label{ex:infectious-disease}
Consider a country that aims to measure the effect of a new public health intervention ({\it e.g.}, encouraging the use of masks or social distancing policies) on the spread of an infectious disease \citep{abaluck2021impact}. To account for spillovers experimentation should be performed at an aggregate level ({\it e.g.}, cities). To facilitate the estimation of cumulative effects it is useful to vary the starting date. 
\end{example}	
	
The primary objective of this paper is to propose experimental designs to optimize the precision of post-experiment estimates of instantaneous and \cmtfinal{lagged} effects of a treatment. 
To operationalize this, the experimenter commits in advance to an estimator which  has a precision associated with each experimental design. The challenge is to find a design that optimizes this precision given the estimator.

We assume all units start in the control (no treatment) state at the initial time period. The design problem is to select, for each unit, the time period to begin treatment. We assume that units cannot switch back to control during the experiment, and thus remain treated once exposed to the treatment. This is a common setting.\footnote{The setting where units can arbitrarily switch between treatment and control is discussed in Sections \ref{sec:experiment-outcome-estimand} and \ref{subsec:reversible-treatment}, and turns out to be a simpler setting.} 
The treatment allocation time can vary across units, leading to a \emph{staggered treatment adoption} or \emph{stepped wedge design}.
    
    \subsection{Summary of Contributions}

    {\bf Non-adaptive experiments. }	
    We first study the design of non-adaptive experiments, where both the number of units and time periods and treatment decisions are determined prior to the start of the experiment. We focus on  the properties of the generalized least squares (GLS) estimator.
    We consider a general version of feasible GLS to estimate instantaneous and lagged treatment effects\footnote{Lagged treatment effect measures the effect of this period's treatment on a future period's outcome.} from observed outcomes that can be non-stationary.
    \cmtrx{The estimator determines the precision of estimated instantaneous and lagged effects. A linear combination of these precisions comprises the objective function for the experimental design problem. }
     Finding the optimal solution is generally NP-hard. We provide the analytical optimality conditions for the design, and propose an algorithm to choose a treatment design based on the optimality conditions. The precision of our selected design approximates the optimal objective (best achievable precision) within a multiplicative factor of $1+O(1/N^2)$ where $N$ is the number of units.
    
	Our solution to the design problem of non-adaptive experiments has two prominent features. First, the fraction of treated units per time period takes an $S$-shaped curve; the treatment rolls out to units more slowly (over time) in the beginning and at the end, but rapidly in the middle, of the experiment. Second, the optimal design imposes this rollout pattern for each stratum, where strata are defined by groups with the same observed and estimated latent covariate values.

	{\bf Adaptive experiments.}
	Next we study the design of adaptive experiments, where the number of units is fixed, but  the experiment can be terminated early and treatment assignment decisions can be adaptively made after each period's data is collected. These experiments are useful when the pre-set duration is more than needed to attain a target precision of treatment effect estimates, and when treatment decisions cannot be optimally made ex ante.  In our main contribution,
	we propose a new algorithm, the  Precision-Guided Adaptive Experiment or PGAE algorithm, that  adaptively terminates the experiment based on the estimated precision of the estimated treatment effect from partially observed results of the experiment. It employs dynamic programming \cmtfinal{to adaptively optimize}  \cmtfinal{the speed of treatment rollout in} subsequent time periods. The resulting adaptive experiment achieves a target precision, using a shorter duration, or equivalently incurring a lower cost, than the non-adaptive experiment. 

    The adaptive nature of the experiment  creates challenges arising from the fact that the outcomes and assignments that occur early in the experiment affect the treatment assignments to units later in the experiment.
    Thus, the treatment assignments are not independent of observed outcomes. We propose an estimation method based on sample splitting that ensures that the estimates obtained by PGAE are consistent and asymptotically converge to a normal distribution. 
     An appealing property of the proposed estimation scheme we propose is that the final treatment effect estimation uses \emph{all} of the data, incurring no efficiency loss compared to an oracle who would have access to the same design at the beginning of the experiment.
     
     Finally, we illustrate the superior performance of our solutions, as compared to benchmarks, for non-adaptive and adaptive experiments through synthetic experiments based on four real data sets about flu occurrence rates, home medical visits, grocery expenditure, and Lending Club loans.

    \subsection{Related Literature}
	Our \cmtfinal{staggered rollout designs of panel experiments} are related to a number of other experimental designs. They are most closely related to the stepped wedge designs in clinical trials \citep{brown2006stepped}. Stepped wedge designs sequentially roll out an intervention to clusters over several periods. Prior work on optimal stepped wedge design \citep{hussey2007design,hemming2015stepped,li2018optimal} studies optimal treatment assignments of clusters under a linear mixed outcome model that has no observed or latent covariates and assumes constant treatment effect over time  with instantaneous effects only. In contrast, we study the optimal design allowing treatment effects to vary over time ({\it i.e.}, with instantaneous and lagged effects). 

    Related to our design, a few other designs proposed recently are suitable for studying time-varying treatment effects. One design is the synthetic control design \cmtfinal{for} panel experiments \cmtfinal{where the design} selects \cmtfinal{units to be treated}, allocates treatment to \cmtfinal{all of} them \cmtfinal{in a single period}, and forms a synthetic treated and control unit for treatment effect estimation \citep{doudchenkodesigning2021,doudchenko2021synthetic,abadie2021synthetic}. Another design is the switchback design \citep{bojinov2020design,xiong2023bias} for a single experimental unit that can arbitrarily switch between treatment and control. In contrast to these two designs, our designs leverage variation in treatment times across units to increase power. The randomized designs proposed in \cite{basse2019minimax} also allow for cross-unit variation in treatment times, but they are studied under a different framework that minimizes the worst-case risk in randomization-based inference.

    When interference between units is a concern, our design can be modified by following a conventional approach to avoid biases  by  aggregating units to a level that \cmtfinal{interference} is no longer  a problem. 
    But we note a growing literature that directly tackles the biases using novel experimental design ideas, such as multiple randomization  \citep{bajari2021multiple,johari2020experimental} and designs \cmtfinal{that perturb treatments near equilibrium outcomes} \citep{wager2021experimenting}. In contrast, our design, by abstracting away from interference, can be used to study the rich dynamics of cumulative effects over time.

    Different from all the aforementioned designs, we additionally study the design and analysis of adaptive experiments. Our proposed PGAE algorithm consists of three components: adaptive treatment decisions, an experiment termination rule, and post-experiment inference. The adaptive treatment decision \cmtfinal{component} relates to the literature on adaptive designs in sequential experiments ({\it e.g.,} \cite{efron1971forcing,bhat2019near,glynn2020adaptive}), online learning and multi-armed bandits ({\it e.g.,} \cite{bubeck2012regret,lattimore2018bandit}). \cmtfinal{Distinct} from this literature, we consider a  \cmtfinal{panel setting}, and our design choices
    allow for experiment termination and ensure that inference is manageable.

    The experiment termination rule \cmtfinal{component} is based on the precision of the treatment effect estimate. \cmtrx{The precision-based termination rule has been used to obtain fixed-volume confidence sets from a sequence of independent random variables {\citep{glynn1992asymptotic,glynn1992asymptoticstopping,singham2012finite}}. We extend the use of this rule to the panel setting.}
    Our proposed PGAE algorithm decides whether to terminate the experiment at every period, which is related to the sequential testing problem, considered by \cite{siegmund1985sequential,wald2004sequential,bertsekas2012dynamic,johari2017peeking,ju2019sequential} among others. The key challenge in sequential testing is to draw valid inference post-experiment. The PGAE algorithm addresses this challenge through a sample-splitting technique. 
    
    We split units into disjoint sets, with each serving a different purpose.
    The idea of sample splitting has been used in the econometrics literature \cmtrx{(\cite{angrist1995split,angrist1999jackknife,athey2016recursive,chernozhukov2018double} among others)}
    for valid inference when machine learning methods are used and overfitting is a concern. \cmtmb{In contrast, we use the \emph{full sample} for the final treatment effect estimation that incurs no loss in estimation efficiency}. However, splitting samples into disjoint sets of units is crucial in decoupling experiment termination from inference in PGAE. In multi-armed bandits and other sequential decision problems, the splitting of \cmtfinal{data in adaptive experiments} has been considered by, among others, \cmtmb{\cite{auer2003using,goldenshluger2013linear,bastani2020online,hamidi2019personalizing}} for consistent estimation of the arm parameters. In contrast to this literature, we repeatedly experiment on the same set of units, and split samples in the unit dimension. \cmtmb{Finally, we note that \cite{lai1982least} showed normal approximation for the estimation error of adaptive least squares under a certain stability condition on the inverse covariance matrix. Recently, stronger results were obtained by leveraging debiasing techniques for OLS  \citep{deshpande2017accurate} and LASSO \citep{deshpande2021online}. In our panel setting, the stability condition of \cite{lai1982least} does not hold; however, the multi-unit nature of the problem allows us to avoid debiasing, even without the stability condition. Moreover, in our setting the experiment stopping rule is adaptively selected.
    }

%% file: section_2.tex
\section{\cmtfinal{Panel Experiments with Staggered Rollouts} and Assumptions}\label{sec:experiment-outcome-estimand}
    In this section, we first introduce \cmtfinal{panel experiments with staggered rollouts} and then describe the two types of experiments that are considered in this paper. We also specify assumptions on the outcome model and treatment designs. Throughout, $[M]$ refers to the set $\{1, 2, \ldots, M\}$, for any positive integer $M$, {and additional notations used in the paper are summarized in Table \ref{tab:notations}}.

    Recall Examples \ref{ex:driver-safety}-\ref{ex:infectious-disease}, and assume
    we are planning to run an experiment to estimate the effect of a treatment of interest (app feature or public health intervention). 
    Let $z_{it}$ in $\{-1,+1\}$ be the treatment variable for unit $i$ at time $t$, for $i, t \in \mathbb{Z}$, where $z_{it} = +1$ means unit $i$ is treated at time $t$ and $z_{it} = -1$ means otherwise. Assume the treatment is not applied to any unit before the experiment starts, that is, $z_{it} \equiv -1$ for all $i$ and $t \leq 0$. The experimental designer decides the treatment assignments in an experiment with $N$ units over $T$ time periods, that is, chooses $z_{it}$ for $i \in [N]$ and $t \in [T]$. All the decision variables can be written in a compact form $Z = [z_{it}]_{(i,t)\in [N] \times [T]}$, which is referred to as the treatment design of the experiment. Different treatment designs lead to different panels of observed outcomes, denoted as $Y = [Y_{it}]_{(i,t)\in [N] \times [T]}$, that affect the precision of treatment effect estimates. The estimation precision, to be defined formally in Section \ref{subsec:objective}, directly impacts the cost of running the experiment since it specifies the required number of units and time periods. Therefore, optimizing the treatment decisions is an integral part of designing \cmtfinal{panel experiments}.
    
    In this paper, we consider two types of \cmtfinal{panel experiments}: non-adaptive experiments and adaptive experiments. For nonadaptive experiments, $N$ and $T$ are fixed, whereas for adaptive experiments, $N$ is fixed, but $T$ is unknown. Non-adaptive experiments enjoy the benefit of simplicity and involve the straightforward construction of statistical tests for the treatment effects. However, non-adaptive experiments can be inefficient in time and cost, as the experiment may run longer than needed to attain a certain precision of treatment effect estimates. In comparison, adaptive experiments can early stop the experiment if needed, at the expense of making statistical inference for treatment effects more challenging. We study the design of non-adaptive experiments in Section \ref{sec:model}, where treatment decisions are made before the experiment starts. Building on our insights from the non-adaptive experiments, we propose an algorithm in Section \ref{sec:sequential-experiment} to run adaptive experiments that make treatment decisions adaptively during the experiment and provide valid post-experiment inference of treatment effects.

\cmtfinal{We focus on a class of panel experiments whose treatment assignments satisfy the irreversible treatment adoption condition. These experiments with treatment adoption times possibly varying across units are referred to as panel experiments with staggered rollouts; the design of these experiments is referred to as the staggered rollout design.}
\begin{assumption}[Irreversible Treatment Adoption]\label{ass:treatment-adoption}
	For all $(i,t)\in [N]\times[T]$,
	$z_{it} \leq z_{i,t+1}$.
\end{assumption}
We mostly focus on the irreversible scenario for two reasons. First, often there are practical constraints
restricting units from switching between control and treatment, such as the policies or programs implemented at the group level ({\it e.g.}, city or state level), or the features that are electronically rolled to the user interface ({\it e.g.}, ride-hailing app feature). Second, when the treated units switch back to the control, they may not return to the original control status. For example, drivers may develop different driving habits via the new feature. In fact, the irreversible pattern is common in observational studies ({\it e.g.}, \cite{card1994minimum,abadie2010synthetic}).
In the experiment design literature, the irreversible pattern appears in the stepped wedge designs of cluster randomized trials in public health \citep{brown2006stepped,hussey2007design,woertman2013stepped,hemming2015stepped}, and in the synthetic control designs \citep{doudchenkodesigning2021,doudchenko2021synthetic,abadie2021synthetic}. 
\begin{remark}[Extension to Reversible Treatments]\label{rem:reversible}
In Section \ref{subsec:reversible-treatment}, we study the design of non-adaptive panel experiments when the treatment can be stopped. We show that optimizing the treatment designs for this case follows from our results in Section \ref{sec:model}.
\end{remark}

	\begin{figure}[t!]
	\begin{subfigure}{0.4\textwidth}
		\centering
		\includegraphics[width=1\linewidth]{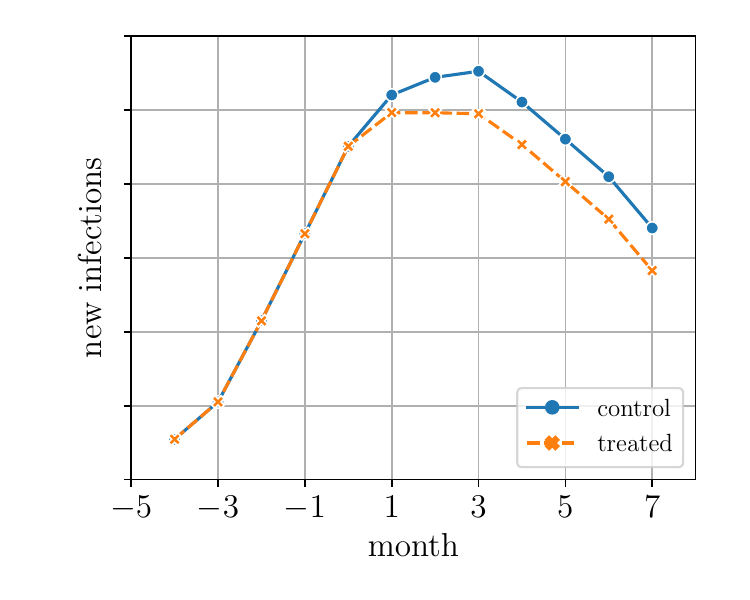}
		\caption{Example 1}\label{fig:cumulative-effect} 
	\end{subfigure}\hfill
	\begin{subfigure}{0.4\textwidth}
		\centering
		\includegraphics[width=1\linewidth]{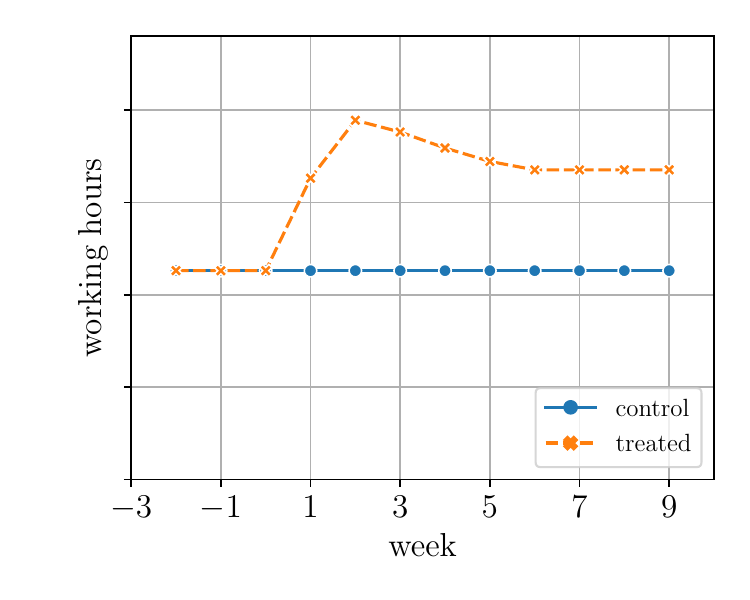}
		\caption{Example 2}\label{fig:wearout-effect} 
	\end{subfigure}
	\vspace{2mm}
	\caption{\textbf{Illustrative examples of cumulative effects.} These figures show the cumulative effects $\tau_0 + \cdots + \tau_j$ as a function of the number of treated periods $j$. Figure \ref{fig:cumulative-effect} illustrates a scenario where the effect of a new public health intervention on new infections accumulates over time ($\tau_0, \tau_1, \tau_2 < 0$). Figure \ref{fig:wearout-effect} illustrates a scenario where the effect of a new app feature on drivers' working hours attenuates over time ($\tau_0, \tau_1 > 0$ and $\tau_2, \tau_3, \tau_4, \tau_5 < 0$). 
	}
	\label{fig:example}
\end{figure}

Next we introduce the potential outcome model. 

\begin{assumption}[Potential Outcome Model]\label{ass:model}
The potential outcomes for unit $i$ at time $t$ can be written as
\cmtrx{\[Y_{it}(z_{i,t-\ell}, \cdots, z_{i,t-1}, z_{it}) \]
for a nonnegative integer $\ell$, where $\ell$ is known.
}

\end{assumption}

Assumption \ref{ass:model} requires that a unit's treatment $z_{it}$ only affects this unit's own outcomes (no cross-unit interference). It further requires that a unit's outcomes are not affected by this unit's future treatment assignments (no anticipation). For example, drivers do not increase their working hours in anticipation of the new app feature. No anticipation is also commonly imposed in the literature \citep{basse2019minimax,bojinov2020design}. 

We allow the potential outcomes to depend on the treatment assignments up to $\ell$ periods in the past. \cmtrx{We assume $\ell$ is known for the design and estimation purposes. Note that there exists a bias-variance tradeoff in the specification of $\ell$ in practice. When the specified $\ell$ is less than the true value, the estimator can be biased. When the specified $\ell$ is more than the true value, the estimator is unbiased, but can be inefficient. See Figure \ref{fig:varying-ell} in Section \ref{ecsec:spec-ell} for an example. }

We define the average instantaneous effect $\tau_0$ and lagged effects $\tau_1,\dots,\tau_\ell$ as 
 \cmtrx{   \[ 
        \tau_j \coloneqq \frac{1}{NT} \sum_{i,t}   \frac{1}{2} \Big[ Y_{it}(-1, \cdots, -1, \underbrace{+1}_{z_{i,t-j}}, +1, \cdots, +1) - Y_{it}(-1, \cdots, -1, \underbrace{-1}_{z_{i,t-j}}, +1, \cdots, +1) \Big]\,,
 \]}
 for all $j\in \{0\} \cup [\ell]$. 
 Here $1/2$ is used for notation simplicity in the specification \eqref{eqn:model-setup} below and remaining sections to account for the scaling of $z_{it}$ that takes value between $\{+1,-1\}$ as opposed to $\{+1,0\}$. Note that $\tau_j$ is the average, over individuals and time periods, of individual-time specific treatment effects that can be heterogeneous in both units and time periods.

Here $\tau_0, \tau_1, \cdots, \tau_\ell$ can take arbitrary values, and therefore allow for general dynamics of cumulative effects over time. To demonstrate this generality, we describe a few scenarios below.
	\begin{itemize}
	    \item The average effect accumulates over time, for example when $\tau_0,\ldots,\tau_\ell$ all have the same sign, as shown in Figure \ref{fig:cumulative-effect}.
	    
	    \item The average effect attenuates over time, for example when $\tau_0$ and $\tau_1$ have a different sign than all of $\tau_2,\ldots,\tau_\ell$, as shown in Figure \ref{fig:wearout-effect}.
	    
	    \item The average effect is constant over time, for example when $\ell = 0$.
	    
	    \item The average cumulative effect is zero, for example when $\sum_{j=0}^\ell \tau_j=0$. After $\ell$ periods, a unit's outcome is the same as the outcome of never being treated. 
	    
	\end{itemize}
	
		In this paper, we are interested in estimating $\bm{\tau} = (\tau_0, \tau_1, \cdots, \tau_\ell)$. Below we study how to choose $Z = [z_{it}]_{(i,t)\in [N]\times[T]}$
		to accurately estimate these parameters.

%% file: section_3.tex
\section{Non-Adaptive Experiments}\label{sec:model}
In this section, we study the \cmtfinal{staggered rollout} design of non-adaptive experiments. In Section \ref{subsec:estimator}, we present an approach for estimating $\tau_0, \tau_1, \cdots, \tau_\ell$.  Next, in Section \ref{subsec:objective}, we define the precision of the proposed estimators. We formulate an optimization problem, where for a given estimator, the solution yields the set of treatment times for each unit that maximizes the precision of the estimator. Finally, we present our solution to the optimization problem in Section \ref{subsec:result-carryover}.

 \subsection{Estimation of Instantaneous and Lagged Effects}\label{subsec:estimator}
    The decision maker needs to consider two objectives when choosing an estimator for $\tau_0, \tau_1, \cdots, \tau_\ell$. First is the statistical properties of this estimator, such as the bias, variance, and mean-squared error (MSE). Second is the feasibility of optimizing units' treatment times based on the properties of this estimator. For example, one can study the optimal treatment assignments based on MSE of a simple estimator \cmtrx{such as the difference-in-means estimator}, but this simple estimator can have a large \cmtrx{variance and} MSE. Alternatively, one can use a sophisticated estimator with a small MSE \cmtrx{such as regularized matrix factorization estimator in Section \ref{sec:algorithm}}, but the functional form of MSE may be very complex, which can lead to an intractable optimization problem in the design phase.

    In this paper, we seek to balance these two objectives. We study the optimization of treatment assignments, for a particular estimator, namely the Generalized Least Squares (GLS) estimator, given a particular model for the data generating process. The GLS estimator for $\tau_0, \tau_1, \cdots, \tau_\ell$ is based on  the following specification
    \begin{equation}\label{eqn:model-setup} 
    Y_{it} = \alpha_i +  \beta_t + \*X_i^\T  \bm{\theta}_t + \tau_0 z_{it} + \tau_1 z_{i,t-1} + \cdots + \tau_{\ell} z_{i,t-\ell}  + \underbrace{\*u_i^\T \*v_t + \varepsilon_{it}}_{e_{it}}\, ,
    \end{equation}
	where $\alpha_i$ and $\beta_t$ are unobserved unit and time fixed effects, $\*X_i  \in \+R^{d_x}$ are observed covariates of dimension $d_x$, $\bm{\theta}_t \in\+R^{d_x}$ are their (non-random) unobserved time-varying coefficients $\bm{\theta}_t$, $\*u_i \in \+R^{d_u}$ are latent (non-random) covariates of dimension $d_u$, $\*v_t \in \+R^{d_u}$ are (random) latent factors, and $\{\varepsilon_{it}\}_{it\in[N]\times[T]}$ are unobserved residuals. We allow $d_x = 0$, that is, \eqref{eqn:model-setup} does not have observed covariates. We also allow $d_u = 0$, that is, \eqref{eqn:model-setup} does not have latent covariates.

The specification with two-way ({\it i.e.}, unit and time) fixed effects and possibly with observed covariates has been widely used in observational studies to estimate treatment effects \citep{angrist2008mostly}.\footnote{There are two reasons. First, $\alpha_i$ and $\beta_t$ can be arbitrarily correlated with the observed explanatory variables $(\*X_i, {z}_{it}, \cdots, {z}_{i,t-\ell})$, so they can capture the unobserved additive unit-specific and time-specific confounders that jointly affect ${z}_{it}$ and $Y_{it}$ \citep{angrist2008mostly}. Second, $\alpha_i$ captures the mean effect of unobservables on unit $i$'s outcome over time, and $\beta_t$ captures the mean effect of unobservables on units' outcomes at time $t$. } Observed covariates $\*X_i$ in \eqref{eqn:model-setup} are time-invariant and are not affected by the treatment. Examples of such covariates are an individual's gender and race or attributes of a city.\footnote{The time-invariance assumption is commonly used in difference-in-differences estimators \citep{card1994minimum} 
and synthetic control estimators \citep{abadie2010synthetic} in observational studies.} Using $\alpha_i$, $\beta_t$ and $\*X_i$ controls for the heterogeneity in units and time periods, that can effectively reduce variance in treatment effect estimation. Note that \eqref{eqn:model-setup} has the interactive latent factor structure $\*u_i^\T \*v_t$ that can capture the multiplicative unobserved effects, and is therefore more general than the specification with additive effects $\alpha_i$ and $\beta_t$ only. \cmtrx{We assume $\*v_t$ is random, but $\*u_i$ is nonrandom for the design purpose. Such an assumption is also imposed in a different literature that estimates the latent factors on large panels  \citep{bai2002determining,bai2003inferential}. }

    	GLS estimates $\tau_0, \cdots, \tau_\ell$, $\alpha_i$, $\beta_t$ and $\bm{\theta}_t$ by minimizing the weighted sum of squared residuals $e_{it}$ in \eqref{eqn:model-setup} (see Section \ref{subsec:gls-carryover} for more details). GLS has two nice properties \cmtfinal{when the optimal weights are chosen, following} the Gauss-Markov Theorem (see Lemma \ref{lemma:gauss-markov} in Section \ref{subsec:gls-carryover}) under the strict exogeneity assumption in Assumption \ref{ass:constant-error} below. First, GLS is the best linear unbiased estimator (BLUE) of $\tau_0, \cdots, \tau_\ell$, meaning that this estimator has the smallest variance among all the linear unbiased estimators. Second, there is an explicit formula for the variance and precision of GLS in terms of $z_{it}$, which makes the treatment design problem tractable.

	\begin{assumption}[Error Structure]\label{ass:constant-error}
	\cmtrx{$\*v_t \in \+R^{d_u}$ is i.i.d. in $t$}
 with  $\+E[\*v_t  \mid \bm{\xi}_{it}] = \bm{0}$ and $\+E[\*v_t \*v_s^\T \mid \bm{\xi}_{it}, \bm{\xi}_{js} ] = \bm\Sigma_{v} \cdot \bm{1}_{t = s}$ for all $i, j, t$ and $s$, where $\bm{\xi}_{it} = (\*X_i, z_{it}, \cdots, z_{i-\ell, t})$. Moreover, $\varepsilon_{it} $ is i.i.d. in $i$ and $t$ with $\+E[\varepsilon_{it} \mid \bm{\xi}_{it}] =0$ and $\+E[\varepsilon_{it} \varepsilon_{js}  \mid \bm{\xi}_{it}, \bm{\xi}_{js} ] = \sigma_\varepsilon^2 \cdot \bm{1}_{i = j, t = s}$ for all $i, j, t$ and $s$. In addition, $\varepsilon_{it}$ is independent of $\*v_s$  for all $i, t$ and $s$.
	\end{assumption}

	\begin{remark}[Machine learning heuristics for $\bm{\tau}$]
	    Instead of focusing on the least squares estimator, one can look at other estimators, for example, machine learning estimators.
	    We provide one such estimator in Section \ref{sec:algorithm} for $\bm{\tau}$ that does not rely on Assumption \ref{ass:constant-error} about $\*v_t$. This type of machine learning-based estimators are typically biased in the estimation of $\bm{\tau}$, but could have a smaller variance than the unbiased estimators. The optimization of $Z$ based on these estimators is generally not tractable, and therefore we do not emphasize them in this paper.
	\end{remark}

		\begin{remark}[Implication of Assumption \ref{ass:constant-error}]
	$\+E[\*v_t] = 0$ holds without loss of generality, as we can always project their mean to $\alpha_i$.\footnote{If $\+E[\*v_t] \neq 0$, let $\alpha^\dagger_i = \alpha_i + \*u_i^\T \+E[\*v_t] $ and $\*v^\dagger_t = \*v_t - \+E[\*v_t]$, and then $\*v^\dagger_t $ has mean zero. } 
	Under Assumption \ref{ass:constant-error}, $e_{it}$ satisfies 
	\[\+E[e_{it}] = 0, \qquad \+E[e_{it}^2] = \*u_i^\T \*\Sigma_{v} \*u_i + \sigma_{\varepsilon}^2, \qquad \+E[e_{it} e_{jt}] = \*u_i^\T \*\Sigma_{v} \*u_j, \qquad \+E[e_{it}e_{js}] = 0, \quad \text{ for } \,\, t \neq s, \]
	implying that $e_{it}$ is correlated in the unit dimension, but it is uncorrelated in the time dimension.
	\end{remark}
\cmtfinal{\begin{remark}[Feasiblity of GLS]
     The weight matrix $\*W$ in GLS is optimal when it is proportional to the inverse covariance matrix of $\bm{e}_t = [e_{it}]_{i \in [N]}$, {\it i.e.}, $\big(\*U \*\Sigma_v \*U^\T + \sigma_\varepsilon^2 \*I_N \big)^\I $ with $\*U = [\*u_i]_{i \in [N]}$. As the optimal weight matrix is unknown, in practice, we can use feasible GLS, where we first use OLS, estimate the optimal weight matrix, and then use GLS with the estimated weight matrix. 
 \end{remark}
 }

	    \begin{remark}[Treatment Effect Estimation with Heterogeneity]\label{rem:hetreogeneous}
    Specifications like \eqref{eqn:model-setup} have been commonly used to estimate average treatment effects in empirical studies in many domains, such as economics \citep{card1994minimum}, operations ({\it e.g.}, \cite{cui2019learning,cachon2019does}),
    and healthcare \citep{abaluck2021impact}. Such specifications do not restrict treatment effects to be homogeneous. In Remark \ref{remark:cate} below, we discuss an alternative estimation approach when heterogeneous treatment effects are the objects of interest.
	\end{remark}

 \subsection{Optimization Problems for Treatment Decisions}\label{subsec:objective}
	
	In this section, we introduce the optimization problem to solve the optimal $Z$ based on the statistical properties of GLS. From a high level perspective, we are interested in precisely estimating $\bm{\tau}$. However, there are $\ell+1$ parameters in $\bm{\tau}$ and it is generally infeasible to find an $Z$ that simultaneously maximizes the precision of each of $\hat{\tau}_0, \cdots, \hat{\tau}_{\ell}$ for $\ell \geq 1$. Instead, one needs to consider an objective function that summarizes the precision of each of $\hat{\tau}_0, \cdots, \hat{\tau}_{\ell}$ into a scalar. There are two categories of objective functions that one may be interested in:
	\begin{itemize}
	    \item Balancing the precision/variance of each of $\hat{\tau}_0, \cdots, \hat{\tau}_{\ell}$
	    \item Maximizing the precision of some linear combination of $\hat{\tau}_0, \cdots, \hat{\tau}_{\ell}$
	\end{itemize}
	Examples of the objective functions related to the first category include \citep{atkinson2007optimum} 
	\begin{itemize}
	    \item \textbf{A-optimal design}: minimizes the trace of $\var(\hat{\bm{\tau}})$
	    \item \textbf{D-optimal design}: minimizes the determinant of $\var(\hat{\bm{\tau}})$
	    \item \textbf{T-optimal design}: maximizes the trace of the inverse of $\var(\hat{\bm{\tau}})$
	\end{itemize}
	An example of the objective functions related to the second category is
\begin{itemize}
    \item \textbf{Cumulative effect} ($\sum_{j=0}^\ell \tau_j$): minimizes $\var\big(\sum_{j = 0}^\ell \hat{\tau}_j\big)$
\end{itemize}


    When $\ell = 0$, the four objective functions mentioned above are equivalent to one another. For general $\ell$, they are not equivalent, and the optimal treatment assignments for these four objectives can be different (but the difference
can be small). 

    In this section, we focus on finding the optimal assignment
    $A = [A_i]_{i \in [N]}$, where $A_i \in [T] \cup \{\infty\}$ denotes the first time that unit $i$ adopts the treatment ($A_i=\infty$ means unit $i$ was never treated\footnote{We use ``$\infty$'' to preserve the ordering that a larger value for $A_i$ implies the treatment is assigned at a later time.}). Because the treatment is irreversible, there is a one-to-one mapping between $(z_{i1}, z_{i2}, \cdots, z_{iT})$ and $A_i$. Given the reduction to the $N$ adoption times we focus on
    analytically solving the T-optimal (or trace-optimal) design:
	 \begin{equation}\label{eqn:obj}
 \cmtrx{\max_{A}  \tr \big(\mathrm{Prec}(\hat{\bm{\tau}}; A)  \big)  ,}
	\end{equation}
 \cmtrx{where $\mathrm{Prec}(\hat{\bm{\tau}}; A)$ is the precision matrix of $\bm{\tau}$ and is defined as the matrix inverse of the variance-covariance matrix of $\bm{\tau}$, $\var(\hat{\bm{\tau}}; A)$, when the assignment $A$ is used.}
	

	T-optimal design was first introduced by \cite{atkinson1975design} to discriminate between two competing regression models ({\it e.g.}, to determine whether $\ell = \ell_0$ or $\ell = \ell_1$ is true, for distinct number of lags $\ell_0$ and $\ell_1$ in our setting). Since then, the T-optimal design has been studied by \cite{atkinson1975optimal,ucinski2005t,wiens2009robust,dette2012t,dette2013robust,dette2015bayesian} and others.  
	
	Solving the T-optimal design is generally challenging, even in some special cases (see Example \ref{example:T-1-ell-0}).
	We provide the explicit optimality conditions for the integer program \eqref{eqn:obj} in Section \ref{subsec:result-carryover}. Based on the optimality conditions, we provide an algorithm on choosing a design in Algorithm \ref{algo:choose-design} in Section \ref{subsubsec:estimate-latent-covariates}. 
	Admittedly, the other three objectives mentioned above would be natural and of practical interest, especially the one that minimizes the variance of the estimated cumulative effect, $\var\big(\sum_{j = 0}^\ell \hat{\tau}_j\big)$. However, analytically solving the other three objectives is generally infeasible, as explained in Remark \ref{rem:alternative-objectives} below. Instead, one could numerically solve the other three objectives in practice. We visualize the numerical solutions for D-optimal design in Figure \ref{fig:carryover-treatment-effect-d-opt} in Section \ref{subsec:d-optimal-design}, which has a similar structure as our solutions to \eqref{eqn:obj}. We empirically show in Section \ref{sec:empirical} and Section \ref{subsubsec:robustness-to-alternative-metrics} that our solutions to \eqref{eqn:obj} outperform the benchmark treatment designs measured by the objective of A-optimal design and $\var\big(\sum_{j = 0}^\ell \hat{\tau}_j\big)$.

	\begin{example}\label{example:T-1-ell-0}
	    If $T=1$, $\ell = 0$, and all the covariates are observed ({\it i.e.}, $d_u = 0$),  \eqref{eqn:obj} coincides with the offline optimization problem in \cite{bhat2019near}, and is equivalent to the MAX-CUT problem and is NP-hard \citep{hayes2002computing,mertens2006easiest}. In this paper, we focus on the case of low-dimensional covariates.\footnote{This makes sense as we allow for latent covariates, which can summarize the information and reduce the dimensionality of (high-dimensional) observed covariates.} Our solution from Algorithm \ref{algo:choose-design} is provably close to the optimal integer solution to \eqref{eqn:obj} for a large $N$. 
	\end{example}

    \begin{remark}[Challenges in alternative objectives]\label{rem:alternative-objectives}
    Note that each entry in $\var(\hat{\bm{\tau}}) $ is a ratio of two polynomial functions of $z_{it}$, where the degrees of numerator and denominator are $2\ell$ and $2(\ell+1)$, respectively, for $\ell \geq 1$. The objective of A-optimal design and $\var\big(\sum_{j = 0}^\ell \hat{\tau}_j\big)$ are both sums of entries in $\var(\hat{\bm{\tau}}) $ ({\it i.e.}, sums of ratios of two higher-order polynomials of $z_{it}$). The objective of D-optimal design is the inverse of a $2(\ell+1)$-th order polynomial function of $z_{it}$. \cmtfinal{The objective functions of both A-optimal and D-optimal design are non-convex. Therefore, using the first order condition only is generally not sufficient to solve the global optimal solution.}
 	\end{remark}

\begin{figure}[t!]
		\centering
		\includegraphics[width=0.4\linewidth]{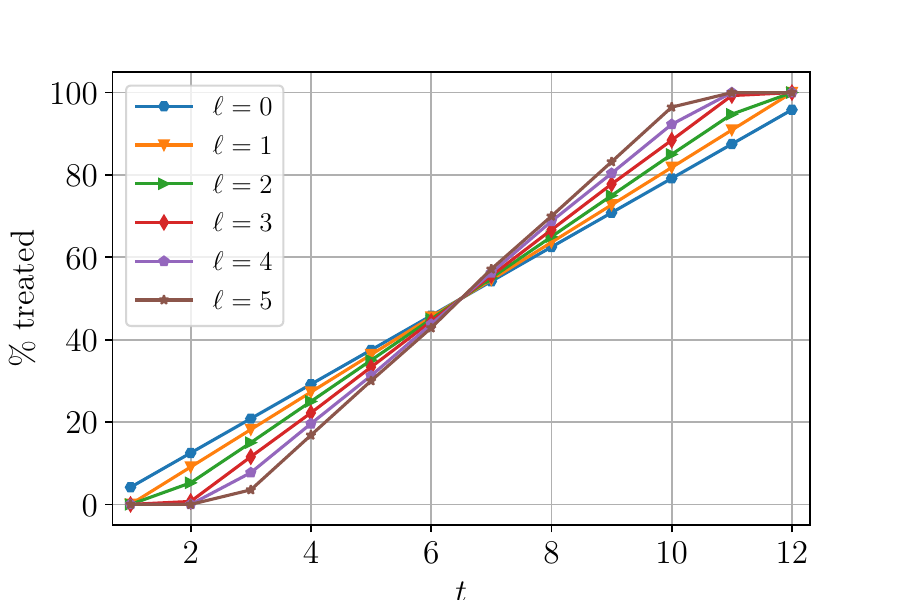}
		\caption{\textbf{{T}-optimal design.} Optimal treated fraction in the design of 12-period non-adaptive experiments with varying $\ell$. }
		\label{fig:carryover-treatment-effect-t-opt}
	\end{figure}

	{\blue 
 \subsection{Optimal Solutions}\label{subsec:result-carryover}
	We provide the optimality conditions for the T-optimal design. The optimality conditions disentangle the effect of different components in \eqref{eqn:model-setup} ({\it i.e.}, two-way fixed effects, observed and latent covariates) on the optimal treatment assignments. This problem is challenging in our setting with multiple units and periods for two reasons. First, different components can potentially affect the optimal design in both unit and time dimensions. Second, the effect of different components may be convoluted and interact with one another.
	
	To build intuition, we start with the solution to a simple specification.
	
	\begin{example}[Two-way fixed effects and $\ell = 0$]\label{example:two-way-fe}
	Suppose $Y_{it} = \alpha_i + \beta_t + \tau_0 z_{it} + \varepsilon_{it}$, and and Assumptions \ref{ass:treatment-adoption}, \ref{ass:model}, and \ref{ass:constant-error} hold. Then the objective function in \eqref{eqn:obj} equals to
\begin{align}\label{eqn:precision-ell-0}
      \mathrm{Prec}(\hat{\tau}_0; A)
    = \frac{N}{\sigma_\varepsilon^2 } \left[ - 2 \bm{b}_T^\T \bm{\omega} -\bm{\omega}^\T \*P_{\bm{1}_{T}} \bm{\omega} \right],
\end{align}
which is a quadratic and concave function of $\bm{\omega} = [\omega_t]_{t \in [T]}$ with $\omega_t = N^{-1} \sum_{i=1}^N z_{it}$, $\*P_{\bm{1}_{T}} = \*I_{T} - \bm{1}_{T} \bm{1}_{T}^\T/T$, and $\*b_T = [b_{t}]$ with $b_{t} = (T+1-2t)/T$. By solving the first-order condition, we can show that any treatment design satisfying
${\omega}^{\ast}_t = -b_t$ for all $t$ is optimal and maximizes the precision. In the optimal solution,  the treated fraction is linear in time. This result is conceptually similar to those in \cite{lawrie2015optimal,girling2016statistical,li2018optimal} that show the optimal treated fraction is linear in time, under a similar specification, but with $\alpha_i$ to be random effects. More intuition for the linear treated fraction is provided in Section \ref{subsec:treated-fraction}.

	\end{example}

Next consider a more general specification with $\ell > 0$, but without $\*X_i$ or $\*u_i$. We can show the objective function in \eqref{eqn:obj} is still quadratic and concave in $\bm{\omega}$, but takes a more complicated form (see Lemma \ref{lemma:simplify-obj} in Section \ref{subsec:separate-quadratic}), leading to nonlinear optimal treated fraction in time. Specifically in Theorem \ref{thm:obs-latent-carryover-model} we show that in the optimal solution, the unit average of $z_{it}$ satisfies 
		\begin{equation}\label{eqn:omega-carryover}
		\omega_{\ell,t}^\ast = \begin{cases}
		-1 & t \leq \lfloor \ell/2 \rfloor \\
		a^{(\ell)}_{t-\lfloor \ell/2 \rfloor} & \lfloor \ell/2 \rfloor < t \leq \ell \\
		-1 + \big(2t - (\ell+1)\big)/(T - \ell)  & \ell < t \leq T - \ell \\
		- \omega^\ast_{\ell,T+1-t} & T - \ell <  t \leq T-\lfloor \ell/2 \rfloor \\
		1 & T-\lfloor \ell/2 \rfloor < t \\
		\end{cases}
		\end{equation}
		where $a^{(\ell)}$ is a vector of length $\ell - \lfloor \ell/2 \rfloor$ defined in Section \ref{subsec:def-A-l-B-l}. $\omega_{\ell,t}^\ast$ has five stages and follows an $S$-shaped curve in time $t$: stage 1:  all units are under control; stage 2: $\omega_t^\ast$ grows non-linearly in time (because of $a^{(\ell)}$); stage 3: $\omega_t^\ast$ grows linearly in time; stage 4: $\omega_t^\ast$ grows non-linearly in time again; stage 5, all units are under treatment. The optimal solution is symmetric with respect to the center ({\it i.e.}, $\left((T+1)/2, 0 \right)$). Figure \ref{fig:carryover-treatment-effect-t-opt} demonstrates $\omega_{\ell,t}^\ast$ for various $\ell$ in a $T=12$ period problem.  More examples of $\omega_{\ell,t}^\ast$ will be provided in Section \ref{subsec:treated-fraction}.
		
	Furthermore, suppose the specification contains at least one of the $\*X_i$ or $\*u_i$ components. A commonly used variance reduction approach in cross-sectional studies is to balance covariates, so that treated and control units are comparable when the two groups have similar covariates ({\it e.g.}, \cite{imbens2015causal}).\footnote{There is a strand of literature in operations research to use discrete optimization to achieve covariate balancing \citep{nikolaev2013balance,bertsimas2015power,bertsimas2019covariate,bhat2019near}, and to use stratified sampling to increase power \citep{fox2000separability,mulvey1983multivariate}.} Similar intuition carries over to the panel setting. We show in Theorem \ref{thm:obs-latent-carryover-model} below that the optimal design balances covariates groups, where groups are defined as the set of units with the same initial treatment time.

	We state the first of our main theorems using the following solution concepts: Let $\mathbb{A}_{\ell} = \big\{ A:  N^{-1} \sum_{i = 1}^N \boldsymbol{1}_{A_i \leq t}  = (1+\omega_{\ell,t}^\ast)/2,\, \forall t\big\}$ be the set of designs satisfying \eqref{eqn:omega-carryover}. When $d_x > 0$, let $\mathbb{A}_{\*X} =  \big\{ A: N^{-1} \sum_{i = 1}^N  \*X_i \boldsymbol{1}_{A_i  =  t}  = \bm{0}_{d_x},\, \forall t \in \{2,\cdots,T\}\big\}$ be the set of designs satisfying covariate balancing conditions (suppose rows in $\*X = [\*X_i]_{i \in [N]}$ are centered). When $d_x = 0$, let $\mathbb{A}_{\*X} = \mathbb{A}^N$ and no conditions need to be imposed related to $\*X_i$. $\mathbb{A}_{\*U}$ is defined similarly as $\mathbb{A}_{\*X}$.
	}

		\begin{theorem}[Optimality Conditions]\label{thm:obs-latent-carryover-model}
		Suppose Assumptions \ref{ass:treatment-adoption}, \ref{ass:model}, and \ref{ass:constant-error} hold, $\hat{\bm{\tau}} $ is \cmtrx{estimated from the infeasible GLS}
		with $\*W \propto \big(\*U \*\Sigma_v \*U^\T + \sigma_\varepsilon^2 \*I_N \big)^\I $,  rows in $\*X$ and $\*U$ are centered $($i.e., $\sum_{i = 1}^N \*X_i = \bm{0}_{d_x}$ and $\sum_{i = 1}^N \*u_i = \bm{0}_{d_u}$$)$, $\*X$ and $\*U$ are orthogonal $($i.e., $\sum_{i = 1}^N \*X_i \*u_i^\T = \mathbf{0}_{d_x \times d_u}$$)$, $\*\Sigma_v = \sigma_\varepsilon^2 \cdot \*I_{d_u}$, and $T > (\ell^3+13\ell^2+7\ell+3)/(8\ell)$.
		 $A$ is an optimal treatment design if $A  \in \mathbb{A}_{\mathrm{opt}} $, where 
		\begin{equation}\label{eqn:carryover-t-optimal-obs-latent-thm}
		\mathbb{A}_{\mathrm{opt}} = \mathbb{A}_{\ell}  \cap \mathbb{A}_{\*X} \cap \mathbb{A}_{\*U}\, .
		\end{equation}
	\end{theorem}

	Theorem \ref{thm:obs-latent-carryover-model} provides the optimality conditions for the T-optimal design when $\tau_0, \cdots, \tau_\ell$ are estimated from GLS using specification \eqref{eqn:model-setup}. The presence of $\alpha_i$ and $\beta_t$ makes the optimal treated fraction grow gradually over time, and the growth rate is determined by $\ell$. We further illustrate how the optimal treated fractions depend on $\alpha_i$, $\beta_t$, and $\ell$ in Section \ref{subsec:treated-fraction}. The presence of $\*X_i$ and $\*u_i$ imposes additional covariate balancing conditions, that can be satisfied if units in each stratum satisfy treated fraction conditions. Here strata are defined as groups of units with the same covariate value.  When covariates are discrete-valued, Example \ref{example:discrete-covariate} below provides a solution that satisfies the optimality conditions. For general cases, we provide guidance on choosing a design based on Theorem \ref{thm:obs-latent-carryover-model} in Section \ref{subsec:choose-a-design}.
\cmtrx{\begin{example}[Discrete $\*X_i$ and $\*u_i$]\label{example:discrete-covariate}
     Suppose $(\*X_i, \*u_i)$ is discrete and can only take $G$ values for a finite $G > 0$, denoted as $\{(x_g, u_g)\}_{g=1}^G$, each with positive probability. Then any $A$ in $\mathbb{A}^{\mathrm{disc}}_{\mathrm{opt}}$ defined below is in $\mathbb{A}_{\mathrm{opt}}$
		\begin{equation}\label{eqn:opt-finite-x-interactive-effect-carryover}
		\mathbb{A}^{\mathrm{disc}}_{\mathrm{opt}} =\bigg\lbrace A: \frac{1}{|\tlo_g|} \sum_{i \in \tlo_g} \boldsymbol{1}_{A_i \leq t}  = \frac{1+\omega_{\ell,t}^\ast}{2}  \,\,\,\, \forall t, g \bigg\rbrace\,,
		\end{equation}
		where $\tlo_g = \{i: (\*X_i, \*u_i) = (x_g, u_{g})\}$.
 \end{example}}
	
    \begin{remark}[Assumptions in Theorem \ref{thm:obs-latent-carryover-model}]
    The assumptions in Theorem \ref{thm:obs-latent-carryover-model} are non-restrictive for the following four reasons. 
	First, the assumption that rows in $\*X$ and $\*U$ are demeaned can be satisfied by projecting the mean onto the unit fixed effects $\bm{\alpha}$. Second, the assumption that $\*X$ and $\*U$ are orthogonal can be satisfied by applying the Gram-Schmidt procedure ({\it i.e.}, QR decomposition) to $\begin{bmatrix}
	 \*X & \*U \end{bmatrix}$ (which is possible as $\bm{\theta}$ is unknown and unrestricted). Third, $\*\Sigma_v = \sigma_\varepsilon^2 \cdot \*I_{d_u}$ is essentially an identification assumption, so that we can uniquely identify $\*u_i$ and $\*v_t$.\footnote{In other words, for arbitrary $\*u_i$ and $\*v_t$, we can right multiply $\*u_i^\T$ by $\sigma_\varepsilon \cdot \*\Sigma_v^{-1/2}$ and left multiply  $\*v_t$ by $\sigma_\varepsilon \cdot \*\Sigma_v^{-1/2}$ so that $v_t$ has variance $\sigma_\varepsilon^2 \cdot \*I_{d_u}$ (conditions $\sum_{i = 1}^N \*u_i = \bm{0}_{d_x}$ and  $\sum_{i = 1}^N \*X_i \*u_i^\T = \mathbf{0}_{d_x,d_u}$  stay valid after this manipulation).} Fourth, the fundamental structure of our problem does not change with these assumptions because $\mathrm{Prec}(\hat{\bm{\tau}})$ is a quadratic function of $z_{it}$ regardless of whether these assumptions are imposed or not. 
    \end{remark}

    	\begin{remark}[Assumption on $T$]
    	The assumption $T > (\ell^3+13\ell^2+7\ell+3)/(8 \ell) $
		is a sufficient condition to show $\omega_{\ell,t}^\ast$ is monotonic, but is not a necessary condition and can potentially be relaxed, based on the numerical solution of treated fractions when this assumption is violated.
	\end{remark}
		
	    \begin{remark}[Magnitude of treatment effects]\label{rem:magnitude-treatment-effect}
    Note that $\mathbb{A}_{\mathrm{opt}} $ and $\mathbb{A}^{\mathrm{disc}}_{\mathrm{opt}}$ do not depend on the value of $\bm{\tau}$. This is because the estimation error $\hat{\bm{\tau}} - \bm{\tau}$ from GLS does not depend on the value of $\bm{\tau}$ under the specification \eqref{eqn:model-setup}. Therefore, $\tr \big(\mathrm{Prec}(\hat{\bm{\tau}})  \big)$ does not depend on the value of $\bm{\tau}$. See \eqref{eqn:precision-ell-0} for an example.
 	\end{remark}

 \subsubsection{Optimal Treated Fractions}\label{subsec:treated-fraction} \texttt{}
 \begin{figure}[t!]
	\centering
	\begin{subfigure}{0.2\textwidth}
		\centering
		\includegraphics[width=1\linewidth]{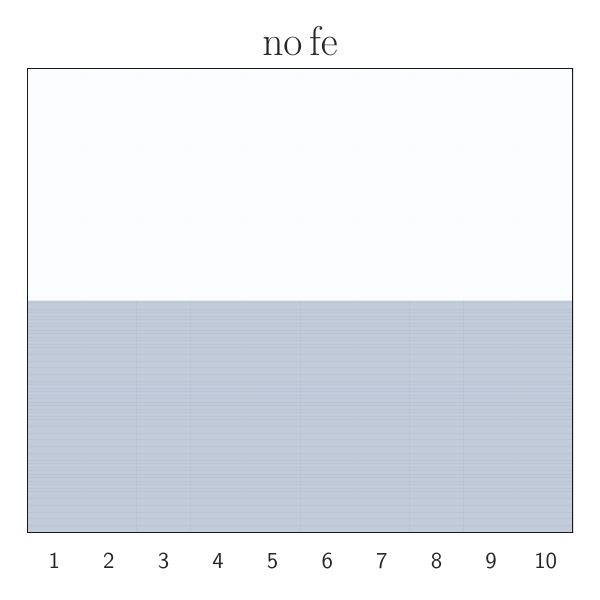}
	\end{subfigure}%
	\begin{subfigure}{0.2\textwidth}
		\centering
		\includegraphics[width=1\linewidth]{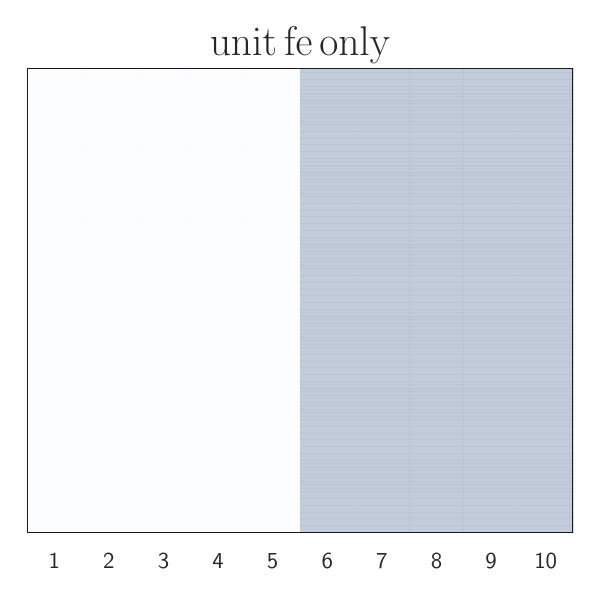}
	\end{subfigure}%
	\begin{subfigure}{0.2\textwidth}
		\centering
		\includegraphics[width=1\linewidth]{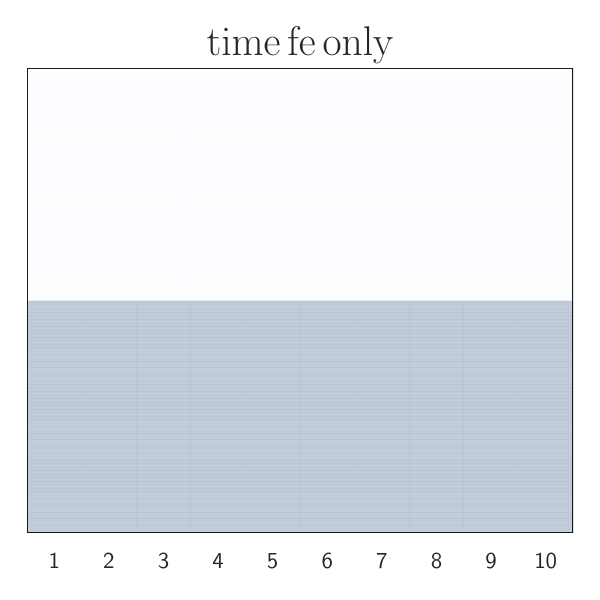}
	\end{subfigure}%
	\begin{subfigure}{0.2\textwidth}
		\centering
		\includegraphics[width=1\linewidth]{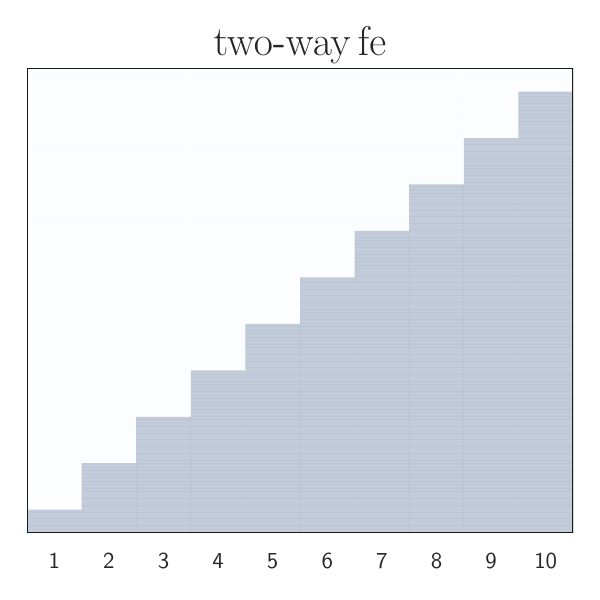}
	\end{subfigure}%
	\begin{subfigure}{0.2\textwidth}
		\centering
		\includegraphics[width=1\linewidth]{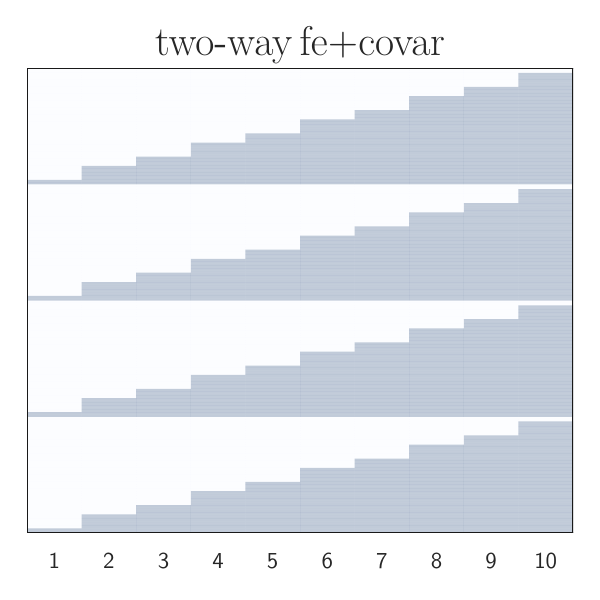}
	\end{subfigure}
	\caption{\textbf{Optimal treatment designs under various specifications.} Examples of T-optimal designs of $10$-period experiments with $\ell = 0$. Darker color denotes treatment, and lighter color denotes control. ``fe'' stands for ``fixed effects'' and ``covar'' stands for ``covariates''
	}
	\label{fig:various-optimal-design}
\end{figure}
	{\blue 
	
	To provide more intuition on the optimal treated fractions, we start with the specification with $\ell = 0$, and with either  $\alpha_i$ or $\beta_t$, but not both. 
	
	\begin{example}[Time Fixed Effects Only and $\ell = 0$]\label{example:time-only}
Suppose $Y_{it} = \beta_t + \tau_0 z_{it} + \varepsilon_{it}$ and Assumptions \ref{ass:treatment-adoption}, \ref{ass:model}, and \ref{ass:constant-error} hold. In this case,
$\mathrm{Prec}(\hat{\tau}_0)  = {N}/{\sigma_\varepsilon^2 } \cdot \big[T - \sum_{t=1}^T \omega_t^2  \big]$.
Any treatment design is optimal if it satisfies $\omega^\ast_t = 0$ for all $t$ and $z_{it} \leq z_{i,t+1}$ for all $i$ and $t$, that is, assigns treatment to 50\% units, while the others functioning as the control, for all time periods. 
\end{example}

\begin{example}[Unit Fixed Effects Only and $\ell = 0$]\label{example:unit-only}
Suppose	$Y_{it} = \alpha_i + \tau_0 z_{it} + \varepsilon_{it}$ and Assumptions \ref{ass:treatment-adoption}, \ref{ass:model}, and \ref{ass:constant-error} hold.  In this case,
$\mathrm{Prec}(\hat{\tau}_0) = {T}/{\sigma_\varepsilon^2 } \cdot \big[N - \sum_{i=1}^N \zeta_i^2 \big]$, 
where $\zeta_i = T^{-1} \sum_{t=1}^T z_{it}$.
Any treatment design is optimal if it satisfies $\omega^\ast_t = -1$ for all $t < (T+1)/2$ and $\omega^\ast_t = 1$ for all $t > (T+1)/2$, that is, allocates treatment to all units at halftime.\footnote{For odd $T$, units can be either treated or untreated at $t = (T+1)/2$.} 
\end{example}

    }

	If the specification has both $\alpha_i$ and $\beta_t$, as in Example \ref{example:two-way-fe}, 
	the optimal treated fractions are intuitively in between those in Examples \ref{example:time-only} and \ref{example:unit-only}. See Figure \ref{fig:various-optimal-design} for the visualization of optimal treated fractions under various specifications.

	The same intuition carries over to $\ell > 0$ and the precision matrix $\mathrm{Prec}(\hat{\bm{\tau}})$ is quadratic and concave in $\omega_t$. The two examples below provide the analytic expression of $\omega^\ast_{\ell,t}$ for $\ell = 1$ and $2$. We provide the expression of $\omega_{\ell,t}^\ast$ for $\ell = 3$ in Example \ref{remark:carryover-example-l3} of Section \ref{subsec:def-A-l-B-l}. 

	\begin{example}[$\ell = 1$]\label{example:ell-1}
	In Theorem \ref{thm:obs-latent-carryover-model}, $\omega^\ast_{\ell,t}$ equals
		$-1 + {2(t-1)}/{(T-1)}$ for all $t$.
	\end{example}
	\begin{example}[$\ell = 2$]\label{example:ell-2}
	In Theorem \ref{thm:obs-latent-carryover-model}, $\omega^\ast_{\ell,t}$ is determined by, 
$\omega_{\ell,1}^\ast = -1$, $\omega_{\ell,2}^\ast = -1 + {2}/{(2T-5)}$, $\omega_{\ell,t}^\ast = -1 + {(2t-3)}/{(T-2)}$ for $t = 3, \cdots, T-2$, $\omega_{\ell,T-1}^\ast = 1 - {2}/{(2T-5)}$, and $\omega_{{\ell,T}}^\ast = 1$.
	\end{example}

\begin{figure}[t!]
		\centering
		\includegraphics[width=0.5\linewidth]{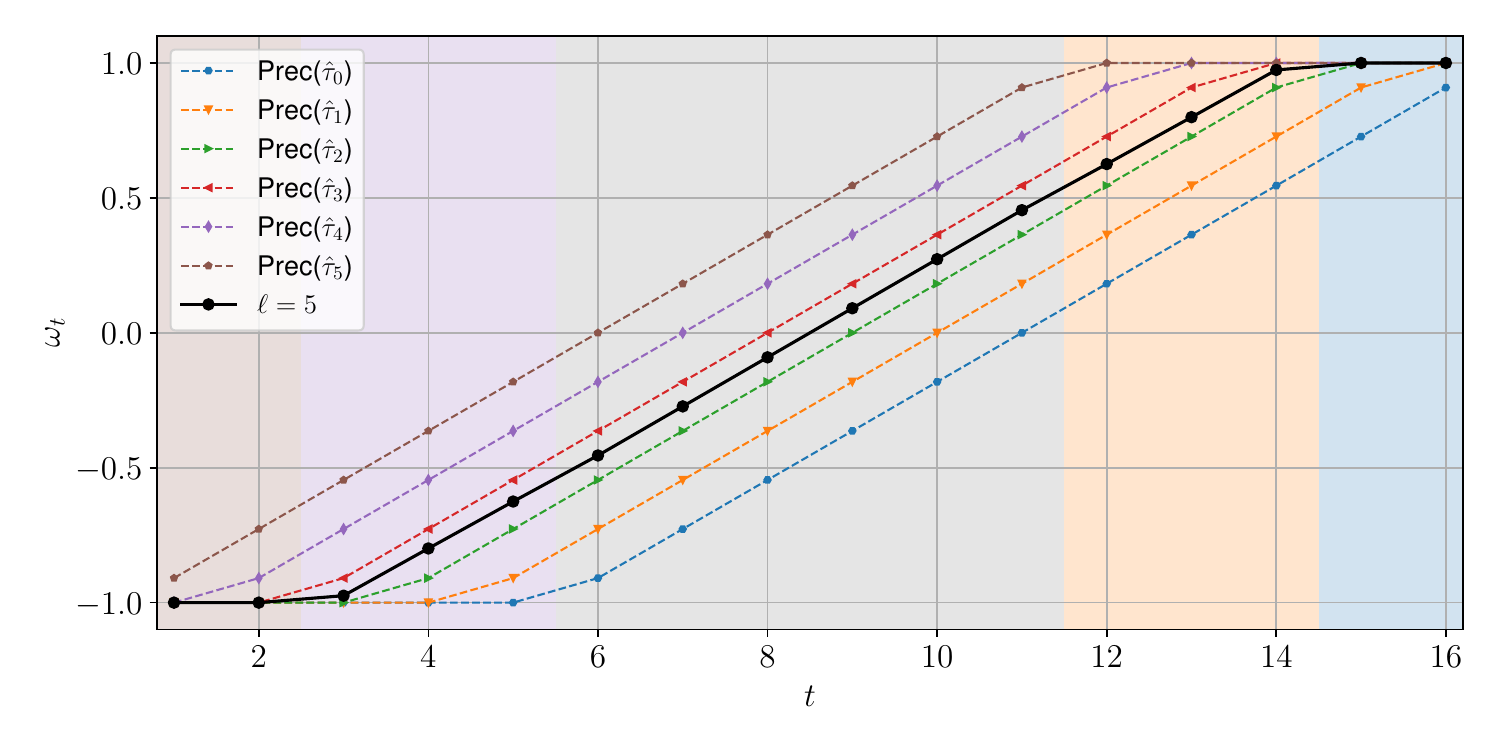}
		\caption{Illustration of the $S$-shaped curve for the $ 16$-period optimal design with $\ell = 5$. For estimating $\tau_j$ only, the dashed lines denote the optimal $\omega_t$ that maximizes $\mathrm{Prec}(\hat{\tau}_j)$ for $j \in \{0,1,\cdots,5\}$. For estimating $\tau_0, \cdots, \tau_5$ simultaneously, the solid line denotes the optimal $\omega_t$ that maximizes $\mathrm{Prec}(\hat{\bm{\tau}})$, which is a convex combination of the dashed lines. The shaded regions from left to right denote the first to fifth stages in the solution $\omega^\ast_{\ell,t}$.  }
		\label{fig:carryover-treatment-effect-t-opt-s-curve}
	\end{figure}

    {\blue \paragraph{Intuition for the $S$-shaped curve of $\omega_{\ell,t}^\ast$.} The objective function in \eqref{eqn:obj} is a sum of $\ell+1$ quadratic functions, one function for each  $\mathrm{Prec}(\hat{\tau}_j)$ with $j \in \{0,1,\cdots,\ell\}$. Let $\omega_{\ell,t}^\ast(j)$ be the unit average of $z_{it}$ in the solution that maximizes $\mathrm{Prec}(\hat{\tau}_j)$ at time $t$, when the duration of lagged effects is $\ell$. This means $\omega^\ast_{\ell,t}$, defined in \eqref{eqn:omega-carryover}, can be written as a convex combination of $\omega_{\ell,t}^\ast(j)$ for all $j$ and $t$. Intuitively, since each $\omega_{\ell,t}^\ast(j)$ is chosen to maximize the precision of a single parameter $\hat{\tau}_j$, it should grow linearly in $t$, like the $\ell=0$ case, except during the first $\ell$ or the last $\ell$ periods that it is truncated to $0$ or $1$ respectively (see Figure \ref{fig:carryover-treatment-effect-t-opt-s-curve}). 
    Equipped with this observation, it is easy to see that $\omega^\ast_{\ell,t}$ grows linearly in $t$ in the middle (the third stage) where all $\omega^\ast_{\ell,t}$ grow linearly. $\omega^\ast_{\ell,t}$ is constant in stage one and five, where all $\omega^\ast_{\ell,t}$ are constant. Moreover, $\omega^\ast_{\ell,t}$ has a non-linear growth in stage two and four, because some of $\omega_{\ell,t}^\ast(j)$ are truncated to $0$ or $1$. 
    }
    
 \subsubsection{Choosing A Treatment Design}\label{subsec:choose-a-design}\texttt{}
	
	{\blue Based on the optimality conditions in Theorem \ref{thm:obs-latent-carryover-model}, we provide Algorithm \ref{algo:choose-design} in Section \ref{subsec:more-details-choose-design} to choose a treatment design.} In this subsection, we outline some general guidance on choosing a treatment design. When the specification does not have covariates ($d_x = d_u = 0$) and $\mathbb{A}_{\ell} $ is not empty, we can randomly choose one from $\mathbb{A}_\ell$ with equal probability. When the specification has observed, discrete-valued covariates ($d_x \neq 0$) and $\mathbb{A}^{\mathrm{disc}}_{\mathrm{opt}}$ is not empty, we can randomly choose one from $\mathbb{A}^{\mathrm{disc}}_\ell$ with equal probability. The random sampling can ensure that the treatment design balances the relevant covariates for the outcomes that are omitted in the specification \eqref{eqn:model-setup} and design of experiments \citep{hayes2017cluster}. 
	
	Note that $\mathbb{A}^{\mathrm{disc}}_{\mathrm{opt}}$ can be empty, if $|\tlo_g|(1+\omega_{\ell,t}^\ast)/(2T)$ is not an integer for each stratum (similarly for $\mathbb{A}_{\ell} $). For this case, Algorithm \ref{algo:choose-design} rounds $|\tlo_g|(1+\omega_{\ell,t}^\ast)/(2T)$ to the nearest integer to obtain a feasible design. As shown in Proposition \ref{prop:rounding-error-obs-cov} in Section \ref{subsubsec:practical-considerations}, the value of $\tr \big(\mathrm{Prec}(\hat{\bm{\tau}})  \big)$ evaluated at this feasible solution is within a factor of $1 + O \Lp {1}/{N^2} \Rp$ of the optimal value of $\tr \big(\mathrm{Prec}(\hat{\bm{\tau}})  \big)$.

	{\blue If observed covariates are continuous, then we can partition units into a small number of strata based on the observed covariate values, and then randomly choose a design that satisfies the treated fraction conditions for each stratum (possibly with rounding). Prior work has suggested keeping the number of strata small to avoid over-stratification \citep{kernan1999stratified}, because over-stratification may lower the precision of the estimated treatment effects \citep{de2008consequences}.  }

	{\blue If there are latent covariates, we suggest using the historical control data for the same set of units for the design of experiments. We can improve precision by using historical data to estimate $\*u_i$, partitioning units into strata ({\it e.g.}, by spectral clustering), and choosing a treatment design based on the estimated $\*u_i$. See Section \ref{subsec:latent-covariates} for more details. }

	\begin{remark}[Conditional Average Treatment Effect (CATE)]\label{remark:cate}
	Suppose we are interested in using observable sources of heterogeneity $\*X_i$ to assess the heterogeneity in treatment effects (\cite{robinson1988root,wager2018estimation} among others). In this case, we may be interested in conditional average instantaneous and lagged effects, denoted by $\tau_j(\*X_i)$. If $\*X_i$ is discrete, the designs satisfying \eqref{eqn:opt-finite-x-interactive-effect-carryover} can maximize the precision of $\hat{\tau}_j(\*X_i)$ for all $j$ and $\*X_i$, following that the optimality conditions for $\omega_{\ell,t}^\ast$ are satisfied within each stratum.
    
	\end{remark}

    {\blue \begin{remark}[Discussion on Adaptive Design]\label{rem:fixed-sample-adaptive-design}
	The treatment assignments studied in this section are chosen and fixed before the experiment starts. For non-adaptive experiments, we do not pursue adaptive designs, where treatment assignments for subsequent periods can vary during the experiment, for the following reason. From Theorem \ref{thm:obs-latent-carryover-model}, the information that matters for the treatment assignments but is unknown before the experiment starts is $\*u_i$. However, the estimation of $\*u_i$ at the beginning of the experiment can be quite noisy, and we do not expect noisy estimates of $\*u_i$ can substantially help the design. See Section \ref{subsubsec:choose-treatment-design} for further elaboration on this.
\end{remark}}

%% file: section_4.tex
\section{Adaptive Experiments}\label{sec:sequential-experiment}

Consider an experiment that is allowed to run $T_{\max}$ periods. The experimenter seeks to achieve a certain precision objective by the end of the experiment. After observing some data during the experiment, the experimenter may find that achieving this objective does not need $T_{\max}$ periods' of experimental data. In this case, the experimenter may want to terminate the experiment early in order to reduce the cost of the experiment. 

In this section, we study the design and analysis of adaptive experiments, where $N$ is fixed, but the experiment duration can vary due to the early termination. Let $\tilde{T} \in [T_{\max}]$ be the duration of the adaptive experiment, which is a random variable and unknown before the adaptive experiment starts. 
At any time period $t$, the experimenter collects data and decides whether to terminate the experiment. If so, the experiment stops, and the realization of $\tilde{T}$ is $t$; otherwise, the experimenter makes treatment decisions for time $t + 1$. For ease of understanding, we present our algorithm and results based on the following simple specification of the observed outcome of unit $i$ at time $s$
\begin{equation}\label{eqn:two-way-direct-model}
    Y_{is}  = \alpha_i +  \beta_s  + \tau z_{is}  + \varepsilon_{is}, \qquad \forall ~ i \in [N], ~ s \in [\tilde{T}].
\end{equation}
For simplicity, we denote instantaneous effect as $\tau$ instead of $\tau_0$ in this section. Later, we discuss how our algorithm and results can be generalized to the specification with $\ell > 0$ and with $\*X_i$ and $\*u_i$. 

Motivated by the objective of maximizing precision in \eqref{eqn:obj}, we consider the following criterion to terminate the experiment if the precision exceeds a target threshold $c$ at time $t$
\begin{equation}\label{eqn:experiment-termination-rule}
    \cmtrx{\mathrm{Prec}(\hat{\tau}; \bm{\omega}) = \frac{Nt}{\sigma_\varepsilon^2} \cdot \underbrace{(- 2 \bm{b}_t^\T \bm{\omega}_{1:t} -\bm{\omega}_{1:t}^\T \*P_{\bm{1}_{t}} \bm{\omega}_{1:t})/t}_{ g_{\tau}(\bm{\omega}, t)}  \geq c\,,}
\end{equation}
\cmtrx{where the expression of $\mathrm{Prec}(\hat{\tau}; \bm{\omega})$ follows from Example \ref{example:two-way-fe} in the case of $\hat{\tau}$ estimated from a panel with $N$ units and $t$ periods. In this section, we parametrize the precision by $\bm{\omega}$ and we optimize over $\bm{\omega}$, as the precision only depends on $\bm{\omega}$ but not whom to treat under specification \eqref{eqn:two-way-direct-model}.} 
This precision-based rule is equivalent to the variance-based rule to terminate the experiment when $\var(\hat{\tau}) \leq 1/c$. This type of termination rules has been used by others in different settings, such as in simulations and in sequential testing on sequentially arrived units without time effects (\cite{chow1965asymptotic,glynn1992asymptotic,singham2012finite} among others). 

There are three key technical challenges in designing and analyzing the experiments that can be terminated early. 
The first challenge concerns adaptively \cmtfinal{choosing the fraction of treated units per period}. 
Recall from Example \ref{example:two-way-fe}, that $\omega_s^\ast =(2s-1-\tilde{T})/\tilde{T}$. But, since $\tilde{T}$ is unknown before the experiment starts, or even during the experiment, \cmtfinal{choosing the optimal fraction of treated units} is non-trivial. To address this challenge, we aim to adaptively improve the treatment decisions as we gather more information about $\tilde{T}$ during the experiment. 

The second challenge concerns implementing the termination rule. As long as we can estimate the critical unknown parameter $\sigma_\varepsilon^2$ in \eqref{eqn:experiment-termination-rule}, we can determine whether to stop the experiment. There are two main difficulties in this task. The first is to have a valid implementation of the termination rule on adaptively collected data. Here, valid implementation means that the precision indeed exceeds threshold $c$ when the experiment terminates. The second is to do it in a way that the next challenge (obtaining valid post-experiment inference of $\tau$) can be manageable. 
Early stopping complicates post-experiment inference because the same data is used to make the decision about stopping and to estimate treatment effects, leading to the well-known bias that can arise when adaptive tests determine whether to terminate experiments (\cite{johari2017peeking} among others).

The third challenge concerns efficient estimation and inference for $\tau$, post-experiment. Based on the adaptive nature of collected data and the implementation of experiment termination rule, we seek to choose a consistent and efficient estimator for $\tau$ that uses as many observations as possible.

We propose the Precision-Guided Adaptive Experiment (PGAE) algorithm in Section \ref{subsec:adaptive-algorithm} to simultaneously tackle these three challenges. PGAE combines ideas from dynamic programming and sample splitting. In Section \ref{subsec:adaptive-theoretical-guarantee},
we prove statistical consistency and asymptotic normality of $\hat\tau$ and $\widehat{\sigma^2}$, estimated by PGAE, paving the way for valid statistical inference for $\tau$. 

\subsection{Estimators}

To start, we first review two existing estimators and then propose a new estimator. All three estimators are extensively used in PGAE. 
Suppose these estimators use the data of units in a set $\mathcal{S}$ over $t$ periods collected so far, where $t$ is small, but set size $|\mathcal{S}|$ can be large. In this subsection, we sub-index the estimators by $\mathcal{S}$ and $t$ to refer to the data used in the estimators.

The first is the \within estimator for $\tau$ \citep{wallace1969use}. The \within estimator of $\tau$, denoted by $\hat{\tau}_{\mathcal{S},t}$, regresses $\dot{Y}_{is}$ on $\dot{z}_{is}$ based on the specification $\dot{Y}_{is} =  \tau \dot{z}_{is}  + \dot{\varepsilon}_{is}$, where for any variables $\{x_{is}\}_{(i,s) \in \mathcal{S} \times [t]}$ ({\it e.g.}, $Y_{is}$ and $z_{is}$), the notation $\dot{x}_{is}$ denotes the \within transformed $x_{is}$ and is defined as
\begin{equation}\label{eqn:within-estimator}
    \dot{x}_{is} = x_{is} - \bar{x}_{i \cdot} - \bar{x}_{\cdot s} + \bar{x}\,,
\end{equation}
in which $\bar{x}_{i \cdot}$, $\bar{x}_{\cdot s}$, and $ \bar{x}$ are averages of $x_{is}$'s over time periods, units, and both of them, respectively. The \within estimator is an efficient estimation approach that does not need to estimate $\alpha_i$ and $\beta_s$, but produces the same estimate of $\tau$ as OLS based on \eqref{eqn:two-way-direct-model} that estimates $\alpha_i$ and $\beta_s$.\footnote{Regressing  $Y_{is}$ on $z_{is}$ and unit and time dummies is the same as GLS with weight matrix $\*W \propto \*I_N$ in \eqref{eqn:gls}. The \within estimator is also called Least-Squares Dummy Variable (LSDV) estimator.
} As shown in Lemma \ref{lemma:asymptotic-tau-sigma}, $\hat{\tau}_{\mathcal{S},t}$ is consistent and asymptotically normal for any finite $t$, when the set size $|\mathcal{S}|$ is large.

The second is the plug-in estimator for $\sigma^2_\varepsilon$, which is used in experiment termination and post-experiment inference, and takes the form of
\begin{equation}
     \estsigmasq_{\mathcal{S},t} = \frac{1}{|\mathcal{S}| \cdot (t-1)} \sum_{i \in \mathcal{S}} \sum_{s = 1}^t \big( \dot{y}_{is} - \hat{\tau}_{\mathcal{S},t} \cdot \dot{z}_{is} \big)^2. \label{eqn:second-moment-sigma-estimator} 
\end{equation}
The factor $1/(t-1)$ is for finite $t$ correction. As shown in Lemma \ref{lemma:asymptotic-tau-sigma}, $\estsigmasq_{\mathcal{S},t}$ is consistent and asymptotically normal for any finite $t$.

The third is a new estimator for the variance of $\varepsilon_{is}^2$, that is, $\xi^2_\varepsilon \coloneqq \+E[(\varepsilon^2_{is} - \sigma^2_\varepsilon)^2]$, which is used to quantify the uncertainty in our estimator for $\sigma^2_\varepsilon$, and takes the form of
\begin{align}
    \estxisq_{\mathcal{S},t} =& \underbrace{\frac{t^2}{(t-1)^2}}_{\substack{\text{correction}\\\text{multiplier}}}  \cdot \underbrace{\frac{1}{|\mathcal{S}| \cdot t}  \sum_{i \in \mathcal{S}}  \left(\sum_{s=1}^t \left[ (\dot{y}_{is} - \hat{\tau}_{\mathcal{S},t} \cdot \dot{z}_{is})^2 - \estsigmasq_{\mathcal{S},t} \right] \right)^2}_{\text{plug-in estimator of } \xi^2_\varepsilon}  - \underbrace{\frac{3t-2}{(t-1)^2} \cdot (\estsigmasq_{\mathcal{S},t})^2}_{\text{correction term}} \, . \label{eqn:fourth-moment-sigma-estimator}
\end{align}
In this estimator, both the correction multiplier and correction term are used to de-bias the plug-in estimator when $t$ is finite. The plug-in estimator is biased, because $(\dot{y}_{is} - \hat{\tau}_{\mathcal{S},t} \cdot \dot{z}_{is})^2$ is not an unbiased estimator of $\sigma^2_{is}$ for each $i$ and $s$, and the bias is squared in the estimation of $\xi_\varepsilon^2$, which cannot be averaged out over $i$. Figure \ref{fig:test-asymptotics} in Section \ref{subsec:finite-sample-lemma} visualizes the bias of the plug-in estimator and shows that $\estxisq_{\mathcal{S},t}$ can correct for the bias in finite samples. Lemma \ref{lemma:asymptotic-tau-sigma} shows that $\estxisq_{\mathcal{S},t}$ is consistent for any finite $t$.

\subsection{Precision-Guided Adaptive Experiment (PGAE) via Dynamic Programming}\label{subsec:adaptive-algorithm}

\begin{figure}[t!]
		\centering
		\includegraphics[width=0.8\linewidth]{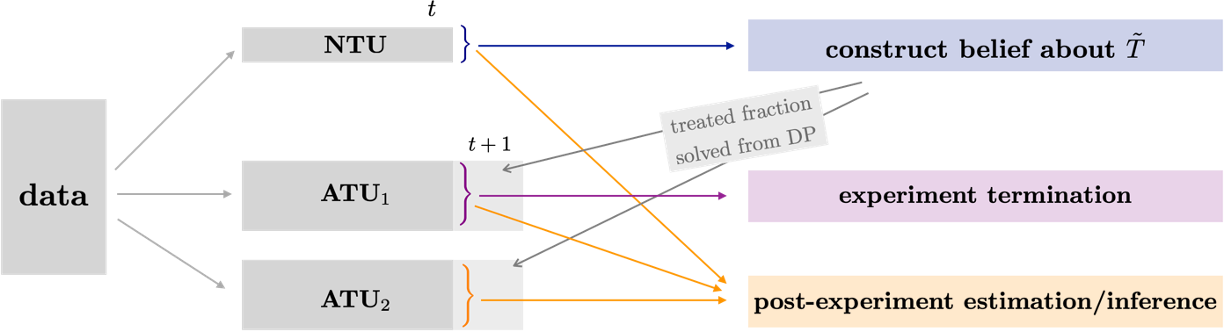}
		\caption{\textbf{Illustration of PGAE.}}
		\label{fig:pgae}
	\end{figure}

PGAE simultaneously addresses the three challenges introduced at the beginning of Section \ref{sec:sequential-experiment}, which is feasible through a careful partitioning of units into disjoint subsets, with each set functioning for a different purpose.
Specifically, we partition units into three mutually disjoint sets $\mathcal{S}_{\fcs}$, $\mathcal{S}_{\ad,1}$ and $\mathcal{S}_{\ad,2}$ that represent a set of non-adaptive treatment units (NTU) and two sets of adaptive treatment units (ATU), respectively. See Figure \ref{fig:pgae} for an illustration.
Let $p_{\fcs} = |\mathcal{S}_{\fcs}|/N$, $p_{\ad,1} = |\mathcal{S}_{\ad,1}|/N$, and $p_{\ad,2} = |\mathcal{S}_{\ad,2}|/N$ be the fractions of units in these three sets. Clearly, $p_{\fcs} + p_{\ad,1} + p_{\ad,2} = 1$. The sets are selected such that $p_{\fcs}$ is small and $p_{\ad,1} = p_{\ad,2}$. 

Before the experiment starts, we initialize the treatment designs of $\mathcal{S}_{\fcs}$, $\mathcal{S}_{\ad,1}$ and $\mathcal{S}_{\ad,2}$ by the optimal design of a $T_{\max}$-period non-adaptive experiment, which is the solution without early stopping. Then the average of $z_{is}$ over $i$ in each set satisfies $\omega_{\mathrm{bm},s} = (2s - 1 - T_{\max})/{T_{\max}}$ for all $s\in[T_{\max}]$ (with rounding if necessary), per Example \ref{example:two-way-fe}. The initial design also serves as the ``benchmark'' design for the adaptive experiment.
 The treatment design of NTU does not change during the experiment, stays equal to $\bm{\omega}_{\mathrm{bm}}$, and observed data from
NTU is used to update the treatment assignments of ATU for subsequent periods, specifically, to improve upon $\bm{\omega}_{\mathrm{bm}}$. One of the two ATU sets ($\mathcal{S}_{\ad,1}$) is used to estimate $\sigma_\varepsilon^2$, and decide whether to terminate the experiment. The other ATU set ($\mathcal{S}_{\ad,2}$) provides another estimate of $\sigma_\varepsilon^2$ for post-experiment inference. As we will see in Section \ref{subsec:adaptive-theoretical-guarantee}, partitioning units into three sets is crucial for decoupling possible correlations due to data reuse and hence obtaining valid statistical inference.

Next, we illustrate how PGAE addresses each of the three challenges.

\paragraph{1. Choosing a treatment design.}
{\blue 
In order to update treatment decisions for ATU, we need to construct a belief about experiment stopping time $\tilde{T}$, using NTU. If we would have known $\sigma_\varepsilon^2$, then we would exactly know the minimum stopping time $\tilde{T}$ such that the precision (using $\bm{\omega}_{\mathrm{bm}}$) is bigger than $c$ in \eqref{eqn:experiment-termination-rule}. However, $\sigma_\varepsilon^2$ is unknown in practice, but it is possible to construct \cmtfinal{the belief} distribution of $\sigma_\varepsilon^2$, which will be explained in detail below. Then we can draw samples of $\sigma^2$ from the \cmtfinal{belief} distribution of $\sigma_\varepsilon^2$. For each sampled  $\sigma^2$, we plug it into the precision expression and find the minimum duration that satisfies the stopping rule \eqref{eqn:experiment-termination-rule}.\footnote{Formally, we find $T$ using the following approach. Let $\sigma_0^2$ be a sampled $\sigma^2$ and let $T_0 = \min \left\{t: Nt / \sigma_0^2 \cdot g_\tau(\bm{\omega}, t) \geq c  \right\}$ be the minimum duration such that the precision exceeds threshold $c$. If $T_0 \in \{t+1, t+2, \cdots, T_{\max}\}$, then we set $T$ as $ T_0$; otherwise, if $T_0 > T_{\max}$, then we set $T$ as $T_{\max}$; otherwise, if $T_0 \leq t$, then we set $T$ as $t+1$. } The minimum duration may vary with the sampled $\sigma^2$. By repeatedly sampling $\sigma^2$ and finding the minimum duration, we can obtain an empirical distribution of $\tilde{T}$, denoted as $P_t(\tilde{T})$. See the pseudocode of the helper function \texttt{estimate\_belief()} of PGAE in Section \ref{subsec:pseudo-code} for more details.

Here is our approach to constructing a \cmtfinal{belief} distribution of $\sigma_\varepsilon^2$. We first estimate $\sigma_\varepsilon^2$ from formula \eqref{eqn:second-moment-sigma-estimator} using data of NTU for $t$ periods. Let $\estsigmasq_{\fcs,t}$ be the estimator. Next, we quantify the uncertainty in $\estsigmasq_{\fcs,t}$. From Lemma \ref{lemma:asymptotic-tau-sigma} below, $\estsigmasq_{\fcs,t}$ is consistent and asymptotically normal with
\begin{align}
    \sqrt{|\mathcal{S}_\fcs| \cdot t} \cdot \frac{\estsigmasq_{\fcs,t} - \sigma_\varepsilon^2 }{\hat{\xi}_{\fcs,t}^{\dagger} }   \xrightarrow{d} \mathcal{N}(0,1)\,, \label{eqn:asymptotic-normality-sigmasq}
\end{align}
where $\hat{\xi}_{\fcs,t}^{\dagger} = \big[\estxisq_{\fcs,t} + (\estsigmasq_{\fcs,t})^2/(t-1) \big]^{1/2}$ and $\estxisq_{\fcs,t}$ is estimated from \eqref{eqn:fourth-moment-sigma-estimator} using data of NTU. Let the left-hand side of \eqref{eqn:asymptotic-normality-sigmasq} be $w$. We have $\sigma_\varepsilon^2 = \estsigmasq_{\fcs,t} - w \cdot {\hat{\xi}^\dagger_{\fcs,t}}/{\sqrt{|\mathcal{S}_\fcs| \cdot t}}$, and we repeatedly use this formula with samples of $w$ from $\mathcal{N}(0,1)$ to obtain a \cmtfinal{belief} distribution of $\sigma_\varepsilon^2$. 

We use a dynamic program to solve the treatment decisions for ATU. In this dynamic program, the state variable is the treated fractions up to time $t$, i.e., $\bm{\omega}_{\ad,1:t}$, and the decision variable is the treated fraction for the next period $\omega_{t+1}$.\footnote{{\blue Example 4.3.4 of \cite{bertsekas2012dynamic} also uses a dynamic program to select a threshold for terminating a sequential hypothesis testing problem, but we use a dynamic program to find the optimal design in subsequent time periods.}}  The payoff in intermediate time periods is zero, and conditional on the realization of $\tilde{T}$, the terminal cost is
$- \tilde{T} \cdot \funfrac\left((\bm{\omega}_{\ad,1:t},
\bm{\omega}_{(t+1):\tilde{T}}), \tilde{T}\right)$, which is equivalent to our objective of maximizing precision. 
Here, in the dynamic program, we aim to solve $\omega_{t+1}$ that minimizes the expected terminal cost. The expectation is taken with respect to the random experiment duration $\tilde{T}$, and we use the empirical distribution $P_t(\tilde{T})$ learned at time $t$ when taking the expectation. Specifically, we solve $\omega_t$ by minimizing
%
\[
\+E_{\tilde{T} \sim P_t(\tilde{T})} \left[ - \tilde{T} \cdot \funfrac\left((\bm{\omega}_{\ad,1:t}, \bm{\omega}_{(t+1):\tilde{T}}), \tilde{T}\right) \right]\,,
\]
subject to the constraint $\omega_{\ad,t} \leq  \omega_{t+1} \leq  \omega_{t+2} \leq \cdots \leq \omega_{T_{\max}} \leq 1$.\footnote{Only $\omega_{t+1}$ is the decision variable. $\omega_{t+2}, \cdots, \omega_{T_{\max}} $ are not decision variables, whose values can vary with $\tilde{T}$.}
Section \ref{subsec:dp-omega} provides more discussion on this dynamic program, and more details about the dynamic program can be found in the pseudocode of the helper function \texttt{update\_treatment\_design()} of PGAE in Section \ref{subsec:pseudo-code}.

}

\paragraph{2. Implementing the termination rule.} We use the data in $\mathcal{S}_{\ad,1}$ to estimate $\sigma^2_\varepsilon$ via formula \eqref{eqn:second-moment-sigma-estimator}. Let the estimator at time $t$ be $\estsigmasq_{\ad,1,t}$.\footnote{For notation simplicity, we denote the estimator as $\estsigmasq_{\ad,1,t}$, as opposed to $\estsigmasq_{\mathcal{S}_{\ad,1},t}$.}
We then plug $\estsigmasq_{\ad,1,t}$ and $\bm{\omega}_{\mathrm{bm},1:t}$ into the termination rule \eqref{eqn:experiment-termination-rule} to estimate precision and decide whether to terminate the experiment. Note that using $\bm{\omega}_{\mathrm{bm},1:t}$ tends to under-estimate the precision (see Proposition \ref{prop:precision-ordering} in Section \ref{subsec:sequential-additional-results}), so that the experiment termination tends to be conservative. If our experiment terminates, we show in Proposition \ref{prop:precision-guarantee} in Section \ref{subsec:sequential-additional-results} that the true precision exceeds $c$ with a high probability.\footnote{Admittedly, it seems natural to use $\bm{\omega}_{\ad,1:t}$ that may yield a more precise estimation of the precision. We do not pursue this route as it is harder to show that the post-experiment inference of $\tau$ is valid and the corresponding precision is indeed larger than $c$, given the complexity of the current proof.}

\paragraph{3. Efficient estimation and valid inference for $\tau$.} In PGAE, we estimate 
$\tau$ using all the data of NTU and ATU over $\tilde{T}$ periods. Let the estimator be $\hat{\tau}_{\all,\tilde T}$. We show in Theorem \ref{theorem:asymptotic-page} below that $\hat{\tau}_{\all,\tilde T}$ is efficient and achieves the optimal convergence rate. But we only use $\mathcal{S}_{\ad,2}$ to estimate $\sigma^2_\varepsilon$ via formula \eqref{eqn:second-moment-sigma-estimator}. Let the estimator be $\estsigmasq_{\ad,2,\tilde T}$, which is consistent, and we use $\estsigmasq_{\ad,2,\tilde T}$ to estimate $\var(\hat{\tau})$.
Note that the consistency of $\sigma^2_\varepsilon$ is sufficient for constructing valid confidence intervals for $\tau$. Therefore the efficiency (determined by the sample size) in the estimation of $\sigma^2_\varepsilon$ is less important. The only tradeoff is that the endpoints of confidence intervals are estimated less precisely, with a larger second-order error term.\footnote{The first-order error term comes from the estimation error of $\hat{\tau}_{\all,\tilde T}$} This explains why we use only $\mathcal{S}_{\ad,2}$ to estimate $\sigma^2_\varepsilon$.

In summary, PGAE takes advantage of all the data collected so far, and then jointly optimizes treatment assignments alongside the choice of whether to continue the experiment. The pseudocode for PGAE is shown in Algorithm \ref{algo:page}, with the pseudocode for the helper functions collected in Section \ref{subsec:pseudo-code}. 

\begin{algorithm}[t!]
 \caption{Precision-Guided Adaptive Experiment (PGAE)}  \label{algo:page}
  \SetAlgoLined\DontPrintSemicolon
  \SetKwFunction{algo}{algo}\SetKwFunction{proc}{proc}
  \SetKwProg{myalg}{Algorithm}{}{}
  \SetKwFunction{pgae}{pgae}
  \SetKwFunction{Fpart}{partition\_initialize}
  \SetKwFunction{estbelief}{estimate\_belief}
  \SetKwFunction{updatedesign}{update\_treatment\_design}
  \SetKwFunction{estvarprec}{estimate\_var\_prec}
  \myalg{\pgae{$N,T_{\max}, \bm{\omega}_{\mathrm{bm}}, c, p_{\fcs}, m, t_0$}}{
  \nl $Z_{\fcs}$, $Z_{\ad,1}$, $Z_{\ad,2} \leftarrow$ \Fpart{$N, p_{\fcs}, T_{\max}$}\;
  \nl Run the experiment for $t_0$ time periods \;
  \nl $\estsigmasq_{\ad,1,t}, \widehat{\Prec}_{\ad,1,t} \leftarrow$ \estvarprec{$Z_{\ad,1,1:t}, Y_{\ad,1,1:t}, \bm{\omega}_{\mathrm{bm}}, N, n_{\ad}, t$} \;
  \nl \While{$\widehat{\Prec}_{\ad,1,t}  < c$ {\rm and} $t < T_{\max}$}{
  \nl $P_t(\cdot) \leftarrow $ \estbelief{$Z_{\fcs,1:t}, Y_{\fcs,1:t}, N, \bm{\omega}_{\mathrm{bm}}, T_{\max}, t, m, c$}\;
  \nl $Z_{\ad,1,t+1}, Z_{\ad,2,t+1} \leftarrow $ \updatedesign{$Z_{\ad,1,1:t}, Z_{\ad,2,1:t}, \omega_{\ad,1:t}, T_{\max}, t$} \;
  \nl $t \leftarrow t+1$, and run the experiment for time period $t+1$\;
  \nl $\estsigmasq_{\ad,1,t}, \widehat{\Prec}_{\ad,1,t} \leftarrow$ \estvarprec{$Z_{\ad,1,1:t}, Y_{\ad,1,1:t}, \bm{\omega}_{\mathrm{bm}}, N, n_{\ad}, t$} \;
  
  }
  \nl $\tilde{T} \leftarrow t$, and $\hat{\tau}_{\all,\tilde{T}} \leftarrow$ \within estimator of $\tau$ from both NTU and ATU \;
  \nl $\estsigmasq_{\ad,2,\tilde{T}}, \widehat{\Prec}_{\ad,2,\tilde{T}} \leftarrow$ \estvarprec{$Z_{\ad,2,1:\tilde{T}}, Y_{\ad,2,1:\tilde{T}}, \bm{\omega}_{\mathrm{bm}}, N, n_{\ad}, \tilde{T}$} \;
  \nl \KwRet $\tilde{T}$, $\hat{\tau}_{\all,\tilde{T}}$ and $\estsigmasq_{\ad,2,\tilde{T}}$\;}{}
\end{algorithm} 

\subsection{Analysis of the Algorithm}\label{subsec:adaptive-theoretical-guarantee}

In this subsection, we present the asymptotic results of the estimated $\tau$ and $\sigma^2_\varepsilon$ in PGAE. These results serve two purposes. The first is to justify our approach in Section \ref{subsec:adaptive-algorithm} to construct a belief about $\sigma^2_\varepsilon$, and specifically to justify \eqref{eqn:asymptotic-normality-sigmasq}. The second is to show that the outputs of PGAE ({\it i.e.}, $\hat{\tau}_{\all,\tilde{T}}$ and $\estsigmasq_{\ad,2,\tilde{T}}$) can be used for valid statistical inference and hypothesis test of treatment effect.

{\blue
To start, we characterize the asymptotic properties of $\hat{\tau}_{\fcs,t}$, $\estsigmasq_{\fcs,t}$ and $\estxisq_{\fcs,t}$, when they are estimated from non-adaptive experimental data, in Lemma \ref{lemma:asymptotic-tau-sigma}. This lemma provides theoretical support for constructing the \cmtfinal{belief} distribution of $\sigma_\varepsilon^2$, and serves as a crucial intermediate step in characterizing asymptotic distributions for the outputs of PGAE.

\begin{lemma}\label{lemma:asymptotic-tau-sigma}
Suppose Assumptions \ref{ass:treatment-adoption} and \ref{ass:model} hold. Under the specification \eqref{eqn:two-way-direct-model}, suppose $\varepsilon_{is}$ is i.i.d. for any $i$ and $s$ with $\+E[\varepsilon_{is}] = 0$, $\+E[\varepsilon^2_{is}] = \sigma_\varepsilon^2$, $\+E[\varepsilon^3_{is}] = 0$,  and $\+E[(\varepsilon^2_{is} - \sigma^2_\varepsilon)^2] = \xi^2_{\varepsilon}$. $\hat{\tau}_{\fcs,t}$ and $\estsigmasq_{\fcs,t}$ are consistent. As $|\mathcal{S}_{\fcs}| \rightarrow \infty$, for any finite $t$, conditional on $Z_\fcs$, we have 
\[\sqrt{|\mathcal{S}_{\fcs}|} \left(\begin{bmatrix} \hat{\tau}_{\fcs,t} \\ \estsigmasq_{\fcs,t} \end{bmatrix}  - \begin{bmatrix} \tau \\ \sigma^2_\varepsilon  \end{bmatrix}\right)  \stackrel{d}{\longrightarrow } \mathcal{N} \left(\begin{bmatrix} 0 \\ 0 \end{bmatrix}, \begin{bmatrix} \sigma_\varepsilon^2/(t \cdot \funfrac(\bm{\omega}_{\fcs, 1:t}, t)) & 0 \\  0 & \xi_{\varepsilon,t}^{\dagger2}/t \end{bmatrix} \right), \]
where $\xi_{\varepsilon,t}^{\dagger2} = \xi^2_\varepsilon + 2 \big(\sigma_\varepsilon^2\big)^2/(t-1)$. Furthermore, 
$\sqrt{|\mathcal{S}_{\fcs}|} \big(\estxisq_{\fcs,t} - \xi_\varepsilon^2\big) = O_p(1)$.
\end{lemma}

Since both $\estsigmasq_{\fcs,t}$ and $\estxisq_{\fcs,t}$ are consistent, $\estxidaggersq_{\fcs,t} = \estxisq_{\fcs,t} + 2 \big(\estsigmasq_{\fcs,t} \big)^2/(t-1)$ is a consistent estimator of $\xi_\varepsilon^{\dagger2}$. From Slutsky's theorem, the asymptotic distribution in \eqref{eqn:asymptotic-normality-sigmasq} holds.

Interestingly, $\hat{\tau}_{\fcs,t}$ and $\estsigmasq_{\fcs,t}$ are asymptotically independent, even though we use $\hat{\tau}_{\fcs,t}$ in the estimation of $\estsigmasq_{\fcs,t}$. The reason of asymptotic independence is as follows. The estimation error of $\hat{\tau}_{\fcs,t}$ is a weighted average of $\varepsilon_{is}$ over $i$ and $s$. The leading term in the estimation error of $\estsigmasq_{\fcs,t} $ is the sum of a weighted average of $\varepsilon^2_{ju} - \sigma_\varepsilon^2$ over $j$ and $u$ and a weighted average of $\varepsilon_{ju} \varepsilon_{jv} $ over $j,u,v$ with $u \neq v$. Note that $\varepsilon_{is}$ is uncorrelated with both $\varepsilon^2_{ju} - \sigma_\varepsilon^2$ and $\varepsilon_{ju} \varepsilon_{jv} $ for all $i,j,s,u,v$, because $\varepsilon_{is}$ is i.i.d. in $i$ and $s$ and has zero first and third moments. Therefore, the leading terms in the estimation errors of $\hat{\tau}_{\fcs,t}$ and $\estsigmasq_{\fcs,t} $ are uncorrelated. For the non-leading terms, they are at a small order of magnitude and do not contribute to the asymptotic covariance. 
As $\hat{\tau}_{\fcs,t}$ and $\estsigmasq_{\fcs,t}$ are jointly asymptotically normal, the zero asymptotic correlation implies asymptotic independence. In Section \ref{subsec:finite-sample-lemma}, we demonstrate the finite sample properties of Lemma \ref{lemma:asymptotic-tau-sigma} and asymptotic independence between $\hat{\tau}_{\fcs,t}$ and $\estsigmasq_{\fcs,t}$.

\begin{remark}\label{remark:lemma-finite-t}
Lemma \ref{lemma:asymptotic-tau-sigma} holds for any finite $t$. In fact, when $t$ grows to infinity, the problem is simpler, because we can show $\plim_{t\rightarrow \infty} \xi^{\dagger 2}_{\varepsilon,t} = \xi^2_\varepsilon$, and the plug-in estimator of $\xi^2_\varepsilon$ mentioned in formula \eqref{eqn:fourth-moment-sigma-estimator} is consistent. In this section, we focus on the challenging case with a finite $t$, because we want to apply Lemma \ref{lemma:asymptotic-tau-sigma} to the estimates on NTU early in the experiment ({\it i.e.}, $t$ is small).
\end{remark}

}

Next we show that $\hat{\tau}_{\all,\tilde{T}}$ and $\estsigmasq_{\ad,2,\tilde{T}}$ from PGAE can be used for valid post-experiment statistical inference and hypothesis testing for $\tau$. To show this, there are two critical steps: (a) show the asymptotic distribution of $\hat{\tau}_{\all,\tilde{T}}$; (b) show that the asymptotic variance of $\hat{\tau}_{\all,\tilde{T}}$ can be consistently estimated using $\estsigmasq_{\ad,2,\tilde{T}}$.

\begin{theorem}\label{theorem:asymptotic-page}
Suppose Assumptions \ref{ass:treatment-adoption} and \ref{ass:model} hold, $p_{\fcs} \in (0,1)$ is a fixed number as $N$ grows. Under the specification \eqref{eqn:two-way-direct-model}, suppose the assumptions about $\varepsilon_{is}$ in Lemma \ref{lemma:asymptotic-tau-sigma} hold and $\varepsilon_{is}$ is bounded with a symmetric distribution around $0$. $\hat{\tau}_{\all,\tilde{T}}$ and $\estsigmasq_{\ad,2,\tilde{T}}$ are consistent.
As $N \rightarrow \infty$,
        \begin{align}
        \sqrt{N} \cdot \begin{bmatrix}
    \big(\tilde{T}\funfrac(\bm{\omega}_{\all,1:\tilde{T}},\tilde{T})/\sigma_\varepsilon^2\big)^{1/2} \cdot \left( \hat{\tau}_{\all,\tilde{T}} - \tau\right) \vspace{0.3cm}  \\ \big(\tilde{T} p_{\ad,2}/\xi^{\dagger 2}_{\varepsilon,\tilde{T}} \big)^{1/2} \cdot \big(\estsigmasq_{\ad,2,\tilde{T}} - \sigma_\varepsilon^2 \big)
        \end{bmatrix}
        \stackrel{d}{\longrightarrow } \mathcal{N} \left(\bm{0}, I_2 \right) \label{eqn:cond-normal}\, .
    \end{align}
    \end{theorem}
{ 

\cmtrx{$\hat{\tau}_{\all,\tilde{T}}$ is consistent for $\tau$ with the optimal convergence rate $\sqrt{N}$. This result is not obvious because $\mathcal{S}_{\fcs}$, $\mathcal{S}_{\ad,1}$ and $\mathcal{S}_{\ad,2}$ are all used to estimate $\tau$, which could potentially lead to two sources of bias in the estimation of $\tau$. The first is that $\hat{\tau}_{\all,\tilde{T}}$ depends on adaptive treatment designs, and the choice of adaptive designs depend on  $\estsigmasq_{\fcs,t}$ and $\estxisq_{\fcs,t}$, and therefore on $\varepsilon_{is}$ for $i \in \mathcal{S}_{\fcs}$. The second is that $\hat{\tau}_{\all,\tilde{T}}$ depends on termination time $\tilde{T}$, where $\tilde{T}$ depends on $\estsigmasq_{\ad,1,t}$ and therefore on $\varepsilon_{is}$ for $i \in \mathcal{S}_{\ad,1}$. Both sources may lead to the violation of the commonly made exogeneity assumption  to show consistency ({\it i.e.}, asymptotic conditional mean of $\varepsilon_{it}$ is zero). With a careful analysis in Lemma \ref{lemma:conditional-mean-variance}, we show this is not the case, and the asymptotic conditional mean is still zero. This is mainly because $\estsigmasq_{\fcs,t}$, $\estxisq_{\fcs,t}$ and $\estsigmasq_{\ad,1,t}$ are all even moments of $\varepsilon_{is}$. With a symmetric distribution of $\varepsilon_{is}$ around $0$, conditioning on the even moments does not change the mean of $\varepsilon_{is}$.}

\cmtrx{
The adaptivity of the design, with the termination time depending on early values of the outcomes, comes at no cost in the estimation of $\tau$ in the following sense. Suppose we run an adaptive experiment, with a distribution of termination times. Now suppose we compare this to a series of non-adaptive experiments with the same distribution of termination times. This series of non-adaptive experiments is not actually feasible, because it depends on values we do not know {\it ex ante}. Nevertheless, this series of experiments does not do better than our proposed adaptive experiment, in the sense that the average of the variances is the same as that for our proposed adaptive experiment. This result is surprising as the adaptive nature in choosing treatment designs and in experiment termination does not affect the estimation efficiency of $\hat{\tau}_{\all,\tilde{T}}$. In Section \ref{subsec:finite-sample-pgae}, we empirically show that the results of Theorem \ref{theorem:asymptotic-page} are valid for a moderate $N$.}

Moreover, as our adaptive treatment decisions seek to maximize $\funfrac(\cdot)$, we expect $\funfrac(\bm{\omega}_{\all,1:\tilde{T}},\tilde{T}) > \funfrac(\bm{\omega}_{\mathrm{bm},1:\tilde{T}},\tilde{T})$ so that $\tau$ can be estimated more efficiently than the benchmark design ({\it i.e.}, $\Prec(\hat{\tau}_{\all,\tilde{T}}) > ({N\tilde{T}}/{\sigma^2_\varepsilon}) \cdot \funfrac(\bm{\omega}_{\mathrm{bm},1:\tilde{T}},\tilde{T})$). This is shown in Proposition \ref{prop:precision-ordering} in Section \ref{subsec:sequential-additional-results} for a large $N$, and is empirically demonstrated in Section \ref{subsec:empirical-sequential} for a moderate $N$.
}

\subsection{Extension to Carryover Model with Covariates}\label{subsec:sequential-carryover}

PGAE and Theorem \ref{theorem:asymptotic-page} can be easily extended to the specification with $\ell > 0$ and with $\*X_i$. We can continue using the \within estimator for $\tau_0, \cdots, \tau_\ell$ by regressing $\dot{Y}_{it}$ on $\dot{z}_{it}, \cdots, \dot{z}_{i,t-\ell}, \dot{\*X}_i$. For the experiment termination rule, we could generalize \eqref{eqn:experiment-termination-rule} to $\tr \big(\mathrm{Prec}(\hat{\bm{\tau}})  \big) \geq c$ or other criteria based on the objectives discussed in Section \ref{subsec:objective}. From Lemma \ref{lemma:simplify-obj} in Section \ref{subsec:separate-quadratic}, the only unknown parameter in the termination rule is $\sigma_\varepsilon^2$. Furthermore, we can partition units into strata based on $\*X_i$. For each stratum, we can then continue using PGAE to construct an empirical distribution about $\tilde{T}$, make adaptive treatment decisions, and sequentially decide whether to terminate the experiment. We can use a similar proof to show that results in Section \ref{subsec:adaptive-theoretical-guarantee} continue to hold with $\hat\tau$ replaced by $\hat\tau_0, \cdots, \hat\tau_\ell$ and $g_{\tau}(\bm{\omega},\tilde{T}) $ replaced by a matrix depending on $\bm{\omega}$ and $\tilde{T}$ (see Lemma \ref{lemma:simplify-obj} for its definition).

{\blue If the specification has $\*u_i$, there are multiple approaches to proceed. First, as a simple solution, we can ignore $\*u_i$ and run PGAE as the case without $\*u_i$. This approach can be shown to be valid under suitable assumptions.\footnote{For example, the suitable assumptions can be $\*u_i$ is mean zero and i.i.d. in $i$, and $\*v_s$ is i.i.d. in $s$.} We can further improve the precision of $\hat{\tau}_{\all,\tilde{T}}$ by re-estimating $\tau$ using GLS post-experiment.

Second, as a more efficient solution, if we have historical data, then we can use it to estimate $\*u_i$, partition units into strata, and run PGAE on each stratum. With abundant historical data, we can precisely estimate $\sigma_\varepsilon^2$, and therefore the minimum duration to achieve a certain precision threshold. For this case, we do not need an adaptive experiment and can instead run the non-adaptive experiment with the estimated minimum duration.}

%% file: section_5.tex
\section{Empirical Applications}\label{sec:empirical}

We run synthetic experiments on multiple real data sets to study our solutions in Sections \ref{sec:model} and \ref{sec:sequential-experiment}.
\footnote{Our code is available at \protect\url{https://github.com/ruoxuanxiong/staggered_rollout_design}.}
  First, we describe the data sets that we study from multiple domains in Section \ref{subsec:data-description}.
Next, in Section \ref{subsec:empirical-fixed-sample-size}, we show that for non-adaptive experiments, our solutions from Section \ref{subsec:result-carryover} require less than 50\% of the sample size to achieve the same treatment effect estimation error as the benchmark designs. For adaptive experiments, in Section \ref{subsec:empirical-sequential}, we show that our adaptive design from PGAE can improve the precision of treatment effect estimation by more than 20\%, on top of the improvements obtained by our non-adaptive designs. 

\subsection{Data Descriptions}\label{subsec:data-description}
Our synthetic experiments are run on four different data sets. The first one, MarketScan Research Databases, is used for the empirical results of this section. As robustness checks, the same results are shown on the remaining three data sets in Section \ref{subsec:description-data-sets}.

\paragraph{MarketScan research databases.} 
These databases contain inpatient and outpatient claim records. Focusing on influenza as the primary diagnosis, there are 21,277 inpatient admissions versus 9,678,572 outpatient visits in the databases. We denote all of these as influenza visits. Our outcome variables are monthly \emph{flu visit occurrence rates} per Metropolitan Statistical Area (MSA) and per thousand patients, which is defined as the ratio of the number of influenza visits among all enrolled patients times $1,000$ for the given month in a given MSA. Moreover, our analysis focuses on the flu peak seasons that are defined as October to April of the next year. We focus on the period from October 2007 to April 2015 as the databases have few observations outside this period. This leaves us with a panel of $185$ MSAs over $56$ months.
See Section \ref{subsec:description-data-sets} for more details.

\paragraph{Other data sets.} The three additional data sets, as described in Section \ref{subsec:description-data-sets}, are home medical visits in 61 cities over 144 weeks, grocery store transactions for 7,130 households over 97 weeks, and Lending Club loans for 956 geographic areas in the US over 139 months.

\subsection{Non-Adaptive Experiments}\label{subsec:empirical-fixed-sample-size}
First, in Section \ref{subsec:experiment-setup}, we discuss the setup of synthetic experiments and evaluation criteria, and then present the results in Section \ref{subsec:fixed-sample-empirical-results}. 

\subsubsection{Setup}\label{subsec:experiment-setup}
\texttt{} \\
Here, we first introduce the benchmark designs as well as different versions of our solution, depending on the specifications of the estimator. Then we explain how the synthetic experiment and treatment effect are generated, and then discuss the evaluation metrics that are used.

	{\bf Treatment designs.}
	We consider the following treatment designs. Illustrations of these designs in Figure \ref{fig:various-optimal-design} of Section \ref{sec:model} can facilitate the reading.
	\begin{enumerate}
		\item \emph{Benchmark treatment designs:}
		\begin{enumerate}
		\item $Z_{\FF}$ (fifty-fifty): $Z_{\FF}$ has 50\% control and 50\% treated units at every time period. More precisely, $Z_{\FF}$ is a rounded solution when starting with $\omega_s = 0$ for all $s$.
			\item $Z_{\BA}$ (before-after): $Z_{\BA}$ has all units in the control state before halftime and all units in the treatment state after halftime. More precisely, $Z_{\BA}$ is a rounded solution that starts with $\omega_s = -1$ for $s < (T+1)/2$ and equals to $\omega_s = 1$ for $s \geq (T+1)/2$.
			\item $Z_{\FFBA}$ (fifty-fifty with before-after): $Z_{\FFBA}$ has all units in the control state before halftime and half of the units in the treatment state after halftime. That is, $Z_{\FFBA}$ is a rounded solution that starts with $\omega_s = -1$ for $s < (T+1)/2$ and has $\omega_s = 0$ for $s \geq (T+1)/2$. $Z_\FFBA$ combines $Z_\FF$ and $Z_\BA$, and has the simultaneous treatment adoption pattern.
		\end{enumerate}
		
		\item \emph{Variations of $Z_\OPT$ from Section \ref{sec:model}:} In order to assess the benefit of various features of our specification \eqref{eqn:model-setup}, we consider three different designs. Each one is a variant of $Z_\OPT$, but with different specifications of the estimator.
		\begin{enumerate}
				\item $Z_{\OPT, \mathrm{linear}}$:
				This design is optimal under the specification \eqref{eqn:model-setup} with $\ell = 0$ and without covariates ($d_x = d_u = 0$). This design can in fact be considered as the ``state-of-the-art'' benchmark design since it is analogous to the optimal stepped wedge designs of \cite{hemming2015stepped} and \cite{li2018optimal} in which the treated fraction increases linearly in time. We sample $\{A_i\}_{i \in [N]}$ from $\mathbb{A}_0$ defined in \eqref{eqn:carryover-t-optimal-obs-latent-thm}, and the sampled $\{A_i\}_{i \in [N]}$ uniquely defines $Z_{\OPT, \mathrm{linear}}$.
				
		\item $Z_{\OPT}$: This design a nonlinear staggered design and is optimal under the specification \eqref{eqn:model-setup} with $\ell > 0$ and without covariates ($d_x = d_u = 0$). We sample $\{A_i\}_{i \in [N]}$ from $\mathbb{A}_{\ell}$ defined in \eqref{eqn:carryover-t-optimal-obs-latent-thm}, and the sampled $\{A_i\}_{i \in [N]}$ uniquely defines $Z_{\OPT}$.

		\item $Z_{\OPT,\mathrm{stratified}}$: This design is also a nonlinear staggered design and is optimal under the specification \eqref{eqn:model-setup} with $\ell > 0$ and with discrete-valued latent covariates ($d_x = 0$, $d_u > 0$). 
		The value of this design is only demonstrated when historical control data is available, which is a realistic assumption in practice. \cmtfinal{In our empirical applications, we first estimate $\*u_i$ by singular value decomposition (SVD) using historical data (see ``evaluation metrics'' below for the construction of historical data). Next we partition units into strata based on estimated $\hat{\*u}_i$ and randomly choose a treatment design that satisfies the conditions in  \eqref{eqn:carryover-t-optimal-obs-latent-thm} for each stratum, where the number of strata varies from $2$ to $4$. See Sections  \ref{subsubsec:estimate-latent-covariates} and \ref{subsubsec:choose-treatment-design} for more details.}  
	
		\end{enumerate}
		
	\end{enumerate}
	
    {\bf Synthetic non-adaptive experimental data.}
    Since we are not aware of any specific experiment that was performed on the data, we assume the data is the control data ({\it i.e.}, original panel data entries are $Y_{is}(-\bm{1}_{\ell+1})$, for all $i$ and $s$). We then create a hypothetical treatment with instantaneous and lagged effects. Given a treatment design $Z$, the observed outcome (in a hypothetical experiment) for unit $i$ at time $s$ would be (recall that $z_{is} \in \{-1,+1\}$)
	\[
	Y_{is} = Y_{is}(-\bm{1}_{\ell+1}) + \sum_{j = 0}^{\ell} (z_{i,s-j} + 1) \cdot \tau_j\,, 
	\]
	where $z_{is} = -1$ for $s \leq 0$. 
    For the results presented in Section \ref{subsec:fixed-sample-empirical-results}, $\ell$ is chosen at 2. We consider other values of $\ell$ in Section \ref{ecsub:varying-ell}. 

	{\bf Evaluation metrics.} Instead of running a single simulation on the entire panel of control data, we select $m$ random sub-blocks of dimension \cmtfinal{$N \times (T_{\mathrm{hist}} + T)$, where the first $T_{\mathrm{hist}}$ periods are historical control data and the synthetic experiment is applied to the last $T$ periods of data}. The estimated $\tau_j$ for $j \in \{0\}  \cup [\ell]$ on $k$-th block are denoted as $\hat \tau^{(k)}_{j} $. 
 For each design, we use the non-adaptive experimental data generated by this design to estimate $\tau_0, \cdots, \tau_\ell$ using GLS with specification \eqref{eqn:model-setup}. 
	As a robustness check, we also compare different estimation methods based on different specifications as well.
	We report the mean and 95\% confidence band of total squared estimation error $\sum_{j = 0}^{\ell} \big( \hat \tau^{(k)}_{j} - \tau_j \big)^2$ which is motivated by the objective of A-optimal design, defined in Section \ref{subsec:objective}. {\blue Note that for GLS, none of the estimation error metrics depend on the actual value of $\tau_0, \cdots, \tau_\ell$, as discussed in Remark \ref{rem:magnitude-treatment-effect}. For illustration purposes, we set the lagged effects to decay linearly in lag, $\tau_{1} = 2 \tau_{0}/3$,  $\tau_{2} =  \tau_{0}/3$, $\tau_j = 0$ for $j > 2$, and the cumulative effect $|\tau_0 + \tau_1 + \tau_2| = 0.1(NT)^{-1}\sum_{i,t} Y_{it}(-\bm{1}_{\ell+1})$. The latter selection means the cumulative effect has a magnitude that is $10\%$ of the average outcome in the panel. We verify that our results are robust to other values of $\bm{\tau}$ and to a much smaller magnitude of the cumulative effect in Figure \ref{fig:varying-effect}. } As a robustness check, we also report the squared estimation error of cumulative effect $\big(\sum_{j = 0}^{\ell} (\hat \tau^{(k)}_{j} - \tau_j) \big)^2$, and metrics related to hypothesis testing, that is, the receiver operating characteristic (ROC) curve and the corresponding area under the curve (AUC) in Section \ref{subsubsec:robustness-to-alternative-metrics}.
	
\subsubsection{Results}\label{subsec:fixed-sample-empirical-results}

\begin{figure}[t!]
	\centering
	\begin{subfigure}{1\textwidth}
		\centering
		\includegraphics[width=1\linewidth]{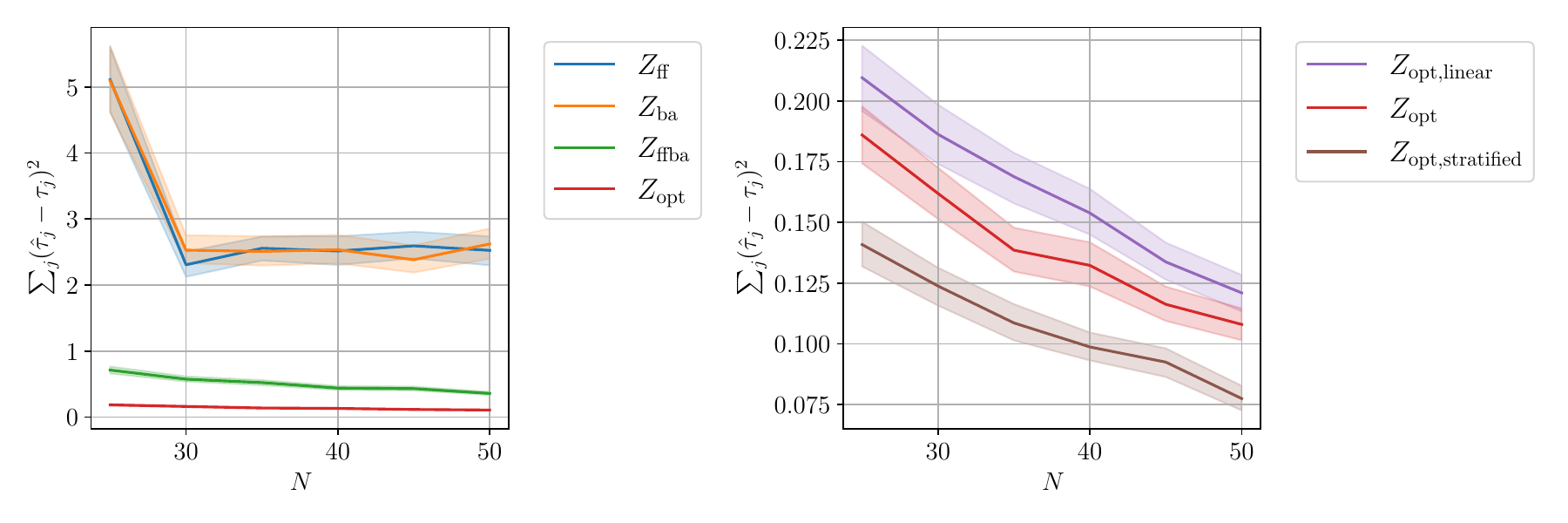}
	\end{subfigure}
	\caption{\textbf{Comparison of various designs in non-adaptive experiments.} These figures show the mean and 95\% confidence band of $\sum_{j}(\hat{\tau}_j - \tau_j)^2$ for various designs, based on 2,000 synthetic non-adaptive experiments with $\ell = 2$, $T = 7$ and varying $N$. The red curve in two figures are identical. The right figure zooms in the left one.
	}
	\label{fig:varying-N-flu}
\end{figure}

	\paragraph{Staggered treatment designs outperform benchmark designs.} The left subplot in Figure \ref{fig:varying-N-flu} shows the total estimation error $\sum_{j}(\hat{\tau}_j - \tau_j)^2$ of our nonlinear staggered design $Z_{\OPT}$ and benchmark designs $Z_\FF$, $Z_\BA$ and $Z_\FFBA$. Both $Z_{\OPT}$ and $Z_\FFBA$ consistently and significantly outperform $Z_\BA$ and $Z_\FF$. 
	The design $Z_{\FFBA}$, as a combination of $Z_\FF$ and $Z_\BA$, performs significantly better, but is still outperformed by $Z_\OPT$. Specifically, by using only 50\% of the sample size ($N=25$ versus $N=50$), $Z_\OPT$ achieves lower estimation error than $Z_\FFBA$.

\paragraph{Nonlinear staggered design outperforms linear staggered design.} The right subplot in Figure \ref{fig:varying-N-flu} compares our nonlinear staggered design $Z_\OPT$ with the linear staggered design $Z_{\OPT,\mathrm{linear}}$. When $\ell > 0$, $Z_{\OPT,\mathrm{linear}}$ requires 10\% more samples than $Z_\OPT$ to achieve the same estimation error. Note that the improvement is solely because of $\ell > 0$. In fact, if $\ell = 0$, the treated fraction of $Z_{\OPT,\mathrm{linear}}$ is optimal. We show this empirically in Figure \ref{fig:instantaneous-effect} of Section \ref{ecsub:varying-ell} by observing that $Z_{\OPT,\mathrm{linear}}$ requires about 5\% fewer samples than $Z_\OPT$ due to the higher variance of the latter.

\paragraph{Stratification further improves upon the staggered treatment design.} The right subplot in Figure \ref{fig:varying-N-flu} additionally compares our nonlinear staggered designs \textit{without} stratification $Z_\OPT$ and \textit{with} stratification $Z_{\OPT,\mathrm{stratified}}$. 
Using $Z_{\OPT,\mathrm{stratified}}$ can further reduce 20\% samples to achieve the same total estimation error. 
Overall, this result suggests the existence of latent covariates in the original data. Therefore, when there are latent covariates, we could use the historical data that contains information about latent covariates to design a stratified experiment.  

\paragraph{Robustness to additional data sets.} Figure \ref{fig:additional-varying-N} in Section \ref{subsubsec:robustness-additional-data-set} shows that the above three findings continue to hold on the other three data sets, as $N$ is varied. Figure \ref{fig:additional-varying-T} in Section \ref{subsubsec:robustness-additional-data-set} shows the above three findings continue to hold on all four data sets, as $T$ is varied. 

{\blue 
\paragraph{Robustness to various specifications of the estimator.} 

We compare the performance of various treatment designs, as the model specification varies in Figure \ref{fig:various-estimation-method} in Section \ref{subsubsec:robustness-to-specification}. We show that the specification with $\alpha_i$, $\beta_t$, and $\*u_i$ significantly outperforms the specification where either $\alpha_i$, $\beta_t$, or $\*u_i$ is absent. Moreover, we show that $Z_{\OPT,\mathrm{stratified}}$ performs best under various specifications. Therefore, both the treatment decisions (design) and specification of the estimator play important roles in reducing the estimation error.}

 \paragraph{Robustness to other evaluation metrics.} The above three findings continue to hold when the evaluation metric is the squared estimation error of cumulative effect, as shown in Figure \ref{fig:varying-N-other-metrics} in Section \ref{subsubsec:robustness-to-alternative-metrics}. Figure \ref{fig:roc-equal-tau} in Section \ref{subsubsec:robustness-to-alternative-metrics} shows the ROC curve of various designs ({\it i.e.}, power vs. significance level), with AUC reported in Table \ref{tab:auc} in Section \ref{subsubsec:robustness-to-alternative-metrics}. Aligned with other metrics, $Z_{\OPT,\mathrm{stratified}}$ has consistently higher power than all other designs.

\subsection{Adaptive Experiments}\label{subsec:empirical-sequential}
In this section, we run synthetic adaptive experiments and evaluate adaptive designs produced by PGAE. We describe the experimental setup in Section \ref{subsec:sequential-setup} and then present the results in Section \ref{subsec:sequential-result}. We show the finite sample properties of Lemma \ref{lemma:asymptotic-tau-sigma} in Section \ref{subsec:finite-sample-lemma}. We also show the finite sample properties of Theorem 
\ref{theorem:asymptotic-page}  in Section \ref{subsec:finite-sample-pgae}\cmtfinal{, which implies the validity of the post-experiment inference using estimates produced by PGAE}.

\subsubsection{Setup}\label{subsec:sequential-setup}
\texttt{} \\
Suppose the adaptive experiment can run for a maximum of $T_{\max}$ periods in total and $\ell = 0$.\footnote{The results are robust to the hypothetical intervention with carryover effects, and are available upon request. } The adaptive experiment is terminated if the estimated precision is larger than threshold $c$. 

{\bf Treatment designs.} Overall, we consider the following three designs
\begin{enumerate}
    \item Adaptive design: The design produced by PGAE, with dimension $N \times \tilde{T}$\cmtfinal{, where $\tilde{T}$ is the actual termination time observed in the adaptive experiment}.
    \item Benchmark design: The initial design applied to $\mathcal{S}_{\fcs}$, $\mathcal{S}_{\ad,1}$, and $\mathcal{S}_{\ad,2}$ with dimension $N \times \tilde{T}$,  where for all $s\in[\tilde{T}]$,  $N^{-1} \sum_i z_{is} = \omega_{\mathrm{bm},s} = (2s - 1 - T_{\max})/{T_{\max}}$ which is optimal when $\tilde{T} = T_{\max}$ (identical to $Z_{\OPT, \mathrm{linear}}$ when $\tilde T = T_{\max}$). 
    \item Oracle design: The optimal design for a $\tilde{T}$-period experiment for $\mathcal{S}_{\fcs}$, $\mathcal{S}_{\ad,1}$, and $\mathcal{S}_{\ad,2}$ with dimension $N \times \tilde{T}$ and $N^{-1} \sum_i z_{is} = ({2s - 1- \tilde{T}})/{\tilde{T}}$ (identical to $Z_{\OPT, \mathrm{linear}}$ when $\tilde{T}$ is known ex ante).
\end{enumerate}

Note that the dimensionality of the three designs is the same, so we can make a fair comparison of the performance of these three designs.

{\bf Synthetic adaptive experimental data.} Similar to the synthetic non-adaptive experiments, we assume the original data does not contain any specific treatment that we study. Given a treatment design $Z$, the observed outcome for unit $i$ at time $s$ ($s \leq \tilde{T}$) is $Y_{is} = Y_{is}(-1) +  \tau_0  (z_{is} + 1)$.

\paragraph{Evaluation metrics.} As before, we randomly select $m$ blocks, each with dimension $N \times T_{\max}$ from the original control data. We report the mean and 95\% confidence band of $\big( \hat{\tau}^{(k)}_{0} - \tau_0 \big)^2$, where $\hat{\tau}^{(k)}_{0}$ is the estimated $\tau_0$ on the synthetic experimental data of dimension $N \times \tilde{T}^{(k)}$ based on the $k$-th block of the original data. $\tilde{T}^{(k)}$ is equal to the value of $\tilde{T}$ for that block; that is, $\tilde{T}$ can vary with $k$. 

\subsubsection{ Results}\label{subsec:sequential-result}
\texttt{}\\
{\blue 
We show an empirical distribution of the termination time $\tilde{T}$ in Figure \ref{fig:experiment-termination-time} and the estimation error of $\hat{\tau}_0$ of various designs in Figure \ref{fig:various-opt-design}. Four observations can be made from these figures.

First, PGAE indeed terminates the experiment early when precision exceeds the threshold. As shown Figure in \ref{fig:experiment-termination-time}, when $T_{\max} > 7$, the experiment is always terminated quite early ($\tilde{T}<T_{\max}/2$). This early stopping does not compromise the estimation error as Figure \ref{fig:various-opt-design} validates that the stopping rule works correctly and the estimation error of the adaptive design always stays below the variance threshold $1/c$. Looking at the results for different values of the threshold $c$, in Figure \ref{fig:experiment-termination-time} and Figure \ref{fig:experiment-termination-time-supp}, we see that the termination time tends to increase with the threshold, which is as expected.

Second, the adaptive design from PGAE consistently reduces the estimation error ({\it i.e.}, improves the precision) compared to the benchmark design ({\it i.e.}, non-adaptive design), where the benchmark design is used as initialization in PGAE. This implies that adaptive treatment decisions in PGAE can be useful in lowering the estimation error post-experiment. The reduction is more substantial for a larger $T_{\max}$. This is because when $T_{\max}$ is larger, the benchmark design is further away from the optimal design; hence there is more room for improvement for the adaptive design. The reduction is more than 20\% for $T_{\max} \geq 14$.

Third, the adaptive design consistently has a larger estimation error than the oracle design. The difference between adaptive and oracle designs is primarily due to the loss in precision from not knowing $\tilde{T}$ before the experiment starts. As the benchmark design differs from the oracle design and the treatment decisions are irreversible, the ``mistakes'' made in early time periods persistently continue to impact later periods. If we seek to narrow down the gap between adaptive and oracle designs, we can increase $N$, so that PGAE can learn the experiment termination time faster and make better treatment decisions early in the experiment.

    \begin{figure}[t!]
    \centering
    \includegraphics[width=1\linewidth]{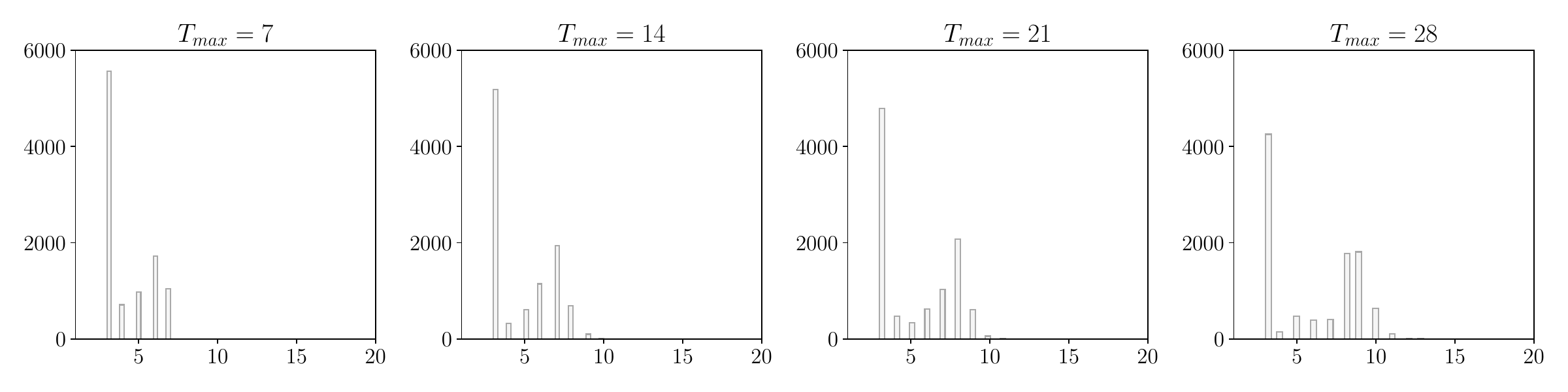}
	\caption{\textbf{Empirical distribution of termination time $\tilde{T}$ for various maximum duration.} This figure shows the histogram of the termination time $\tilde{T}$ for $T_{\max} \in \{7,14,21,28\}$ and $N=50$, based on 10,000 adaptive experiments. 
	In PGAE, NTU and each set of ATU have $10$ and $20$ units, respectively. $\tau_0$ is chosen at the minus 10\% of the average monthly flu occurrence rate, {\it i.e.}, $\tau_0 = -0.1 (NT_{\max})^{-1} \sum_{i,s} Y_{is}(-1)$. The experiment termination threshold is $c = 0.015 \cdot N/\tau_0^2 = 13.83$. The results in this figure are robust to the choice of $c$, as shown in Figure \ref{fig:experiment-termination-time-supp}.
	}
	\label{fig:experiment-termination-time}
\end{figure}

Finally, we note that there is a different trend in how the estimation error varies with $T_{\max}$ for different designs. For the benchmark design, the estimation error generally increases with $T_{\max}$. This is because as $T_{\max}$ increases, the benchmark design deviates more from the oracle design. For the oracle design, the estimation error consistently decreases with $T_{\max}$. This is because $\tilde{T}$ tends to increase with $T_{\max}$, as shown in Figure \ref{fig:experiment-termination-time}, and as a result, the precision of $\hat{\tau}_0$ using the oracle design increases with $\tilde{T}$.\footnote{The precision of $\hat{\tau}_0$ using the oracle design equals to $N (4\tilde{T}^2 - 1)/(3  \tilde T \sigma^2_\varepsilon) $, which increases with $\tilde{T}$.} For the adaptive design, since the algorithm stops as the estimated precision reaches the fixed threshold $c$, we expect the estimation error to generally stay flat for various $T_{\max}$. But since the precision estimation is not exact, and is actually a conservative one, there is no specific pattern for fluctuations in the estimation error.
    \begin{figure}[t!]
    \centering
    \includegraphics[width=0.6\linewidth]{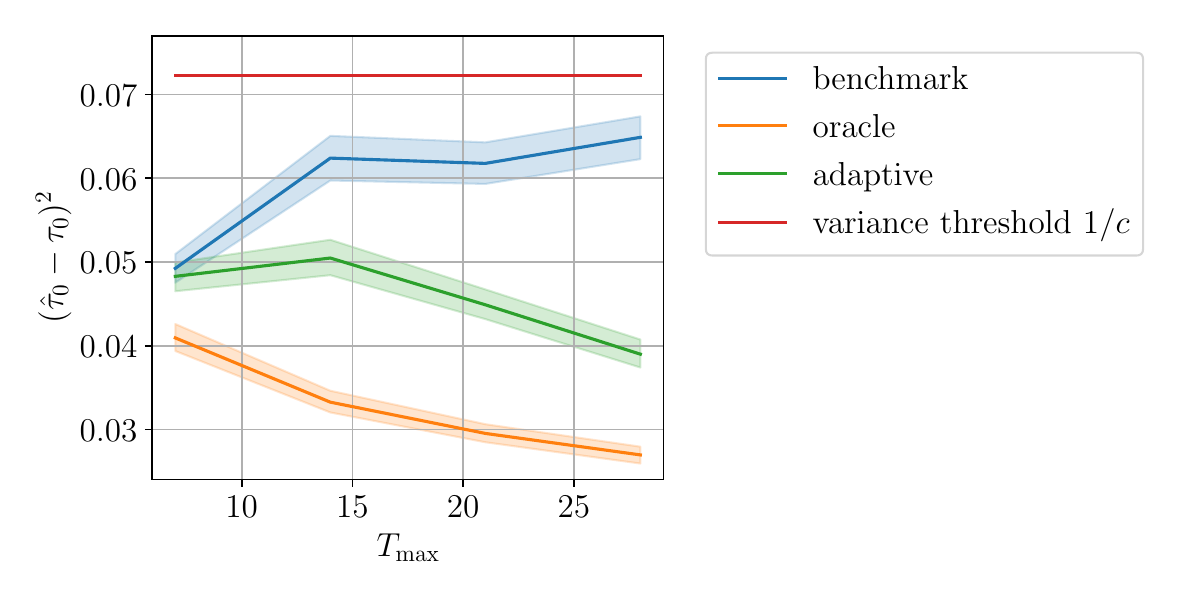}
	\caption{\textbf{Comparison of various designs in adaptive experiments.} This figure shows the mean and 95\% confidence band of $(\hat{\tau}_0 - \tau_0)^2$ for adaptive, benchmark, and oracle designs, based on 10,000 synthetic adaptive experiments. As the precision threshold is $c = 13.83$, the corresponding variance threshold $\var(\hat{\tau})$ is $1/c=0.072$. 
	The results in this figure are robust to the choice of $c$, as shown in Figure \ref{fig:experiment-termination-time-supp}.
	}
	\label{fig:various-opt-design}
\end{figure}

}

%% file: section_6.tex
\section{Concluding Remarks}\label{sec:conclusion}
	In this paper, we study the optimal design of \cmtfinal{staggered rollout experiments}. \cmtfinal{These experiments} are particularly useful for studying \cmtfinal{the impact of} treatments that have causal effects on both current and future outcomes. Our goal is to optimally make treatment decisions for every unit at every time period, in anticipation of most precisely estimating the average instantaneous and lagged effects. This optimization problem can reduce the sample size requirement and directly minimize the opportunity cost of the experiment in practice. We first study the non-adaptive experiments, where the sample size is fixed and treatment decisions are made pre-experiment. We provide a near-optimal solution to the optimization problem.  We further study adaptive experiments, where the experiments can be stopped early if needed. We propose the Precision-Guided Adaptive Experimentation (PGAE) algorithm for adaptive experiments. PGAE makes adaptive treatment decisions and allows for valid post-experiment inference. Finally, synthetic experiments on multiple data sets show that our proposed solutions for non-adaptive and adaptive experiments reduce the opportunity cost of the experiments by over 50\%, compared to non-adaptive design benchmarks.

%% file: appendix_A.tex
\renewcommand{\theHsection}{A\arabic{section}}

\begin{longtable}{ll}
        	{\bf Notation} & {\bf Description} \\
         \hline \\
         $[M]$ & $\{1, 2, \cdots, M\}$ for any positive integer $M$ \\
        $\bm{x}_{a:b}$ &  $(x_a,\ldots,x_b)$ for integers $a$ and $b$, $1\le a<b\le n$ \\
        $\tau_0$ & average instantaneous effect \\
        $\tau_j$ & average $j$-th period lagged effect \\
        $\bm{\tau}$ & $\begin{pmatrix} \tau_\ell & , & \cdots, & \tau_0   \end{pmatrix}$ \\
        $\*{y}_t$ & $\big(Y_{1t} , Y_{2t} , \cdots , Y_{Nt}\big)^\T \in \+R^{N \times 1}$\\
        $\*z_t$ & $\big(z_{1t} , z_{2t} , \cdots , z_{Nt}\big)^\T \in \{-1,1\}^{N \times 1}$\\
        $\bm{y}$ & $\big(
	\*y_1^\T , \*y_2^\T , \cdots , \*y_T^\T\big)^\T \in \+R^{NT\times 1}$\\
 $\bm{z}$ & $\big(\*z_1^\T , \*z_2^\T , \cdots , \*z_T^\T\big)^\T \in \+R^{NT\times 1}$ \\
 $Y$ & $\big(
	\*y_1 , \*y_2 , \cdots , \*y_T\big) \in \+R^{N\times T}$  \\ 
 $Z$ & $\big(
	\*z_1 , \*z_2 , \cdots , \*z_T\big) \in \{-1,1\}^{N\times T}$ \\
 $\omega_t$ & equals $N^\I \sum_{i = 1}^N z_{it}$ \\
 $\zeta_i$ & equals $T^\I \sum_{t = 1}^T z_{it}$ \\
 $A_i$ & first time that unit $i$ adopts the treatment, $A_i \in [T]  \cup \{\infty\}$ \\
 $A$ & $(A_1, \cdots, A_N^\T)$ \\
 $\alpha_i$ & unobserved unit fixed effect \\
 $\beta_t$ & unobserved time fixed effect \\
 $\*X_i$ & observed covariates of dimension $d_x$ \\
 $\*u_i$ & observed covariates of dimension $d_u$ \\
 $\bm{\theta}_t$ & unobserved coefficients of $\*X_i$ at time $t$ \\
 $\*u_t$ & unobserved coefficients of $\*v_t$ at time $t$ \\
 $\varepsilon_{it}$ & residual of unit $i$ at time $t$ \\
 $\sigma_\varepsilon^2$ & equals $\+E[\varepsilon^2_{it}]$ \\
 $\xi_\varepsilon^2$ & equals $\+E[(\varepsilon^2_{it} - \sigma_\varepsilon^2)^2]$ \\
 $e_{it}$ & equals $\varepsilon_{it} + \*u_i^\T \*v_t$ \\
 $\bm{1}_N$ & a vector of all ones of dimension $N$ \\
 $\*I_N$ & identity matrix of dimension $N \times N$\\
 $\hat{\tau}_j$ & estimate of $\tau_j$ \\
 $\hat{\bm{\tau}}$ & estimate of $\bm{\tau}$ \\
 $\xrightarrow{d}$ & convergence in distribution\\
 $\xrightarrow{p}$ & 
 convergence in probability \\
 \multicolumn{2}{l}{For a set of random variables $\tilde{X}_n$ and constants $a_n$,} \\
 $\tilde{X}_n = O_p(a_n)$ & the set of values $\tilde{X}_n/a_n$ is stochastically bounded \\ 
 $\tilde{X}_n = o_p(a_n)$ &  the set of values $\tilde{X}_n/a_n$ converges to zero in probability \\ 
 $A\mathrm{Cov}$ & asymptotic covariance\\
 \hline
 \caption{Mathematical notations.}
 \label{tab:notations}
    \end{longtable}

    \newpage

    \setcounter{tocdepth}{2}

    \tableofcontents

    \newpage
  
  \section{Supplementary Material for Non-Adaptive Experiments}\label{sec:constant-appendix}

		\subsection{Supplementary Material for Generalized Least Squares}\label{subsec:gls-carryover}
		
		    {\blue GLS estimates $\tau_0, \cdots, \tau_\ell$, $\alpha_i$, $\beta_t$, and $\bm{\theta}_t$ in \eqref{eqn:model-setup} by solving the following optimization problem }
    \begin{align} \label{eqn:gls}
        \min_{\bm{\tau}, \bm{\alpha}, \bm{\beta}_{(\ell+1):T}, \bm{\theta}_{(\ell+1):T} } \quad & \sum_{t=\ell+1}^{T} \bm{e}_t^\T   \cdot \*W^{-1} \cdot  \bm{e}_t   \\
       \nonumber \mathrm{s.t.} \quad & \bm{e}_t =  \*y_t - \bm{\alpha} - \beta_t \cdot \bm{1}_N  - \*X \bm{\theta}_t   - \tau_0 \cdot \*z_t - \cdots  -  \tau_\ell \cdot \*z_{t - \ell} \\
      \nonumber  & \alpha_i = 0 \text{ for } i > N-d_x-1 
    \end{align}
	where $\bm{\alpha} = \begin{pmatrix}
\alpha_1, &  \cdots, & \alpha_N
	\end{pmatrix} $, $\bm{\beta}_{(\ell+1):T} = \begin{pmatrix}
\beta_{\ell+1}, &  \cdots, & \beta_T
	\end{pmatrix}  $, $ \bm{\theta}_{(\ell+1):T} = \begin{pmatrix}
	\begin{array}{c|c|c}
	     \bm{\theta}_{\ell+1} &  \cdots & \bm{\theta}_T
	\end{array}
	\end{pmatrix}$, $\*X = \begin{pmatrix}
	\begin{array}{c|c|c|c}
	     \*X_1 & \*X_2 & \cdots & \*X_N
	\end{array}
	\end{pmatrix}^\T  \in \+R^{N \times d_x}$, and $\*W$ is a positive definite weighting matrix. The solution to \eqref{eqn:gls} is denoted as $\big(\hat{\bm{\tau}}, \hat{\bm{\alpha}}, \hat{\bm{\beta}}_{(\ell+1):T}, \hat{\bm{\theta}}_{(\ell+1):T}\big)$. {\blue In \eqref{eqn:gls}, we use observed outcomes from time $\ell+1$ to $T$, because these are the time periods that we have access to the current and past $\ell$ periods' treatment assignments. The second constraint, $\alpha_i$ is equal to zero for the last $d_x+1$ units, ensures that all parameters can be uniquely identified. Below we provide a few special cases of how the estimation problem is simplified when a simpler specification is used.
	
	\begin{example}[No latent covariates]
	    If there are no latent covariates ($d_u = 0$) in \eqref{eqn:model-setup}, then the optimal $\*W$ is proportional to $ \*I_N$. The objective function in    \eqref{eqn:gls} is then simplified to $\sum_{t =\ell+ 1}^T \*e_t^\T \*e_t$, which is the case where GLS and ordinary least squares (OLS) coincide. 
	\end{example}
	\begin{example}[No observed covariates]
	    If there are no observed covariates ($d_x = 0$), then the two constraints in \eqref{eqn:gls} are simplified to $\bm{e}_t =  \*y_t - \bm{\alpha} - \beta_t \cdot \bm{1}_N   - \tau_0 \cdot \*z_t - \cdots  -  \tau_\ell \cdot \*z_{t - \ell} $ and only $\alpha_N$ is enforced to be $0$. 
	\end{example}
	\begin{example}[No time or unit fixed effects]
	    If \eqref{eqn:model-setup} does not include time fixed effects, then the two constraints in \eqref{eqn:gls} are simplified to $\bm{e}_t =  \*y_t - \bm{\alpha}  - \*X \bm{\theta}_t  - \tau_0 \cdot \*z_t - \cdots  -  \tau_\ell \cdot \*z_{t - \ell} $ and $ \alpha_i = 0 \text{ for } i > N-d_x $. If \eqref{eqn:model-setup} does not include unit fixed effects, then \eqref{eqn:gls} only has one constraint, that is, $\bm{e}_t =  \*y_t  - \beta_t \cdot \bm{1}_N  - \*X \bm{\theta}_t   - \tau_0 \cdot \*z_t - \cdots  -  \tau_\ell \cdot \*z_{t - \ell} $. 
	\end{example}
	}

	When \eqref{eqn:model-setup} has latent covariates ($d_u > 0$), the optimal $\*W$ is  proportional to the inverse covariance matrix of $\bm{e}_t$, which is $\big(\*U \*\Sigma_v \*U^\T + \sigma_\varepsilon^2 \*I_N \big)^\I$ under Assumption \ref{ass:constant-error} below and $\*U = \begin{pmatrix}
	\begin{array}{c|c|c|c}
	     \*u_1 & \*u_2 & \cdots & \*u_N
	\end{array}
	\end{pmatrix}^\T  \in \+R^{N \times d_u}$. As $\*U$, $\*\Sigma_v$, and $\sigma_\varepsilon^2 $ are unknown in practice, we can use feasible generalized least squares (FGLS). FGLS first solves \eqref{eqn:gls} using identity matrix as the weighting matrix, then estimates $\*U$, $\*\Sigma_v$, and $\sigma_\varepsilon^2 $, and lastly uses $\big(\hat{\*U} \hat{\*\Sigma}_v \hat{\*U}^\T + \hat\sigma_\varepsilon^2 \*I_N \big)^\I$ as the weighting matrix to solve \eqref{eqn:gls} again.

		\begin{lemma}[Gauss-Markov Theorem]\label{lemma:gauss-markov}
		Consider a linear model $\bm{y} = \*X \bm\beta + \text{noise}$ with $\+E[\text{noise} \mid \*X] = 0$ and $\Cov[\text{noise} \mid \*X] = \bm\Omega$. If $\bm\Omega = \sigma^2 \bm I$, then the ordinary least squares estimator
		$\hat{\bm\beta} = (\*X^\T \*X)^\I \*X^\T \bm{y}$
		is the best linear unbiased estimator (BLUE). Otherwise, when $\bm\Omega$ is not a multiple of the identity matrix, the generalized least squares estimator $\hat{\bm\beta} = (\*X^\T \bm\Omega^\I \*X)^\I \*X^\T \bm\Omega^\I \bm{y}$
		is BLUE. Here, ``best'' means the estimator has the lowest variance among all unbiased linear estimators.
	\end{lemma}
		
		\subsection{Supplementary Material for Theorem \ref{thm:obs-latent-carryover-model}}\label{subsec:def-A-l-B-l}
		The $a^{(\ell)}$ in Theorem \ref{thm:obs-latent-carryover-model} is defined as
		\[a^{(\ell)} =  (M^{(\ell)})^{-1} b^{(\ell)}, \]
		where 
		$M^{(\ell)}$ and $b^{(\ell)}$ are defined as
		
		\begin{align}
		M^{(\ell)} =& \begin{bmatrix}
		\lfloor \ell/2 \rfloor+1 \\ & \lfloor \ell/2 \rfloor + 2 \\ && \ddots \\ &&& \ell 
		\end{bmatrix} - \frac{1}{T - \ell} \begin{bmatrix}
		\ell - \lfloor \ell/2 \rfloor & \ell  - 1 - \lfloor \ell/2 \rfloor & \ell  - 2 - \lfloor \ell/2 \rfloor & \cdots & 1 \\
		\ell  - 1 - \lfloor \ell/2 \rfloor & \ell  - 1 - \lfloor \ell/2 \rfloor & \ell  - 2 - \lfloor \ell/2 \rfloor & \cdots & 1 \\ 
		\vdots & \vdots & \vdots & \ddots & \vdots \\
		1 & 1 & 1 & \cdots &  1
		\end{bmatrix} \label{eqn:A-ell}
		\end{align}
		
		\begin{align}
		    b^{(\ell)} =& - \begin{bmatrix}
		\lfloor \ell/2 \rfloor+1 \\ \vdots \\ \ell - 1 \\ \ell
		\end{bmatrix}  + \frac{1}{T - \ell} \begin{bmatrix}
		(\lfloor \ell/2 \rfloor+1)^2  \\ \vdots \\  (\ell-1)^2 \\  \ell^2
		\end{bmatrix} - \frac{1}{T - \ell} \begin{bmatrix} 
		\sum_{l = 1}^{\ell - \lfloor \ell/2 \rfloor} (\lfloor \ell/2 \rfloor  + 1 - l) \\
		\vdots \\  2 \lfloor \ell/2 \rfloor - 1\\  \lfloor \ell/2 \rfloor
		\end{bmatrix} \label{eqn:b-ell}
		\end{align}
		
		Below we provide the expression of $\omega_{\ell,t}^\ast$ for $\ell = 3$, which has five stages. The example below complements the examples of $\omega_{\ell,t}^\ast$ for $\ell = 0, 1$ and $2$ in Examples \ref{example:two-way-fe}, \ref{example:ell-1} and \ref{example:ell-2}.
		
		\begin{example}[$\ell = 3$]\label{remark:carryover-example-l3}
		In Theorem \ref{thm:obs-latent-carryover-model}, $\omega^\ast_{\ell,t}$ takes the form of
			\begin{eqnarray*}
				&& \omega_1^\ast = -1, \quad \omega_2^\ast = -1 + \frac{6}{6T^2 - 44T + 79}, \quad \omega_3^\ast = -1 + \frac{12(T-4)}{6T^2 - 44T + 79}, \\
				&& \omega_t^\ast = -1 + \frac{2t-4}{T-3} \,\, \text{ for } t = 4, \cdots, T-3, \\
				&& \omega_{T-2}^\ast = 1 - \frac{12(T-4)}{6T^2 - 44T + 79}, \quad \omega_{T-1}^\ast = 1- \frac{6}{6T^2 - 44T + 79}, \quad \omega_T^\ast = 1.
			\end{eqnarray*} 
		\end{example}
	
	    \subsection{D-Optimal Treatment Design}\label{subsec:d-optimal-design}
	    We consider the D-optimal design that minimizes the determinant of $\var(\hat{\bm{\tau}})$. Note that 
	    $\var(\hat{\bm{\tau}}) = \mathrm{Prec}(\hat{\bm{\tau}})^{-1}$ and $\det(\var(\hat{\bm{\tau}})) = 1/\det(\mathrm{Prec}(\hat{\bm{\tau}})^{-1}) $. Minimizing $\var(\hat{\bm{\tau}})$ is equivalent to minimizing $1/\det(\mathrm{Prec}(\hat{\bm{\tau}}))$ and equivalent to:
	\begin{equation}\label{eqn:carryover-d-opt-transform}
	\min_{\{A_i\}_{i \in [N]}} -\det \big( \mathrm{Prec}(\hat{\bm{\tau}}) \big).
	\end{equation}
	Below we consider solving \eqref{eqn:carryover-d-opt-transform} for the specification $Y_{it}= \alpha_i +  \beta_t + \tau_0 z_{it} + \tau_1 z_{i,t-1} + \cdots + \tau_{\ell} z_{i,t-\ell} +  \varepsilon_{it}$.
	If the specification has covariates (either $d_x > 0$ or $d_u > 0$), then the role of covariates in the optimality conditions of \eqref{eqn:carryover-d-opt-transform} is identical to that for the T-optimal design in Theorem \ref{thm:obs-latent-carryover-model}.
	
	To solve \eqref{eqn:carryover-d-opt-transform}, we first need to write every entry in $\mathrm{Prec}(\hat{\bm{\tau}})$ as a function of  $z_{it}$:

	\begin{equation}\label{eqn:d-optimal-precision}
	    \begin{aligned}
	        \mathrm{Prec}(\hat{\bm{\tau}})_{jm}  = \begin{cases} 
					- \frac{N}{\sigma_\varepsilon^2} \Ls \sum_{t = j}^{T - \ell-1+j} \omega_t^2  - \frac{1}{T - \ell} \Lp  \sum_{t=j}^{T - \ell-1+j} \omega_t \Rp^2 + \frac{T - \ell}{2} \sum_{t = 1}^T (\upsilon_{T+1-t}^{(j,j)} - \upsilon_{T-t}^{(j,j)}) \omega_t  \Rs &  j = m\\
					 - \frac{N}{\sigma_\varepsilon^2} \left[ \sum_{t=j}^{T - \ell-1+j} \omega_t \omega_{t+m-j} -  \frac{1}{T - \ell} \Lp  \sum_{t=j}^{T - \ell-1+j} \omega_t \Rp  \Lp \sum_{t=m}^{T - \ell-1+m} \omega_t \Rp  \right.  \\
					\quad \left.  + \frac{T - \ell}{2} \sum_{t = 1}^T (\upsilon_{T+1-t}^{(j,m)} - \upsilon_{T-t}^{(j,m)}) \omega_t -  \sum_{t = j}^{m-1} (\omega_t  -  \omega_{T - \ell+t}) \right]    & j < m\\
					\mathrm{Prec}(\hat{\bm{\tau}})_{mj} & j > m \\
				\end{cases}
	    \end{aligned}
	\end{equation}
	where $\omega_t = \frac{1}{N} \sum_{i = 1}^N z_{it}$ and $\upsilon_t^{(j,m)}$  is defined as
	\begin{eqnarray*}
		\upsilon_t^{(j,m)} = \begin{cases}
			1 &  t \leq \ell+1-m\\
			-\Lp -1 + \frac{2(t-1-\ell+m)}{T - \ell} \Rp & \ell+1-m < t \leq \ell+1-j \\
			\Lp  -1 + \frac{2(t-1-\ell+m)}{T - \ell} \Rp \Lp  -1 + \frac{2(t-1-\ell+j)}{T - \ell} \Rp & \ell+1-j < t \leq T+1-m \\
			\Lp  -1 + \frac{2(t-1-\ell+j)}{T - \ell} \Rp & T+1-m < t \leq T+1-j \\
			1 & T+1-j < t \\
		\end{cases}
	\end{eqnarray*}
	
		Note that each entry in $\mathrm{Prec}(\hat{\bm{\tau}})$ is a quadratic function of $\omega_t$. Based on the Leibniz formula for determinants,  $\det \big( \mathrm{Prec}(\hat{\bm{\tau}}) \big)$ is a linear combination of $2(\ell+1)$ products of $\ell+1$ distinct elements in $\mathrm{Prec}(\hat{\bm{\tau}})$ (recall that $\mathrm{Prec}(\hat{\bm{\tau}})$ is an $(\ell+1) \times (\ell +1)$ matrix. Therefore, $\mathrm{Prec}(\hat{\bm{\tau}})$ is the $2(\ell+1)$-th degree polynomial function of $z_{it}$. 
		
\begin{figure}[t!]
			\centering
			\includegraphics[width=0.5\linewidth]{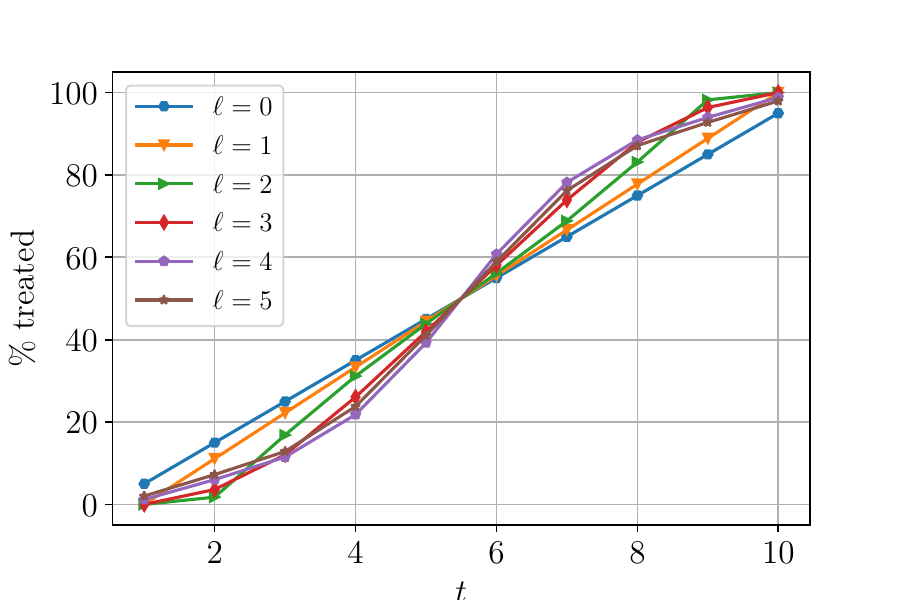}
			\caption{{D}-optimal treatment design: Optimal treated proportion ($(1+\omega_t)/2$) at each period for a $T$-period treatment design and various $\ell$, where $T = 10$. Different colors represent different $\ell$.}
			\label{fig:carryover-treatment-effect-d-opt}
		\end{figure}

		It is generally infeasible to analytically solve \eqref{eqn:carryover-d-opt-transform} for $\ell > 1$. We therefore use the off-the-shelf software to find the optimal $\omega_t$. We show the optimal solution to \eqref{eqn:carryover-d-opt-transform} for $T=10$ in Figure \ref{fig:carryover-treatment-effect-d-opt}. Similar to the T-optimal design,  the optimal $\omega_t$ is symmetric with respect to the center ({\it i.e.}, $((T+1)/2,0)$). Also similar to the T-optimal design, if $\ell$ is larger, then the optimal $\omega_t$ is generally smaller at the beginning, increases at a faster rate in the middle, and is generally larger in the end. 
	
\subsection{Reversible Treatment Adoption Case}\label{subsec:reversible-treatment}

There are two ways to look at the reversible treatment adoption case: first, each unit is treated at most once; second, units can be treated more than once. Surprisingly, the first case is a special case of the irreversible pattern introduced in Section \ref{sec:experiment-outcome-estimand}, as shown in \ref{subsubsec:one-time-treatment}.  For the second case, we derive analogous results as those for the irreversible pattern in \ref{subsubsec:general-reversible-treatment}.

\subsubsection{Equivalent Representation of One-time Treatment}\label{subsubsec:one-time-treatment}
\texttt{} \\
For the one-time treatment, such as the participation of a job-training or skill-improvement program, we can still use specification \eqref{eqn:model-setup} to estimate how the effect of this one-time treatment varies with the time. However, we need to use a different definition of $z_{it}$ in specification \eqref{eqn:model-setup}: $z_{it} = +1$ denotes that unit $i$ has been treated up to time $t$, and $z_{it} = -1$ denotes otherwise.

An alternative approach is to use the definition of treatment variable as in the main text: Let $w_{it} = \{0,1\}$ denote whether unit $i$ is treated at time $t$ ($w_{it} = 1$) or not ($w_{it} = 0$). In the case of one-time treatment, at most one of $w_{it}, \cdots, w_{i,t-\ell}$ is $1$. Consider the following specification using $w_{it}$:
\begin{equation}\label{eqn:model-setup-one-time}
    Y_{it}  = \tilde\alpha_i +  \tilde\beta_t + \*X_i^\T  \bm{\theta}_t + \*u_i^\T \*v_t + \tilde\tau_0 w_{it} + \tilde\tau_1 w_{i,t-1} + \cdots + \tilde\tau_{\ell} w_{i,t-\ell}  + \varepsilon_{it}\,.
\end{equation}
$\tilde\tau_j$ can be interpreted as the treatment effect given that a unit was treated $j$-period ago. If $|\tilde \tau_\ell| < \cdots < |\tilde \tau_1| < |\tilde \tau_0| $, then the effect of the one-time treatment attenuates over time.

We can show that specification \eqref{eqn:model-setup-one-time} using $w_{it}$ is equivalent to specification \eqref{eqn:model-setup} using $z_{it}$ with the change of variables:
\begin{align*}
    \tilde\tau_j = 2 \sum_{k = 0}^j \tau_k \qquad \tilde\alpha_i = \alpha_i -\sum_{k = 0}^\ell \tau_k \qquad \tilde\beta_t = \beta_t.
\end{align*}

\subsubsection{General Reversible Treatment Pattern}\label{subsubsec:general-reversible-treatment}
\texttt{} \\
Next we consider the case where a unit can arbitrarily switch between treatment and control ({\it i.e.}, Assumption \ref{ass:treatment-adoption} is violated). Below we provide the optimality conditions of T-optimal design for this case. 

\begin{proposition}\label{prop:rev-treat}
			Suppose Assumption \ref{ass:model} holds, and the treatment is reversible. Let $\omega_t =  \sum_{i = 1}^N z_{it}/N $ and $\zeta_i = \sum_{i = 1}^T z_{it} /T$.
			\begin{enumerate}
				\item Suppose $\tau$ is estimated from the least squares estimator under the specification $Y_{it} =\beta_t+ \tau z_{it} + \varepsilon_{it}$. If $\varepsilon_{it}$ is i.i.d. in $i$ and $t$ with mean $0$ and variance $\sigma_\varepsilon^2$, then any treatment design is optimal if it satisfies
				\[\omega_t = 0.\]
				\item Suppose $\tau$ is estimated from the least squares estimator under the specification $Y_{it} =\alpha_i+ \tau z_{it} + \varepsilon_{it}$. If $\varepsilon_{it}$ is i.i.d. in $i$ and $t$ with mean $0$ and variance $\sigma_\varepsilon^2$, then any treatment design is optimal if it satisfies
				\[\zeta_i = 0.\]
				\item Suppose $\tau$ is estimated from the least squares estimator under the specification $Y_{it} =\alpha_i+ \beta_t +\tau z_{it} + \varepsilon_{it}$. If $\varepsilon_{it}$ is i.i.d. in $i$ and $t$ with mean $0$ and variance $\sigma_\varepsilon^2$, then any treatment design is optimal if it satisfies
				\[\omega_t = 0, \quad \zeta_i = 0.\]
				\item Suppose $\tau$ is estimated from the least squares estimator under the specification $\alpha_i +  \beta_t +  \*X^\T_i  \bm{\theta}_t+  \*u^\T_i \*v_t + \tau_0 z_{it} + \tau_1 z_{i,t-1} + \cdots + \tau_{\ell} z_{i,t-\ell} + \varepsilon_{it}$ for $\ell \geq 0$. Suppose the assumptions in  Theorem \ref{thm:obs-latent-carryover-model} hold, and both $\*X_i$ and $\*u_i$ are discrete-valued. Let $\omega_{g,t} = \frac{1}{|\tlo_g|} \sum_{i \in \tlo_g} z_{it}$, where $\tlo_g = \{i: X_i = x_g, u_i = u_{0,g}\}$. Any treatment design is optimal if it satisfies
				\[\omega_{g,t} = 0, \, \text{ for all $t$ and $g$}, \quad \zeta_i^{(j)} = 0, \, \text{ for all $i$ and $j$},\]
    where $\zeta_i^{(j)} = \sum_{t=j}^{T - \ell+j-1} z_{it} / (T - \ell)$
			\end{enumerate}
		\end{proposition}
		
		\proof{Proof of Proposition \ref{prop:rev-treat}}
		\begin{enumerate}
		    \item Under the specification  $Y_{it} =\beta_t+ \tau z_{it} + \varepsilon_{it}$, the precision of $\hat{\tau}$ equals  
		    \[\Prec(\hat{\tau}) = NT - N \sum_{t=1}^T \omega_t^2.  \]
		    The precision is maximized at $\omega_t = 0$.
		    \item Under the specification $Y_{it}= \alpha_i + \tau z_{it} + \varepsilon_{it}$, the precision of $\hat{\tau}$ equals  
		    \[\Prec(\hat{\tau}) = NT - T \sum_{i = 1}^N \zeta_i^2. \]
		    The precision is maximized at $\zeta_i = 0$.
		    \item  Under the specification $Y_{it}= \alpha_i + \beta_t + \tau z_{it} + \varepsilon_{it}$, from the proof of Lemma \ref{lemma:simplify-obj}, the precision of $\hat{\tau}$ equals 
		    \[\Prec(\hat{\tau}) = NT - \left(T \sum_{i = 1}^{N} \zeta_i^2 -\frac{N}{T} \Lp\sum_{t=1}^T \omega_t  \Rp^2 + N \sum_{t = 1}^T \omega_t^2 \right). \]
		    The precision is maximized at $\zeta_i = 0$ and $\omega_t = 0$, given that $\sum_{t = 1}^T \omega_t^2 - \frac{1}{T}  \Lp\sum_{t=1}^T \omega_t  \Rp^2$ is minimized at $\omega_t = 0$.
		    \item Under the specification $Y_{it} = \alpha_i +  \beta_t +  \*X^\T_i  \bm{\theta}_t+  \*u^\T_i \*v_t + \tau_0 z_{it} + \tau_1 z_{i,t-1} + \cdots + \tau_{\ell} z_{i,t-\ell} + \varepsilon_{it}$, the role of observed and latent covariates in the optimal design is the same as that in Theorem \ref{thm:obs-latent-carryover-model}.   The major difference with Theorem \ref{thm:obs-latent-carryover-model} is the optimal treated fraction conditions. Note that the precision matrix $\Prec(\hat{\bm{\tau}})$ takes the form of (see Equation \eqref{eqn:precision-matrix}), 
      \[\Prec\left( \hat{\bm{\tau}}\right) =  \bm{Z}_\ell^\T  \bm{\Sigma}^\I_e  ( \bm{\Sigma}_e - \bm{\Gamma} (\bm{\Gamma}^\T \bm{\Sigma}^\I_e  \bm{\Gamma})^\I \bm{\Gamma}^\T) \bm{\Sigma}^\I_e  \bm{Z}_\ell  \]
where the $(j,j)$-th entry ({\it i.e.}, diagonal entry) in $\Prec\left( \hat{\bm{\tau}}\right)$ takes the form of (see Equations \eqref{eqn:precision-entries-1} and \eqref{eqn:precision-entries-2} in Lemma \ref{eqn:carryover-separate-obj}, and the proof of Lemma \ref{lemma:simplify-obj})
\begin{align*}
    & N(T-\ell) -   \Bigg[  N \sum_{t=j}^{T - \ell+j-1} \omega_t^2   - \frac{N}{T - \ell}  \Lp \sum_{t=j}^{T - \ell+j-1} \omega_t  \Rp^2    + (T - \ell) \sum_{i = 1}^N (\zeta_i^{(j)})^2 \Bigg] \\ & - \frac{N}{\sigma_\varepsilon^2} \sum_{j = 1}^{\ell + 1} f_{j,\*X}(Z) - \frac{N}{\sigma_\varepsilon^2} \sum_{j = 1}^{\ell + 1} f_{j,\*U}(Z)
\end{align*}
$f_{j,\*X}(Z)$ and $f_{j,\*U}(Z)$ are defined in Lemma \ref{lemma:simplify-obj}.
The $(j,j)$-th entry in $\Prec(\hat{\bm{\tau}})$ is maximized at $\omega_t = 0$ for all $t$ and $\zeta_i^{(j)} = 0$ for all $i$. Moreover both $f_{j,\*X}(Z)$ and $f_{j,\*U}(Z)$ are minimized at $\omega_{g,t} = 0$. The T-optimal design maximizes the trace of the precision matrix, which is satisfied by the solution provided in Lemma \ref{prop:rev-treat}.4. 
		\end{enumerate}
		We conclude the proof of Lemma \ref{prop:rev-treat}. \Halmos
		\endproof

		    \subsection{A Rounding Approach for A Feasible Solution}\label{subsubsec:practical-considerations}
    
    When there does not exist a feasible solution in $\mathbb{A}^{\mathrm{disc}}_{\mathrm{opt}}$,  we suggest using the nearest integer rounding rule as follows to obtain a feasible $\{A_i\}_{i \in [N]}$.

	\textit{Nearest integer rounding rule:} If the number of treated units for each stratum suggested by Theorem \ref{thm:obs-latent-carryover-model}, {\it i.e.}, is not an integer,
 \[\frac{|\tlo_g|(1+\omega_{\ell,t}^\ast)}{2T}\]
  we suggest rounding it to an integer using the nearest integer rule. We separate the number of treated units into the integer and decimal parts
 \[\frac{|\tlo_g|(1+\omega_{\ell,t}^\ast)}{2T} = \underbrace{\lfloor \frac{|\tlo_g|(1+\omega_{\ell,t}^\ast)}{2T} \rfloor }_{N_{\mathrm{treated},g,t}^{\mathrm{int}}} + \underbrace{\frac{|\tlo_g|(1+\omega_{\ell,t}^\ast)}{2T} - \lfloor \frac{|\tlo_g|(1+\omega_{\ell,t}^\ast)}{2T} \rfloor }_{N_{\mathrm{treated},g,t}^{\mathrm{dec}}} \]
 The nearest integer rounding rule works as follows
\begin{itemize}
    \item If the decimal part $N_{\mathrm{treated},g,t}^{\mathrm{dec}} < 0.5$, or if $N_{\mathrm{treated},g,t}^{\mathrm{dec}} = 0.5$ with $t < T/2$, then the rounded number of treated units is 
    \[\frac{1}{|\tlo_g|} \sum_{i \in \tlo_g} \boldsymbol{1}_{A^\rnd_i \leq t} = N_{\mathrm{treated},g,t}^{\mathrm{int}}  \]
    \item Otherwise, the rounded number of treated units is
    \[\frac{1}{|\tlo_g|} \sum_{i \in \tlo_g} \boldsymbol{1}_{A^\rnd_i \leq t} = N_{\mathrm{treated},g,t}^{\mathrm{int}} + 1  \]
\end{itemize}

Let $Z^\rnd$ be the treatment design satisfying the rounded number of treated units for each stratum. Let $Z^{\mathrm{int}\ast}$ be the optimal integer solution to optimization problem \eqref{eqn:obj}. $ \tr \big(\mathrm{Prec}(\hat{\bm{\tau}})  \big)_{Z} $ be the objective function \eqref{eqn:obj} ({\it i.e.} $\tr \big(\mathrm{Prec}(\hat{\bm{\tau}})  \big)$) evaluated at $Z$.
 
	The following proposition bounds the difference between $ \tr \big(\mathrm{Prec}(\hat{\bm{\tau}})  \big)_{Z^\rnd} $ and $ \tr \big(\mathrm{Prec}(\hat{\bm{\tau}})  \big)_{Z^{\mathrm{int}^\ast}} $.
	\begin{proposition}
		\label{prop:rounding-error-obs-cov}
		Suppose the assumptions in Theorem \ref{thm:obs-latent-carryover-model} hold, $d_u = 0$ and $N$ is even, and $x_{k,\max} = \max_{g} |x_{gk}|$ is finite, where $x_{gk}$ is the $k$-th coordinate of $x_g$ for all strata $g$. We have
		\[ \tr \big(\mathrm{Prec}(\hat{\bm{\tau}})  \big)_{Z^\rnd} = \tr \big(\mathrm{Prec}(\hat{\bm{\tau}})  \big)_{Z^{\mathrm{int}\ast}}  \cdot\left(1+ O\left(\frac{1+ \sum_{k=1}^{d_x} x_{k,\max}^2}{N^2_{\min}} \right) \right).\]
	\end{proposition}

    Since the probability of each realization of $\*X_i$ is bounded away from 0 and $x_{k,\max}$ is finite,  we have $O\left((1+ \sum_{k=1}^{d_x} x_{k,\max}^2)/N^2_{\min} \right)  = O \Lp 1/N^2 \Rp$ as $N \rightarrow \infty$. However, if $G$ is large compared with $N$ (or $\*X_i$ takes infinitely many values), $ \tr \big(\mathrm{Prec}(\hat{\bm{\tau}})  \big)_{Z^\rnd} $ could be much larger than $ \tr \big(\mathrm{Prec}(\hat{\bm{\tau}})  \big)_{Z^{\mathrm{int}\ast}} $. In this case, we could instead partition units into only a few ({\it e.g.}, 2 or 3) groups based on their covariates' values using $K$-means or some other methods. Within each group, the treated fraction conditions are satisfied. 
    
    {\blue 
    \subsection{Supplementary Details in Choosing a Treatment Design}\label{subsec:more-details-choose-design}
    
    In Algorithm \ref{algo:choose-design}, we provide a procedure to randomly choose a treatment design for each stratum based on the treated fraction conditions. If there are no covariates (in the outcome model), then this is the case where we only have one stratum that consists of all experimental units. 
    
    We want to highlight that Algorithm \ref{algo:choose-design} allows for the case where the treatment fraction conditions in Theorem \ref{thm:obs-latent-carryover-model} can not be exactly satisfied for any treatment design. For this case, Algorithm \ref{algo:choose-design} uses the nearest integer rounding algorithm in \ref{subsubsec:practical-considerations} in choosing a treatment design.

    	\begin{algorithm}[htp]
 \caption{Choose a treatment design for each stratum $g$}  \label{algo:choose-design}
 \SetKwInOut{Input}{Inputs}
 \nl \Input{$|\mathcal{O}_g|$, $[\omega_{\ell,t}^\ast]_{t\in [T]}$} 
  \nl \For{$t = 1,\cdots, T $}{
  \nl $N_{\mathrm{treated},g,t}^{\mathrm{int}} \leftarrow \lfloor |\tlo_g|(1+\omega_{\ell,t}^\ast)/2 \rfloor$\;
  \nl $N_{\mathrm{treated},g,t}^{\mathrm{dec}} \leftarrow |\tlo_g|(1+\omega_{\ell,t}^\ast)/2-  N_{\mathrm{treated},g,t}^{\mathrm{int}} $ \;
  \nl \eIf{$N_{\mathrm{treated},g,t}^{\mathrm{dec}} < 0.5 $ or $ N_{\mathrm{treated},g,t}^{\mathrm{dec}} = 0.5 $ with $t < T/2$}{
  \nl $N_{g,t} \leftarrow N_{\mathrm{treated},g,t}^{\mathrm{int}} $ \;}{
  \nl $N_{g,t} \leftarrow N_{\mathrm{treated},g,t}^{\mathrm{int}} + 1 $ \;
  } 
    }
  \nl $f(\cdot)  \leftarrow $ a random function that shuffles $\{1, 2, \cdots, |\mathcal{O}_g| \}$\;
  \nl $Z_g \leftarrow [-1]^{|\mathcal{O}_g| \times T}$ \;
  \nl \For{$i = 1, \cdots, |\mathcal{O}_g|$}{
  \nl \For{$t = 1, \cdots, T$}{
  \nl \eIf{$f(i) \leq N_{g,t}$}{
  $z_{g,it} \leftarrow 1$ \;
  }{
  $z_{g,it} = -1$ \;
  }
  }
  }
  \nl \KwRet $Z_g $\;
\end{algorithm}

\subsection{Latent Covariates}\label{subsec:latent-covariates}

\subsubsection{Estimation of latent covariates}\label{subsubsec:estimate-latent-covariates}
\texttt{}
 
If there are latent covariates, we suggest, whenever possible, using the historical control data for the same set of units in designing the experiment. This is because historical data has information about $\*u_i$. Below we provide a heuristic approach to using historical data to find a better treatment design.

We can apply singular value decomposition (SVD) to the historical data and take the top singular vectors as the estimated $\*u_i$, denoted as $\hat{\*u}_i$. We can use cross-validation to choose the number of singular vectors.
Specifically, we can partition the historical data in the time dimension into training and validation sets. Then we estimate the singular vectors on the training set, and design synthetic experiments on the validation set based on various numbers of singular vectors. Finally, we can select the optimal number of singular vectors whose experimental design has the lowest estimation error of treatment effects on the validation set.

We suggest using SVD with cross-validation for the following reason. Since the time horizon $T$ of the experiment is typically short ({\it e.g.}, one flu season with $T = 7$ in our empirical application, and short $T$ is the regime where statistical power is a concern), it is typically sufficient to account for the first one or two latent factors in the feasible GLS estimator. Then it is natural to account for the top a few latent factors in the design of experiments, where SVD with cross-validation seems to be a reasonable approach. 

If historical data have missing observations, or if both the historical data and experimental data have a long time horizon (where it is reasonable to estimate more latent factors), then we could use low-rank matrix estimation with nuclear norm regularization to estimate latent covariates from the historical data.

\subsubsection{Non-adaptive treatment design based on latent covariates}\label{subsubsec:choose-treatment-design}
\texttt{}

We can first estimate latent covariates by SVD from historical data, and then treat $\hat{\*u}_i$ as ``observed'' covariates in the design of non-adaptive experiments. If $\hat{\*u}_i$ is continuous, then we can partition units into strata based on $\hat{\*u}_i$ by applying $k$-means clustering on the largest singular vectors, similar to observed continuous-valued covariates. Here the number of clusters $k$ can be chosen by cross-validation. Each cluster of units from $k$-means clustering is a stratum, and we can apply Algorithm \ref{algo:choose-design} to randomly choose a treatment design for each stratum. 

The approach of stratifying based on top singular vectors is conceptually similar to spectral clustering, which enjoys nice theoretical and empirical properties (see {\it e.g.}, \cite{ng2001spectral}) and works well in our empirical applications. Admittedly, it is possible to stratify based on the ideas of other clustering algorithms ({\it e.g.}, density-based clustering, mean-shift clustering), and we leave this for future work.

Note that it seems possible to use the currently available experimental data to improve the estimation precision of $\hat{\*u}_i$ and treatment decisions for subsequent experimental periods. However, we do not pursue this route for two reasons.

First, if historical data does not have many periods, then $\hat{\*u}_i$ can be quite noisy early in the experiment. We may not want to use noisy estimates of $\*u_i$ for treatment decisions, because the mistakes we make in the early periods (due to very noisy estimates of $\*u_i$) can carry over to later periods. We have verified this in numerical simulations but did not include them in the paper, given the paper is quite lengthy already.

Second, if historical data has many periods, which allows us to precisely estimate $\*u_i$, then having a few more experimental periods can only marginally improve the precision of $\hat{\*u}_i$ and treatment allocations. This is because the convergence rate of $\hat{\*u}_i$ is $\sqrt{T}$ \citep{bai2003inferential}.

    \subsection{Separable Quadratic Representation}\label{subsec:separate-quadratic}
    
    In this subsection, we state a critical lemma that shows $\tr\left(\mathrm{Prec}(\hat{\bm{\tau}})\right) $ can be decomposed into three separable quadratic functions: the first one does not depend on $\*X$ and $\*U$, the second one only depends on $\*X$, and the third one only depends on $\*U$. The solutions to each of these three show the effects of fixed effects, observed covariates, and latent covariates on the optimal treatment assignments, respectively. If a solution simultaneously optimizes all three quadratic functions, then this solution is an optimal solution to \eqref{eqn:obj}.
    
    \begin{lemma}[Separable Quadratic Representation]\label{lemma:simplify-obj}
		Suppose Assumptions \ref{ass:treatment-adoption}, \ref{ass:model}, and \ref{ass:constant-error} hold. Suppose $\hat{\bm{\tau}} $ is estimated from the infeasible GLS with $\*W= [w_{ij}] \propto \big(\*U \*\Sigma_v \*U^\T + \sigma_\varepsilon^2 \*I_N \big)^\I $,  rows in $\*X$ and $\*U$ are demeaned, that is, $\sum_{i = 1}^N \*X_i = \bm{0}_{d_x}$ and $\sum_{i = 1}^N \*u_i = \bm{0}_{d_u}$,  $\sum_{i = 1}^N \*X_i \*u_i^\T = \mathbf{0}_{d_x,d_u}$, and $\*\Sigma_v = \sigma_\varepsilon^2 \cdot \*I_{d_u}$.
		Let $T_\ell = T - \ell$ and $j_\ell = j + T_\ell - 1$ for all $j$. Then $\tr \big(\mathrm{Prec}(\hat{\bm{\tau}})  \big) $  takes the form of
		\begin{equation}\label{eqn:carryover-quadratic}
		\begin{aligned}
		\tr \big(\mathrm{Prec}(\hat{\bm{\tau}})  \big)  =& - \frac{N}{\sigma_{\varepsilon}^2} \sum_{j = 1}^{\ell+1} 
		\bigg(  \underbrace{ (\bm{\omega}_{j:j_\ell})^\T \*P_{\bm{1}_{T_\ell}}  \bm{\omega}_{j:j_\ell}  + 2 \bm{b}_{\ell}^\T \bm{\omega}_{j:j_\ell} }_{f_{j,\bm{1}}(Z)} +  \underbrace{\textstyle  \sum_{k=1}^{d_x}  (\bm{\omega}_{j:j_\ell}^{x_k})^\T \*P_{\bm{1}_{T_\ell}}  \bm{\omega}_{j:j_\ell}^{x_k}  }_{f_{j,\*X}(Z)} + \underbrace{\textstyle \frac{1}{N} \bm{z}_{j:j_\ell}^\T \*M_{\*U} \bm{z}_{j:j_\ell} }_{f_{j,\*U}(Z)} \bigg),
		\end{aligned}
		\end{equation}
		where $\bm{\omega}_{j:j_\ell} = \big({\omega}_j, \cdots, {\omega}_{j_\ell}\big)^\T$ with ${\omega}_t = \frac{1}{N} \sum_{i=1}^N z_{it} $, $\bm{\omega}_{j:j_\ell}^{x_k} = \big(	{\omega}^{x_k}_j, \cdots, {\omega}^{x_k}_{j_\ell} \big)^\T$ with ${\omega}^{x_k}_t =\frac{1}{N} \sum_{i=1}^N X_{ik} z_{it}$, 
		$\bm{z}_{j:j_\ell} = \big(\*z_{j}^\T, \cdots, \*z_{j_\ell}^\T \big)^\T \in \{-1,+1\}^{N T_\ell\times 1}$,  $\*P_{\bm{1}_{T_\ell}} = \*I_{T_\ell} - \frac{1}{T_\ell} \bm{1}_{T_\ell} \bm{1}_{T_\ell}^\T$, $\*b_\ell = [b_{\ell,t}] \in [-1,1]^{T_\ell}$ with $b_{\ell,t} = \frac{T_\ell+2-2t}{T_\ell} $, and $\*M_{\*U} = \*P_{\bm{1}_{T_\ell}} \otimes \*U (\*I_{d_u} + \*U^\T \*U)^\I \*U^\T$.\footnote{``$\otimes$'' is the Kronecker product.}
	\end{lemma}

    From Lemma \ref{lemma:simplify-obj}, maximizing $	\tr \big(\mathrm{Prec}(\hat{\bm{\tau}})  \big)$ is equivalent to 
	\begin{equation}\label{eqn:objective-reduced}
	 \min_{\{A_i\}_{i \in [N]} }  \underbrace{\textstyle \sum_{j = 1}^{\ell+1}   f_{j,\bm{1}}(Z)  }_{f_{\bm{1}}(Z) }  + \underbrace{\textstyle \sum_{j = 1}^{\ell+1}   f_{j,\*X}(Z) }_{f_{\*X}(Z) }  + \underbrace{\textstyle \sum_{j = 1}^{\ell+1}   f_{j,\*U}(Z)  }_{f_{\*U}(Z) }. 
	\end{equation}

    If we can find an $\{A_i\}_{i \in [N]}$ that simultaneously minimizes each of $f_{\bm{1}}(Z)$, $f_{\*X}(Z)$, and $f_{\*U}(Z)$, then this $\{A_i\}_{i \in [N]}$ minimizes \eqref{eqn:objective-reduced}. In the following, we separately analyze the three sub-problems of \eqref{eqn:objective-reduced}: $\min_{\{A_i\}_{i \in [N]} }  f_{\bm{1}}(Z) $, $\min_{\{A_i\}_{i \in [N]} }  f_{\*X}(Z) $, and $\min_{\{A_i\}_{i \in [N]} }  f_{\*U}(Z) $.

	We first consider the sub-problem $\min_{\{A_i\}_{i \in [N]} }  f_{\bm{1}}(Z) $. The solution to this problem characterizes the effect of the presence of two-way fixed effects, $\alpha_i$ and $\beta_t$ on the optimal treatment assignments. Note that $f_{\bm{1}}(Z)$ is a sum of quadratic and linear terms:
	\begin{equation}\label{eqn:f1w-carryover-obj}
	   f_{\bm{1}}(Z) =\sum_{j = 1}^{\ell+1} \big( (\bm{\omega}_{j:j_\ell})^\T \*P_{\bm{1}_{T_\ell}}  \bm{\omega}_{j:j_\ell}  + 2 \bm{b}_{\ell}^\T \bm{\omega}_{j:j_\ell} \big).
	\end{equation}
	It is possible to provide the analytical solution to $\min_{\{A_i\}_{i \in [N]} }  f_{\bm{1}}(Z) $ based on its first order condition. The analytical solution is provided in \eqref{eqn:omega-carryover}.

	Next we consider the other two sub-problems: $\min_{\{A_i\}_{i \in [N]} }  f_{\*X}(Z) $ and $\min_{\{A_i\}_{i \in [N]} }  f_{\*U}(Z) $. The solution to these two problems characterizes the effects of the presence of $\*X_i$ and/or $\*u_i$ on the optimal treatment assignments, respectively. 
	Note that both $f_{\*X}(Z) $ and $f_{\*U}(Z) $ are sums of quadratic functions of $\{\*z_t\}_{t=1}^T$ that do not have linear terms. The Hessian of $f_{\*X}(Z) $ and $f_{\*U}(Z) $ are both semidefinite, following the definition of two matrices in $f_{\*X}(Z) $ and $f_{\*U}(Z) $, {\it i.e.}, $\*P_{\bm{1}_{T_\ell}} = \*I_{T_\ell} - \frac{1}{T_\ell} \bm{1}_{T_\ell} \bm{1}_{T_\ell}^\T$ and $\*M_{\*U} = \*P_{\bm{1}_{T_\ell}} \otimes \*U (\*I_{d_u} +  \*U^\T \*U)^\I \*U^\T$. Then the minimum possible value of $f_{\*X}(Z) $ and $f_{\*U}(Z) $ is 0. As shown in Theorem \ref{thm:obs-latent-carryover-model} below, the minimum value of $f_{\*X}(Z) $ can be achieved if $\{A_i\}_{i \in [N]} \in \mathbb{A}_{\*X}$, and the minimum value of $f_{\*U}(Z)$ can be achieved if $\{A_i\}_{i \in [N]} \in \mathbb{A}_{\*U}$.

}

\clearpage

%% file: appendix_B.tex
\section{Proof of Results for Non-Adaptive Experiments}\label{sec:constant-proof}

    We can combine $\beta_t$ with $\bm{\theta}_t$ in the specification \eqref{eqn:model-setup}, that is, 
		\[Y_{it}= \alpha_i + \underbrace{\begin{bmatrix}
			1 & \*X^\T_i
			\end{bmatrix}}_{\tilde{\*X}_i^\T}  \begin{bmatrix}
		\beta_t \\ \bm{\theta}_t
		\end{bmatrix} + \tau_0 z_{it} + \tau_1 z_{i,t-1} + \cdots + \tau_{\ell} z_{i,t-\ell} + \underbrace{\*u_i^\T \*v_t  + \varepsilon_{it}}_{e_{it}}.\]
		Denote $p \coloneqq d_x + 1$, and then $\tilde{\*X}_i \in \+R^{p}$. Denote  $\zeta_i = \frac{1}{T} \sum_{t = 1}^T z_{it} $ for all $i$ and $\tilde \omega_t = \frac{1}{N} \sum_{i=1}^N \tilde{\*X}_i z_{it} \in \+R^{p}$ for all $t$.

    We write the potential outcomes from time $\ell+1$ to $T$ into a vectorized form, and then we have 
    \[\bm{y}_{(\ell+1):T} = \begin{bmatrix} \bm{z}_{1:(T-\ell)} &  \cdots &  \bm{z}_{\ell:(T-1)} & \bm{z}_{(\ell+1):T} & 
    \bm{\Gamma}
    \end{bmatrix} \begin{bmatrix}
    \bm{\tau} \\ \bm{\alpha}_{1:(N-p)} \\ \bm{\beta}_{(\ell+1):T} \\ \bm{\theta}_{(\ell+1):T}
    \end{bmatrix} + \bm{e}_{(\ell+1):T}, \]
    where $\hat{\bm{\tau}} = \begin{pmatrix} \tau_\ell & , & \cdots, & \tau_0   \end{pmatrix}$,
    
    \[ \bm{\Gamma} = 
		\begin{bmatrix}
		\tilde{\*I}_{N-p} & \bm{1}_N & \*X  &  &  &  \\
		\tilde{\*I}_{N-p} & & &    \bm{1}_N  &  \*X & \\
		\vdots & & & & & &  \ddots \\
		\tilde{\*I}_{N-p} & & & &   & & & & \bm{1}_N & \*X\\
		\end{bmatrix}  = \begin{bmatrix}
		\tilde{\*I}_{N-p} & \tilde{\*X} \\
		\tilde{\*I}_{N-p} & & \tilde{\*X} \\
		\vdots & & & \ddots \\
		\tilde{\*I}_{N-p} & & & & \tilde{\*X} \\
		\end{bmatrix}\in \+R^{(N(T - \ell)) \times (N+(T - \ell-1)p)}, \]
		
		$\tilde{\*I}_{N-p} = \begin{bmatrix} \*I_{N-p} & \mathbf{0}_{N-p,p} \end{bmatrix}^\T  \in \+R^{N \times (N-p)}$, $\*I_{N-p}$ is an identity matix of dimension $(N-p)\times (N-p)$ and $\mathbf{0}_{N-p,p}$ is a matrix of $0$. Note that we restrict $\bm\alpha_{(N-p+1):N} = 0$ such that all other $\alpha_i$ and $\beta_t$ can be uniquely identified.
		
		Let 
		\[\bm{Z}_\ell = \begin{bmatrix}
		 \bm{z}_{(\ell+1):T} & \bm{z}_{\ell:(T-1)} & \cdots & \bm{z}_{1:(T-\ell)}
		\end{bmatrix}. \]
		
		Then the precision of the estimated $\big( \hat{\bm{\tau}}, \hat{\bm{\alpha}}, \hat{\bm{\beta}}_{(\ell+1):T}, \hat{\bm{\theta}}_{(\ell+1):T} \big)$ from \eqref{eqn:gls}
    \begin{align}\label{eqn:vcov-matrix}
        \var\left(\begin{bmatrix}
        \hat{\bm{\tau}} \\ \hat{\bm{\alpha}}_{1:(N-p)} \\ \hat{\bm{\beta}}_{(\ell+1):T} \\ \hat{\bm{\theta}}_{(\ell+1):T}
        \end{bmatrix} \right) = \left(\begin{bmatrix}
        \bm{Z}_\ell^\T \\ \bm{\Gamma}^\T 
        \end{bmatrix} 
        \cdot \bm{\Sigma}^\I_e \cdot
        \begin{bmatrix}
        \bm{Z}_\ell & \bm{\Gamma}
        \end{bmatrix} \right)^\I \, ,
    \end{align}
    where $\bm{\Sigma}_e = \diag(\bm{\Psi}, \bm{\Psi}, \cdots, \bm{\Psi}) \in \+R^{(N(T - \ell)) \times (N(T - \ell))}$ and $\bm{\Psi} = \*U \*\Sigma_v \*U^\T + \sigma_\varepsilon^2 \*I_N$ from Assumption \ref{ass:constant-error}.

    From block matrix inversion, we have 
    \begin{align}\label{eqn:precision-matrix}
        \Prec\left( \hat{\bm{\tau}}\right) ^\I = \var\left( \hat{\bm{\tau}}\right) = \left(\bm{Z}_\ell^\T  \bm{\Sigma}^\I_e  ( \bm{\Sigma}_e - \bm{\Gamma} (\bm{\Gamma}^\T \bm{\Sigma}^\I_e  \bm{\Gamma})^\I \bm{\Gamma}^\T) \bm{\Sigma}^\I_e  \bm{Z}_\ell \right)^\I 
    \end{align}

    \subsection{Proof of Lemma \ref{lemma:simplify-obj}}

    To prove the separable quadratic representation of $\Prec\left( \hat{\bm{\tau}}\right)$, we first state and prove a useful lemma.  
    \begin{lemma}\label{eqn:carryover-separate-obj}
    Suppose the assumptions in Lemma \ref{lemma:simplify-obj} hold and $\sigma_\varepsilon^2 = 1$. For the entries in $\Prec\left( \hat{\bm{\tau}}\right)$ in Equation \eqref{eqn:precision-matrix}, we have 
    \begin{enumerate}
        \item The $(j,m)$-th entry in $\bm{Z}_\ell^\T \bm{\Sigma}_e  \bm{Z}_\ell$ equals
        \begin{equation}\label{eqn:precision-entries-1}
            \sum_{t=1}^{T-\ell} \bm{z}_{j-1+t}^\T \left( \*I_N - \*U(\*I_k + \*U^\T \*U)^\I \*U^\T  \right) \bm{z}_{m-1+t} 
        \end{equation}
        \item The $(j,m)$-th entry in $\bm{Z}_\ell^\T  \bm{\Sigma}^\I_e  \bm{\Gamma} \cdot (\bm{\Gamma}^\T \bm{\Sigma}^\I_e  \bm{\Gamma})^\I \cdot \bm{\Gamma}^\T \bm{\Sigma}^\I_e  \bm{Z}_\ell $ equals
        \begin{equation}\label{eqn:precision-entries-2}
            \begin{aligned}
                & N \sum_{t=j}^{T - \ell+j-1} \tilde \omega_t^\T \tilde \omega_{t+m-j}   - \frac{N}{T - \ell}  \Lp \sum_{t=j}^{T - \ell+j-1} \tilde \omega_t^\T  \Rp  \Lp \sum_{t=m}^{T - \ell+m-1} \tilde \omega_t \Rp  \\ & + (T - \ell) (\zeta^{(j)})^\T \Lp \*I_N - \*U (\*I_k + \*U^\T \*U)^\I \*U^\T \Rp \zeta^{(m)} 
            \end{aligned}
        \end{equation}
        where $\tilde \omega_t = \frac{1}{N} \sum_{i = 1}^N \tilde{\*X}_i z_{it} \in \+R^p$ and $\zeta^{(j)} = \frac{1}{T - \ell} \sum_{t=j}^{T - \ell+j-1} \bm{z}_t \in \+R^{N}$.
    \end{enumerate}

    \end{lemma}
    
    \proof{Proof of Lemma \ref{eqn:carryover-separate-obj}}
    \texttt{}

     \textbf{Step 1: Prove Lemma \ref{eqn:carryover-separate-obj}.1}
     
    Since $\bm{\Sigma}_e = \diag(\bm{\Psi}, \bm{\Psi}, \cdots, \bm{\Psi}) \in \+R^{(N(T - \ell)) \times (N(T - \ell))}$ and $\bm{\Psi} = \*U \*\Sigma_v \*U^\T +\*I_N $ following the assumption that $\Sigma_v = \sigma_\varepsilon^2 \cdot \*I_k$ and $\sigma_\varepsilon^2 = 1$, we have 
         
         \[\bm{\Psi}^\I  = \frac{1}{\sigma_\varepsilon^2 } (\*I_N + \*U \*U^\T)^\I = \left( \*I_N - \*U(\*I_k + \*U^\T \*U)^\I \*U^\T  \right)\in \+R^{N \times N}. \]
         Then 
         \[\bm{Z}_{\ell,j}^\T \bm{\Sigma}_e  \bm{Z}_{\ell,m} = \sum_{t=1}^{T-\ell} \bm{z}_{j-1+t}^\T \bm{\Psi} \bm{z}_{m-1+t}.  \]

    \textbf{Step 2: Prove Lemma \ref{eqn:carryover-separate-obj}.2.}
    
    We show the $(j,m)$-th entry in $\bm{Z}_\ell^\T  \bm{\Sigma}^\I_e  \bm{\Gamma} \cdot (\bm{\Gamma}^\T \bm{\Sigma}^\I_e  \bm{\Gamma})^\I \cdot \bm{\Gamma}^\T \bm{\Sigma}^\I_e  \bm{Z}_\ell $. This consists of the following three steps. 

    \textbf{Step 2.1: Provide the expression of $\bm{Z}_{\ell,j}^\T \bm{\Sigma}_e^\I \bm{\Gamma}$ for all $j$.}
    
    $\bm{Z}_{\ell}^\T \bm{\Sigma}_e^\I \bm{\Gamma}$ has
		\begin{align*}
		    \bm{Z}_{\ell,j}^\T \bm{\Sigma}_e^\I \bm{\Gamma} =&  \bm{Z}_{\ell,j}^\T \begin{bmatrix} \bm{\Psi}  \tilde{\*I}_{N-p} & \bm{\Psi} \tilde{\*X} \\ \bm{\Psi} \tilde{\*I}_{N-p} & & \bm{\Psi} \tilde{\*X} \\ \vdots  & & & \ddots  \\ \bm{\Psi} \tilde{\*I}_{N-p} && & &\Psi \tilde{\*X} \end{bmatrix} \\ =&
		\begin{bmatrix} \sum_{t=1}^{T - \ell} \bm{z}_{j-1+t}^\T \bm{\Psi} \tilde{\*I}_{N-p},  & \bm{z}_j^\T \bm{\Psi} \tilde{\*X}, & \cdots, &  \bm{z}_{T - \ell+j-1}^\T \bm{\Psi} \tilde{\*X} \end{bmatrix} \\ =& \begin{bmatrix}
		(\phi^{(j)})^\T & (\iota^{(j)})^\T
		\end{bmatrix}, 
		\end{align*}
		where
  \[(\phi^{(j)})^\T = \sum_{t=1}^{T - \ell} \bm{z}_{j-1+t}^\T \bm{\Psi} \tilde{\*I}_{N-p}  = (\sum_{t=1}^{T - \ell} \bm{z}_{j-1+t})^\T \bm{\Psi} \tilde{\*I}_{N-p} \]
  and
  \[(\iota^{(j)})^\T = \begin{bmatrix}
		\bm{z}_j^\T \bm{\Psi} \tilde{\*X}, & \cdots, &  \bm{z}_{T - \ell+j-1}^\T \bm{\Psi} \tilde{\*X}
		\end{bmatrix}. \]
		
		Since $\zeta^{(j)} = \frac{1}{T - \ell} \sum_{t=j}^{T - \ell+j-1} \bm{z}_t \in \+R^{N}$, we have
		\[\phi^{(j)} = T (\zeta^{(j)})^\T \bm{\Psi} \tilde{\*I}_{N-p}.\]

		Note that $\*U$ and $\tilde{\*X}$ are orthogonal (from the assumptions in Lemma \ref{lemma:simplify-obj}), we have
		\begin{align}\label{eqn:psi-simplify}
		    \bm{\Psi} \tilde{\*X} = \tilde{\*X} \qquad \text{and} \qquad \tilde{\*X}^\T \bm{\Psi} \tilde{\*X} = N \cdot \*I_p 
		\end{align}
		
		Then $$(\iota^{(j)})^\T = \begin{bmatrix}
		\bm{z}_j^\T \tilde{\*X}, & \cdots, &  \bm{z}_{T - \ell+j-1}^\T  \tilde{\*X}
		\end{bmatrix} = \begin{bmatrix}
		N \tilde \omega_j^\T, & \cdots, & N \tilde \omega_{T - \ell+j-1}^\T
		\end{bmatrix} = N \tilde{\bm{\omega}}_{j:j_\ell}^\T  \in \+R^{(T-\ell)p},$$
		where $\tilde \omega_t = \frac{1}{N} \sum_{i = 1}^N \tilde{\*X}_i z_{it} \in \+R^p$, and $\tilde{\bm{\omega}}_{j:j_\ell}^\T $ is defined as $\begin{bmatrix}
		\tilde \omega_j^\T, & \cdots, &  \tilde \omega_{T - \ell+j-1}^\T
		\end{bmatrix} $.
		
		In summary, 
		\[\bm{Z}_{\ell,j}^\T \bm{\Sigma}_e^\I \bm{\Gamma} = \begin{bmatrix}
		T (\zeta^{(j)})^\T \bm{\Psi} \tilde{\*I}_{N-p}, & N \tilde{\bm{\omega}}_{j:j_\ell}^\T
		\end{bmatrix}. \]
		
    \textbf{Step 2.2: Provide the expression of $(\bm{\Gamma}^\T \bm{\Sigma}_e \bm{\Gamma})^\I$.}
  
	Using block matrix inverse, we decompose  $(\bm{\Gamma}^\T \bm{\Sigma}_e^\I \bm{\Gamma} )^\I$ as 
		\[(\bm{\Gamma}^\T \bm{\Sigma}_e \bm{\Gamma})^\I =  \begin{bmatrix} \bm{\Xi}_{11} & \bm{\Xi}_{12} \\ \bm{\Xi}_{21} & \bm{\Xi}_{22} \end{bmatrix}  \in \+R^{(N+(T - \ell-1)p)) \times (N+(T - \ell-1)p)},  \]
		where
  \begin{align*}
      \bm{\Xi}_{11} =& \*M \\
      \bm{\Xi}_{12} =&  - \*M \tilde{\*M} \\
      \bm{\Xi}_{21} =& \bm{\Xi}_{12}^\T \\
      \bm{\Xi}_{22} =& \bar{\*M} +\tilde{\*M}^\T \*M \tilde{\*M}
  \end{align*}
 with
		\begin{align*}
		    \*M =&   \frac{1}{T - \ell} \Lp \tilde{\*I}_{N-p}^\T \bm{\Psi} \tilde{\*I}_{N-p} -\tilde{\*I}_{N-p}^\T \bm{\Psi}\tilde{\*X} (\tilde{\*X}^\T \bm{\Psi}  \tilde{\*X})^\I \tilde{\*X}^\T \bm{\Psi} \tilde{\*I}_{N-p}  \Rp^\I  \\
      =& \frac{1}{T - \ell} \Lp  \tilde{\*I}_{N-p}^\T \bm{\Psi} \tilde{\*I}_{N-p} - \frac{1}{N} \tilde{\*X} \tilde{\*X}^\T \Rp^\I \in \+R^{(N-p) \times (N-p)}  
		\end{align*}
  and 
  \begin{align*}
      \tilde{\*M} =& \begin{bmatrix}\tilde{\*I}_{N-p}^\T \bm{\Psi} \tilde{\*X} (\tilde{\*X}^\T \bm{\Psi}  \tilde{\*X})^\I, & \cdots,  & \tilde{\*I}_{N-p}^\T \bm{\Psi}\tilde{\*X} (\tilde{\*X}^\T \bm{\Psi}  \tilde{\*X})^\I \end{bmatrix} = \begin{bmatrix}
				\frac{1}{N} \tilde{\*X} & \cdots & \frac{1}{N} \tilde{\*X}
			\end{bmatrix} \in \+R^{(N-p)\times ((T - \ell)p)} 
  \end{align*}
  and 
  \begin{align*}
      \bar{\*M} =& \diag((\tilde{\*X}^\T \bm{\Psi} \tilde{\*X})^\I, (\tilde{\*X}^\T \bm{\Psi} \tilde{\*X})^\I,
			\cdots, (\tilde{\*X}^\T \bm{\Psi} \tilde{\*X})^\I) = \frac{1}{N} \*I_{((T - \ell)p)} \in \+R^{((T - \ell)p) \times ((T - \ell)p)}
  \end{align*}

		and we use \eqref{eqn:psi-simplify} in the simplification.

		We can further simplify $\*M$ using the Woodbury matrix identity
		\begin{align*}
		    \*M = \frac{1}{T - \ell} \Ls \Lp \tilde{\*I}_{N-p}^\T \bm{\Psi} \tilde{\*I}_{N-p} \Rp^\I + \Lp \tilde{\*I}_{N-p}^\T \bm{\Psi} \tilde{\*I}_{N-p} \Rp^\I \tilde{\*X} \Lp N - \tilde{\*X}^\T \Lp \tilde{\*I}_{N-p}^\T \bm{\Psi} \tilde{\*I}_{N-p} \Rp^\I  \tilde{\*X} \Rp^\I \tilde{\*X}^\T \Lp \tilde{\*I}_{N-p}^\T \bm{\Psi} \tilde{\*I}_{N-p} \Rp^\I   \Rs 
		\end{align*}
		where 
		\begin{align*}
		    \Lp \tilde{\*I}_{N-p}^\T \bm{\Psi} \tilde{\*I}_{N-p} \Rp^\I = \Lp \*I_{N-p} - \*U_{(1)} (\*I_k + \*U^\T \*U)^\I \*U_{(1)}^\T \Rp^\I = \*I_{N-p} + \*U_{(1)} (\*I_k + \*U_{(2)}^\T \*U_{(2)})^\I \*U_{(1)}^\T,
		\end{align*}
  with $\*U = \begin{bmatrix}
		\*U_{(1)}^\T & \*U_{(2)}^\T  
		\end{bmatrix}^\T$ and 
  \begin{align*}
      \*U_{(1)} =& \begin{bmatrix}
		\*u_1 & \*u_2 & \cdots & \*u_{N-p}
		\end{bmatrix}^\T \in \+R^{(N-p) \times k} \\
  \*U_{(2)} =& \begin{bmatrix}
		\*u_{N-p+1}  & \cdots & \*u_{N}
		\end{bmatrix}^\T \in \+R^{p \times k}
  \end{align*}

    \textbf{Step 2.3: Provide the expression of $\bm{Z}_{\ell,j}^\T  \bm{\Sigma}^\I_e  \bm{\Gamma} \cdot (\bm{\Gamma}^\T \bm{\Sigma}^\I_e  \bm{\Gamma})^\I \cdot \bm{\Gamma}^\T \bm{\Sigma}^\I_e  \bm{Z}_{\ell,m}$. }
    
  We combine steps (a) and (b) to calculate $\bm{Z}_{\ell,j}^\T \bm{\Sigma}_e^\I \bm{\Gamma} (\bm{\Gamma}^\T \bm{\Sigma}_e^\I \bm{\Gamma} )^\I \bm{\Gamma}^\T \bm{\Sigma}_e^\I  \bm{Z}_{\ell,m}$ for $1\leq j, m \leq \ell+1$.
		
		From step (a), it is equivalent to calculating each term in
		\[\begin{bmatrix}
		(\phi^{(j)})^\T & (\iota^{(j)})^\T
		\end{bmatrix}  \begin{bmatrix} \bm{\Xi}_{11} & \bm{\Xi}_{12} \\ \bm{\Xi}_{21} & \bm{\Xi}_{22} \end{bmatrix} \begin{bmatrix}
		\phi^{(m)} \\ \iota^{(m)}
		\end{bmatrix} = (\phi^{(j)})^\T \bm{\Xi}_{11} \phi^{(m)} + (\phi^{(j)})^\T \bm{\Xi}_{12} \iota^{(m)} +  (\iota^{(j)})^\T \bm{\Xi}_{21} \phi^{(m)} +  (\iota^{(j)})^\T \bm{\Xi}_{22} \iota^{(m)} \]
		
		Each term has 
		\begin{eqnarray*}
			(\phi^{(j)})^\T \bm{\Xi}_{11} \phi^{(m)} &=& (T - \ell)^2 (\zeta^{(j)})^\T \bm{\Psi} \tilde{\*I}_{N-p} \*M \tilde{\*I}_{N-p}^\T \bm{\Psi} \zeta^{(m)} \\
			(\phi^{(j)})^\T \bm{\Xi}_{12} \iota^{(m)} &=& - N (T - \ell) (\zeta^{(j)})^\T\Psi \tilde{\*I}_{N-p} \*M \tilde{\*M} \tilde{\bm{\omega}}_{m:m_\ell} = -(T - \ell)  (\zeta^{(j)})\Psi \tilde{\*I}_{N-p} \*M \tilde{\*X} \Lp \sum_{t=m}^{T - \ell+m-1} \tilde{{\omega} }_{m-1+t} \Rp  \\
			(\iota^{(j)})^\T \bm{\Xi}_{21} \phi^{(m)} &=& -N (T - \ell) \tilde{\bm{\omega}}_{j:j_\ell}^\T \tilde{\*M}^\T \*M \bm{\Psi} \tilde{\*I}_{N-p} \zeta^{(m)} = - (T - \ell) \Lp \sum_{t=j}^{T - \ell+j-1} \tilde \omega_t \Rp  \tilde{\*X}^\T \*M \bm{\Psi} \tilde{\*I}_{N-p}^\T \zeta^{(m)} \\
			(\iota^{(j)})^\T \bm{\Xi}_{22} \iota^{(m)} &=& N \sum_{t=j}^{T - \ell+j-1} \tilde \omega_{t}^\T \tilde \omega_{t+m-j}  + \Lp \sum_{t=j}^{T - \ell+j-1} \tilde \omega_t^\T \Rp  \tilde{\*X}^\T \*M \tilde{\*X}  \Lp \sum_{t=m}^{T - \ell+m-1} \tilde \omega_t \Rp
		\end{eqnarray*}
		
		where we use  $\zeta^{(j)} = \frac{1}{T - \ell} \sum_{t=j}^{T - \ell+j-1} \bm{z}_t$ and $\tilde{\bm{\omega}}_{j:j_\ell} = \begin{bmatrix}
		\tilde \omega_j^\T & \cdots \tilde \omega_{T - \ell+j-1}^\T
		\end{bmatrix} $.

		We partition $\tilde{\*X}$ as $\tilde{\*X} = \begin{bmatrix}
			\tilde{\*X}_{(1)}^\T & \tilde{\*X}_{(2)}^\T  
			\end{bmatrix}^\T$, where 
   \begin{align*}
       \tilde{\*X}_{(1)} \coloneqq&  \begin{bmatrix}
			\tilde{\*X}_1 & \tilde{\*X}_2 & \cdots & \tilde{\*X}_{N-p}
			\end{bmatrix}^\T \in \+R^{(N-p) \times p} \\
   \tilde{\*X}_{(2)} \coloneqq& \begin{bmatrix}
			\tilde{\*X}_{N-p+1}  & \cdots & \tilde{\*X}_{N} \end{bmatrix}^\T \in \+R^{p \times p}
   \end{align*}
    We can simplify $(\phi^{(j)})^\T \bm{\Xi}_{11} \phi^{(m)}, (\phi^{(j)})^\T \bm{\Xi}_{12} \iota^{(m)}, (\iota^{(j)})^\T \bm{\Xi}_{21} \phi^{(m)}$ and $(\iota^{(j)})^\T \bm{\Xi}_{22} \iota^{(m)}$  by calculating the following terms
			\begin{align*}
			&	\tilde{\*I}_{N-p} (\tilde{\*I}_{N-p}^\T \bm{\Psi} \tilde{\*I}_{N-p} )^\I \tilde{\*I}_{N-p}^\T = \begin{bmatrix}
					\*I_{N-p} + \*U_{(1)} (\*I_k + \*U_{(2)}^\T \*U_{(2)})^\I \*U_{(1)}^\T & \bm{0} \\ \bm{0}^\T & \bm{0}
				\end{bmatrix} \in \+R^{N \times N} 
			\end{align*}
			and
			\begin{align*}
			    \bm{\Omega} \coloneqq& \*\Psi \tilde{\*I}_{N-p} ( \tilde{\*I}_{N-p}^\T \bm{\Psi} \tilde{\*I}_{N-p} )^\I \tilde{\*I}_{N-p}^\T  \*\Psi \\
				=&  \begin{bmatrix}
					\*I_{N-p}  & \bm{0} \\ - \*U_{(2)} (\*I_k + \*U_{(2)}^\T \*U_{(2)})^\I \*U_{(1)}^\T & \bm{0}
				\end{bmatrix} \*\Psi \\ =& \begin{bmatrix}
					\*I_{N-p} - \*U_{(1)}(\*I_k +  \*U^\T \*U)^\I \*U_{(1)}^\T & -\*U_{(1)} (\*I_k + \*U^\T \*U)^\I \*U_{(2)}^\T \\
					- \*U_{(2)} (\*I_k +  \*U^\T \*U)^\I \*U_{(1)}^\T & \*U_{(2)} (\*I_k + \*U_{(2)}^\T \*U_{(2)})^\I \*U_{(1)}^\T \*U_{(1)} (\*I_k +  \*U^\T \*U)^\I \*U_{(2)}^\T
				\end{bmatrix} \in \+R^{N \times N}  
			\end{align*}
			and
			\begin{align*}
			    (\tilde{\*I}_{N-p}^\T \bm{\Psi} \tilde{\*I}_{N-p} )^\I \tilde{\*X}_{(1)} =&\tilde{\*X}_{(1)} - \*U_{(1)} (\*I_k + \*U_{(2)}^\T \*U_{(2)} )^\I \*U_{(2)}^\T \tilde{\*X}_{(2)}  \in \+R^{(N-p) \times p} \\
				 \tilde{\*I}_{N-p} (\tilde{\*I}_{N-p}^\T \bm{\Psi} \tilde{\*I}_{N-p} )^\I \tilde{\*X}_{(1)}  =& \begin{bmatrix}
					\tilde{\*X}_{(1)} - \*U_{(1)} (\*I_k + \*U_{(2)}^\T \*U_{(2)} )^\I \*U_{(2)}^\T \tilde{\*X}_{(2)}   \\ 0
				\end{bmatrix}  \in \+R^{N \times p} 
			\end{align*}
			and
			\begin{align*}
			&	\bm{\delta} \coloneqq \tilde{\*X}_{(1)}^\T (\tilde{\*I}_{N-p}^\T \bm{\Psi} \tilde{\*I}_{N-p} )^\I \tilde{\*X}_{(1)}  = N \*I_p - (\tilde{\*X}_{(2)}^\T \tilde{\*X}_{(2)}  - \tilde{\*X}_{(2)}^\T \*U_{(2)} (\*I_k + \*U_{(2)}^\T \*U_{(2)} )^\I \*U_{(2)}^\T \tilde{\*X}_{(2)} ) \in \+R^{p \times p} \\
		&		\bm{\gamma} \coloneqq \*\Psi \tilde{\*I}_{N-p} (\tilde{\*I}_{N-p}^\T \bm{\Psi} \tilde{\*I}_{N-p} )^\I \tilde{\*X}_{(1)}  = \begin{bmatrix}
					\tilde{\*X}_{(1)} \\  \*U_{(2)} (\*I_k + \*U_{(2)}^\T \*U_{(2)} )^\I \*U_{(2)}^\T \tilde{\*X}_{(2)} 
				\end{bmatrix}\in \+R^{N \times p}.
			\end{align*}
	
		From the definition of $\zeta_i^{(j)} $ and $\omega_t$, we have $(T - \ell) \sum_{i = 1}^N \zeta_i^{(j)} = N \sum_{t=j}^{T - \ell-1+j} \omega_t$ for $j = 1, 2, \cdots, \ell+1$ and $ \sum_{i = 1}^{N-1} \zeta_i^{(j)} = \frac{N}{T - \ell} \sum_{t=j}^{T - \ell-1+j} \omega_t - \zeta_N^{(j)} $. More generally, we have $(T - \ell) \sum_{i = 1}^N \tilde{\*X}_i \zeta_i^{(j)} = N \sum_{t=j}^{T - \ell-1+j}  \tilde{\omega}_t$. 
		Using these properties, 
		\begin{eqnarray*}
			(\phi^{(j)})^\T \bm{\Xi}_{11} \phi^{(m)} &=& (T - \ell)  (\zeta^{(j)})^\T \Lp \*\Omega +  \bm{\gamma} (N \*I_p - \bm{\delta})^\I \bm{\gamma}^\T  \Rp \zeta^{(m)} \\
			(\phi^{(j)})^\T \bm{\Xi}_{12} \iota^{(m)} &=&  -(T - \ell)  (\zeta^{(j)})^\T \Lp  \bm{\gamma} (N \*I_p - \bm{\delta})^\I \tilde{\*X}^\T \Rp \zeta^{(m)}  \\
			(\iota^{(j)})^\T \bm{\Xi}_{21} \phi^{(m)} &=&  - (T - \ell) (\zeta^{(j)})^\T \Lp   \tilde{\*X}  (N \*I_p - \bm{\delta})^\I \bm{\gamma}^\T \Rp \zeta^{(m)}  \\
			(\iota^{(j)})^\T \bm{\Xi}_{22} \iota^{(m)} &=& N \sum_{t=j}^{T - \ell+j-1} \tilde \omega_{t}^\T \tilde \omega_{t+m-j}  + (T - \ell)  (\zeta^{(j)})^\T \Lp   \tilde{\*X} (N \*I_p - \bm{\delta})^\I  \tilde{\*X}^\T \Rp \zeta^{(m)} \\ && - \frac{N}{T - \ell}  \Lp \sum_{t=j}^{T - \ell+j-1} \tilde \omega_t^\T  \Rp  \Lp \sum_{t=m}^{T - \ell+m-1} \tilde \omega_t \Rp
		\end{eqnarray*}

		We sum these four terms together and obtain
		\begin{eqnarray*}
				  N \sum_{t=j}^{T - \ell+j-1} \tilde \omega_{t}^\T \tilde \omega_{t+m-j}  - \frac{N}{T - \ell}  \Lp \sum_{t=j}^{T - \ell+j-1} \tilde \omega_t^\T  \Rp  \Lp \sum_{t=m}^{T - \ell+m-1} \tilde \omega_t \Rp \\  + (T-\ell)  (\zeta^{(j)})^\T \Lp \*\Omega + \Lp \bm{\gamma} - \tilde{\*X} \Rp  (N \*I_p - \bm{\delta})^\I \Lp \bm{\gamma} - \tilde{\*X} \Rp^\T  \Rp \zeta^{(m)}.
			\end{eqnarray*}
			with
			\begin{eqnarray*}
				&&  \*\Omega + \Lp \bm{\gamma} - \tilde{\*X} \Rp  (N \*I_p - \bm{\delta})^\I \Lp \bm{\gamma} - \tilde{\*X} \Rp \\
				&=&  \*\Omega + \begin{bmatrix}
					0 & \bm{0} \\ \bm{0}^\T & \*I_p - \*U_{(2)} (\*I_k + \*U_{(2)}^\T \*U_{(2)} )^\I \*U_{(2)}^\T 
				\end{bmatrix} \\
				&=& \*I_N - \*U (\*I_k + \*U^\T \*U)^\I \*U^\T,
			\end{eqnarray*}
			following $\*U^\T \*U = \*U_{(1)}^\T \*U_{(1)} + \*U_{(2)}^\T \*U_{(2)}$ and
			\begin{eqnarray*}
				&& \*U_{(2)} (\*I_k + \*U_{(2)}^\T \*U_{(2)})^\I \*U_{(1)}^\T \*U_{(1)} (\*I_k +  \*U^\T \*U)^\I \*U_{(2)}^\T  - \*U_{(2)} (\*I_k + \*U_{(2)}^\T \*U_{(2)} )^\I \*U_{(2)}^\T  \\
				&=& \*U_{(2)} (\*I_k + \*U_{(2)}^\T \*U_{(2)})^\I ( \*U_{(1)}^\T \*U_{(1)} - \*I_k- \*U^\T \*U )(\*I_k +  \*U^\T \*U)^\I \*U_{(2)}^\T \\
				&=& - \*U_{(2)} (\*I_k +  \*U^\T \*U)^\I \*U_{(2)}^\T.
			\end{eqnarray*}
			In summary  $\bm{Z}_{\ell,j}^\T \bm{\Sigma}_e^\I \bm{\Gamma} (\bm{\Gamma}^\T \bm{\Sigma}_e^\I \bm{\Gamma} )^\I \bm{\Gamma}^\T \bm{\Sigma}_e^\I  \bm{Z}_{\ell,m}$ equals
			\begin{align*}
			    N \sum_{t=j}^{T - \ell+j-1} \tilde \omega_{t}^\T \tilde \omega_{t+m-j} -  \frac{N}{T - \ell}   \Lp \sum_{t=j}^{T - \ell+j-1} \tilde \omega_t^\T  \Rp  \Lp \sum_{t=m}^{T - \ell+m-1} \tilde \omega_t \Rp 
			 + (T - \ell)  (\zeta^{(j)})^\T \Lp I_N - \*U (\*I_k + \*U^\T \*U)^\I \*U^\T \Rp \zeta^{(m)}.
			\end{align*}

    \Halmos
    \endproof

    Next we prove Lemma \ref{lemma:simplify-obj}. In this proof, we can simultaneously obtain Equation \eqref{eqn:d-optimal-precision} for the D-optimal design.
    
    	\proof{Proof of Lemma \ref{lemma:simplify-obj} and Equation \eqref{eqn:d-optimal-precision}}
     \texttt{} 
		
		From Lemma \ref{eqn:carryover-separate-obj}, when $\sigma_\varepsilon^2 = 1$, the $(j,m)$-th entry in $\Prec(\hat{\bm{\tau}})$ is $ \bm{Z}_\ell^\T  \bm{\Sigma}^\I_e  ( \bm{\Sigma}_e - \bm{\Gamma} (\bm{\Gamma}^\T \bm{\Sigma}^\I_e  \bm{\Gamma})^\I \bm{\Gamma}^\T) \bm{\Sigma}^\I_e  \bm{Z}_\ell$ and equals
		\begin{eqnarray*}
			&& \bm{Z}_{\ell,j}^\T  \bm{\Sigma}_e^\I \bm{Z}_{\ell,m} - \bm{Z}_{\ell,j}^\T \bm{\Sigma}_e^\I \bm{\Gamma} (\bm{\Gamma}^\T \bm{\Sigma}_e^\I \bm{\Gamma})^\I \bm{\Gamma}^\T \bm{\Sigma}_e^\I \bm{Z}_{\ell,m}  \\
			&=& \sum_{t=j}^{T - \ell+j-1} \bm{z}_{j-1+t}^\T \left( \*I_N - \*U(\*I_k + \*U^\T \*U)^\I \*U^\T  \right) \bm{z}_{m-1+t} \\
			&& - N \sum_{t=j}^{T - \ell+j-1} \tilde \omega_t^\T \tilde \omega_{t+m-j} +  \frac{N}{T - \ell}  \Lp \sum_{t=j}^{T - \ell+j-1} \tilde \omega_t^\T  \Rp  \Lp \sum_{t=m}^{T - \ell+m-1} \tilde \omega_t \Rp  \\ &&  - (T - \ell)  (\zeta^{(j)})^\T \Lp I_N - \*U (\*I_k + \*U^\T \*U)^\I \*U^\T \Rp \zeta^{(m)} \\
			&=& \underbrace{\sum_{t=j}^{T - \ell+j-1} \bm{z}_{j-1+t}^\T \bm{z}_{m-1+t}  - \Bigg[ (T - \ell)  (\zeta^{(j)})^\T    \zeta^{(m)}   -  \frac{N}{T - \ell}  \Lp \sum_{t=j}^{T - \ell+j-1} \omega_t  \Rp  \Lp \sum_{t=m}^{T - \ell+m-1} \omega_t \Rp + N \sum_{t=j}^{T - \ell+j-1} \omega_t \omega_{t+m-j}  \Bigg] }_{\coloneqq a^{(j,m)}} \\
			&& + \underbrace{N \cdot  \sum_{q=1}^{d_x} \Bigg[  -\sum_{t=j}^{T - \ell+j-1} \omega_t^{x_q} \omega^{x_q}_{t+m-j}   + \frac{1}{T - \ell}  \Lp \sum_{t=j}^{T - \ell+j-1} \omega^{x_q}_t  \Rp  \Lp \sum_{t=m}^{T - \ell+m-1} \omega^{x_q}_t \Rp  \Bigg] }_{\coloneqq b^{(j,m)}} \\
			&& + \underbrace{(T - \ell)  (\zeta^{(j)})^\T  \*U (\*I_k + \*U^\T \*U)^\I \*U^\T  \zeta^{(m)}  - \sum_{t=j}^{T - \ell+j-1} \bm{z}_t^\T U (\*I_k +  \*U^\T \*U)^\I U^\T \bm{z}_{t+m-j}  }_{\coloneqq c^{(j,m)}},   
		\end{eqnarray*}
		where $\omega_t = \frac{1}{N} \sum_{i = 1}^N z_{it}$ and $\omega_t^{x_q} = \frac{1}{N} \sum_{i = 1}^N X_{iq} z_{it}$ for $q = 1, \cdots, d_x$.

		When there are no covariates, we only have the term $a^{(j,m)}$. We can write $\sum_{t=j}^{T - \ell+j-1} \bm{z}_{j-1+t}^\T \bm{z}_{m-1+t} $ and $(\zeta^{(j)})^\T    \zeta^{(m)} $ in $a^{(j,m)}$ in terms of $\omega_1, \cdots, \omega_T$.
		
		First, for the term $\sum_{t=j}^{T - \ell+j-1} \bm{z}_{j-1+t}^\T \bm{z}_{m-1+t} $ and $(\zeta^{(j)})^\T    \zeta^{(m)} $,  if $j = m$, then  $\sum_{t=j}^{T - \ell+j-1} \bm{z}_{j-1+t}^\T \bm{z}_{j-1+t} = N(T - \ell)$. if $j \neq m$, suppose $j < m$, then we have 
		\[\sum_{t=j}^{T - \ell+j-1} \bm{z}_{j-1+t}^\T \bm{z}_{m-1+t} = N \Ls (T - \ell) + \sum_{t = j}^{m-1} (\omega_t  -  \omega_{T - \ell+t})\Rs.  \]
		
		Second, let us write $(\zeta^{(j)})^\T \zeta^{(m)}$ in terms of $\omega_1, \cdots, \omega_T$.
		Recall the definition $\zeta^{(j)}_i = \frac{1}{T - \ell} \sum_{t=j}^{T - \ell-1+j} z_{1t} $, there are $T+1$ different values that $\zeta_i^{(j)} \zeta_i^{(m)}$ can take, denoted as $\upsilon_0^{(j,m)}, \upsilon_1^{(j,m)}, \cdots, \upsilon_T^{(j,m)}$, where $\upsilon_t^{(j,m)}$ denotes the value of $\zeta_i^{(j)} \zeta_i^{(m)}$ when unit $i$ starts to get  the treatment at time period $T+1-t$ (and $\upsilon_0^{(j,m)}$ represents the value of $\zeta_i^{(j)} \zeta_i^{(m)}$ when unit $i$ stays in the control group for all time periods). Without loss of generality, we assume $j \leq m$ and have
		\begin{eqnarray*}
			\upsilon_t^{(j,m)} = \begin{cases}
				1 &  t \leq \ell+1-m\\
				-\Lp -1 + \frac{2(t-1-\ell+m)}{T - \ell} \Rp & \ell+1-m < t \leq \ell+1-j \\
				\Lp  -1 + \frac{2(t-1-\ell+m)}{T - \ell} \Rp \Lp  -1 + \frac{2(t-1-\ell+j)}{T - \ell} \Rp & \ell+1-j < t \leq T+1-k \\
				\Lp  -1 + \frac{2(t-1-\ell+j)}{T - \ell} \Rp & T+1-m < t \leq T+1-j \\
				1 & T+1-j < t \\
			\end{cases}
		\end{eqnarray*}
		Given $\omega_t$, there are $\frac{N(1+ \omega_1)}{2}, \frac{N(1+\omega_2)}{2}, \cdots \frac{N(1+\omega_T)}{2}$ treated units in time period $1, 2, \cdots, T$. It is equivalent to having  $\frac{N(1+\omega_1)}{2}, \frac{N(\omega_2 - \omega_1)}{2}, \cdots,\frac{N(\omega_T-\omega_{T-1})}{2}$ untreated units to start the treatment in time period $1, 2, \cdots, T$ and leaving $\frac{N(1-\omega_T)}{2}$ units in the control group in the end. 
		\begin{eqnarray*}
			(\zeta^{(j)})^\T \zeta^{(m)} = \sum_{i = 1}^{N} \zeta_i^{(j)} \zeta_i^{(m)}  &=& N \left[ \frac{1 + \omega_1}{2} \cdot \upsilon^{(j,m)}_T + \frac{ \omega_2 -  \omega_1}{2} \upsilon^{(j,m)}_{T-1} + \cdots  + \frac{1 - \omega_T}{2} \cdot \upsilon^{(j,m)}_0 \right] \\
			&=& N \left[ 1 + \frac{\upsilon^{(j,m)}_T - \upsilon^{(j,m)}_{T-1}}{2}  \omega_1 + \frac{\upsilon^{(j,m)}_{T-1} - \upsilon^{(j,m)}_{T-2}}{2}  \omega_2  + \cdots + \frac{\upsilon^{(j,m)}_1 - \upsilon^{(j,m)}_0}{2} \omega_T  \right],
		\end{eqnarray*}
		following $\upsilon^{(j,m)}_0 = \upsilon^{(j,m)}
		_T = 1$.  

		We plug the expression of $\sum_{t=j}^{T - \ell+j-1} \bm{z}_{j-1+t}^\T \bm{z}_{m-1+t} $ and $(\zeta^{(j)})^\T \zeta^{(m)} $ into $a^{(j,m)}$ and multiply $a^{(j,m)}$ by $1/\sigma_\varepsilon^2$ to account for $\sigma_\varepsilon^2 \neq 1$, then we obtain Equation \eqref{eqn:d-optimal-precision}.
		
		To show Lemma \ref{lemma:simplify-obj}, we plug the expression of $\sum_{t=j}^{T - \ell+j-1} \bm{z}_{j-1+t}^\T \bm{z}_{j-1+t} $ and $(\zeta^{(j)})^\T \zeta^{(j)} $ into $a^{(j,j)}$, and multiply $a^{(j,j)}$, $b^{(j,j)}$ and $c^{(j,j)}$ by $1/\sigma_\varepsilon^2$ to account for $\sigma_\varepsilon^2 \neq 1$
		then we have 
  {\small 
		\begin{align*}
		    & \tr \left( \bm{Z}_\ell^\T  \bm{\Sigma}^\I_e  ( \bm{\Sigma}_e - \bm{\Gamma} (\bm{\Gamma}^\T \bm{\Sigma}^\I_e  \bm{\Gamma})^\I \bm{\Gamma}^\T) \bm{\Sigma}^\I_e  \bm{Z}_\ell\right) \\
		    =&\frac{1}{\sigma_\varepsilon^2}  \sum_{j = 1}^{\ell+1} \left(a^{(j,j)} + b^{(j,j)} + c^{(j,j)} \right) \\
		  =&  -   \frac{N}{\sigma_\varepsilon^2} \cdot  \sum_{j = 1}^{\ell+1} \underbrace{\Ls \sum_{t = j}^{T - \ell-1+j} \omega_t^2  - \frac{1}{T - \ell} \Lp  \sum_{t=j}^{T - \ell-1+j} \omega_t  \Rp^2 +  \sum_{t = j}^{T - \ell-1+j} \frac{2(T - \ell-1+2j-2t)}{T - \ell}  \omega_t   \Rs}_{f_{j,\bm{1}}(Z)}  \\
		    & -  \frac{N}{\sigma_\varepsilon^2} \cdot   \sum_{j = 1}^{\ell+1} \underbrace{\sum_{k=1}^{d_x} \Bigg[  \sum_{t=j}^{T - \ell+j-1} \left(\frac{1}{N} \sum_{i = 1}^N X_{ik} z_{it} \right)^2  - \frac{1}{T - \ell}  \Lp \frac{1}{N} \sum_{t=j}^{T - \ell+j-1} \sum_{i = 1}^N X_{ik} z_{it}  \Rp^2  \Bigg]}_{f_{j,\*X}(Z)}  \\
		    & - \frac{N}{\sigma_\varepsilon^2} \sum_{j = 1}^{\ell+1} \underbrace{\Bigg[ \frac{1}{N} \sum_{t=j}^{T - \ell+j-1} \bm{z}_t^\T \*U (\*I_k +  \*U^\T \*U)^\I \*U^\T \bm{z}_{t} -  \frac{1}{N(T - \ell)} \bigg( \sum_{t=j}^{T - \ell+j-1} \bm{z}_t \bigg)^\T  \*U (\*I_k + \*U^\T \*U)^\I \*U^\T   \bigg( \sum_{t=j}^{T - \ell+j-1} \bm{z}_t \bigg)  \Bigg] }_{f_{j,\*U}(Z)},
		\end{align*} }
		where $f_{j,\*X}(Z) $ can be written as $ \sum_{k=1}^{d_x}  (\bm{\omega}_{j:j_\ell}^{x_k})^\T \*P_{\bm{1}_{T_\ell}}  \bm{\omega}_{j:j_\ell}^{x_k} $, and $f_{j,\*U}(Z) $ can be written as $f_{j,\*U}(Z) = \frac{1}{N} \bm{z}_{j:j_\ell}^\T\*M_{\*U}  \bm{z}_{j:j_\ell} $ with $\*M_{\*U} = \*P_{\bm{1}_{T-\ell}} \otimes \*U (\*I_{d_u} + \*U^\T \*U)^\I \*U^\T$.

		\Halmos
		\endproof

		\subsection{Proof of Theorem \ref{thm:obs-latent-carryover-model}}

		\proof{Proof of Theorem \ref{thm:obs-latent-carryover-model}}
		
		As described in Section \ref{subsec:separate-quadratic}, if we can find a design that can separately minimize $f_{j,\bm{1}}(Z)$, $f_{j,\*X}(Z)$, $f_{j,\*U}(Z)$, then this design can maximize the precision $\Prec(\hat{\bm{\tau}})$.

		Let us first consider the design that minimizes $f_{j,\bm{1}}(Z)$. We can write it out as
		\begin{align}\label{eqn:fj1z}
		    f_{j,\bm{1}}(Z) = \sum_{j = 1}^{\ell+1} \Ls \sum_{t = j}^{T - \ell-1+j} \omega_t^2  - \frac{1}{T - \ell} \Lp  \sum_{t=j}^{T - \ell-1+j} \omega_t \Rp^2 +  \sum_{t = 1}^{T - \ell-1+j} \frac{2(T - \ell-1+2j-2t)}{T - \ell}  \omega_t  \Rs.
		\end{align}
		
		The Lagrangian of $f_{j,\bm{1}}(Z)$ is 
		\begin{align*}
			\mathcal{L}(\bm{\omega}, \bm{\lambda}, \bm{\kappa}, \bm{\iota}) &= \sum_{j = 1}^{\ell+1} \Ls \sum_{t = j}^{T - \ell-1+j} \omega_t^2  - \frac{1}{T - \ell} \Lp  \sum_{t=j}^{T - \ell-1+j} \omega_t \Rp^2 +  \sum_{t = 1}^T \frac{2(T - \ell-1+2j-2t)}{T - \ell}  \omega_t  \Rs \\
			&  + \sum_{t=1}^T \lambda_t (-1 - \omega_t) + \sum_{t=1}^T \kappa_t (\omega_t - 1) + \sum_{t=1}^{T-1} \iota_t ( \omega_t - \omega_{t+1}). 
		\end{align*}

		The KKT conditions of $\mathcal{L}(\bm{\omega}, \bm{\lambda}, \bm{\kappa}, \bm{\iota})$ are
		\begin{align}
		\frac{\partial \mathcal{L}}{\partial \omega_t } =& t \omega_t - \frac{\sum_{j=1}^t s_j}{T - \ell}  + \frac{(T - \ell - t)t}{T - \ell} - \lambda_t + \kappa_t + \iota_t - \iota_{t-1} = 0, \quad t \leq \ell \label{eqn:kkt-small-t} \\
		\frac{\partial \mathcal{L}}{\partial \omega_t } =& (\ell+1) \omega_t - \frac{\sum_{j=1}^{\ell+1} s_j }{T - \ell} + \frac{(\ell+1)(T+1-2t)}{T - \ell}  - \lambda_t + \kappa_t + \iota_t - \iota_{t-1} = 0, \quad \ell <  t \leq T - \ell \label{eqn:kkt-medium-t} \\
		\frac{\partial \mathcal{L}}{\partial \omega_t } =& (T+1-t) \omega_t - \frac{\sum_{j=1}^{T+1-t} s_j}{T - \ell}  + \frac{(T - \ell - 1)(T+1-t)}{T - \ell}  - \lambda_t + \kappa_t + \iota_t - \iota_{t-1} = 0, \, t >  T - \ell  \label{eqn:kkt-large-t} \\
		\nonumber & \lambda_t (-1 - \omega_t) = 0, \quad \kappa_t (\omega_t - 1) = 0, \quad \iota_t ( \omega_t - \omega_{t+1}) = 0 \\
		\nonumber & -1 \leq \omega_t \leq 1, \quad \omega_t \leq \omega_{t+1},\quad  \lambda_t \geq 0, \quad \kappa_t \geq 0, \quad \iota_t \geq 0
		\end{align}
		where $s_j = \sum_{t=j}^{T - \ell-1+j} \omega_t$ for $j = 1, \cdots, \ell+1$ and $\iota_0 = 0$.
		
		The Hessian of $f(\bm{\omega})$ is positive semi-definite. Any solution that satisfies the KKT conditions is optimal. 
		
		First, we can show the optimal solution is symmetric with respect to the origin. The proof is as follows. If $\bm{\omega}^\ddagger$ is the optimal solution that minimizes \eqref{eqn:fj1z}, then we can show $\bm{\omega}^\dagger = \begin{bmatrix}
		-\omega^\ddagger_T & -\omega^\ddagger_{T-1} & \cdots & -\omega^\ddagger_1
		\end{bmatrix}$ has the same  value in the objective function as $\bm{\omega}^\ddagger$ because $\sum_{j = 1}^{\ell+1} \sum_{t = 1}^{T - \ell-1+j} \frac{2(T - \ell-1+2j-2t)}{T - \ell}  \omega_t $ in $f(\bm{\omega})$ is symmetric with respect to the origin and similarly for the other two terms in $f(\bm{\omega})$. Since \eqref{eqn:fj1z} is convex, 
		$f\Lp \frac{\bm{\omega}^\ddagger + \bm{\omega}^\dagger }{2} \Rp \leq \frac{1}{2} \big[ f(\bm{\omega}^\ddagger) + f(\bm{\omega}^\dagger)  \big]  = f(\bm{\omega}^\ddagger).$
		Then if $\bm{\omega}^\ddagger$ is optimal, $\bm{\omega}^\dagger = \bm{\omega}^\ddagger$. 
		
		Now we can focus on the $\omega$ that satisfies $\begin{bmatrix}
		\omega_1 & \omega_2 & \cdots & \omega_T 
		\end{bmatrix} = \begin{bmatrix}
		-\omega_T & -\omega_{T-1} & \cdots & -\omega_1
		\end{bmatrix}$. From the definition of $s_j = \sum_{t=j}^{T - \ell-1+j} \omega_t$, we have $s_j = -s_{\ell+1-j}$. If $\ell$ is even, $s_{\ell/2+1} = 0$.

		Now we are going to verify $\bm{\omega}^\ast = \begin{bmatrix}
		\omega_1^\ast & \omega_2^\ast & \cdots & \omega_T^\ast
		\end{bmatrix}$ defined in Equation \eqref{eqn:omega-carryover} satisfies the KKT conditions with feasible $\bm{\lambda}, \bm{\kappa}, \bm{\iota}$.

        \textbf{Case 1: $\omega_t^\ast$ for $\ell < t \leq T - \ell$.}
        
        $\omega_t^\ast = -1 + \frac{2t - (\ell+1)}{T - \ell}  $ satisfies Equation \eqref{eqn:kkt-medium-t} with $\lambda_t = \kappa_t = \iota_t = 0$ and $\iota_{\ell} = 0$.
	
        \textbf{Case 2: $\omega_t^\ast$ for $t \leq \ell$.}
        
        Given $\omega_t = - \omega_{T+1-t}$. We can simplify $s_j$  to
			\begin{eqnarray*}
				s_j = \begin{cases} 
					\sum_{t = j}^{\ell+1-j} \omega_t & \text{ for } j = 1, \cdots, \lfloor (\ell+1)/2 \rfloor \\
					\sum_{t = T-j}^{T - \ell+j} \omega_t & \text{ for } j = \lfloor (\ell+1)/2 \rfloor+1, \cdots, \ell+1 \, .  
				\end{cases}
			\end{eqnarray*}
			As an example, when $\ell = 2$,  we have $s_1 = \omega_1 + \omega_2$, $s_2 = 0$ and $s_3 = \omega_{T-1}+\omega_T$; when $\ell = 3$,  we have $s_1 = \omega_1 + \omega_2 + \omega_3$, $s_2 = \omega_2$, $s_3 = \omega_{T-1}$ and  $s_4 = \omega_{T-2}+\omega_{T-1}+\omega_T$. Furthermore, $s_j + s_{\ell+2-j} = 0$ for $1 \leq j \leq  \ell+1$. Using this property, for $\lfloor \ell/2 \rfloor < t \leq \ell$, we have $\sum_{j = 1}^t s_j = \sum_{j = 1}^{\ell+1-t} s_j$.
			
			Next we show when $ \omega_t = -1$ for $t \leq \lfloor \ell/2 \rfloor$, there exist some $\omega_t$  for $ \lfloor \ell/2 \rfloor < t \leq \ell$ and some feasible $\lambda_t, \kappa_t, \iota_t$ that satisfy Equation \eqref{eqn:kkt-small-t}. 
			
			When  $\omega_t = -1$ for $t \leq \lfloor \ell/2 \rfloor$, then for $ \lfloor \ell/2 \rfloor < t \leq \ell$, $\sum_{j=1}^t s_j = \sum_{j = 1}^{\ell+1-t} s_j = \big[ \sum_{j = 1}^{\ell+1-t} (\lfloor \ell/2 \rfloor + 1 - j) \big] + \min(\ell+1-t, \ell - \lfloor \ell/2 \rfloor) \cdot \omega_{\lfloor \ell/2 \rfloor+1} + \cdots + \min(\ell+1-t, 2) \cdot   \omega_{\ell-1} + \min(\ell+1-t, 1) \cdot \omega_{\ell}$. As an example, when $\ell = 2$, $s_2=-1+\omega_2$; when $\ell = 3$, $s_1+s_2 = -1+2\omega_2+\omega_3$, $s_1+s_2+s_3 = -1 + \omega_2 + \omega_3$. We can rewrite Equation \eqref{eqn:kkt-small-t} for  $ \lfloor \ell/2 \rfloor < t \leq \ell$ in a vectorized form as the following (we will consider Equation \eqref{eqn:kkt-small-t} for $t \leq \lfloor \ell/2 \rfloor$ in the later part of this proof)
			\begin{equation} 
			\begin{bmatrix}
			\frac{\partial \mathcal{L}}{\partial \omega_{\lfloor \ell/2 \rfloor+1} }\\ \vdots \\  \frac{\partial \mathcal{L}}{\partial  \omega_{\ell} }
			\end{bmatrix} = A^{(\ell)} \begin{bmatrix}
			\omega_{\lfloor \ell/2 \rfloor+1} \\ \vdots \\   \omega_{\ell}
			\end{bmatrix} - b^{(\ell)} - \begin{bmatrix}
			\lambda_{\lfloor \ell/2 \rfloor+1} \\ \vdots \\ \lambda_{\ell}
			\end{bmatrix} + \begin{bmatrix}
			\kappa_{\lfloor \ell/2 \rfloor+1} \\ \vdots \\ \kappa_{\ell}
			\end{bmatrix} + \begin{bmatrix}
			\iota_{\lfloor \ell/2 \rfloor+1} \\ \vdots \\ \iota_{\ell}
			\end{bmatrix} - \begin{bmatrix}
			\iota_{\lfloor \ell/2 \rfloor} \\ \vdots \\ \iota_{\ell-1}
			\end{bmatrix} = 0 \label{eqn:kkt-small-t-simplified},
			\end{equation}
			where $A^{(\ell)}$ and $b^{(\ell)}$ are defined in Equation \eqref{eqn:A-ell} and Equation \eqref{eqn:b-ell}.
			When $\begin{bmatrix} \omega_{\lfloor \ell/2 \rfloor+1}  & \cdots &  \omega_{\ell} \end{bmatrix}^\T  = (A^{(\ell)})^{-1} b^{(\ell)}$, Equation \eqref{eqn:kkt-small-t-simplified} holds with $\lambda_t = \kappa_t = \iota_t = 0$ for $t = \lfloor \ell/2 \rfloor + 1, \cdots, \ell$ and $\iota_{\lfloor \ell/2 \rfloor} = 0$. The remaining step is to verify the constraints $-1 \leq \omega_t \leq -1 + \frac{\ell+1}{T - \ell}$ and $\omega_t \leq \omega_{t+1}$ hold if $\begin{bmatrix} \omega_{\lfloor \ell/2 \rfloor+1}  & \cdots &  \omega_{\ell} \end{bmatrix}^\T  = (A^{(\ell)})^{-1} b^{(\ell)}$.
			
		\textbf{Step 1: Show $- A^{(\ell)} \bm{1} \leq b^{(\ell)} \leq (-1 + \frac{\ell+1}{T - \ell}) A^{(\ell)}  \bm{1}$.}
  
  Note that the diagonal entries in $A^{(\ell)}$ are positive while the off-diagonal entries in $A^{(\ell)}$ are negative, then $A_{t^\prime, :}^{(\ell)} \bm{\omega}_{(\lfloor \ell/2 \rfloor+1): L}$ is increasing in $\omega_{t}$ and decreasing in $\omega_s$ for $t^\prime = 1, \cdots, \ell - \lfloor \ell/2 \rfloor$, $t = t^\prime+ \lfloor \ell/2 \rfloor$ and $s \neq t^\prime+ \lfloor \ell/2 \rfloor $. If $- A^{(\ell)} \bm{1} \leq b^{(\ell)} \leq (-1 + \frac{\ell+1}{T - \ell}) A^{(\ell)}  \bm{1}$  hold, then $\omega_t$ defined in $\begin{bmatrix} \omega_{\lfloor \ell/2 \rfloor+1}  & \cdots &  \omega_{\ell} \end{bmatrix}^\T  = (A^{(\ell)})^{-1} b^{(\ell)}$  is between $-1$ and $-1 + \frac{\ell+1}{T - \ell}$ for $t = \lfloor \ell/2 \rfloor + 1, \cdots, \ell$.
				
				First, let us show $- A^{(\ell)} \bm{1} \leq b^{(\ell)} $, which is equivalent to showing every entry in $A^{(\ell)} \bm{1} + b^{(\ell)} $ is non-negative, that is, for $t^\prime = 1, \cdots, \ell - \lfloor \ell/2 \rfloor$, $( A^{(\ell)} \bm{1})_{t^\prime} +  b^{(\ell)}_{t^\prime} \geq 0$. If $\ell$ is even, $\sum_{l=1}^{\ell-t} (\ell - \lfloor \ell/2 \rfloor + 1 - l) = \frac{t(\ell+1-t)}{2}$ and $\sum_{l=1}^{\ell-t} ( \lfloor \ell/2 \rfloor  + 1 - l) = \frac{t(\ell+1-t)}{2} $. Let $t = t^\prime + \lfloor \ell/2 \rfloor$. We have 
				\[ ( A^{(\ell)} \bm{1})_{t^\prime} +  b^{(\ell)}_{t^\prime}  = t - \frac{1}{T - \ell} \frac{t(\ell+1-t)}{2} - t + \frac{t^2}{T - \ell} - \frac{1}{T - \ell} \frac{t(\ell+1-t)}{2} = \frac{t(2T - \ell-1)}{T - \ell} \geq 0. \]
				If $\ell$ is odd, $\sum_{j=1}^{\ell-t} (\ell - \lfloor \ell/2 \rfloor + 1 - j) = \frac{(t+1)(\ell+1-t)}{2}$ and $\sum_{j=1}^{\ell-t} ( \lfloor \ell/2 \rfloor  + 1 - j) = \frac{(t-1)(\ell+1-t)}{2} $. We have 
				\[ ( A^{(\ell)} \bm{1})_{t^\prime} +  b^{(\ell)}_{t^\prime}  = t - \frac{1}{T - \ell} \frac{(t+1)(\ell+1-t)}{2} - t + \frac{t^2}{T - \ell} - \frac{1}{T - \ell} \frac{(t-1)(\ell+1-t)}{2} = \frac{t(2T - \ell-1)}{T - \ell} \geq 0. \]
				
				Second, let us show $ b^{(\ell)} \leq (-1 + \frac{\ell+1}{T - \ell}) A^{(\ell)}  \bm{1}$, which is equivalent to showing every entry in $ (1 - \frac{\ell+1}{T - \ell}) A^{(\ell)}  \bm{1} + b^{(\ell)} $ is non-positive, that is, for $t^\prime = 1, \cdots, \ell - \lfloor \ell/2 \rfloor$, $\big( A^{(\ell)} (1-\frac{\ell+1}{T - \ell}) \bm{1} \big)_{t^\prime} +  b^{(\ell)}_{t^\prime} \leq 0$. 
				If $\ell$ is even
				\[ \Big( A^{(\ell)} (1-\frac{\ell+1}{T - \ell}) \bm{1} \Big)_{t^\prime} +  b^{(\ell)}_{t^\prime} = \frac{t(2T - \ell-1)}{T - \ell} - \frac{\ell+1}{T - \ell} \Big(t - \frac{t(\ell+1-t)}{2(T - \ell)}\Big) = \frac{t(T - \ell-1)}{T - \ell} \Big(2- \frac{1}{2}\frac{\ell+1}{T - \ell}\Big) < 0 \]
				following $t(T - \ell-1) < 0$ and $2- \frac{1}{2}\frac{\ell+1}{T - \ell} > 0$. If $\ell$ is odd, 
				\[ \Big( A^{(\ell)} (1-\frac{\ell+1}{T - \ell}) \bm{1} \Big)_{t^\prime} +  b^{(\ell)}_{t^\prime} = \frac{t(2T - \ell-1)}{T - \ell} - \frac{\ell+1}{T - \ell} \Big(t - \frac{(t+1)(\ell+1-t)}{2(T - \ell)}\Big) = \frac{t(T - \ell-1)}{T - \ell} \Big(2- \frac{t+1}{2t}\frac{\ell+1}{T - \ell}\Big) < 0 \]
				following $t(T - \ell-1) < 0$ and $2- \frac{t+1}{2t}\frac{\ell+1}{T - \ell} > 0$.

			\textbf{Step 2: Show $- b^{(\ell)}_{t^\prime}/ (A^{(\ell)} \bm{1})_{t^\prime}$ is non-decreasing in $t^\prime$ for $t^\prime = 1, \cdots, \ell - \lfloor \ell/2 \rfloor$. } 
   
   Note that the diagonal entries in $A^{(\ell)}$ are positive while the off-diagonal entries in $A^{(\ell)}$ are negative, then $A_{t^\prime, :}^{(\ell)} \bm{\omega}_{(\lfloor \ell/2 \rfloor+1): L}$ is increasing in $\omega_{t}$ and decreasing in $\omega_s$ for $t^\prime = 1, \cdots, \ell - \lfloor \ell/2 \rfloor$, $t = t^\prime+ \lfloor \ell/2 \rfloor$ and $s \neq t^\prime+ \lfloor \ell/2 \rfloor $. If $- b^{(\ell)}_{t^\prime}/ (A^{(\ell)} \bm{1})_{t^\prime}$ is non-decreasing in $t^\prime$, then  $\omega_t$ is non-decreasing in $t$, where $t = t^\prime+ \lfloor \ell/2 \rfloor$. 
				
				Let $c_{t^\prime}$ be the   $c_{t^\prime}$ that satisfies $(A^{(\ell)} (-1 + \frac{c_{t^\prime}}{T - \ell}) \bm{1})_{t^\prime} =  b^{(\ell)}_{t^\prime}$ and let $t = t^\prime+ \lfloor \ell/2 \rfloor$. If $\ell$ is even, we have
				\begin{eqnarray*}
					&& \frac{t(2T - \ell-1)}{T - \ell}  = \frac{c_{t^\prime}}{T - \ell} \Lp  t - \frac{1}{T - \ell} \frac{t(\ell+1-t)}{2} \Rp \\
					&\Leftrightarrow& 2T - \ell-1 = c_{t^\prime} \frac{t+2T-3\ell-1}{2(T - \ell)}.
				\end{eqnarray*}
				Since $\frac{\partial (2T - \ell-1)}{\partial t}= 2$, $\frac{\partial \frac{t+2T-3\ell-1}{2(T - \ell)}}{\partial t} = \frac{1}{2(T - \ell)}$, and $2 > \frac{1}{2(T - \ell)}$, we have $c_{t^\prime}$ increases in $t$ and $t^\prime$.  This implies $- b^{(\ell)}_{t^\prime}/ (A^{(\ell)} \bm{1})_{t^\prime}$ is non-decreasing in $t^\prime$ for even $\ell$. 
				
				If $\ell$ is odd, we have 
				\begin{eqnarray*}
					&& \frac{t(2T - \ell-1)}{T - \ell}  = \frac{c_{t^\prime}}{T - \ell} \Lp  t - \frac{1}{T - \ell} \frac{(t+1)(\ell+1-t)}{2} \Rp \\
					&\Leftrightarrow& 2T - \ell-1 = c_{t^\prime} \Lp 1 + \frac{(t+1)(T - \ell-1)}{2t(T - \ell)} \Rp.
				\end{eqnarray*}
				Since $\frac{\partial (2T - \ell-1)}{\partial t}= 2$, $\frac{\partial \frac{t+2T-3\ell-1}{2(T - \ell)}}{\partial t} \leq \frac{\ell+3}{\ell+1} \frac{1}{T - \ell}$, and $2 > \frac{\ell+3}{(T - \ell)(\ell+1)}$, we have $c_{t^\prime}$ increases in $t$ and $t^\prime$. This again implies $- b^{(\ell)}_{t^\prime}/ (A^{(\ell)} \bm{1})_{t^\prime}$ is non-decreasing in $t^\prime$ for odd $\ell$. 
			
			We have verified that for  $ \lfloor \ell/2 \rfloor < t \leq \ell$,  $\omega_t$ defined in $\begin{bmatrix} \omega_{\lfloor \ell/2 \rfloor+1}  & \cdots &  \omega_{\ell} \end{bmatrix}^\T  = (A^{(\ell)})^{-1} b^{(\ell)}$ satisfies the KKT conditions. The remaining step is to verify for $t \leq \lfloor \ell/2 \rfloor$,  $\omega_t$ defined as $\omega_t = -1$  satisfies the KKT conditions. When $\omega_t = -1$, constraints $-1 \leq \omega_t \leq 1$, $\omega_t \leq \omega_{t+1}$ for $t \leq \lfloor \ell/2 \rfloor$  and $\omega_{ \lfloor \ell/2 \rfloor}\leq \omega+_{ \lfloor \ell/2 \rfloor+1}$ are satisfied. We only need to verify that we can find  feasible $\lambda_t, \kappa_t, \iota_t$ to satisfy Equation  \eqref{eqn:kkt-small-t}. Since $\omega_t = -1$, from complementary slackness, $\kappa_t = 0$. Plug $\omega_t = -1$ into Equation \eqref{eqn:kkt-small-t}, we have 
			\begin{eqnarray*}
				\lambda_1  - \iota_1 &=& -\frac{1+ s_1}{T - \ell} \\
				\lambda_t  - \iota_t + \iota_{t-1} &=& -\frac{t^2 + \sum_{j=1}^t s_j}{T - \ell} \text{ for } t = 2, \cdots \lfloor \ell/2 \rfloor - 1 \\
				\lambda_t   + \iota_{t-1} &=& -\frac{t^2 + \sum_{j=1}^t s_j}{T - \ell} \text{ for } t =  \lfloor \ell/2 \rfloor 
			\end{eqnarray*}
			We only need to verify $-\frac{t^2 + \sum_{j=1}^t s_j}{T - \ell} \geq 0$ for $t = \lfloor \ell/2 \rfloor $ as for the other conditions, $\lambda_1 - \iota$ and $\lambda_t  - \iota_t + \iota_{t-1}$ can take any value by properly choosing $\lambda_t$ and $\iota$. 
			
			Note that $\frac{1}{T - \ell} \sum_{j=1}^{\lfloor \ell/2 \rfloor} s_j = \frac{1}{T - \ell} \sum_{j = 1}^{L +1 - \lfloor \ell/2 \rfloor} s_j = (\ell+1 - \lfloor \ell/2 \rfloor ) (\omega_{\ell+1 - \lfloor \ell/2 \rfloor} +1) - \frac{(\ell+1- \lfloor \ell/2 \rfloor )^2}{T - \ell}$. Furthermore, if we can show  $\omega_{\ell+1 - \lfloor \ell/2 \rfloor} +1 \leq \frac{\ell+1}{T - \ell} \frac{1}{\ell+1 - \lfloor \ell/2\rfloor}$ for even $\ell$ and $\omega_{\ell+1 - \lfloor \ell/2 \rfloor} +1 \leq \frac{\ell+1}{T - \ell} \frac{2}{\ell+1 - \lfloor \ell/2\rfloor}$ for odd $\ell$, then we have 
			\begin{eqnarray*}
				&& - \frac{\lfloor \ell/2 \rfloor^2}{T - \ell} - (\ell+1 - \lfloor \ell/2 \rfloor ) (\omega_{\ell+1 - \lfloor \ell/2 \rfloor} +1) + \frac{(\ell+1- \lfloor \ell/2 \rfloor )^2}{T - \ell} \\
				&=& \frac{(\ell+1- 2\lfloor \ell/2 \rfloor) (\ell+1)}{T - \ell} - (\ell+1 - \lfloor \ell/2 \rfloor ) (\omega_{\ell+1 - \lfloor \ell/2 \rfloor} +1) \geq 0
			\end{eqnarray*}
			and therefore $-\frac{t^2 + \sum_{j=1}^t s_j}{T - \ell} \geq 0$.
			
			Next is to show ``$\omega_{\ell+1 - \lfloor \ell/2 \rfloor} +1 \leq \frac{\ell+1}{T - \ell} \frac{1}{\ell+1 - \lfloor \ell/2\rfloor}$ for even $\ell$ and $\omega_{\ell+1 - \lfloor \ell/2 \rfloor} +1 \leq \frac{\ell+1}{T - \ell} \frac{2}{\ell+1 - \lfloor \ell/2\rfloor}$ for odd $\ell$.'' Denote $c^u_t \coloneqq -1 + \frac{\ell+1}{T - \ell} \frac{t^\prime}{\lfloor (\ell+1)/2 \rfloor+1} $. If we can show $\omega_t \leq -1 + \frac{\ell+1}{T - \ell} \frac{t^\prime}{\lfloor (\ell+1)/2 \rfloor+1} \coloneqq c^u_t$ for $t = t^\prime + \lfloor \ell/2 \rfloor$, then it implies ``$\omega_{\ell+1 - \lfloor \ell/2 \rfloor} +1 \leq \frac{\ell+1}{T - \ell} \frac{1}{\ell+1 - \lfloor \ell/2\rfloor}$ for even $\ell$ and $\omega_{\ell+1 - \lfloor \ell/2 \rfloor} +1 \leq \frac{\ell+1}{T - \ell} \frac{2}{\ell+1 - \lfloor \ell/2\rfloor}$ for odd $\ell$.'' Note that the diagonal entries in $A^{(\ell)}$ are positive while the off-diagonal entries in $A^{(\ell)}$ are negative, then $A_{t^\prime, :}^{(\ell)} \bm{\omega}_{(\lfloor \ell/2 \rfloor+1): L}$ is increasing in $\omega_{t}$ and decreasing in $\omega_s$ for $t = t^\prime+ \lfloor \ell/2 \rfloor$ and $s \neq t^\prime+ \lfloor \ell/2 \rfloor $.  We only need to show $ (A^{(\ell)} c^u)_{\ell-\lfloor \ell/2 \rfloor} \geq b^{(\ell)}_{\ell-\lfloor \ell/2 \rfloor}$, where $c^u = \begin{bmatrix}
			c^u_{\lfloor \ell/2 \rfloor+1} & \cdots & c^u_{\ell- \lfloor \ell/2 \rfloor}
			\end{bmatrix}^\T $. If $\ell$ is even, and when $T > \frac{\ell^2 + 11\ell + 2}{8}$,
			\[  -(A^{(\ell)} c^u)_{\ell-\lfloor \ell/2 \rfloor} + b^{(\ell)}_{\ell-\lfloor \ell/2 \rfloor} = \frac{\ell(\ell-1)}{T - \ell} - \frac{\ell(\ell+1)}{T - \ell} \Lp \frac{\ell}{\ell+2} - \frac{4}{T - \ell} \Rp = - \frac{\ell}{T - \ell} \Lp \frac{2}{\ell+2} - \frac{\ell+1}{4(T - \ell)} \Rp  < 0. \]
			If $\ell$ is odd, and when $T > \frac{\ell^3+13\ell^2+7\ell+3}{8\ell}$ (note that $\frac{\ell^3+13\ell^2+7\ell+3}{8\ell} > \frac{\ell^2 + 11\ell + 2}{8}$), 
			\[-(A^{(\ell)} c^u)_{\ell-\lfloor \ell/2 \rfloor} + b^{(\ell)}_{\ell-\lfloor \ell/2 \rfloor} =  \frac{\ell(\ell-1)}{T - \ell} - \frac{\ell(\ell+1)}{T - \ell} \frac{\ell+1}{\ell+3} + \frac{(\ell+1)^2}{4(T - \ell)^2} = - \frac{1}{T - \ell} \Lp \frac{2\ell}{\ell+3} - \frac{(L+2)^2}{4(T - \ell)} \Rp < 0.\]

    \textbf{Case 3: $\omega_t^\ast$ for $t > T - \ell$.}	
 
    This is a symmetric case of $\omega_t^\ast$ for $t < \ell$. The proof of $\omega_t^\ast$ for $t > T - \ell$ carries over to this case. 
    
	Combining three cases together,	we have verified that the $\bm{\omega}^\ast$ defined in Equation \eqref{eqn:carryover-t-optimal-obs-latent-thm} satisfies the KKT conditions and the Hessian of \eqref{eqn:fj1z} is positive semi-definite, then  $\bm{\omega}^\ast$ is an optimal solution that minimizes $f_{j,\bm{1}}(Z)$.

    Next is to find a solution that minimizes $f_{j,\*X}(Z)$. Note that $f_{j,\*X}(Z) $ can be written as $ \sum_{k=1}^{d_x}  (\bm{\omega}_{j:j_\ell}^{x_k})^\T \*P_{\bm{1}_{T_\ell}}  \bm{\omega}_{j:j_\ell}^{x_k} $. $\*P_{\bm{1}_{T_\ell}}$ is a positive semi-definite matrix with one eigenvalue to be $0$ and the corresponding eigenvector to be $\bm{1}$. Therefore, $(\bm{\omega}_{j:j_\ell}^{x_k})^\T \*P_{\bm{1}_{T_\ell}}  \bm{\omega}_{j:j_\ell}^{x_k}  \geq 0$  for all $\bm{z}$ and the minimum value is attained when $\*X^\T \*z_t$ is the same for all $t$, or equivalently $\frac{1}{N} \sum_{i = 1}^N X_{i} z_{it} = \mu_X$ for some $\mu_X \in \+R^{d_x}$.

    Finally is to find a solution that minimizes $f_{j,\*U}(Z)$. Note that $f_{j,\*U}(Z) $ can be written as $f_{j,\*U}(Z) = \frac{1}{N} \bm{z}_{j:j_\ell}^\T\*M_{\*U}  \bm{z}_{j:j_\ell} $ with $\*M_{\*U} = \*P_{\bm{1}_{T-\ell}} \otimes \*U (\*I_{d_u} + \*U^\T \*U)^\I \*U^\T$. Similar to $f_{j,\*X}(Z)$,  $\bm{z}^\T M_U \bm{z} \geq 0$ for all $\bm{z}$ and the minimum value is  attained when $ \*U^\T \*z_t$  is the same for all $t$, or equivalently, $\frac{1}{N} \sum_{i = 1}^N u_{i} z_{it} = \mu_U$ for some $\mu_U \in \+R^{d_u}$.
		
	Combining the optimality conditions for $f_{j,\bm{1}}(Z)$ $f_{j,\*X}(Z)$ and $f_{j,\*U}(Z)$. A solution is optimal if it satisfies
		\begin{eqnarray}
		\frac{1}{N} \sum_{i = 1}^N z_{it} = \omega_t^\ast, \quad \quad \frac{1}{N} \sum_{i = 1}^N X_i z_{it} = \mu_X , \quad \frac{1}{N} \sum_{i = 1}^N u_{i} z_{it} = \mu_U, \quad  \text{ for all } t.
		\end{eqnarray}
		
		We, therefore, finish the proof of Theorem \ref{thm:obs-latent-carryover-model}.
		\Halmos
		\endproof

		\subsection{Proof of Proposition \ref{prop:rounding-error-obs-cov}}

				\proof{Proof of Proposition \ref{prop:rounding-error-obs-cov}}
				Let $Z^\ast$ be the optimal solution to \eqref{eqn:obj} with relaxed constraint $z_{it} \in [-1,+1]$, and therefore $Z^\ast$ is feasible and satisfies all the conditions in Theorem \ref{thm:obs-latent-carryover-model}. Note that 
				\[ \tr \big(\mathrm{Prec}(\hat{\bm{\tau}})  \big)_{Z^\ast}  \geq \tr \big(\mathrm{Prec}(\hat{\bm{\tau}})  \big)_{Z^{\mathrm{int}\ast}} \geq \tr \big(\mathrm{Prec}(\hat{\bm{\tau}})  \big)_{Z^\rnd}. \]
				We can provide a bound of $\tr \big(\mathrm{Prec}(\hat{\bm{\tau}})  \big)_{Z^{\mathrm{int}\ast}} - \tr \big(\mathrm{Prec}(\hat{\bm{\tau}})  \big)_{Z^\rnd}$ by bounding 
				\[ \tr \big(\mathrm{Prec}(\hat{\bm{\tau}})  \big)_{Z^\ast}  - \tr \big(\mathrm{Prec}(\hat{\bm{\tau}})  \big)_{Z^\rnd},\]
				which is what we are going to do in the following.
				
				Note that when $d_u = 0$, 
				\[\tr \big(\mathrm{Prec}(\hat{\bm{\tau}})  \big)_{Z}  = - \frac{N}{\sigma_{\varepsilon}^2} \left(f_{\bm{1}}(Z) + f_{\*X}(Z) \right), \]
				where 
				\begin{align*}
				 f_{\bm{1}}(Z)   =&   \sum_{j = 1}^{\ell+1} \underbrace{\Ls \sum_{t = j}^{T - \ell-1+j} \omega_t^2  - \frac{1}{T - \ell} \Lp  \sum_{t=j}^{T - \ell-1+j} \omega_t  \Rp^2 +  \sum_{t = j}^{T - \ell-1+j} \frac{2(T - \ell-1+2j-2t)}{T - \ell}  \omega_t   \Rs}_{f_{j,\bm{1}}(Z)}  \\
		     f_{\*X}(Z) =&    \sum_{j = 1}^{\ell+1} \underbrace{\sum_{k=1}^{d_x} \Bigg[  \sum_{t=j}^{T - \ell+j-1} \left(\frac{1}{N} \sum_{i = 1}^N X_{ik} z_{it} \right)^2  - \frac{1}{T - \ell}  \Lp \frac{1}{N} \sum_{t=j}^{T - \ell+j-1} \sum_{i = 1}^N X_{ik} z_{it}  \Rp^2  \Bigg]}_{f_{j,\*X}(Z)}.
				\end{align*}
				We can bound $\tr \big(\mathrm{Prec}(\hat{\bm{\tau}})  \big)_{Z^\ast}  - \tr \big(\mathrm{Prec}(\hat{\bm{\tau}})  \big)_{Z^\rnd}$ by bounding $f_{\bm{1}}(Z^\rnd) - f_{\bm{1}}(Z^\ast)$
				and $f_{\*X}(Z^\rnd) - f_{\*X}(Z^\ast)$.

				Let us bound the gap between $f_{\bm{1}}(Z^\rnd)$ and $f_{\bm{1}}(Z^\ast)$.

				Note that $\omega_t$ is bounded between $-1$ and $+1$ and the eigenvalues of the Hessian of $f_{j,\bm{1}}(Z) $ are either $1$ or $0$. Therefore, for all $j$,
				\[f_{j,\bm{1}}(Z) = O(T - \ell ). \]
				Next let us bound the difference between $\omega_t^\rnd$ and $\omega^\ast_t$. We introduce the notation $\omega_{g,t}$:
				
				\[\omega_{g,t} = \frac{1}{|\tlo_g|} \sum_{i \in \tlo_g} \underbrace{(2 \cdot \boldsymbol{1}_{A_i \leq t} - 1)}_{z_{it}} \]
				be the treated fraction of units in stratum $g$ scaled between $-1$ and $+1$. Let $\omega_{g,t}^\rnd$ be the value of $\omega_{g,t}$ evaluated at the rounded feasible solution $\{A^\rnd_i\}_{i=1}^N$.

				Since we use the nearest rounding rule to get a feasible $Z^\rnd$, we have $|\omega^\rnd_{g,t} -  \omega^\ast_{g,t}| \leq \frac{1}{|\tlo_g|}$ for all $t$ and $g$, and therefore,
		\[|w^\rnd_{\ell,t} - w^\ast_{\ell,t}| = |\sum_{g = 1}^G p_g (\omega^\rnd_{g,t} - \omega^\ast_{g,t}) | \leq \sum_{g = 1}^G p_g |\omega^\rnd_{g,t} -\omega^\ast_{g,t}|  \leq \sum_{g = 1}^G \frac{p_g }{N_{\min}} = O\left(\frac{1}{N_{\min}} \right). \]
		Let $\delta_t = w^\rnd_{t} - w^\ast_t$.
		The difference between $f_{j,\bm{1}}(Z^\rnd)+f_{\ell+1-j,\bm{1}}(Z^\rnd)$ and $f_{j,\bm{1}}(Z^\ast)+f_{\ell+1-j,\bm{1}}(Z^\ast)$ equals
		\begin{align*}
		    & \big(f_{j,\bm{1}}(Z^\rnd)+f_{\ell+1-j,\bm{1}}(Z^\rnd) \big) - \big(f_{j,\bm{1}}(Z^\ast)+f_{\ell+1-j,\bm{1}}(Z^\ast) \big) \\
		    =& \underbrace{2 \left( \sum_{t = j}^{T - \ell-1+j} \omega^\ast_{\ell,t} \delta_t + \sum_{t = \ell+1-j}^{T - j} \omega^\ast_{\ell,t} \delta_t \right)}_{a_1} \\
      & - \underbrace{\frac{2}{T-\ell} \left[ \left( \sum_{t = j}^{T - \ell-1+j} \omega^\ast_{\ell,t} \right) \left( \sum_{t = j}^{T - \ell-1+j} \delta_t \right) +  \left( \sum_{t = \ell+1-j}^{T - j} \omega^\ast_{\ell,t} \right) \left( \sum_{t = \ell+1-j}^{T - j} \delta_t \right) \right]}_{a_2}  \\
		    & + \underbrace{ \left( \sum_{t = j}^{T - \ell-1+j} \delta^2_t + \sum_{t = \ell+1-j}^{T - j}  \delta^2_t \right)}_{a_3}  -  \underbrace{\frac{1}{T-\ell} \left[  \left( \sum_{t = j}^{T - \ell-1+j} \delta_t \right)^2 +   \left( \sum_{t = \ell+1-j}^{T - j} \delta_t \right)^2 \right]}_{a_4}   \\
		    & + \underbrace{\left(\sum_{t = j}^{T - \ell-1+j}  b_{\ell,t} \delta_t + \sum_{t = \ell+1-j}^{T - j}b_{\ell,t} \delta_t  \right)}_{a_5} 
		\end{align*}
		Since $\omega^\ast_{\ell,t} = - \omega^\ast_{\ell,T+1-t}$ for all $t$, we have $\omega^\ast_{\ell,t} \delta_t  + \omega^\ast_{\ell,T+1-t} \delta_{T+1-t} = 0$ and $b_{\ell,t} \delta_t  + b_{\ell,T+1-t} \delta_{T+1-t} = 0$ following the property of our rounding algorithm. Therefore, assuming $N$ is even, we have
		\[ a_1 = 0 \qquad a_2 = 0 \qquad  a_5 = 0 \]
		and 
        \begin{align*}
		    & \big(f_{j,\bm{1}}(Z^\rnd)+f_{\ell+1-j,\bm{1}}(Z^\rnd) \big) - \big(f_{j,\bm{1}}(Z^\ast)+f_{\ell+1-j,\bm{1}}(Z^\ast) \big) = a_3 + a_4 = O\left(\frac{2 (T-\ell)}{N^2_{\min}} \right).
		\end{align*}
		We sum $j$ together and obtain
		\[ f_{\bm{1}}(Z^\rnd) - f_{\bm{1}}(Z^\ast) = O\left(\frac{(\ell+1) (T-\ell)}{N^2_{\min}} \right). \]
		Next let us bound the gap between $\sum_{j = 1}^{\ell+1} f_{j,\*X}(Z^\rnd)$ and $\sum_{j = 1}^{\ell+1} f_{j,\*X}(Z^\ast)$. Then for each covariate $X_{ik}$, 
				\[|w^{x_k,\rnd}_{\ell,t} - w^{x_k,\ast}_{\ell,t}| = |\sum_{g = 1}^G p_g x_{gk} (\omega^\rnd_{g,t} - \omega^\ast_{g,t}) | \leq \sum_{g = 1}^G p_g x_{j,\max} |\omega^\rnd_{g,t} -\omega^\ast_{g,t}|  \leq \sum_{g = 1}^G \frac{p_g x_{k,\max}}{N_{\min}} = O\left(\frac{x_{k,\max}}{N_{\min}} \right). \]
				We use a similar procedure as above and obtain that for covariate $X_{ik}$, 
				\begin{align*}
		    & \big(f_{j,\*X_{k}}(Z^\rnd)+f_{\ell+1-j,\*X_{k}}(Z^\rnd) \big) - \big(f_{j,\*X_{k}}(Z^\ast)+f_{\ell+1-j,\*X_{k}}(Z^\ast) \big)  = O\left(\frac{2 x_{k,\max}^2 (T-\ell)}{N^2_{\min}} \right).
		\end{align*}
		We sum over $k$ (all covariates) and $j$ (all treatment effects) together and obtain
		\[  f_{\*X}(Z^\rnd) - f_{\*X}(Z^\ast) = O\left(\frac{(\ell+1) (T-\ell) \sum_{k=1}^{d_x} x_{k,\max}^2}{N^2_{\min}} \right). \]
		Since both $f_{\*X}(Z^\ast)$ and $ f_{\*X}(Z^\ast)$ are at the order of $O((\ell+1) (T-\ell))$, we have 
		\begin{align*}
		   f_{\bm{1}}(Z^\rnd)+  f_{\*X}(Z^\rnd) = \left(  f_{\bm{1}}(Z^\rnd)+  f_{\*X}(Z^\rnd)\right) \cdot\left(1+ O\left(\frac{1+ \sum_{k=1}^{d_x} x_{k,\max}^2}{N^2_{\min}} \right) \right)
		\end{align*}
		and from the definition of $\tr \big(\mathrm{Prec}(\hat{\bm{\tau}})  \big)_{Z} $ we have 
		\[ \tr \big(\mathrm{Prec}(\hat{\bm{\tau}})  \big)_{Z^\rnd} = \tr \big(\mathrm{Prec}(\hat{\bm{\tau}})  \big)_{Z^\ast}  \cdot\left(1+ O\left(\frac{1+ \sum_{k=1}^{d_x} x_{k,\max}^2}{N^2_{\min}} \right) \right).\]
		Together with $\tr \big(\mathrm{Prec}(\hat{\bm{\tau}})  \big)_{Z^\ast}  \geq \tr \big(\mathrm{Prec}(\hat{\bm{\tau}})  \big)_{Z^{\mathrm{int}\ast}}$, we have
		\[ \tr \big(\mathrm{Prec}(\hat{\bm{\tau}})  \big)_{Z^\rnd} = \tr \big(\mathrm{Prec}(\hat{\bm{\tau}})  \big)_{Z^{\mathrm{int}\ast}}  \cdot\left(1+ \mathcal{O}\left(\frac{1+ \sum_{k=1}^{d_x} x_{k,\max}^2}{N^2_{\min}} \right) \right).\]
  This concludes the proof of Proposition \ref{prop:rounding-error-obs-cov}.
\Halmos
		\endproof

%% file: appendix_C.tex
\section{Supplementary Material for Adaptive Experiments}\label{sec:more-detail-sequential}

{\blue 
\subsection{Dynamic Program to Solve $\omega_{t+1}$}\label{subsec:dp-omega}
In this section, we provide supplementary details on the dynamic program used to make adaptive treatment decisions for ATU.
In this dynamic program, let the dynamic system be
\[x_{s+1} = \tilde{f}(x_s, \omega_{s+1}, \eta_{s+1}), \qquad \forall  s \leq T_{\max}-1, \]
where $x_s$ is the state of the system and summarizes the information up to time $s$ with the definition provided below, $\omega_{s+1}$ is our decision variable at time $s$, $\eta_{s+1} \in \{0,1\}$ is a random variable with $\eta_{s+1} = 0$ indicating that the experiment has been terminated at time $s+1$, and $\eta_{s+1} = 1$ denoting otherwise. The sequence of random variables $\{\eta_s\}_{s \in [T_{\max}]}$ are sampled as follows. We first sample a termination time $\tilde T$ from $P_t(\tilde T)$, and $\xi_s$ equals to
\[\eta_s = \bm{1}_{s \leq \tilde T}, \qquad \forall  s \in [T_{\max}]. \]

Given $\omega_{s}$ and $\eta_{s}$, we define $x_s$ as a tuple with two elements
\[x_s = \bigg( \underbrace{\vphantom{\sum_{q = 1}^s \xi_q}\bm{\omega}_{1:s}}_{x_{s1}},\,\,\,\, \underbrace{\sum_{q = 1}^s \eta_q}_{x_{s2}} \bigg). \]
The first element $x_{s1}$ is $\bm{\omega}_{1:s} = (\omega_1, \cdots, \omega_s)$. The second element $x_{s2}$ is the number of periods that the experiment has run up to time $s$. If $s > \tilde T$, then  $x_{s2} = \tilde T$. Given the definition of $x_s$, the function $\tilde{f}(\cdot)$ works as follows.  $\tilde{f}(\cdot)$ appends $\omega_s$ to $x_{s1}$ to obtain $x_{s+1,1}$, and adds $\xi_{s+1}$ to $x_{s2}$ to obtain $x_{s+1,2}$. 

Finally we define the cost function for the dynamic program. Let $h_{T_{\max}}(x_{T_{\max}})$ be the terminal cost incurred at the end of the experiment, which is defined as
\[h_{T_{\max}}(x_{T_{\max}})= - \tilde T\cdot \funfrac(\bm{\omega}, \tilde T), \]
where $\tilde T = x_{T_{\max}2} = \sum_{q=1}^{T_{max}} \eta_q$. The definition of cost is aligned with our objective to maximize the estimation precision post experiment. We do not have an intermediate cost in this dynamic program.

We can formulate the dynamic program as an optimization of the expected cost
\[\+E_{\{\eta_s\}_{s \in [T_{\max}]}} \left[ h_{T_{\max}}(x_{T_{\max}}) \right], \]
where the expectation is with respect to the joint distribution of $\{\xi_s\}_{s \in [T_{\max}]}$. This expected value is equivalent to that in Section \ref{subsec:adaptive-algorithm}. The optimization is with respect to the decision variables $\omega_{t+1}, \cdots \omega_{T_{\max}}$ subject to the constraints
\[ \omega_{\ad,t} \leq  \omega_{t+1} \leq  \omega_{t+2} \leq \cdots \leq \omega_{T_{\max}}  \leq 1. \]
The optimal $\omega_{t+1}$ is then used to make treatment decisions for ATU, such that the average of $z_{i,t+1}$ over $i$ equals to $\omega_{t+1}^\ast$. 
}

\subsection{Additional Results}\label{subsec:sequential-additional-results}

\begin{proposition}\label{prop:precision-ordering}
Suppose the Assumptions in Lemma \ref{lemma:asymptotic-tau-sigma} hold. Suppose $\tilde T < T_{\max}$. For a sufficiently large $N$, conditional on $\tilde T = T$, 
\[\mathrm{Prec}(\hat{\tau}_{\all,T}) > \frac{N T}{\sigma^2_{\varepsilon}} \funfrac(\bm{\omega}_{\mathrm{bm},1:T}, T). \]
\end{proposition}

\begin{proposition}\label{prop:precision-guarantee}
    Suppose the Assumptions in Lemma \ref{lemma:asymptotic-tau-sigma} hold, and $\varepsilon_{is}$ is i.i.d. bounded $\sigma$-sub-gaussian random variables with mean zero and variance $\sigma^2_\varepsilon$. If the estimated precision exceeds the threshold $c_0$, that is, $NT/\estsigmasq_{\ad,1,T} \cdot \funfrac(\bm{\omega}, T) \geq c_0 + \delta$, then the corresponding true precision with the same $\bm{\omega}$ exceeds the threshold $c_0$, $NT/\sigma_\varepsilon^2 \cdot \funfrac(\bm{\omega}, T) \geq c_0 $ with probability at least 
    \begin{align}\label{eqn:precision-probability}
        1 - \left(4 + 2T \right) \exp\left(-N  p_{\ad,1} \cdot T c_1^2/(2\sigma^2) \right) = 1 - O\left(  \exp\left(-N^2 T^2 \right) \right)
    \end{align}
    for 
    \begin{align*}
        c_1 = \frac{\sqrt{\left(1+\sigma_\varepsilon \sqrt{2/(T-1)}  \right)^2 + 4 \Delta (2T-1)/(T-1) } -  \left(1+\sigma \sqrt{2/(T-1)}  \right)}{2  (2T-1)/(T-1) }
    \end{align*}
    with 
    \begin{align*}
        \Delta = NT\left(\frac{1}{c_0 + \delta} - \frac{1}{c_0}  \right)\cdot \funfrac(\bm{\omega}_{\mathrm{bm}}, T)
    \end{align*}
\end{proposition}

\proof{Proof of Proposition \ref{prop:precision-ordering}}
Recall that the definition of $\mathrm{Prec}(\hat{\tau}_{\all,T})$ is 
\[\mathrm{Prec}(\hat{\tau}_{\all,T}) = \frac{N T}{\sigma^2_{\varepsilon}} \funfrac(\bm{\omega}_{\all,1:T}, T).  \]
Then proving this proposition is equivalent to proving 
\[\funfrac(\bm{\omega}_{\all,T}, T) > \funfrac(\bm{\omega}_{\mathrm{bm},1:T}, T). \]
Since $\bm{\omega}_{\all,1:T}$ is the average of $Z_{it}$ over $i$ in both ATU and NTU for every $t$, we have  
\[\bm{\omega}_{\all,1:T} = p_{\fcs} \bm{\omega}_{\mathrm{bm},1:T}  + (1- p_{\fcs}) \cdot \bm{\omega}_{\ad,1:T},  \]
where $\omega_{\ad,t+1}$ is solved from a dynamic program based on the empirical distribution of experiment duration, $P_t(\cdot)$. When $N$ is sufficiently large (as assumed in the proposition), the confidence interval for $\estsigmasq_{\fcs,t}$ used to obtain  $P_t(\cdot)$ is sufficiently narrow. Then $P_t(\cdot)$ has probability one at $T$ and zero elsewhere for all $t$. Using this property, and the cost function of the dynamic program, $\omega_{\ad,t+1}$ is the solution that minimizes
\[
\+E_{\tilde{T} \sim P_t(\tilde{T})} \left[ - 
\tilde{T} \cdot \funfrac\left((\bm{\omega}_{\ad,1:t}, \bm{\omega}_{(t+1):\tilde{T}}), \tilde{T}\right) \right] = - T \cdot \funfrac\left((\bm{\omega}_{\ad,1:t}, \bm{\omega}_{(t+1):T}), T\right),  \]
for all $t \geq t_0$ ($t_0$ is the first period to make adaptive treatment decisions for subsequent periods). For the first $t_0$ periods, $\bm{\omega}_{\ad,1:t_0} = \bm{\omega}_{\mathrm{bm}, 1;t_0}$. Therefore, $\bm{\omega}_{\mathrm{bm}}$ is also a feasible solution for the dynamic program. However, $\bm{\omega}_{\ad}$ is solution with lower cost (when $T < T_{\max}$), and then
\[\funfrac(\bm{\omega}_{\ad}, T) > \funfrac(\bm{\omega}_{\mathrm{bm}}, T) \]
Recall that for a generic $\bm{\omega}$, $\funfrac(\bm{\omega}, T) = - 2 \*b^\T \bm{\omega} - \bm{\omega}^\T P_{\bm{1}_T} \bm{\omega}$, which is a concave function of $\bm{\omega}$. Therefore, 
\begin{align*}
    \funfrac(\bm{\omega}_{\all}, T) >& p_\fcs \funfrac(\bm{\omega}_{\mathrm{bm}}, T) + (1-p_\fcs) \funfrac(\bm{\omega}_{\ad}, T) \\ >& \funfrac(\bm{\omega}_{\mathrm{bm}}, T).
\end{align*}
This completes the proof. \halmos

\proof{Proof of Proposition \ref{prop:precision-guarantee}}
If 
$NT/\estsigmasq_{\ad,1,T} \cdot \funfrac(\bm{\omega}, T) \geq c_0 + \delta$, then this implies 
\[\estsigmasq_{\ad,1,T} \leq NT/(c_0 + \delta) \cdot \funfrac(\bm{\omega}, T).   \]
Similarly, if $NT/\sigma_\varepsilon^2 \cdot \funfrac(\bm{\omega}, T) \geq c_0 $, then this implies 
\[ \sigma_\varepsilon^2 \leq NT/c_0 \cdot \funfrac(\bm{\omega}, T) \]
Then showing Proposition \ref{prop:precision-guarantee} is equivalent to showing that 
\[\sigma_\varepsilon^2 - \estsigmasq_{\ad,1,T} \leq NT\left(\frac{1}{c_0 + \delta} - \frac{1}{c_0}  \right)\cdot \funfrac(\bm{\omega}, T) \]
with at least the probability in \ref{eqn:precision-probability}.

From the decomposition of the estimation error of $\estsigmasq_{\ad,1,T}$ (following the proof of Lemma \ref{lemma:asymptotic-tau-sigma-general}), we have 
\begin{align*}
    \sigma_\varepsilon^2 - \estsigmasq_{\ad,1,T} =& \underbrace{\frac{1}{N  p_{\ad,1} \cdot T} \sum_{i \in \mathcal{S}_{\ad,1}, 1 \leq s \leq T}\left[ \sigma_\varepsilon^2 - \varepsilon_{is}^2 \right] }_{x_1} + \underbrace{\frac{1}{N  p_{\ad,1} \cdot T (T-1) } \sum_{i \in \mathcal{S}_{\ad,1}, 1 \leq s, u \leq t: s\neq u}( - \varepsilon_{is} \varepsilon_{iu})}_{x_2}  \\
    & + \underbrace{\frac{1}{T-1} \sum_s \bar{\varepsilon}^2_{\cdot,s} }_{x_3}  - \frac{T}{T-1} \bar{\varepsilon}^2 + \underbrace{(\hat{\tau}_{\ad,1,T} - \tau)^2}_{x_4}  \cdot \underbrace{\frac{1}{N(t-1) p_{\ad,1}} \sum_{i,s} \dot{z}_{is}^2 }_{ \leq 1}
\end{align*}
If we can show for any $c_1, c_2, c_3$ and $c_4$, there exist $p_1, p_2, p_3$ and $p_4$ such that  
\begin{enumerate}
    \item $x_1 \leq c_1$ with probability at least $1 - p_1$
    \item $x_2 \leq c_2$ with probability at least $1 - p_2$
    \item $x_3 \leq c_3$ with probability at least $1 - p_3$
    \item $x_4 \leq c_4$ with probability at least $1 - p_4$,
\end{enumerate}
then by union bound, we have 
\begin{align*}
    \sigma_\varepsilon^2 - \estsigmasq_{\ad,1,T} \leq x_1 + x_2 + x_3 + x_4 \leq c_1 + c_2 + c_3 + c_4
\end{align*}
with probability at least $1 - p_1 - p_2 - p_3 - p_4$.

For $x_1$, it is an average of $N p_{\ad,1} \cdot T$ i.i.d $\sigma$-sub-gaussian random variables with mean $0$. By Hoeffding's inequality, we have 
\[\*P(x_1 \geq c_1) \leq \exp\left(-N  p_{\ad,1} \cdot T c_1^2/(2\sigma^2) \right) = p_1.\]

For $x_2$, note that it can be written as
\[\frac{1}{N  p_{\ad,1} \cdot T (T-1) } \sum_{i \in \mathcal{S}_{\ad,1}, 1 \leq s, u \leq t: s\neq u}( - \varepsilon_{is} \varepsilon_{iu}) = \frac{1}{N p_{\ad,1}} \sum_{i \in \mathcal{S}_{\ad,1}} \underbrace{\frac{1}{T(T-1)} \sum_{1 \leq s, u \leq t: s\neq u} ( - \varepsilon_{is} \varepsilon_{iu})}_{\text{mean 0 variance $\frac{2}{T(T-1) } \sigma_\varepsilon^4$}} \]
following the proof of Lemma \ref{lemma:asymptotic-tau-sigma-general}, implying that it is an average of $N p_{\ad,1}$ i.i.d $2/\sqrt{T(T-1)} \cdot \sigma^2$-sub-gaussian random normal variables with mean $0$. By Hoeffding's inequality, we have 
\[\*P(x_2 \geq c_2) \leq \exp\left(-N  p_{\ad,1} \cdot T (T-1) c_2^2/(4\sigma^4) \right)= p_2.  \]

For $x_3$, if $|\bar{\varepsilon}_{\cdot,s}| \leq 
\left(\frac{T-1}{T} c_3\right)^{1/2}$ for all $s$, then $x_3 \leq c_3$. As $\bar{\varepsilon}_{\cdot,s} = 1/(N p_{\ad,1}) \sum_{i \in \mathcal{S}_{\ad,1}} \varepsilon_{is}$. By Hoeffding's inequality, we have 
\[\*P\left(|\bar{\varepsilon}_{\cdot,s}| \geq \left(\frac{T-1}{T} c_3\right)^{1/2} \right) \leq 2 \exp\left(-N  p_{\ad,1} \cdot (T-1) c_3/(2\sigma^2) \right)\]
and by union bound
\[\*P\left(x_3 \geq c_3 \right) \leq 2 T  \exp\left(-N  p_{\ad,1} \cdot (T-1) c_3/(2\sigma^2) \right)= p_3.\]
For $x_4$, recall that $\hat{\tau}$ takes the form of (see the proof of Lemma \ref{lemma:asymptotic-tau-sigma-general})
\[\hat{\tau}_{\ad,1,T} - \tau =\frac{1}{N p_{\ad,1} T} \sum_{i,t}  \left(\frac{1}{N p_{\ad,1} T} \sum_{i,t} \dot{z}_{it}^2 \right)^\I \dot{z}_{it} \varepsilon_{it} \]
Conditional on $Z_{\ad,1}$, $\hat{\tau}_{\ad,1,T} - \tau$ is a sum of $N p_{\ad,1} T$ independent $ \left(1/(N p_{\ad,1} T) \sum_{i,t} \dot{z}_{it}^2 \right)^\I |\dot{z}_{it}| \sigma$ random variables with mean $0$ over $i$ and $t$. Then by Hoeffding's inequality, we have 
\begin{align*}
    \*P\left((\hat{\tau}_{\ad,1,T} - \tau)^2 \geq c_4 \right) =& \*P\left(|\hat{\tau} - \tau| \geq  c_4^{1/2} \right) \\
    \leq& 2 \exp\left(-N^2  p^2_{\ad,1} T^2 \cdot  c_4/\left(2 \left(1/(N p_{\ad,1} T) \sum_{i,t} \sum_{i,t} \dot{z}_{it}^2 \right)^\I \dot{z}_{it}^2 \sigma^2\right) \right) \\
    =& 2 \exp\left(-N  p_{\ad,1} T \cdot  c_4/\left(2 \sigma^2\right) \right)= p_4
\end{align*}
Now we consider the choice of $c_1$, $c_2$, $c_3$ and $c_4$, such that 
\[c_1 + c_2 + c_3 + c_4 \leq NT\left(\frac{1}{c_0 + \delta} - \frac{1}{c_0}  \right)\cdot \funfrac(\bm{\omega}, T) = \Delta \]
and $1 - p_1 - p_2 - p_3 - p_4$ is as large as possible. Based on the expression of $c_1$, $c_2$, $c_3$ and $c_4$, we match the exponential terms in $p_1$, $p_2$, $p_3$ and $p_4$ (which is the dominating term), set
$p_1 = p_2 = p_3/(2T) = p_4/2$, or equivalently, 
\[c_1^2 = (T-1)/(2\sigma_\varepsilon^2) c_2^2 = (T-1)/T c_3 = c_4, \]
and solve the $c_1$, $c_2$, $c_3$ and $c_4$ such that
\[c_1 + c_2 + c_3 + c_4 = \Delta. \]
Then 
\[c_1 = \frac{\sqrt{\left(1+\sigma_\varepsilon \sqrt{2/(T-1)}  \right)^2 + 4 \Delta (2T-1)/(T-1) } -  \left(1+\sigma \sqrt{2/(T-1)}  \right)}{2  (2T-1)/(T-1) } \]
and
\begin{align*}
    1 - p_1 - p_2 - p_3 - p_4 = 1 - \left(4 + 2T \right) p_1 =1 - \left(4 + 2T \right) \exp\left(-N  p_{\ad,1} \cdot T c_1^2/(2\sigma^2) \right).
\end{align*}
Note that $\Delta = O(NT)$. Then $c_1 = O(\sqrt{NT})$ and 
\[1 - \left(4 + 2T \right) \exp\left(-N  p_{\ad,1} \cdot T c_1^2/(2\sigma^2) \right) = 1 - O\left(  \exp\left(-N^2 T^2 \right) \right)\]
This concludes the proof of Proposition \ref{prop:precision-guarantee}.
\halmos

\subsection{Pseudo Code for Functions in PGAE}\label{subsec:pseudo-code}

\begin{algorithm}
  \DontPrintSemicolon
  \SetKwFunction{Fpart}{partition\_initialize}
  \SetKwProg{Fn}{Function}{:}{}
  \Fn{\Fpart{$N, p_{\fcs}, T_{\max}$}}{
        \nl Randomly partition units into three sets, $\mathcal{S}_{\fcs}$, $\mathcal{S}_{\ad,1}$, and $\mathcal{S}_{\ad,2}$, that satisfy $\mathcal{S}_\fcs \cup \mathcal{S}_{\ad,1} \cup \mathcal{S}_{\ad,2}  =  \{1, \cdots, N\} $,  $|\mathcal{S}_{\ad,1}| = |\mathcal{S}_{\ad,2}|= \lfloor N (1-p_{\fcs})/2 \rfloor$ and $|\mathcal{S}_\fcs|= N - |\mathcal{S}_{\ad,1}| - |\mathcal{S}_{\ad,2}|$\;
        \nl $\bm{\omega}_{\mathrm{bm}} \leftarrow \left[(2s - 1 - T_{\max})/T_{\max} \right]_{s\in [T_{\max}]} $ \;
  \nl $Z_{\fcs} \leftarrow$ treatment design for $\mathcal{S}_{\fcs}$ with $ \bm{1}^\T Z_{\fcs,s}=|\mathcal{S}_\fcs| \cdot \bm{\omega}_{\mathrm{bm}}$ (subject to rounding)\;
  \nl $Z_{\ad,1} \leftarrow$ treatment design for $\mathcal{S}_{\ad,1}$ with $\bm{1}^\T Z_{\ad,1,s} =|\mathcal{S}_{\ad,1}| \cdot \bm{\omega}_{\mathrm{bm}}$ (subject to rounding)\;
  \nl  $Z_{\ad,2} \leftarrow$ treatment design for $\mathcal{S}_{\ad,2}$ with $\bm{1}^\T Z_{\ad,2,s}=|\mathcal{S}_{\ad,2}| \cdot \bm{\omega}_{\mathrm{bm}}$ (subject to rounding)\;
\nl \KwRet $Z_{\fcs}$, $Z_{\ad,1}$, $Z_{\ad,2}$\;}
\end{algorithm}

\begin{algorithm}
  \DontPrintSemicolon
  \SetKwFunction{estbelief}{estimate\_belief}
  \SetKwProg{Fn}{Function}{:}{}
  \Fn{\estbelief{$Z, Y, N, \bm{\tilde{\omega}}, T_{\max}, t, m, c$}}{
  \nl $n \leftarrow $ number of rows in $Z$  \;
        \nl $\hat{\tau} \leftarrow$ within estimator from $Y$ and $Z$\;
  \nl $\estsigmasq \leftarrow (n (t-1))^\I \sum_{i=1}^n \sum_{s = 1}^t \big( \dot{y}_{is} - \hat{\tau} \dot{z}_{is} \big)^2$ \;
\nl $\estxisq \leftarrow t \cdot n^{-1} \cdot (t-1)^{-2} \sum_{i =1}^n  \big[ \sum_{s = 1}^t \left[ \big( \dot{y}_{is} - \hat{\tau}\dot{z}_{is} \right] \big)^2 - \estsigmasq \big]^2 - (3t-2) \cdot ((t-1)^{-2} \cdot (\estsigmasq)^2 $ \;
\nl $\widehat{\xi^{\dagger2}} = \estxisq + 2/(t-1) \cdot (\estsigmasq)^2$ \;
\nl $P_t(T) \leftarrow 0$ for $T = 1, \cdots, T_{\max}$ \;
\nl \For{$j = 1, \cdots, m$}{
\nl $\samplesigmasq \leftarrow $ draw from $ \mathcal{N} \big(\estsigmasq, \widehat{\xi^{\dagger2}} /(n t) \big)$\;
\nl $\tilde{T}\leftarrow$ minimum $T$ such that $NT/\samplesigmasq \cdot \funfrac(\tilde{\bm{\omega}}_{1:T}, T) \geq c$  \;
\nl $\tilde{T}\leftarrow \max( \min(\tilde{T}, T_{\max}),t)$  \;
\nl $P_t(\tilde{T}) \leftarrow P_t(\tilde{T}) + 1/m$ \;
}
\nl \KwRet $P_t(\cdot)$\;}
\end{algorithm}

\begin{algorithm}
  \DontPrintSemicolon
  \SetKwFunction{updatedesign}{update\_treatment\_design}
  \SetKwProg{Fn}{Function}{:}{}
  \Fn{\updatedesign{$P_t(\cdot) , Z_1, Z_2, \tilde{\bm{\omega}}, T_{\max}, t$}}{
  $n_1$, $n_2 \leftarrow $ number of rows in $Z_1$, $Z_2$ \;
  \nl $\omega^\ast_{t+1} \leftarrow$  optimal $\omega_{t+1}$ that minimizes the terminal cost of the dynamic program $\+E_{T \sim P_t(T)} \left[ - T \cdot \funfrac\left((\bm{\omega}_{\ad,1:t}, \bm{\omega}_{(t+1):T}), T\right) \right]$ subject to $\tilde{\omega}_t \leq \omega_{t+1} \leq \omega_{t+1} \leq \cdots \leq \omega_{T_{\max}} \leq 1$ \;
\nl $Z_{1,t+1}, Z_{2,t+1} \leftarrow$ randomly assign treatment to $n_1 (\omega_{t+1} - \tilde{\omega}_{t}) $ and $n_2 (\omega_{t+1} - \tilde{\omega}_{t}) $ control units at time $t$ in $Z_1$ and $ Z_2$ respectively \;
  \nl \KwRet $Z_{1,t+1}, Z_{2,t+1}$\;}
\end{algorithm}

\begin{algorithm}
  \DontPrintSemicolon
  \SetKwFunction{estvarprec}{estimate\_var\_prec}
  \SetKwProg{Fn}{Function}{:}{}
  \Fn{\estvarprec{$Z, Y, \bm{\omega}, N, n, t$}}{
        \nl $\hat{\tau} \leftarrow$ within estimator from $Y$ and $Z$\;
  \nl $\estsigmasq \leftarrow (n (t-1))^\I \sum_{i =1}^n \sum_{s = 1}^t \big( \dot{y}_{is} - \hat{\tau} \dot{z}_{is} \big)^2$ \;
  \nl $\Prec \leftarrow (N t/\estsigmasq) \cdot \funfrac(\bm{\omega}_{1:t},t)  $\;
  \nl \KwRet $\estsigmasq, \Prec$\;}
\end{algorithm}

%% file: appendix_D.tex
\section{Proof of Results for Adaptive Experiments}\label{sec:sequential-proof}

We start with a lemma that provides an expansion of the average of products of two within transformed variables, $\dot{a}_{it} $ and $\dot{b}_{it}$. This lemma will be used to show Lemma \ref{lemma:asymptotic-tau-sigma} and Theorem \ref{theorem:asymptotic-page}.
\begin{lemma}\label{lemma:within-average}
    For any $\{a_{it}\}_{(i,t)\in [N_o]\times [T_o]}$ and $\{b_{it}\}_{(i,t)\in \in [N_o]\times [T_o]}$, we have 
    \begin{align}
    \frac{1}{N_o T_o}\sum_{i, t} \dot{a}_{it} \dot{b}_{it} = \overline{ab} - \frac{1}{N_o} \sum_i \bar{a}_{i,\cdot} \bar{b}_{i,\cdot} - \frac{1}{T_o} \sum_t \bar{a}_{\cdot,t} \bar{b}_{\cdot,t} + \bar{a} \bar{b},\label{eqn:within-average}
\end{align}
where $\overline{ab} = \frac{1}{N_o T_o} \sum_{i,t} a_{it} b_{it}$, $\bar{a}_{i,\cdot} = \frac{1}{T} \sum_t a_{it}$, $\bar{a}_{\cdot,t} = \frac{1}{N_o} \sum_i a_{it}$, and $\bar{a} = \frac{1}{N_o T_o} \sum_{i,t} a_{it}$. $\bar{b}_{i,\cdot}$, $\bar{b}_{\cdot,t}$ and $\bar{b}$ are similarly defined. 
\end{lemma}

\begin{proof}{Proof of Lemma \ref{lemma:within-average}}
We can prove this lemma by using the definition of $\dot{a}_{it}$ and $\dot{b}_{it}$, and writing the average of products (of $\dot{a}_{it} $ and $\dot{b}_{it}$) as multiple averages of products based on the definition of $\dot{a}_{it}$ and $\dot{b}_{it}$
    \begin{align*}
    \frac{1}{N_o T_o}\sum_{i, t} \dot{a}_{it} \dot{b}_{it} =& \frac{1}{N_o T_o}\sum_{i, t} \left(a_{it} - \bar{a}_{i,\cdot} -  \bar{a}_{\cdot,t} + \bar{a}\right) \left(b_{it} - \bar{b}_{i,\cdot} -  \bar{b}_{\cdot,t} + \bar{b}\right)  \\
    =& \frac{1}{N_o T_o}\sum_{i, t} a_{it} b_{it} - \underbrace{\frac{1}{N_o T_o}\sum_{i, t} a_{it}  \bar{b}_{i,\cdot}}_{\frac{1}{N_o }\sum_{i} \bar{a}_{i,\cdot} \bar{b}_{i,\cdot}}  - \underbrace{\frac{1}{N_o T_o}\sum_{i, t} a_{it}  \bar{b}_{\cdot,t}}_{\frac{1}{T_o}\sum_{t} \bar{a}_{\cdot,t}  \bar{b}_{\cdot,t}}  + \underbrace{\frac{1}{N_o T_o}\sum_{i, t} a_{it}  \bar{b}}_{\bar{a} \bar{b}}  \\
    & - \underbrace{\frac{1}{N_o T_o}\sum_{i, t}  \bar{a}_{i,\cdot} b_{it}}_{\frac{1}{N_o }\sum_{i} \bar{a}_{i,\cdot} \bar{b}_{i,\cdot}} + \underbrace{\frac{1}{N_o T_o}\sum_{i, t}  \bar{a}_{i,\cdot} \bar{b}_{i,\cdot}}_{\frac{1}{N_o }\sum_{i} \bar{a}_{i,\cdot} \bar{b}_{i,\cdot}}  + \underbrace{\frac{1}{N_o T_o}\sum_{i, t}  \bar{a}_{i,\cdot} \bar{b}_{\cdot,t}}_{\bar{a}  \bar{b}}  - \underbrace{\frac{1}{N_o T_o}\sum_{i, t}  \bar{a}_{i,\cdot} \bar{b}}_{\bar{a}  \bar{b}}  \\
    & - \underbrace{\frac{1}{N_o T_o}\sum_{i, t}  \bar{a}_{\cdot,t} b_{it}}_{\frac{1}{T_o}\sum_{t} \bar{a}_{\cdot,t}  \bar{b}_{\cdot,t}}  + \underbrace{\frac{1}{N_o T_o}\sum_{i, t}  \bar{a}_{\cdot,t} \bar{b}_{i,\cdot}}_{\bar{a}  \bar{b}}  + \underbrace{\frac{1}{N_o T_o}\sum_{i, t}  \bar{a}_{\cdot,t} \bar{b}_{\cdot,t}}_{\frac{1}{T_o}\sum_{t} \bar{a}_{\cdot,t}  \bar{b}_{\cdot,t}}  - \underbrace{\frac{1}{N_o T_o}\sum_{i, t}  \bar{a}_{\cdot,t} \bar{b}}_{\bar{a}  \bar{b}}  \\
    & + \underbrace{\frac{1}{N_o T_o}\sum_{i, t} \bar{a} b_{it}}_{\bar{a}  \bar{b}}  - \underbrace{\frac{1}{N_o T_o}\sum_{i, t} \bar{a}  \bar{b}_{i,\cdot}}_{\bar{a}  \bar{b}}  - \underbrace{\frac{1}{N_o T_o}\sum_{i, t} \bar{a}  \bar{b}_{\cdot,t}}_{\bar{a}  \bar{b}}  + \underbrace{\frac{1}{N_o T_o}\sum_{i, t} \bar{a}  \bar{b}}_{\bar{a}  \bar{b}}  \\
    =& \overline{ab} - \frac{1}{N_o} \sum_i \bar{a}_{i,\cdot} \bar{b}_{i,\cdot} - \frac{1}{T_o} \sum_t \bar{a}_{\cdot,t} \bar{b}_{\cdot,t} + \bar{a} \bar{b}
\end{align*}
This finishes the proof of this lemma. \halmos
\end{proof}

\subsection{Proof of Lemma \ref{lemma:asymptotic-tau-sigma-general}}

We first state a more general version of Lemma \ref{lemma:asymptotic-tau-sigma} in Lemma \ref{lemma:asymptotic-tau-sigma-general}. Then we prove Lemma \ref{lemma:asymptotic-tau-sigma-general}, and Lemma \ref{lemma:asymptotic-tau-sigma} follows from Lemma \ref{lemma:asymptotic-tau-sigma-general}.

\begin{lemma}\label{lemma:asymptotic-tau-sigma-general}
Suppose $\hat{\tau}$ is the \within estimator of $\tau$, $\estsigmasq$ and $\estxisq$ are estimators of $\sigma^2$ and $\xi^2$ using the formula \eqref{eqn:second-moment-sigma-estimator} and formula \eqref{eqn:fourth-moment-sigma-estimator} from any experimental data with $N_o$ units and $T_o$ periods, whose treatment design is selected before the experiment starts. 
Suppose $\varepsilon_{is}$ is i.i.d. for any $i$ and $s$ with $\+E[\varepsilon_{is}] = 0$, $\+E[\varepsilon^2_{is}] = \sigma_\varepsilon^2$, $\+E[\varepsilon^3_{is}] = 0$,  and $\+E[(\varepsilon^2_{is} - \sigma^2_\varepsilon)^2] = \xi^2_{\varepsilon}$. As $N_o \rightarrow \infty$, for any  $T_o$, conditional on the treatment design $Z$, we have 
\[\sqrt{N_o T_o} \left(\begin{bmatrix} \hat{\tau} \\ \estsigmasq \end{bmatrix}  - \begin{bmatrix} \tau \\ \sigma^2_\varepsilon  \end{bmatrix}\right)  \xrightarrow{d} \mathcal{N} \left(\begin{bmatrix} 0 \\ 0 \end{bmatrix}, \begin{bmatrix} \sigma_\varepsilon^2/\funfrac(\bm{\omega}, T_o) & 0 \\  0 & \xi_\varepsilon^{\dagger2} \end{bmatrix} \right), \]
where $\xi_\varepsilon^{\dagger2} = \xi^2_\varepsilon + \frac{1}{T_o-1} \sigma_\varepsilon^4$ and $\bm{\omega} = \bm{1}^\T Z/N_o$. Furthermore, $\estxisq$ is consistent and asymptotically normal with 
\begin{align*}
    \sqrt{N_o} (\estxisq - \xi_\varepsilon^2) \xrightarrow{d} \mathcal{N}(0, \+E[f_\xi(\varepsilon_{i1}, \cdots, \varepsilon_{i,T_o})^2])
\end{align*}
where 
\begin{align*}
    f_\xi(\varepsilon_{i1}, \cdots, \varepsilon_{i,T_o}) =&  \frac{T_o}{(T_o - 1)^2}   \Big( \sum_{t} ( \varepsilon_{it}^2 - \sigma^2_\varepsilon)^2  + \sum_{t\neq s} ( \varepsilon_{it}^2 - \sigma^2_\varepsilon) ( \varepsilon_{is}^2 - \sigma^2_\varepsilon) \Big)  - \frac{2}{(T_o - 1)^2}  \sum_{t,s,u} ( \varepsilon_{it}^2 - \sigma^2_\varepsilon) \varepsilon_{is} \varepsilon_{iu}  \\
    &+ \frac{T_o^3}{(T_o - 1)^2} \left(\frac{1}{T_o} \sum_t \varepsilon_{it} \right)^4  - \frac{\sigma_\varepsilon^2}{T_o} \sum_{t} (\varepsilon_{it}^2 - \sigma_\varepsilon^2 ) - \frac{\sigma_\varepsilon^2}{ T_o(T_o-1)} \sum_{t\neq s} \varepsilon_{it} \varepsilon_{is} - \xi_\varepsilon^2 -\frac{3T_o-2}{(T_o-1)^2} \sigma_\varepsilon^4.
\end{align*}
In addition, $\hat{\tau}$, $\estsigmasq$ and $\estxisq$ are jointly asymptotically normal.
\end{lemma}

\begin{proof}{Proof of Lemma \ref{lemma:asymptotic-tau-sigma-general}}

\textbf{Step 1.1: Show the consistency of $\hat{\tau}$.}

The estimation error of $\hat{\tau}$ from the \within estimator is 
\begin{align*}
    \hat{\tau} - \tau = \left(\frac{1}{N_o T_o} \sum_{i, t} \dot{z}^2_{it} \right)^\I \frac{1}{N_o T_o}\sum_{i, t} \dot{z}_{it} \dot{\varepsilon}_{it}. 
\end{align*}
We first show the consistency of $\hat{\tau}$. This is equivalent to showing that the estimation error converges to zero in probability as $N_o \rightarrow \infty$.

The first term in the estimation error of $\hat{\tau}$ equals to
\begin{align*}
    \frac{1}{N_o T_o} \sum_{i,t} \dot{z}^2_{it}  =& - \frac{1}{T_o} \bm{\omega}^\T \*P_{\bm{1}_{T_o}}\bm{\omega} - \frac{2}{T_o} \*b_0^\T \bm{\omega} = \funfrac(\bm{\omega}, T_o),
\end{align*}
following Lemma \ref{lemma:simplify-obj} (without observed and latent covariates, and $\ell = 0$). This term is nonzero with a properly chosen $\bm{\omega}$.
The second term in the estimation error of $\hat{\tau}$ has the following decomposition from Lemma \ref{lemma:within-average}
\begin{align*}
\frac{1}{N_o T_o} \sum_{i,t} \dot{z}_{it} \dot{\varepsilon}_{it}  =& \underbrace{\frac{1}{N_o T_o} \sum_{i,t} z_{it} \varepsilon_{it}}_{a_1}  - 
 \underbrace{\frac{1}{N_o} \sum_{i} \bar{z}_{i,\cdot} \bar{\varepsilon}_{i,\cdot}}_{a_2} - \underbrace{\frac{1}{T_o} \sum_{t} \bar{z}_{\cdot,t} \bar{\varepsilon}_{\cdot,t}}_{a_3}  + \underbrace{\bar{z} \bar{\varepsilon}}_{a_4} 
\end{align*}
We can show that
\begin{enumerate}
    \item $a_1 \xrightarrow{p} 0$
    \item $a_2 \xrightarrow{p} 0$
    \item $a_3 \xrightarrow{p} 0$ (follows from $\bar{\varepsilon}_{\cdot,t} \xrightarrow{p} 0$ from the law of large numbers)
    \item $a_4 \xrightarrow{p} 0$ (follows from $\bar{\varepsilon} \xrightarrow{p} 0$ from the law of large numbers)
\end{enumerate}

\begin{proof}{Proof of $a_1 \xrightarrow{p} 0$.}
Note that the mean of $a_1$ is zero 
\[\+E_{\varepsilon}\left[\frac{1}{N_o T_o} \sum_{i,t} z_{it} \varepsilon_{it} \right] = \frac{1}{N_o T_o} \sum_{i,t} z_{it} \+E\left[\varepsilon_{it} \right] = 0. \]
The variance of $a_1$ is (using the property that $\varepsilon_{it}$ is i.i.d in $i$ and $t$)
\begin{align*}
\var_{\varepsilon}\left(\frac{1}{N_o T_o} \sum_{i,t} z_{it} \varepsilon_{it} \right) = \frac{1}{N^2_o T^2_o} \sum_{i,t} z^2_{it} \var(\varepsilon_{it})  = O\left( \frac{1}{N_o}\right).
\end{align*}
From Chebyshev's inequality, we have $a_1 \xrightarrow{p} 0$.
   \halmos 
\end{proof}

\begin{proof}{Proof of $a_2 \rightarrow 0$.}
   The mean of  $a_2$ is zero
   \[\+E_{\varepsilon}\left[\frac{1}{N_o } \sum_{i} \bar{z}_{i,\cdot} \bar{\varepsilon}_{i,\cdot} \right] = \frac{1}{N_o } \sum_{i,t} \left(\bar{z}_{i,\cdot} \cdot  \frac{1}{T_o} \sum_t \+E\left[ \bar{\varepsilon}_{it}\right] \right)= 0. \]
   The variance of  $a_2$ is (using the property that $\varepsilon_{it}$ is i.i.d in $i$ and $t$)
   \begin{align*}
       \var_{\varepsilon}\left(\frac{1}{N_o } \sum_{i} \bar{z}_{i,\cdot} \bar{\varepsilon}_{i,\cdot} \right) = \frac{1}{N_o^2} \sum_i \bar{z}_{i,\cdot}^2 \var\left( \bar{\varepsilon}_{i,\cdot}\right) = \frac{1}{N_o^2 T_o^2} \sum_i \left(\bar{z}_{i,\cdot}^2 \sum_t \var\left( \varepsilon_{it}\right)\right)= O\left( \frac{1}{N_o}\right).
   \end{align*}
   From Chebyshev's inequality, we have $a_2 \xrightarrow{p} 0$.
   \halmos
\end{proof}
Therefore we have shown that all four terms in the decomposition of $\var_{\varepsilon}\left(\frac{1}{N_o T_o} \sum_{i,t} z_{it} \varepsilon_{it} \right)$ converges to zero in probability. This implies the consistency of $\hat{\tau}$.

\textbf{Step 1.2: Show the asymptotic normal distribution of $\hat{\tau}$.}

For the estimation error of $\hat{\tau}$, we have 
\begin{flalign*}
    \hat{\tau} - \tau =& \left(\frac{1}{N_o T_o} \sum_{i, t} \dot{z}^2_{it} \right)^\I \frac{1}{N_o T_o}\sum_{i, t} \dot{z}_{it} \dot{\varepsilon}_{it} \\
    =& \left(\frac{1}{N_o T_o} \sum_{i, t} \dot{z}^2_{it} \right)^\I \frac{1}{N_o T_o}\sum_{i, t} \dot{z}_{it} \varepsilon_{it}  & \tag{by the proof of Lemma \ref{lemma:within-average}} \\
    =&  \frac{1}{N_o T_o}\sum_{i, t} \underbrace{\left(\frac{1}{N_o T_o} \sum_{i, t} \dot{z}^2_{it} \right)^\I \left(z_{it} - \frac{1}{N_o} \sum_i \bar{z}_{i,\cdot} - \frac{1}{T_o} \sum_t \bar{z}_{\cdot,t} + \bar{z}\right)}_{\coloneqq \zeta_{it}}\varepsilon_{it} & \tag{ by the definition of $\dot{z}_{it}$} \\
    =& \frac{1}{N_o T_o}\sum_{i, t} \zeta_{it} \varepsilon_{it},
\end{flalign*}
where $\zeta_{it}$
is a function of $Z$ and is independent of $\varepsilon_{it}$ (as $Z$ is chosen before the experiment starts). 

Since $\varepsilon_{it} \stackrel{\iid}{\sim} (0, \sigma_{\varepsilon}^2)$, the estimation error of $\hat{\tau}$ is an average of $N_o T_o$ independent terms $\zeta_{it} \varepsilon_{it}$. We can apply Lindeberg-Feller CLT to $\hat{\tau}$ as long as the 
 condition in Lindeberg-Feller CLT 
holds. Note that the condition in Lindeberg-Feller CLT for the estimation error of $\hat{\tau}$ takes the form of 
\begin{align*}
   & \max_{i,t}\frac{\sigma_\varepsilon^2 \zeta_{it}^2}{\sum_{i,t} \sigma_\varepsilon^2 \zeta_{it}^2 } \rightarrow 0 \qquad \text{ as } N_o \rightarrow \infty \\
    \Leftrightarrow&  \max_{i,t}\frac{\dot{z}_{it}^2}{\sum_{i,t} \dot{z}_{it}^2 } \rightarrow 0 \qquad \text{ as } N_o \rightarrow \infty \\
     \Leftrightarrow&  \max_{i,t}\frac{\dot{z}_{it}^2}{N_o T_o \funfrac(\bm{\omega}, T_o)} \rightarrow 0 \qquad \text{ as } N_o \rightarrow \infty
\end{align*}
where the last line holds as $\dot{z}_{it}$ is bounded and $\funfrac(\bm{\omega}, T_o)$ is bounded away from 0. 

Then we can apply Lindeberg-Feller CLT and show that
\[\sqrt{N_o T_o} (\hat{\tau} - \tau) \xrightarrow{d} \mathcal{N}(0, \sigma_{\varepsilon}^2 /\funfrac(\bm{\omega}, T_o))\,, \]
where the asymptotic variance is $\sigma_{\varepsilon}^2 /\funfrac(\bm{\omega}, T_o)$ following that
\begin{align*}
    \mathrm{AVar}(\hat{\tau}) =& \left(\frac{1}{N_o T_o} \sum_{i, t} \dot{z}^2_{it} \right)^\I \+E_\varepsilon\left[ N_o T_o \left(\frac{1}{N_o T_o}\sum_{i, t} \dot{z}_{it} \varepsilon_{it}\right)^2 \right]\left(\frac{1}{N_o T_o} \sum_{i, t} \dot{z}^2_{it} \right)^\I \\
    =& \sigma_\varepsilon^2 \cdot  \left(\frac{1}{N_o T_o} \sum_{i, t} \dot{z}^2_{it} \right)^\I = \sigma_{\varepsilon}^2 /\funfrac(\bm{\omega}, T_o).
\end{align*}

\textbf{Step 2.1: Show the consistency of $\estsigmasq$.}

We can decompose $\estsigmasq$ as 
\begin{flalign}
  \nonumber  \estsigmasq =& \frac{1}{N_o(T_o-1)} \sum_{i,t} (\dot{y}_{it} - \hat{\tau} \dot{z}_{it})^2 = \frac{1}{N_o(T_o-1)} \sum_{i,t} (\underbrace{\dot{y}_{it} - \tau \dot{z}_{it}}_{\dot{\varepsilon}_{it}}  - (\hat{\tau} - \tau) \dot{z}_{it})^2 \\
  \nonumber  =& \frac{1}{N_o(T_o-1)}\sum_{i,t} \dot{\varepsilon}^2_{it} - 2 (\hat{\tau} - \tau) \cdot \underbrace{\frac{1}{N_o(T_o-1)} \sum_{i,t}\dot{z}_{it} \dot{\varepsilon}_{it}}_{\frac{\hat{\tau} - \tau}{N_o(T_o-1)}  \sum_{i,t} \dot{z}_{it}^2 }  + (\hat{\tau} - \tau)^2 \cdot \frac{1}{N_o(T_o-1)} \sum_{i,t} \dot{z}_{it}^2  \label{eqn:exact-decomposition-sigma-sq}
    \\
    =& \frac{1}{N_o(T_o-1)}\sum_{i,t} \dot{\varepsilon}^2_{it} - \underbrace{(\hat{\tau} - \tau)^2}_{O_p\left( \frac{1}{N_o} \right)} \cdot \frac{1}{N_o(T_o-1)} \sum_{i,t} \dot{z}_{it}^2  \\  \nonumber =&  \frac{1}{N_o(T_o-1)}\sum_{i,t} \dot{\varepsilon}^2_{it} + O_p\left( \frac{1}{N_o} \right).  
\end{flalign}

We can further decompose the leading term of
$\estsigmasq$ as 
\begin{flalign*}
    & \frac{1}{N_o(T_o-1)} \sum_{i,t} \dot{\varepsilon}^2_{it} \\ =& \frac{1}{N_o(T_o-1)} \sum_{i,t} \varepsilon_{it}^2 - \frac{T_o}{N_o(T_o-1)} \sum_i \bar{\varepsilon}^2_{i,\cdot}  - \frac{1}{T_o-1} \sum_t \underbrace{\bar{\varepsilon}^2_{\cdot,t}}_{O_p\left( \frac{1}{N_o} \right)}  + \frac{T_o}{T_o-1} \underbrace{\bar{\varepsilon}^2}_{O_p\left( \frac{1}{N_o} \right)}  & \tag{by Lemma \ref{lemma:within-average}} \\
    =& \frac{1}{N_o(T_o-1)} \sum_{i,t} \varepsilon_{it}^2 - \frac{T_o}{N_o(T_o-1)} \sum_i \left( \frac{1}{T_o^2} \sum_{t} \varepsilon_{it}^2 + \frac{1}{T_o^2} \sum_{t \neq s} \varepsilon_{it} \varepsilon_{is} \right) + O_p\left( \frac{1}{N_o} \right) & \tag{by expanding $\bar{\varepsilon}^2_{i,\cdot}$} \\
    =& \sigma_\varepsilon^2 +  \frac{1}{N_o T_o} \sum_{i,t} (\varepsilon_{it}^2 - \sigma_\varepsilon^2 ) - \frac{1}{N_o T_o(T_o-1)} \sum_{i, t\neq s} \varepsilon_{it} \varepsilon_{is}  +  O_p\left( \frac{1}{N_o} \right) & \tag{add and subtract $\sigma_\varepsilon^2$}
\end{flalign*}

The estimation error of $\estsigmasq$ is $\estsigmasq - \sigma_\varepsilon^2$, whose leading terms are $\frac{1}{N_o T_o} \sum_{i,t} (\varepsilon_{it}^2 - \sigma_\varepsilon^2 )$ and $\frac{1}{N_o T_o(T_o-1)} \sum_{i, t\neq s} \varepsilon_{it} \varepsilon_{is} $. Note that $\varepsilon_{it}$ is i.i.d. in $i$ and $t$. Therefore, $\varepsilon_{it}^2 - \sigma_\varepsilon^2 $ is i.i.d. in $i$ and $t$. We can then apply the law of large numbers to the first leading term $\frac{1}{N_o T_o} \sum_{i,t} (\varepsilon_{it}^2 - \sigma_\varepsilon^2 )$, and show it converges to zero in probability. Furthermore, $\sum_{t\neq s} \varepsilon_{it} \varepsilon_{is}$ is i.i.d. in $i$. We can also apply the law of large numbers to the second leading term $\frac{1}{N_o T_o(T_o-1)} \sum_{i, t\neq s} \varepsilon_{it} \varepsilon_{is} $, and show it converges to zero in probability.

Since both leading terms converge to zero in probability, we finish the proof of consistency of $\estsigmasq$.

\textbf{Step 2.2: Show the asymptotic normal distribution of $\estsigmasq$.}

We can write the estimation error of $\estsigmasq$ as 
\begin{align*}
    \estsigmasq -\sigma_\varepsilon^2 = \frac{1}{N_o} \sum_i \left[ \frac{1}{T_o} \sum_t (\varepsilon_{it}^2 - \sigma_\varepsilon^2 )  - \frac{1}{T_o (T_o-1)} \sum_{t\neq s} \varepsilon_{it} \varepsilon_{is}\right] +  O_p\left( \frac{1}{N_o} \right) 
\end{align*}
As $\varepsilon_{it}$ is i.i.d. in $i$ and $t$, $\frac{1}{T_o} \sum_t (\varepsilon_{it}^2 - \sigma_\varepsilon^2 )  - \frac{1}{T_o (T_o-1)} \sum_{t\neq s} \varepsilon_{it} \varepsilon_{is}$ is i.i.d. in $i$ with mean zero (following that $\+E[\varepsilon_{it}^2 -\sigma_\varepsilon^2 ] = 0$ and $\+E[\varepsilon_{it} \varepsilon_{is}] = 0$ for $t \neq s$. Then $\estsigmasq$ is asymptotically normal following the standard CLT. Next, we compute the asymptotic variance of $\estsigmasq$, which is equivalent to computing the following term 
\begin{flalign*}
    & \var\left(\frac{1}{T_o} \sum_t (\varepsilon_{it}^2 - \sigma_\varepsilon^2 )  - \frac{1}{T_o (T_o-1)} \sum_{t\neq s} \varepsilon_{it} \varepsilon_{is} \right) \\
    =& \underbrace{\var\left(\frac{1}{T_o} \sum_t (\varepsilon_{it}^2 - \sigma_\varepsilon^2 ) \right)}_{a_1}  - 2 \underbrace{\mathrm{Cov}\left(\frac{1}{T_o} \sum_t (\varepsilon_{it}^2 - \sigma_\varepsilon^2 ),  \frac{1}{T_o (T_o-1)} \sum_{t\neq s} \varepsilon_{it} \varepsilon_{is} \right)}_{a_2} +\underbrace{\var\left( \frac{1}{T_o (T_o-1)} \sum_{t\neq s} \varepsilon_{it} \varepsilon_{is}\right)}_{a_3}  \\
    =& \frac{1}{T_o} \xi^2_\varepsilon + \frac{2}{T_o(T_o-1)} \sigma_\varepsilon^2 
\end{flalign*}
following that term $a_1$ has
\begin{flalign*}
    a_1 = \var\left(\frac{1}{T_o} \sum_t (\varepsilon_{it}^2 - \sigma_\varepsilon^2 ) \right) =& \frac{1}{T^2_o} \sum_{t,s} \mathrm{Cov}\left(\varepsilon_{it}^2 - \sigma_\varepsilon^2 , \varepsilon_{is}^2 - \sigma_\varepsilon^2  \right) \\
    =& \frac{1}{T^2_o} \sum_{t} \var\left(\varepsilon_{it}^2 - \sigma_\varepsilon^2 \right) & \tag{following $\varepsilon_{it}$ is i.i.d.}  \\
    =& \xi^2_\varepsilon,
\end{flalign*}
and term $a_2$ has 
\begin{flalign*}
  a_2 = & \mathrm{Cov}\left(\frac{1}{T_o} \sum_t (\varepsilon_{it}^2 - \sigma_\varepsilon^2 ),  \frac{1}{T_o (T_o-1)} \sum_{t\neq s} \varepsilon_{it} \varepsilon_{is} \right) 
 \\ =& \frac{1}{T^2_o (T_o-1)} \sum_{t, u, s: u \neq s} \mathrm{Cov}\left(\varepsilon_{it}^2 - \sigma_\varepsilon^2 ,  \varepsilon_{iu} \varepsilon_{is} \right) \\
 =& \frac{1}{T^2_o (T_o-1)} \sum_{t, u, s: u \neq s} \+E\left[(\varepsilon_{it}^2 - \sigma_\varepsilon^2 ) \cdot \varepsilon_{iu} \varepsilon_{is} \right] & \tag{both $\varepsilon_{it}^2 - \sigma_\varepsilon^2$ and $\varepsilon_{iu} \varepsilon_{is}$ have mean $0$} \\
 =& 0 ,
\end{flalign*}
where the last line follows that at least one of $\varepsilon_{iu}$ and $\varepsilon_{is}$ differs from $\varepsilon_{it}$ given that $u \neq s$, and therefore $\+E\left[(\varepsilon_{it}^2 - \sigma_\varepsilon^2 ) \cdot \varepsilon_{iu} \varepsilon_{is} \right] = 0$. Term $a_3$ has 
\begin{flalign*}
    a_3 =& \var\left( \frac{1}{T_o (T_o-1)} \sum_{t\neq s} \varepsilon_{it} \varepsilon_{is}\right) \\
    =& \frac{1}{T_o^2 (T_o - 1)^2} \sum_{t,s,u,v:t\neq s, u\neq v}\+E[\varepsilon_{it} \varepsilon_{is}\varepsilon_{iu} \varepsilon_{iv}]  \\
    =& \frac{1}{T_o^2 (T_o - 1)^2} \cdot 2 T_o(T_o-1) \+E[\varepsilon^2_{it} \varepsilon^2_{is}] & \tag{$2 T_o(T_o-1)$ terms with mean nonzero} \\
    =& \frac{2}{T_o(T_o-1)} \sigma_\varepsilon^4 
\end{flalign*}
where the third line follows that $2T_o(T_o-1)$ terms in the sum (in the second line) equal to $\+E[\varepsilon^2_{it} \varepsilon^2_{is}]$ and the remaining terms in the sum equal to $0$. The reason for the  $2T_o(T_o-1)$ terms is that for each pair of $(t,s)$, there are two choices of $(u,v)$ such that $\+E[\varepsilon_{it} \varepsilon_{is}\varepsilon_{iu} \varepsilon_{iv}] = \+E[\varepsilon^2_{it} \varepsilon^2_{is}]$. One choice is $t = u$ and $s = v$. The other choice is $t = v$ and $s = u$. Since there are $T_o(T_o - 1)$ pairs of $(t,s)$, there are $2T_o(T_o-1)$ terms 
in the sum equal to $\+E[\varepsilon^2_{it} \varepsilon^2_{is}]$.

With the asymptotic variance of $\estsigmasq$, we have 
\[    \sqrt{N_o} (\estsigmasq - \sigma_\varepsilon^2) \xrightarrow{d} \mathcal{N}\left(0,\frac{1}{ T_o} \xi^2_\varepsilon + \frac{2}{  T_o(T_o-1)}\sigma_\varepsilon^4 \right). \]

Multiplying both side by $\sqrt{T_o}$, we have 
\begin{align*}
    \sqrt{N_o T_o} (\estsigmasq - \sigma_\varepsilon^2) \xrightarrow{d} \mathcal{N}\left(0, \xi^2_\varepsilon + \frac{2}{T_o-1}\sigma_\varepsilon^4 \right).
\end{align*}

\textbf{Step 3: Show the joint asymptotic normal distribution of $\hat{\tau}$ and $\estsigmasq$.}

Based on Steps 1.2 and 2.2, we have 
\begin{align*}
    \begin{bmatrix}
        \hat{\tau} \\ \\ \estsigmasq 
    \end{bmatrix} - \begin{bmatrix}
       \tau \\ \\ \sigma_\varepsilon^2 
    \end{bmatrix} = \frac{1}{N_o} \sum_i \begin{bmatrix}
        \frac{1}{T_o}\sum_{t} \zeta_{it} \varepsilon_{it} \\ \\
        \frac{1}{T_o} \sum_t (\varepsilon_{it}^2 - \sigma_\varepsilon^2 )  - \frac{1}{T_o (T_o-1)} \sum_{t\neq s} \varepsilon_{it} \varepsilon_{is}
    \end{bmatrix} + \begin{bmatrix}
        0 \\  \\ O_p\left( \frac{1}{N_o} \right) 
    \end{bmatrix}
\end{align*}
We can use the same procedure as Step 1.2 to show that the conditions in multivariate Lindeberg-Feller CLT hold and $\hat{\tau}$ and $\estsigmasq$ are jointly asymptotically normal. Then the next step is to compute the asymptotic covariance between $\hat{\tau} - \tau$ and $\estsigmasq - \sigma^2_\varepsilon$. We separate this task into two sub-tasks. The first sub-task is to compute the asymptotic covariance between $\hat{\tau} - \tau$ and the leading terms of $\estsigmasq - \sigma^2_\varepsilon$ (which is at the order of $O_p\left(1/\sqrt{N_o}\right)$). The second sub-task is to compute the asymptotic variance between $\hat{\tau} - \tau$ and the non-leading terms of $\estsigmasq - \sigma^2_\varepsilon$. 

We first consider the second sub-task, which is a simpler task. Note that  the non-leading terms of $\sqrt{N_o} \left(\estsigmasq - \sigma^2_\varepsilon\right)$ is at the order of $O_p\left(1/\sqrt{N_o}\right)$, and the order of $\sqrt{N_o}(\hat{\tau} - \tau)$ is $O_p(1)$. Therefore, their product is at the order of $O_p\left(1/\sqrt{N_o}\right) = o_p(1)$. Equivalently, the asymptotic covariance between non-leading terms of $\estsigmasq - \sigma^2_\varepsilon$ and $\hat{\tau} - \tau$ is $0$. 

Next we consider the first sub-task. There are two leading terms in $\estsigmasq - \sigma^2_\varepsilon$. The first one is $\frac{1}{N_o T_o} \sum_{i,t} (\varepsilon_{it}^2 - \sigma_\varepsilon^2 )$. The second one is $ \frac{1}{N_o T_o(T_o-1)} \sum_{i, t\neq s} \varepsilon_{it} \varepsilon_{is}$. 

For the asymptotic covariance between $\hat{\tau} - \tau$ and $\frac{1}{N_o T_o} \sum_{i,t} (\varepsilon_{it}^2 - \sigma_\varepsilon^2 )$, we have 

\begin{align*}
     & \+E\left[\left( \frac{1}{N_o T_o} \sum_{i,t} \zeta_{it} \varepsilon_{it} \right) \left(  \frac{1}{N_o T_o} \sum_{i,t} (\varepsilon_{it}^2 - \sigma_\varepsilon^2 ) \right) \right] \\
    =& \frac{1}{N^2_o T^2_o} \sum_{i,t} \zeta_{it} \+E\left[\varepsilon_{it} (\varepsilon_{it}^2 - \sigma_\varepsilon^2) \right] + \frac{1}{N^2_o T^2_o} \sum_{(i,t) \neq (j,s)}  \zeta_{it} \+E\left[ \varepsilon_{it} (\varepsilon_{js}^2 - \sigma_\varepsilon^2) \right] 
    = 0
\end{align*}
following that $\+E\left[\varepsilon_{it} (\varepsilon_{it}^2 - \sigma_\varepsilon^2) \right] = \+E[\varepsilon_{it}^3] - \+E[\varepsilon_{it}] \cdot\sigma_\varepsilon^2 = 0$ and $\+E\left[\varepsilon_{it} (\varepsilon_{js}^2 - \sigma_\varepsilon^2) \right] = \+E[\varepsilon_{it}] \cdot \+E[\varepsilon_{js}^2 - \sigma_\varepsilon^2] = 0$. Therefore the asymptotic covariance between these two terms is $0$. 

For the asymptotic covariance between $\hat{\tau} - \tau$ and $\frac{1}{N_o T_o (T_o-1)} \sum_{i, t\neq s} \varepsilon_{it} \varepsilon_{is} $, we have
\begin{align*}
        & \+E\left[\left( \frac{1}{N_o T_o} \sum_{i,t} \zeta_{it} \varepsilon_{it} \right) \left( \frac{1}{N^2_o T^2_o (T_o-1)} \sum_{i, t\neq s} \varepsilon_{it} \varepsilon_{is}  \right) \right]
    = \frac{1}{N_o T_o (T_o-1)} \sum_{i,j, t, u \neq s} \zeta_{it} 
\+E \left[\varepsilon_{it} \varepsilon_{ju} \varepsilon_{js} \right]= 0
\end{align*}
following that $\+E \left[\varepsilon_{it} \varepsilon_{ju} \varepsilon_{js} \right] = 0$ for any $i, j, u, t, s$ with $u \neq s$ (at least one of $\varepsilon_{ju} $ and $\varepsilon_{js}$ differs from $\varepsilon_{it}$, and $\varepsilon_{it}$ is i.i.d. in $i$ and $t$). Therefore the asymptotic covariance between these two terms is $0$. 

We have finished the two sub-tasks, and concluded that asymptotic covariance between $\hat{\tau} - \tau$ and $\estsigmasq - \sigma_\varepsilon^2$ is $0$. Therefore we have finished the proof of the following joint asymptotic normal distribution of $\hat{\tau}$ and $\estsigmasq$ 
\[\sqrt{N_o T_o} \left(\begin{bmatrix} \hat{\tau} \\ \estsigmasq \end{bmatrix}  - \begin{bmatrix} \tau \\ \sigma^2_\varepsilon  \end{bmatrix}\right)  \xrightarrow{d} \mathcal{N} \left(\begin{bmatrix} 0 \\ 0 \end{bmatrix}, \begin{bmatrix} \sigma_\varepsilon^2/\funfrac(\bm{\omega}, T_o) & 0 \\  0 & \xi_\varepsilon^{\dagger2} \end{bmatrix} \right). \]

\textbf{Step 4.1: Show the consistency of $\estxisq$.}

Recall the definition of $\estxisq$, 
\begin{align*}
    \estxisq =& \frac{T_o}{N_o(T_o-1)^2}  \sum_i \Big(\sum_{t} \big[ (\dot{y}_{it} - \hat{\tau} \dot{z}_{it})^2 - \estsigmasq \big] \Big)^2 - \frac{3T_o-2}{(T_o-1)^2} (\estsigmasq)^2.
\end{align*}
From Step 2.1, we have shown $\estsigmasq - \sigma_\varepsilon^2 = O_p(1/\sqrt{N_o})$, and then we have 
\begin{flalign*}
    (\estsigmasq)^2 =&  \sigma^4_\varepsilon + 2\sigma^2_\varepsilon (\estsigmasq - \sigma^2_\varepsilon) + (\estsigmasq - \sigma^2_\varepsilon)^2 \\
    =&\sigma^4_\varepsilon +O_p\left(\frac{1}{\sqrt{N_o}}\right)
\end{flalign*}

If we can show
\begin{align}\label{eqn:xi-proof-step-1}
        \frac{1}{N_o T_o} \sum_i \Big( \sum_{t} \big[ (\dot{y}_{it} - \hat{\tau} \dot{z}_{it})^2 - \estsigmasq \big] \Big)^2 = \frac{(T_o-1)^2}{T_o^2} \xi_\varepsilon^2 + \frac{3T_o-2}{T_o^2}\sigma^4_\varepsilon +O_p\left(\frac{1}{\sqrt{N_o}}\right)
    \end{align}
then we have
\begin{align*}
    \sqrt{N_o} \left(\estxisq - \xi^2_\varepsilon \right) =& \sqrt{N_o} \left( \frac{T_o^2}{(T_o - 1)^2}\frac{1}{N_o T_o}  \sum_i \Big(\sum_{t} \big[ (\dot{y}_{it} - \hat{\tau} \dot{z}_{it})^2 - \estsigmasq \big] \Big)^2 - \frac{3T_o-2}{(T_o-1)^2} (\estsigmasq)^2 - \xi^2_\varepsilon \right) \\
    =& \sqrt{N_o} \left( \frac{T_o^2}{(T_o - 1)^2}\left[\frac{(T_o-1)^2}{T_o^2} \xi_\varepsilon^2 + 
  \frac{3T_o-2}{T_o^2}\sigma^4_\varepsilon \right]  - \frac{3T_o-2}{(T_o-1)^2}\sigma^4_\varepsilon - \xi^2_\varepsilon +O_p\left(\frac{1}{\sqrt{N_o}}\right) \right) \\ =& O_p(1).
\end{align*}

Let us first prove \eqref{eqn:xi-proof-step-1}
\begin{flalign*}
   & \frac{1}{N_o T_o} \sum_i \Big(\sum_{t} \big( (\dot{y}_{it} - \hat{\tau} \dot{z}_{it})^2 - \estsigmasq \big) \Big)^2 \\
   =& \frac{1}{N_o T_o} \sum_i \Big(\sum_{t} \big( (\dot{y}_{it} - \hat{\tau} \dot{z}_{it})^2 - \sigma_\varepsilon^2 - (\estsigmasq - \sigma_\varepsilon^2) \big) \Big)^2 \\
   =& \frac{1}{N_o T_o} \sum_i \Big(\sum_{t} \big( (\dot{y}_{it} - \hat{\tau} \dot{z}_{it})^2 - \sigma_\varepsilon^2 \big) \Big)^2 - \frac{2 (\estsigmasq - \sigma_\varepsilon^2)}{N_o} \sum_i \Big(\sum_{t} \big( (\dot{y}_{it} - \hat{\tau} \dot{z}_{it})^2 - \sigma_\varepsilon^2 \big) \Big) + T_o (\estsigmasq - \sigma_\varepsilon^2)^2 \\
   =& \underbrace{\frac{1}{N_o T_o} \sum_i \Big(\sum_{t} \big( (\dot{y}_{it} - \hat{\tau} \dot{z}_{it})^2 - \sigma_\varepsilon^2 \big) \Big)^2}_{a_1}- T_o \underbrace{(\estsigmasq - \sigma_\varepsilon^2)^2}_{O_p\left(\frac{1}{N_o} \right)}
\end{flalign*}
where the last line follows the definition of $\estsigmasq$ and the order $\estsigmasq - \sigma_\varepsilon^2 = O(1/\sqrt{N_o})$.

For $a_1$, we have 
\begin{flalign}
   \nonumber a_1 =& \frac{1}{N_o T_o} \sum_i \Big(\sum_{t} \big( (\dot{\varepsilon}_{it} - (\hat{\tau} - \tau) \dot{z}_{it})^2 - \sigma_\varepsilon^2 \big) \Big)^2 \\
   \nonumber =& \frac{1}{N_o T_o} \sum_i \Big(\sum_{t} \big( \dot{\varepsilon}^2_{it} - \sigma_\varepsilon^2 - 2(\hat{\tau} - \tau) \dot{z}_{it} + \underbrace{(\hat{\tau} - \tau)^2 }_{O_p\left( \frac{1}{N_o}\right)} \dot{z}^2_{it}\big) \Big)^2 \\
    \nonumber =& \frac{1}{N_o T_o} \sum_i \Big(\sum_{t} \big( \dot{\varepsilon}^2_{it} - \sigma_\varepsilon^2 - 2(\hat{\tau} - \tau) \dot{z}_{it} \big) \Big)^2+ O_p\left( \frac{1}{N_o}\right)\\
   \nonumber =& \frac{1}{N_o T_o} \sum_i \Big(\sum_{t} \big( \dot{\varepsilon}^2_{it} - \sigma_\varepsilon^2 \big) \Big)^2 + O_p\left( \frac{1}{N_o}\right) & \tag{by $\sum_t \dot{z}_{it} = 0$} 
\end{flalign}
To further decompose $a_1$, we prove the the following lemma. 

\begin{lemma}\label{lemma:decompose-xi}
    In the setting of Lemma \ref{lemma:asymptotic-tau-sigma-general}, we have the following decomposition 
    \[  \frac{1}{N_o T_o} \sum_i \Big(\sum_{t} \big( \dot{\varepsilon}^2_{it} - \sigma_\varepsilon^2 \big) \Big)^2 =  \frac{1}{N_o T_o} \sum_i \Big( \sum_{t} [ ( \varepsilon_{it}^2 - \sigma^2_\varepsilon) -  \bar{\varepsilon}_{i,\cdot}^2] \Big)^2 + O_p\left(\frac{1}{N_o}\right).\]
\end{lemma}

\begin{proof}{Proof of Lemma \ref{lemma:decompose-xi}}
We first decompose the following summation over $t$
    \begin{align*}
& \sum_t \big( \dot{\varepsilon}^2_{it} - \sigma_\varepsilon^2 \big) = \sum_t \big( (\varepsilon_{it} - \bar{\varepsilon}_{i,\cdot} - \bar{\varepsilon}_{\cdot,t} + \bar{\varepsilon})^2 - \sigma_\varepsilon^2 \big)\\ =& \sum_t (\varepsilon_{it}^2 - \sigma_\varepsilon^2) + T_o \bar{\varepsilon}_{i,\cdot}^2 + \sum_t \bar{\varepsilon}_{\cdot,t}^2 + T_o \bar{\varepsilon}^2  - 2 T_o \bar{\varepsilon}_{i,\cdot}^2 - 2 \sum_t \varepsilon_{it} \bar{\varepsilon}_{\cdot,t} + 2T_o \bar{\varepsilon}_{i,\cdot}\bar{\varepsilon} + 2T_o \bar{\varepsilon}_{i,\cdot}\bar{\varepsilon} -2 T_o \bar{\varepsilon}_{i,\cdot} \bar{\varepsilon} - 2T_o \bar{\varepsilon}^2 \\
=& \sum_t (\varepsilon_{it}^2 - \sigma_\varepsilon^2) - T_o \bar{\varepsilon}_{i,\cdot}^2 + \underbrace{\sum_t \bar{\varepsilon}_{\cdot,t}^2}_{O_p\left(\frac{1}{N_o} \right)} - T_o \underbrace{\bar{\varepsilon}^2}_{O_p\left(\frac{1}{N_o} \right)} - 2 \sum_t \varepsilon_{it} \bar{\varepsilon}_{\cdot,t}+ 2T_o \bar{\varepsilon}_{i,\cdot}\bar{\varepsilon}  \\
=& \sum_t (\varepsilon_{it}^2 - \sigma_\varepsilon^2) - T_o \bar{\varepsilon}_{i,\cdot}^2 - 2 \sum_t \varepsilon_{it} \bar{\varepsilon}_{\cdot,t}+ 2T_o \bar{\varepsilon}_{i,\cdot}\bar{\varepsilon} + O_p\left(\frac{1}{N_o} \right)
\end{align*}
Using this decomposition, we have the following decomposition
\begin{align*}
    &\frac{1}{N_o T_o} \sum_i \Big(\sum_{t} \big( \dot{\varepsilon}^2_{it} - \sigma_\varepsilon^2 \big) \Big)^2 =\frac{1}{N_o T_o} \sum_i \Big(  \sum_t (\varepsilon_{it}^2 - \sigma_\varepsilon^2) - T_o \bar{\varepsilon}_{i,\cdot}^2 - 2 \sum_t \varepsilon_{it} \bar{\varepsilon}_{\cdot,t}+ 2T_o \bar{\varepsilon}_{i,\cdot}\bar{\varepsilon}\Big)^2 + O_p\left(\frac{1}{N_o} \right) \\
    \stackrel{(a)}{=}&  \frac{1}{N_o T_o} \sum_i \Big( \sum_{t} [ ( \varepsilon_{it}^2 - \sigma^2_\varepsilon) -  \bar{\varepsilon}_{i,\cdot}^2] \Big)^2 \\
    & -\frac{2}{T_o}  \Big(  \sum_{t,s} \underbrace{\Big(\frac{1}{N_o} \sum_{i} (\varepsilon_{it}^2 - \sigma_\varepsilon^2)\varepsilon_{is} \Big) }_{O_p\left(\frac{1}{\sqrt{N_o}}\right)}\underbrace{\bar{\varepsilon}_{\cdot,s} }_{O_p\left(\frac{1}{\sqrt{N_o}} \right)}- T_o \sum_s \underbrace{\Big( \frac{1}{N_o}\sum_{i} \bar{\varepsilon}_{i,\cdot}^2 \varepsilon_{is}  \Big)}_{O_p\left(\frac{1}{\sqrt{N_o}}\right)} \underbrace{\bar{\varepsilon}_{\cdot,s}}_{O_p\left(\frac{1}{\sqrt{N_o}}\right)}\Big)  \\
     & + 2 \underbrace{\bar{\varepsilon}}_{O_p\left(\frac{1}{\sqrt{N_o}}\right)} \Big( \underbrace{\frac{1}{N_o}\sum_{i,t} (\varepsilon_{it}^2 - \sigma_\varepsilon^2) \bar{\varepsilon}_{i,\cdot} }_{O_p\left(\frac{1}{\sqrt{N_o}}\right)}- \underbrace{\frac{T_o}{N_o} \sum_i \bar{\varepsilon}_{i,\cdot}^3 \Big)}_{O_p\left(\frac{1}{\sqrt{N_o}}\right)} + O_p\left(\frac{1}{N_o}\right) \\
     =&  \frac{1}{N_o T_o} \sum_i \Big( \sum_{t} [ ( \varepsilon_{it}^2 - \sigma^2_\varepsilon) -  \bar{\varepsilon}_{i,\cdot}^2] \Big)^2 + O_p\left(\frac{1}{N_o}\right)
\end{align*}
where the order of each term in (a) can be shown using $\varepsilon_{it}$ is i.i.d. in $i$ and $t$ with $\+E[\varepsilon_{it}] = \+E[\varepsilon_{it}^3] = 0$ and $\+E[\varepsilon^2_{it}] = \sigma_\varepsilon^2$.  \halmos
\end{proof}

Then we use Lemma \ref{lemma:decompose-xi} to decompose $a_1$, 
\begin{flalign}
   \nonumber a_1 =& \frac{1}{N_o T_o} \sum_i \Big( \sum_{t} [ ( \varepsilon_{it}^2 - \sigma^2_\varepsilon) -  \bar{\varepsilon}_{i,\cdot}^2] \Big)^2 + O_p\left(\frac{1}{N_o} \right)   \\ 
  =& \underbrace{\frac{1}{N_o T_o} \sum_i \Big(\sum_{t} ( \varepsilon_{it}^2 - \sigma^2_\varepsilon)  \Big)^2 }_{a_2} - \underbrace{\frac{2}{N_o} \sum_{i,t} ( \varepsilon_{it}^2 - \sigma^2_\varepsilon) \cdot \bar{\varepsilon}_{i,\cdot}^2}_{a_3} + \underbrace{\frac{T_o}{N_o} \sum_i \bar{\varepsilon}_{i,\cdot}^4 }_{a_4} + O_p\left(\frac{1}{N_o} \right),\label{eqn:xi-proof-i}
\end{flalign}

Below we further decompose each of $a_2$, $a_3$ and $a_4$. For the term $a_2$ in \eqref{eqn:xi-proof-i}, we have
\begin{align*}
    a_2 = \frac{1}{N_o T_o} \sum_i \Big( \sum_{t} ( \varepsilon_{it}^2 - \sigma^2_\varepsilon)^2  + \sum_{t\neq s} ( \varepsilon_{it}^2 - \sigma^2_\varepsilon) ( \varepsilon_{is}^2 - \sigma^2_\varepsilon) \Big) =  \xi_\varepsilon^2 + O_p\left(\frac{1}{\sqrt{N_o}} \right).
\end{align*}
following the independence between $\varepsilon_{it}$ and $\varepsilon_{is}$. 

For the term $a_3$ in \eqref{eqn:xi-proof-i}, we have 
\begin{flalign*}
    a_3=& \frac{2}{N_o T_o^2} \sum_{i,t,s,u} ( \varepsilon_{it}^2 - \sigma^2_\varepsilon) \varepsilon_{is} \varepsilon_{iu} \\ =& \frac{2}{N_o T_o^2} \sum_{i,t} ( \varepsilon_{it}^2 - \sigma^2_\varepsilon) \varepsilon_{it}^2 +O_p\left(\frac{1}{\sqrt{N_o}} \right) &\tag{by $\+E[( \varepsilon_{it}^2 - \sigma^2_\varepsilon) \varepsilon_{is} \varepsilon_{iu}]$ if $s\neq t$ or $u \neq t$} \\
    =& \frac{2}{T_o} \xi_\varepsilon^2 +O_p\left(\frac{1}{\sqrt{N_o}} \right) & \tag{by $\xi_\varepsilon^2 = \+E[\varepsilon_{it}^4] - \sigma_\varepsilon^4$}
\end{flalign*}

Fo the term $a_4$ in \eqref{eqn:xi-proof-i}, we have 
\begin{flalign*}
   a_4 =& \frac{T_o}{N_o} \sum_i \left(\frac{1}{T_o} \sum_t \varepsilon_{it} \right)^4 \\
   \stackrel{(a)}{=}& \frac{1}{N_o T_o^3} \sum_{i,t} \varepsilon_{it}^4 + \frac{3}{N_o T_o^3} \sum_{i,t \neq s} \varepsilon_{it}^2 \varepsilon_{is}^2 + O_p\left(\frac{1}{\sqrt{N_o}}\right) & \\ \stackrel{(b)}{=}& \frac{1}{T_o^2} (\xi_\varepsilon^2 + \sigma_\varepsilon^4) + \frac{3(T_o-1)}{T_o^2} \sigma_\varepsilon^4 + O_p\left(\frac{1}{\sqrt{N_o}}\right).
\end{flalign*}
where (a) follows from the law of large numbers and $\+E[\varepsilon_{it}\varepsilon_{is}\varepsilon_{iu}\varepsilon_{iv}] = 0$ if one of $t,s,u,v$ differs from the rest; (b) follows from the law of large numbers and $\+E[\varepsilon_{it}^4] = \xi_\varepsilon^2 + \sigma_\varepsilon^4$. 

Back to \eqref{eqn:xi-proof-i}, we sum $a_2$, $a_3$ and $a_4$, and have
\begin{align*}
    a_1=& \frac{(T_o-1)^2}{T_o^2} \xi_\varepsilon^2 + \frac{3T_o-2}{T_o^2}\sigma^4_\varepsilon +O_p\left(\frac{1}{\sqrt{N_o}}\right).
\end{align*}
This concludes the proof of \eqref{eqn:xi-proof-step-1} and the proof of $\sqrt{N_o} \left(\estxisq - \xi^2_\varepsilon \right) = O_p(1)$.

\textbf{Step 4.2: Show the asymptotic normal distribution of $\estxisq$.}

From Step 4.1, the estimation error of $\estxisq$ can be decomposed as
\begin{align*}
    &\estxisq - \xi^2_\varepsilon = \frac{1}{N_o} \sum_i  \left[ \frac{T_o}{(T_o - 1)^2}   \Big( \sum_{t} ( \varepsilon_{it}^2 - \sigma^2_\varepsilon)^2  + \sum_{t\neq s} ( \varepsilon_{it}^2 - \sigma^2_\varepsilon) ( \varepsilon_{is}^2 - \sigma^2_\varepsilon) \Big)  - \frac{2}{(T_o - 1)^2}  \sum_{t,s,u} ( \varepsilon_{it}^2 - \sigma^2_\varepsilon) \varepsilon_{is} \varepsilon_{iu} \right.  \\
    & \left. + \frac{T_o^3}{(T_o - 1)^2} \left(\frac{1}{T_o} \sum_t \varepsilon_{it} \right)^4  - \frac{\sigma_\varepsilon^2}{T_o} \sum_{t} (\varepsilon_{it}^2 - \sigma_\varepsilon^2 ) - \frac{\sigma_\varepsilon^2}{ T_o(T_o-1)} \sum_{t\neq s} \varepsilon_{it} \varepsilon_{is} - \xi_\varepsilon^2 -\frac{3T_o-2}{(T_o-1)^2} \sigma_\varepsilon^4 \right] + O_p\left(\frac{1}{N_o} \right) \\
    =& \frac{1}{N_o} \sum_i f_\xi(\varepsilon_{i1}, \cdots, \varepsilon_{i,T_o})+ O_p\left(\frac{1}{N_o} \right)
\end{align*}
where we denote the term in square bracket as $f_\xi(\varepsilon_{i1}, \cdots, \varepsilon_{i,T_o})$.
As $\varepsilon_{it}$ is i.i.d. in $i$ and $t$, $f_\xi(\varepsilon_{i1}, \cdots, \varepsilon_{i,T_o})$ is i.i.d. in $i$. We can apply the standard CLT to $\estxisq$, and then $\estxisq$ is asymptotically normal with the asymptotic variance $\+E\left[f_\xi(\varepsilon_{i1}, \cdots, \varepsilon_{i,T_o})^2 \right]$. 

Similar to Step 3, we can apply multivariate Lindeberg-Feller CLT to $\hat{\tau}$, $\estsigmasq$ and $\estxisq$, and obtain the joint asymptotical normality of $\hat{\tau}$, $\estsigmasq$ and $\estxisq$.
\halmos

\end{proof}

\subsection{Proof of Theorem \ref{theorem:asymptotic-page}}

In this section, we prove Theorem \ref{theorem:asymptotic-page}. Before we start, we first state and prove a useful lemma that shows the asymptotic conditional mean and second moment of $\varepsilon_{it}$. This lemma will be an important intermediate step to show the asymptotic distribution of $\hat{\tau}_{\all, T}$ and $\estsigmasq_{\ad,2,T}$.

In the proof of this section, we introduce a random variable $\tilde{N}$ to denote the number of units in the adaptive experiment that can grow to infinity. $N$ denotes any (deterministic) realization of $\tilde{N}$. Furthermore, we let $\bm{\omega}_{\all, 1:T}(N)$ be the unit average of $Z = [Z_\fcs^\T~ Z_{\ad,1}^\T~ Z_{\ad,2}^\T]^\T \in \{-1,+1\}^{N \times T}$ in the experiment with $N$ units over $T$ periods. Similarly we index $\hat{\tau}_{\fcs,t}(N)$, $\hat{\tau}_{\ad,1,t}(N)$, $\estsigmasq_{\fcs,t}(N)$, $\estxisq_{\fcs,t}(N)$ and $\estsigmasq_{\ad,1,t}(N)$ by $N$ to specify the number of units to estimate the corresponding estimator.

For notation simplicity, we introduce 
\[\bm{\upvarphi}^\prime_T(N) = \left(\estsigmasq_{\fcs,2}(N), \estxisq_{\fcs,2}(N), \estsigmasq_{\ad,1,2}(N), \cdots, \estsigmasq_{\fcs,T}(N), \estxisq_{\fcs,T}(N), \estsigmasq_{\ad,1,T}(N) \right) \]
that includes all the information used to make the adaptive treatment decisions and to make the experiment termination decision. From Lemma \ref{lemma:asymptotic-tau-sigma-general}, each entry in $\bm{\upvarphi}^\prime_T(N)$ is consistent, and then we let 
\[\bar{\bm{\upvarphi}}^\prime  = \left(\sigma_\varepsilon^2, \xi_\varepsilon^2, \sigma_\varepsilon^2, \cdots, \sigma_\varepsilon^2, \xi_\varepsilon^2, \sigma_\varepsilon^2 \right) \]
that is the limit of $\bm{\upvarphi}^\prime_T(N)$ as $N$ grows to infinity. Moreover, we concatenate $\bm{\upvarphi}^\prime_T(N)$ with  $\bm{\omega}_{\all, 1:T}(N)$ and concatenate $\bar{\bm{\upvarphi}}^\prime$ with $\bm{\omega}_{\all, 1:T}$ to encompass the information in $\bm{\omega}_{\all, 1:T}(N)$ that affects the asymptotic distribution of $\hat{\tau}_{\all,1:T}$
\begin{align*}
    \bm{\upvarphi}_T(N) =& \left(\bm{\omega}_{\all, 1:T}(N),  \bm{\upvarphi}^\prime_T(N) \right) \\
    \bar{\bm{\upvarphi}} =& \left(\bm{\omega}_{\all, 1:T}, \bar{\bm{\upvarphi}}^\prime\right).
\end{align*}

\begin{lemma}\label{lemma:conditional-mean-variance}
In the setting of Theorem \ref{theorem:asymptotic-page}, 
the asymptotic conditional mean and second moment of $\varepsilon_{it}$ satisfy
\begin{enumerate}
    \item For any $j \in \mathcal{S}_{\fcs} \cup \mathcal{S}_{\ad,1}$, 
\begin{align}
    \+E\left[\varepsilon_{ju} \mid \tilde{N} = N, \tilde{T} = T, \bm{\upvarphi}_T(N) = \bar{\bm{\upvarphi}} + \bm{\epsilon}, Z \right]  =& 0 \label{eqn:lemma-7-3-1-1}
\end{align}
  For any $\bm{\epsilon} = (\epsilon_{\omega}, \bm{\epsilon}^\prime)$, and for any $\delta$, there exist $N_0$, such that for $N > N_0$,
  \begin{align}
  &\sup_{j \in \mathcal{S}_{\fcs} \cup \mathcal{S}_{\ad,1}, s, u} \left|\+E\left[\varepsilon_{ju} \varepsilon_{js}  \mid \tilde{N} = N, \tilde{T} = T, |\bm{\upvarphi}_T(N) - \bar{\bm{\upvarphi}}| \leq {\bm{\epsilon}}, Z \right] - \sigma_\varepsilon^2\right| < C \norm{\bm{\epsilon}^\prime}_2 + \delta  \label{eqn:lemma-7-3-1-2} \\
 & \sup_{j \in \mathcal{S}_{\fcs} \cup \mathcal{S}_{\ad,1}, s, u: s\neq u} \left|\+E\left[\varepsilon_{ju} \varepsilon_{js}  \mid \tilde{N} = N, \tilde{T} = T, |\bm{\upvarphi}_T(N) - \bar{\bm{\upvarphi}}| \leq {\bm{\epsilon}}, Z \right] \right| < C \norm{\bm{\epsilon}^\prime}_2 + \delta \label{eqn:lemma-7-3-1-3} \\
   &   \sup_{j,k \in \mathcal{S}_{\fcs} \cup \mathcal{S}_{\ad,1}, s, u: j \neq k} \left|\+E\left[N\varepsilon_{ju} \varepsilon_{ks}  \mid \tilde{N} = N, \tilde{T} = T, |\bm{\upvarphi}_T(N) - \bar{\bm{\upvarphi}}| \leq {\bm{\epsilon}}, Z \right] \right| < C \norm{\bm{\epsilon}^\prime}_2 + \delta   \label{eqn:lemma-7-3-1-4}
  \end{align}
    \item For $i \not \in \mathcal{S}_{\fcs} \cup \mathcal{S}_{\ad,1}$, 
    \begin{align}
        &\+E\left[\varepsilon_{is} \mid \tilde{N} = N, \tilde{T} = T, \bm{\upvarphi}_T(N) = \bar{\bm{\upvarphi}} + \bm{\epsilon}, Z \right]  =  0 \label{eqn:lemma-7-3-2-1} \\
        &\+E\left[\varepsilon^2_{is} \mid \tilde{N} = N, \tilde{T} = T, \bm{\upvarphi}_T(N) = \bar{\bm{\upvarphi}} + \bm{\epsilon}, Z \right]  = \sigma_\varepsilon^2 \label{eqn:lemma-7-3-2-2} \\
        &\+E\left[\varepsilon_{is} \varepsilon_{ju}  \mid \tilde{N} = N, \tilde{T} = T, \bm{\upvarphi}_T(N) = \bar{\bm{\upvarphi}} + \bm{\epsilon}, Z \right]  = 0 & (i,t) \neq (j,s)\label{eqn:lemma-7-3-2-3}
    \end{align}
\end{enumerate}
\end{lemma}

Now we prove Lemma \ref{lemma:conditional-mean-variance}.

\begin{proof}{Proof of Lemma \ref{lemma:conditional-mean-variance}.}

\textbf{Step 1: Show Lemma \ref{lemma:conditional-mean-variance}.1.}

Given that  $Z_{\fcs}$ is selected to satisfy the treated fraction condition $(2s-1-T_{\max})/T_{\max}$ at time $s$, we can partition the  $Np_{\fcs}$ NTUs into $T+1$ disjoint sets, $\mathcal{K}_{\fcs,1}, \cdots, \mathcal{K}_{\fcs,T+1}$, where units in set $\mathcal{K}_g$ are first treated at time $g$ for $g \leq T$ and units in set $\mathcal{K}_{\fcs,T+1}$ are not treated until the end of the experiment. Similarly, we partition the $Np_{\ad,1}$ ATUs into $T+1$ disjoint sets, $\mathcal{K}_{\ad,1,1} \cdots, \mathcal{K}_{\ad,1,T+1}$. 

If both $i$ and $k$ are in the same set $\mathcal{K}_{\fcs,g}$ or $\mathcal{K}_{\ad,1,g}$ for some $g$, then $\dot{z}_{iu} = \dot{z}_{ku}$ for any $u$. 
Using this property, if both $i$ and $k$ are NTU, then $\varepsilon_{iu}$ and $\varepsilon_{ku}$ are exchangeable in $\hat{\tau}_{\fcs,t}(N) $ based on the expression of $\hat{\tau}_{\fcs,t}(N) $
\[ \hat{\tau}_{\fcs,t}(N) - \tau = \left( \sum_{i,s} \dot{z}_{is}^2 \right)^\I \sum_{i,s} \dot{z}_{is} \varepsilon_{is}. \]
Moreover, $\varepsilon_{iu}$ and $\varepsilon_{ku}$ are exchangeable in $\estsigmasq_{\fcs,t}(N)$ and $\estxisq_{\fcs,t}(N)$ based on the expression of $\estsigmasq_{\fcs,t}(N)$ 
\begin{align*}
    &\estsigmasq_{\fcs,t}(N) = \frac{1}{N(t-1) p_{\fcs}} \sum_{i \in \mathcal{S}_{\fcs}, 1 \leq s \leq t} \left(\dot{y}_{is} - \hat{\tau}_{\fcs,t}(N) \cdot \dot{z}_{is} \right)^2 \\
 =& \frac{1}{N(t-1) p_{\fcs}} \sum_{i,s} \varepsilon_{is}^2 - \frac{t}{N(t-1) p_{\fcs}} \sum_i \bar{\varepsilon}^2_{i,\cdot}  - \frac{1}{t-1} \sum_s \bar{\varepsilon}^2_{\cdot,s}  + \frac{t}{t-1} \bar{\varepsilon}^2 - (\hat{\tau}_{\fcs,t}(N) - \tau)^2 \cdot \frac{1}{N(t-1) p_{\fcs}} \sum_{i,s} \dot{z}_{is}^2  
\end{align*}
and the expression of $\estxisq_{\fcs,t}(N)$
\begin{align*}
    \estxisq_{\fcs,t}(N) - \xi^2_\varepsilon =&  \frac{t^2}{(t - 1)^2}\frac{1}{N p_{\fcs} t}  \sum_i \Big(\sum_{s} \big[ (\dot{y}_{is} - \hat{\tau}_{\fcs,t}(N) \cdot \dot{z}_{is})^2 - \estsigmasq_{\fcs,t}(N) \big] \Big)^2 \\ & - \frac{3t-2}{(t-1)^2} (\estsigmasq_{\fcs,t}(N))^2 - \xi^2_\varepsilon. 
\end{align*}
As we do not use $i$ and $k$ to estimate $\hat{\tau}_{\ad,1,t}(N)$, $\estsigmasq_{\ad,1,t}(N)$ and $\estxisq_{\ad,1,t}(N)$, $i$ and $k$ are also exchangeable in $\hat{\tau}_{\ad,1,t}(N)$, $\estsigmasq_{\ad,1,t}(N)$ and $\estxisq_{\ad,1,t}(N)$. Similarly, we can show that if $i$ and $k$ are both ATU and in the same set $\mathcal{K}_{\ad,1,g}$ for some $g$, $\varepsilon_{iu}$ and $\varepsilon_{ku}$ are exchangeable in $\hat{\tau}_{\ad,1,t}(N)$, $\estsigmasq_{\ad,1,t}(N)$, $\estxisq_{\ad,1,t}(N)$, $\hat{\tau}_{\fcs,t}(N)$, $\estsigmasq_{\fcs,t}(N)$ and $\estxisq_{\fcs,t}(N)$ for all $t$.

We are going to use the set $\mathcal{K}_{\fcs,g}$ and $\mathcal{K}_{\ad,1,g}$ and this exchangeability property to show \eqref{eqn:lemma-7-3-1-2}, \eqref{eqn:lemma-7-3-1-3} and \eqref{eqn:lemma-7-3-1-4}. But before we start, let us first prove \eqref{eqn:lemma-7-3-1-1}.

We first show a useful lemma that is used to prove \eqref{eqn:lemma-7-3-1-1}.

\begin{lemma}\label{lemma:show-lemma-7-3-1-1}
    In the setting of Lemma \ref{lemma:asymptotic-tau-sigma-general}, let $g(\cdot) : \+R^{NT} \rightarrow \+R^p$ be a function that maps $\bm{\varepsilon} = [\varepsilon_{it}]_{(i,t) \in [N] \times [T]}$ to a $p$-dimensional nonnegative vector. Suppose $g(\bm{\varepsilon}) = g(-\bm{\varepsilon})$. For any $i$, conditional on $g(\bm{\varepsilon}) = \*g_0$ for any $\*g_0 \in \+R^p$, we have 
    \[\+E\left[\varepsilon_{it} \mid g(\bm{\varepsilon})  \right] = 0\]
\end{lemma}

\begin{proof}{Proof of Lemma \ref{lemma:show-lemma-7-3-1-1}}
Let the density of $\varepsilon_{it}$ be $f(\varepsilon_{it})$, let the density of $g(\bm{\varepsilon})$ be $f(g(\bm{\varepsilon}))$, and let the conditional density of $\varepsilon_{it}$ on $g(\bm{\varepsilon})$ be $f(\varepsilon_{it} \mid g(\bm{\varepsilon}))$. The conditional expectation equals
\begin{align*}
    &\+E\left[\varepsilon_{it} \mid g(\bm{\varepsilon}) = \*g_0\right] = \int_{-\infty}^{\infty} \varepsilon_{it} f(\varepsilon_{it} \mid g(\bm{\varepsilon}) = \*g_0) d\varepsilon_{it} \\
    =& \int_{-\infty}^{\infty} \varepsilon_{it} \frac{f(\varepsilon_{it}) f(g(\bm{\varepsilon} = \*g_0) \mid \varepsilon_{it})}{f(g(\bm{\varepsilon}))}  d\varepsilon_{it}  \\
    =& \frac{1}{f(g(\bm{\varepsilon}) = \*g_0)} \left[ \int_0^\infty \varepsilon_{it} f(\varepsilon_{it}) f(g(\bm{\varepsilon}) = \*g_0 \mid \varepsilon_{it}) d\varepsilon_{it} +\int_0^\infty -\varepsilon_{it} f(-\varepsilon_{it}) f(g(\bm{\varepsilon} ) = \*g_0\mid -\varepsilon_{it}) d\varepsilon_{it} \right] 
\end{align*}
As $\varepsilon_{it}$ has a symmetric distribution $f(\varepsilon_{it}) =f(-\varepsilon_{it}) $. Let $\bm{\varepsilon}_{-(it)}$ be the vector of $\bm{\varepsilon}$ excluding $\varepsilon_{it}$.  The conditional density $f(g(\bm{\varepsilon} ) = \*g_0 \mid \varepsilon_{it})$ has
\begin{flalign*}
    f(g(\bm{\varepsilon} ) = \*g_0 \mid \varepsilon_{it}) =& \int f(g(\bm{\varepsilon})  = \*g_0 \mid \bm{\varepsilon}) \cdot f(\bm{\varepsilon} \mid \varepsilon_{it}) d \bm{\varepsilon}_{-(it)} \\
    =& \int \bm{1}(g(\bm{\varepsilon})  = \*g_0) \cdot f(\bm{\varepsilon}_{-(it)}) d \bm{\varepsilon}_{-(it)} & \tag{$\varepsilon_{it}$ is i.i.d.} \\
    =& \int \bm{1}(g(-\bm{\varepsilon})  = \*g_0) \cdot f(\bm{\varepsilon}_{-(it)}) d \bm{\varepsilon}_{-(it)} & \tag{ by $g(\bm{\varepsilon}) = g(-\bm{\varepsilon})$}\\
    =& \int \bm{1}(g(-\bm{\varepsilon})  = \*g_0) \cdot f(-\bm{\varepsilon}_{-(it)}) d \bm{\varepsilon}_{-(it)} & \tag{$\varepsilon_{it}$ has a symmetric distribution}\\
    =& \int \bm{1}(g(\bm{-\varepsilon})  = \*g_0) \cdot f(-\bm{\varepsilon} \mid - \varepsilon_{it}) d \bm{\varepsilon}_{-(it)} & \tag{$\varepsilon_{it}$ is i.i.d.} \\
    =& f(g(\bm{\varepsilon} ) = \*g_0 \mid -\varepsilon_{it}) 
\end{flalign*}
Then the conditional expectation has 
\begin{align*}
    \+E\left[\varepsilon_{it} \mid g(\bm{\varepsilon}) \right] =& \frac{1}{f(g(\bm{\varepsilon}) = \*g_0)} \left[ \int_0^\infty \varepsilon_{it} f(\varepsilon_{it}) f(g(\bm{\varepsilon}) = \*g_0 \mid \varepsilon_{it}) d\varepsilon_{it} +\int_0^\infty -\varepsilon_{it} f(\varepsilon_{it}) f(g(\bm{\varepsilon} ) = \*g_0\mid \varepsilon_{it}) d\varepsilon_{it} \right] = 0.
\end{align*}
We conclude the proof of Lemma \ref{lemma:show-lemma-7-3-1-1}.
    \halmos
\end{proof}

Now we apply Lemma \ref{lemma:show-lemma-7-3-1-1} to prove \eqref{eqn:lemma-7-3-1-1}. Let $g(\bm{\varepsilon}) = \bm{\upvarphi}^\prime_T(N)$. For such $g(\bm{\varepsilon})$, the condition $g(\bm{\varepsilon}) = g(-\bm{\varepsilon})$ is satisfied, based on the expression of $\hat{\tau}_{\fcs,t}(N)$,  $\estsigmasq_{\fcs,t}(N)$ and $\estxisq_{\fcs,t}(N)$ and $\estsigmasq_{\ad,1,t}(N)$. 
Applying Lemma \ref{lemma:show-lemma-7-3-1-1}, we have 
\[\+E\left[\varepsilon_{it} \mid \tilde{N} = N, \tilde{T} = T, \bm{\upvarphi}^\prime_T(N) = \bar{\bm{\upvarphi}}^\prime + \bm{\epsilon}^\prime \right] = 0.\]
Note that $\bm{\upvarphi}_T(N)$ is  $\bm{\upvarphi}^\prime_T(N)$ concatenated with $\bm{\omega}_{\all,1:T}(N)$. $\bm{\upvarphi}^\prime_T(N)$ encompasses all the information in $\bm{\omega}_{\all,1:T}(N)$ that is relevant to $\varepsilon_{it}$. Furthermore, $Z_{\ad}$ is randomly chosen subject to the treated fraction constraints $\bm{\omega}_{\ad,1:T}(N)$. We can then write both $Z$ and $Z_{\ad}$ as a function of  $(\estsigmasq_{\fcs,2}(N),  \estxisq_{\fcs,2}(N), \cdots, \estsigmasq_{\fcs,T}(N), \estxisq_{\fcs,T}(N))$ and some random variable $\eta$ that determines which units are treated and which are not subject to the treated fraction constraints. By definition, $\eta$ is independent of $(\estsigmasq_{\fcs,2}(N),  \estxisq_{\fcs,2}(N), \cdots, \estsigmasq_{\fcs,T}(N), \estxisq_{\fcs,T}(N))$ and $\varepsilon_{ju}$ for any $j$ and $u$. Using $\eta$, we have 
\begin{align*}
    & \+E\left[\varepsilon_{ju} \mid \tilde{N} = N, \tilde{T} = T, \bm{\upvarphi}_T(N) = \bar{\bm{\upvarphi}} + \bm{\epsilon}, Z \right]  \\
    =& \+E\left[\varepsilon_{it} \mid \tilde{N} = N, \tilde{T} = T, \bm{\upvarphi}^\prime_T(N) = \bar{\bm{\upvarphi}}^\prime + \bm{\epsilon}^\prime \right] = 0.
\end{align*}
This concludes the proof of \eqref{eqn:lemma-7-3-1-1}.

To show \eqref{eqn:lemma-7-3-1-2}, \eqref{eqn:lemma-7-3-1-3} and \eqref{eqn:lemma-7-3-1-4}, we first show two useful lemmas.

\begin{lemma}\label{lemma:show-lemma-7-3-1-2-4}
    In the setting of Lemma \ref{lemma:asymptotic-tau-sigma-general}, let $g(\cdot) : \+R^{NT} \rightarrow \+R^p$ be a function that maps $\bm{\varepsilon} = [\varepsilon_{it}]_{(i,t) \in [N] \times [T]}$ to a $p$-dimensional vector, and let $h(\cdot) : \+R^{N} \rightarrow \+R^q$ be a function that maps $\bm{\varepsilon}_i = [\varepsilon_{it}]_{t \in  [T]}$ to a $q$-dimensional vector. Let $\mathcal{S}$ be the set of indices, where for any $i$ and $j$ in $\mathcal{S}$, $\bm{\varepsilon}_i $ and $\bm{\varepsilon}_j $ are exchangeable in $g(\bm{\varepsilon})$. For all $i \in \mathcal{S}$, conditional on $g(\bm{\varepsilon}) = \*g_0$ and $\sum_{i \in \mathcal{S}} h(\bm{\varepsilon}_i) = \*h_0$ for any $\*g_0 \in \+R^p$ and $\*h_0 \in \+R^q$,
    \begin{align}
        \+E\left[h(\bm{\varepsilon}_i) \mid  \sum_{k \in \mathcal{S}} h(\bm{\varepsilon}_k) = \*h_0, g(\bm{\varepsilon}) = \*g_0  \right] = \frac{\*h_0}{|\mathcal{S}|}.
    \end{align}
\end{lemma}

\begin{proof}{Proof of Lemma \ref{lemma:show-lemma-7-3-1-2-4}}
    Based on the condition in Lemma \ref{lemma:show-lemma-7-3-1-2-4}, if $i$ and $j$ in $\mathcal{S}$, then 
    \[  \+E\left[h(\bm{\varepsilon}_i) \mid  \sum_{k \in \mathcal{S}} h(\bm{\varepsilon}_k) = \*h_0, g(\bm{\varepsilon}) = \*g_0  \right]  =  \+E\left[h(\bm{\varepsilon}_j) \mid  \sum_{k \in \mathcal{S}} h(\bm{\varepsilon}_k) = \*h_0, g(\bm{\varepsilon}) = \*g_0  \right]. \]
    Using this property, we have 
    \begin{align*}
        \sum_{i \in \mathcal{S}}  \+E\left[h(\bm{\varepsilon}_i) \mid  \sum_{k \in \mathcal{S}} h(\bm{\varepsilon}_k) = \*h_0, g(\bm{\varepsilon}) = \*g_0  \right] =& \*h_0 \\
        \Leftrightarrow	 |\mathcal{S}| \+E\left[h(\bm{\varepsilon}_i) \mid  \sum_{k \in \mathcal{S}} h(\bm{\varepsilon}_k) = \*h_0, g(\bm{\varepsilon}) = \*g_0  \right] =& \*h_0
    \end{align*}
    We conclude the proof of Lemma \ref{lemma:show-lemma-7-3-1-2-4}. \halmos
\end{proof}

\begin{lemma}\label{lemma:conditional-expectation}
    Suppose $\{X_n\}$ and $\{Y_n\}$ are sequences of bounded random variables.  As $n \rightarrow \infty$, $\{X_n\}$ and $\{Y_n\}$ converge to a joint normal distribution 
    \[\sqrt{n}\left( \begin{bmatrix}
        X_n \\ Y_n
    \end{bmatrix} - \begin{bmatrix}
       \mu_X \\ \mu_Y
    \end{bmatrix}\right) \xrightarrow{d} \mathcal{N} \left(\begin{bmatrix}
        \bm{0} \\ \bm{0}
    \end{bmatrix}, \begin{bmatrix}
        \Sigma_{XX} & \Sigma_{XY} \\ \Sigma_{YX} & \Sigma_{YY}
    \end{bmatrix} \right) \,\, . \]
    Then for any $y \in \+R^q$, the asymptotic conditional expectation of  $X$ on $Y_n = y$ is
    \[ \lim_{n \rightarrow \infty} \+E\left[X_n \mid Y_n = y \right]= \mu_X + \Sigma_{XY} \Sigma_{YY}^\I (y - \mu_Y). \]
\end{lemma}

\begin{proof}{Proof of Lemma \ref{lemma:conditional-expectation}}
    Let $Z_n =X_n  + B Y_n $, where $B = - \Sigma_{XY} \Sigma_{YY}^\I$. Then the conditional covariance between $Y_n$ and $Z_n$ is 
    \begin{align*}
        \lim_{n \rightarrow \infty}\mathrm{Cov}(Z_n, Y_n) =& \lim_{n \rightarrow \infty} \mathrm{Cov}(X_n, Y_n) +  \lim_{n \rightarrow \infty}  \mathrm{Cov}(B Y_n, Y_n) \\
        =& \Sigma_{XY}  - \Sigma_{XY} \Sigma_{YY}^\I \Sigma_{YY} = \bm{0} 
    \end{align*}
    As $Z_n$ is a linear combination of $X_n$ and $Y_n$, we can show that $Y_n$ and $Z_n$ are joint asymptotically normal. As they are asymptotically uncorrelated, $Y_n$ and $Z_n$ are asymptotically independent. As the asymptotic expectation of $Z_n$ is $ \lim_{n \rightarrow \infty}\+E\left[Z_n  \right] = \mu_X + B \mu_Y$, we have 
    \begin{align*}
        \lim_{n \rightarrow \infty}\+E\left[X_n \mid Y_n = y \right] =& \lim_{n \rightarrow \infty}\+E\left[Z_n - B Y_n \mid Y_n = y \right] \\
        =& \lim_{n \rightarrow \infty}\+E\left[Z_n \mid Y_n = y \right] - B \cdot \lim_{n \rightarrow \infty}\+E\left[ Y_n \mid Y_n = y \right] \\
        =& \mu_X + B \mu_Y - B y \\
        =& \mu + \Sigma_{XY} \Sigma_{YY}^\I (\mu_Y - y).
    \end{align*}
    This concludes the proof of Lemma \ref{lemma:conditional-expectation}. \halmos
\end{proof}

Next, let us consider the asymptotic conditional second moment of $\varepsilon_{ju}$. 
Suppose $j$ is in $\mathcal{K}$, where $\mathcal{K} \in \{\mathcal{K}_{\fcs,1}, \cdots, \mathcal{K}_{\fcs,T+1}, \mathcal{K}_{\ad,1,1}, \cdots, \mathcal{K}_{\ad,1,T+1}\}$. Let
\begin{align*}
    h(\bm{\varepsilon}_i) =& \varepsilon_{iu}^2 - \sigma_\varepsilon^2  \\
    h_{\mathcal{K}u}(\bm{\varepsilon}) =& \frac{1}{|\mathcal{K}|} \sum_{j \in \mathcal{K}} [\varepsilon_{iu}^2 - \sigma_\varepsilon^2]
\end{align*}
By applying Lemma \ref{lemma:show-lemma-7-3-1-2-4}, the conditional second moment equals
\begin{flalign*}
    & \+E\left[\varepsilon^2_{ju} \mid \bm{\upvarphi}^\prime_T(N) \right] \\
    =&  \+E\left[ \+E\left[\varepsilon^2_{ju} \mid h_{\mathcal{K}u}(\bm{\varepsilon}), \bm{\upvarphi}^\prime_T(N)  \right] \mid \bm{\upvarphi}^\prime_T(N) \right] \\
    =& \sigma_\varepsilon^2 + \+E\left[  h_{\mathcal{K}u}(\bm{\varepsilon})  \mid \bm{\upvarphi}^\prime_T(N) \right]
\end{flalign*}
As $N \rightarrow \infty$ and $T_{\max}$ is fixed, the number of units in $\mathcal{K}$ grows to infinity, that is, $|\mathcal{K}| \rightarrow \infty$. The randomness of $h_{\mathcal{K}u}(\bm{\varepsilon})$ and $\bm{\upvarphi}^\prime_T(N)$ comes from $\varepsilon_{it}$, where $\varepsilon_{it}$ is i.i.d. in $i$ and $t$. We can apply multivariate CLT to $h_{\mathcal{K}u}(\bm{\varepsilon})$ and $\bm{\upvarphi}^\prime_T(N)$  (similar to Step 4.2 in the proof of Lemma \ref{lemma:asymptotic-tau-sigma-general}), and show that $\sigma_\varepsilon^2$ and $\bm{\upvarphi}^\prime_T(N)$ are consistent and joint asymptotically normal with 
\begin{equation}\label{eqn:joint-asymptotic-normal-h-g}
    \begin{aligned}
        & \sqrt{N p_{\fcs}} \left(\begin{bmatrix}
    h_{\mathcal{K}u}(\bm{\varepsilon}) \\ \bm{\upvarphi}^\prime_T(N)
\end{bmatrix} - \begin{bmatrix}
    0 \\ 
    \bar{\bm{\upvarphi}}^\prime
\end{bmatrix} 
\right) \\ \xrightarrow{d} &\mathcal{N} \left(\begin{bmatrix}
    0 \\ \bm{0}_{3(T-1)}
\end{bmatrix}, \begin{bmatrix}
    A\mathrm{Var}(h_{\mathcal{K}u}(\bm{\varepsilon}))& A\mathrm{Cov} (h_{\mathcal{K}u}(\bm{\varepsilon}), \bm{\upvarphi}^\prime_T(N)^\T) \\ A\mathrm{Cov} (\bm{\upvarphi}^\prime_T(N), h_{\mathcal{K}u}(\bm{\varepsilon})) & A\mathrm{Var}(\bm{\upvarphi}^\prime_T(N))
\end{bmatrix} \right) \,\,. 
    \end{aligned} 
\end{equation}
We can then apply Lemma \ref{lemma:conditional-expectation}, as $\varepsilon_{ju}$ is bounded, we have for any $j$ and $u$, 
\begin{align*}
    & \lim_{N \rightarrow \infty} \+E\left[\varepsilon^2_{ju}  \mid \tilde{N} = N, \tilde{T} = T, \bm{\upvarphi}^\prime_T(N) - \bar{\bm{\upvarphi}}^\prime = \bm{\epsilon}^\prime\right] - \sigma_\varepsilon^2 \\
    =& \lim_{N \rightarrow \infty} \+E\left[\varepsilon^2_{ju} \mid \bm{\upvarphi}^\prime_T(N) - \bar{\bm{\upvarphi}}^\prime = \bm{\epsilon}^\prime\right]  - \sigma_\varepsilon^2
    \tag{$\bm{\upvarphi}^\prime_T(N)$ encompasses all the information in $\tilde{T}$} \\
    =& \lim_{N \rightarrow \infty} \+E\left[h_{\mathcal{K}u}(\bm{\varepsilon}) \mid \bm{\upvarphi}^\prime_T(N) - \bar{\bm{\upvarphi}}^\prime = \bm{\epsilon}^\prime\right]  \\
    =& A\mathrm{Cov} (h_{\mathcal{K}u}(\bm{\varepsilon}), \bm{\upvarphi}^\prime_T(N)^\T)  \cdot A\mathrm{Var}(\bm{\upvarphi}^\prime_T(N))^\I \cdot \bm{\epsilon}^\prime \,. 
\end{align*}
Equivalently for any $\tilde{\delta}$, there exists some $N_0$ such that for $N > N_0$, and for any $j \in \mathcal{K}$ and $u$, 
\begin{align*}
    & \left| \+E\left[\varepsilon^2_{ju} \mid \tilde{N} = N, \tilde{T} = T, \bm{\upvarphi}^\prime_T(N) - \bar{\bm{\upvarphi}}^\prime = \bm{\epsilon}^\prime\right] - \sigma_\varepsilon^2 \right| \\
    < & \left| A\mathrm{Cov} (h_{\mathcal{K}u}(\bm{\varepsilon}), \bm{\upvarphi}^\prime_T(N)^\T)  \cdot A\mathrm{Var}(\bm{\upvarphi}^\prime_T(N))^\I \cdot \bm{\epsilon}^\prime \right|  + {\delta}.
\end{align*}
Then we have for any ${\delta}$, there exists some $N_0$ such that for $N > N_0$
\begin{align*}
    & \sup_{j \in \mathcal{S}_{\fcs} \cup \mathcal{S}_{\ad,1},u} \left|\+E\left[\varepsilon^2_{ju} \mid \tilde{N} = N, \tilde{T} = T, \bm{\upvarphi}^\prime_T(N) - \bar{\bm{\upvarphi}}^\prime = \bm{\epsilon}^\prime\right] - \sigma^2_\varepsilon \right| \\
    <& \sup_{\mathcal{K},u}  \left| A\mathrm{Cov} (h_{\mathcal{K}u}(\bm{\varepsilon}), \bm{\upvarphi}^\prime_T(N)^\T)  \cdot A\mathrm{Var}(\bm{\upvarphi}^\prime_T(N))^\I \cdot \bm{\epsilon}^\prime \right|  + {\delta}  \\
    \leq& \left(\sup_{\mathcal{K},u}  \norm{A\mathrm{Cov} (h_{\mathcal{K}u}(\bm{\varepsilon}), \bm{\upvarphi}^\prime_T(N)^\T)}_2 \right) \cdot \norm{A\mathrm{Var}(\bm{\upvarphi}^\prime_T(N))^\I}_2 \cdot \norm{\bm{\epsilon}^\prime}_2 + {\delta}  \\
    <& C \norm{\bm{\epsilon}^\prime}_2 +  {\delta}.
\end{align*}
for some constant $C$.

Following the same argument as the proof of \eqref{eqn:lemma-7-3-1-1}, $\bm{\upvarphi}^\prime_T(N) $ encompasses all the information in $\bm{\omega}_{\all,1:T}(N)$ and $Z$ that is relevant to $\varepsilon_{ju}$. Therefore we have for any ${\delta}$, there exists some $N_0$ such that for $N > N_0$, 
\begin{align*}
&\sup_{j \in \mathcal{S}_{\fcs} \cup \mathcal{S}_{\ad,1}, u} \left|\+E\left[\varepsilon_{ju}^2  \mid \tilde{N} = N, \tilde{T} = T, |\bm{\upvarphi}_T(N) - \bar{\bm{\upvarphi}}| \leq {\bm{\epsilon}}, Z \right] - \sigma_\varepsilon^2\right| \\
=&\sup_{j \in \mathcal{S}_{\fcs} \cup \mathcal{S}_{\ad,1},u} \left|\+E\left[\varepsilon^2_{ju} \mid \tilde{N} = N, \tilde{T} = T, |\bm{\upvarphi}^\prime_T(N) - \bar{\bm{\upvarphi}}^\prime| \leq \bm{\epsilon}^\prime\right] - \sigma^2_\varepsilon \right| \\
<& \sup_{\tilde{\bm{\epsilon}}^\prime: |\tilde{\bm{\epsilon}}^\prime| < \bm{\epsilon}^\prime } C \norm{\tilde{\bm{\epsilon}}^\prime}_2 +  {\delta} \\
=& C \norm{{\bm{\epsilon}}^\prime}_2 +  {\delta}
\end{align*} 
We conclude the proof of \eqref{eqn:lemma-7-3-1-2}.

Next let us consider the conditional covariance between $\varepsilon_{ju}$ and $\varepsilon_{js}$ for $u \neq s$. Suppose $j$ is in $\mathcal{K}$, where $\mathcal{K} \in \{\mathcal{K}_{\fcs,1}, \cdots, \mathcal{K}_{\fcs,T+1}, \mathcal{K}_{\ad,1,1}, \cdots, \mathcal{K}_{\ad,1,T+1}\}$. Let 
\begin{align*}
    h(\bm{\varepsilon}_j) =& \varepsilon_{js}\varepsilon_{ju} \\
    h_{\mathcal{K}su}(\bm{\varepsilon}) =& \frac{1}{|\mathcal{K}|} \sum_{j \in \mathcal{K}} \varepsilon_{js}\varepsilon_{ju} \,.
\end{align*}
For this definition of $h_{\mathcal{K}su}(\bm{\varepsilon})$, we can similarly show a joint asymptotic normal distribution of $h_{\mathcal{K}su}(\bm{\varepsilon})$ and $\bm{\upvarphi}^\prime_T(N)$ as \eqref{eqn:joint-asymptotic-normal-h-g}, and we can use a similar approach as the proof of  \eqref{eqn:lemma-7-3-1-2} to show that for any $\delta$, there exist $N_0$ such that for any $N > N_0$, we have
\begin{align*}
    & \sup_{j \in \mathcal{S}_{\fcs} \cup \mathcal{S}_{\ad,1}, s,u} \left|\+E\left[\varepsilon_{ju} \varepsilon_{js}  \mid \tilde{N} = N, \tilde{T} = T, |\bm{\upvarphi}_T(N) - \bar{\bm{\upvarphi}}| \leq {\bm{\epsilon}}, Z \right]\right| \\
    =& \sup_{j \in \mathcal{S}_{\fcs} \cup \mathcal{S}_{\ad,1}, s,u} \left|\+E\left[\varepsilon_{ju} \varepsilon_{js}  \mid \tilde{N} = N, \tilde{T} = T, |\bm{\upvarphi}^\prime_T(N) - \bar{\bm{\upvarphi}}^\prime| \leq {\bm{\epsilon}}^\prime \right]\right| \\
    \leq& \sup_{\tilde{\bm{\epsilon}}^\prime: |\tilde{\bm{\epsilon}}^\prime| \leq \bm{\epsilon}^\prime} \left(\sup_{\mathcal{K},s,u}  \norm{A\mathrm{Cov} (h_{\mathcal{K}su}(\bm{\varepsilon}), \bm{\upvarphi}^\prime_T(N)^\T)}_2 \right) \cdot \norm{A\mathrm{Var}(\bm{\upvarphi}^\prime_T(N))^\I}_2 \cdot \norm{\bm{\epsilon}^\prime}_2 + {\delta}   \\
    \leq& C \norm{\bm{\epsilon}^\prime}_2 + \delta 
\end{align*}
for some constant $C$. 
This concludes the proof of \eqref{eqn:lemma-7-3-1-3}. 

Finally, let us consider the asymptotic covariance between $\varepsilon_{ju}$ and $\varepsilon_{ks}$ for $k \neq j$. Suppose $j$ is in $\mathcal{K}$
\begin{align*}
    h(\bm{\varepsilon}) =& \varepsilon_{ju}\varepsilon_{ks} \\
    h^\prime_{\mathcal{K}su}(\bm{\varepsilon}) =& \frac{1}{|\mathcal{K}|} \sum_{j \in \mathcal{K}}  \varepsilon_{ju}\varepsilon_{k_j s}
\end{align*}
where $\{k_l\}_{l \in \mathcal{K}}$ is a randomly selected sequence of indices that satisfy $k_l \in [N]$, satisfy $k_l \neq k_{l^\prime}$ and $k_l \neq m$ for any $l^\prime, m \in \mathcal{K}$ with $l^\prime \neq l$, and satisfy $k_l = k$ for $l = j$. Given this sequence, we have $\varepsilon_{ju}\varepsilon_{k_j s} $ to be i.i.d. for any $j \in \mathcal{K}$. Then $h^\prime_{\mathcal{K}su}(\bm{\varepsilon}) $ is asymptotically normal. We can then show a joint asymptotic normal distribution of $h^\prime_{\mathcal{K}su}(\bm{\varepsilon}) $ and $\bm{\upvarphi}^\prime_T(N)$. In this joint distribution, we note that the covariance between $h^\prime_{\mathcal{K}su}(\bm{\varepsilon}) $ and $\bm{\upvarphi}^\prime_T(N)$ has 
\[\mathrm{Cov} \left(h_{\mathcal{K}su}(\bm{\varepsilon}), \bm{\upvarphi}^\prime_T(N)^\T \right) = O \left({1}/{N} \right) \]
because $h_{\mathcal{K}su}(\bm{\varepsilon})$ is an average of residuals of two different units and the weight of $h_{\mathcal{K}su}(\bm{\varepsilon})$ in $\bm{\upvarphi}^\prime_T(N)$ is at the order of $O(1/(N p_{\fcs}))$. As $\varepsilon_{ju}$ is bounded, we have 
\begin{align*}
    & \sup_{i,j \in \mathcal{S}_{\fcs} \cup \mathcal{S}_{\ad,1},s,u} \left|\+E\left[N\varepsilon_{iu} \varepsilon_{js}  \mid \tilde{N} = N, \tilde{T} = T, |\bm{\upvarphi}_T(N) - \bar{\bm{\upvarphi}}| \leq {\bm{\epsilon}}, Z \right]\right| \\
    =& \sup_{i,j \in \mathcal{S}_{\fcs} \cup \mathcal{S}_{\ad,1},s,u} \left|\+E\left[N\varepsilon_{iu} \varepsilon_{js}  \mid \tilde{N} = N, \tilde{T} = T, |\bm{\upvarphi}^\prime_T(N) - \bar{\bm{\upvarphi}}^\prime| \leq {\bm{\epsilon}}^\prime \right]\right| \\
    \leq& \sup_{\tilde{\bm{\epsilon}}^\prime: |\tilde{\bm{\epsilon}}^\prime| \leq \bm{\epsilon}^\prime} \left(\sup_{\mathcal{K},s,u}  \norm{A\mathrm{Cov} (Nh_{\mathcal{K}jsu}(\bm{\varepsilon}), \bm{\upvarphi}^\prime_T(N)^\T)}_2 \right) \cdot \norm{A\mathrm{Var}(\bm{\upvarphi}^\prime_T(N))^\I}_2 \cdot \norm{\bm{\epsilon}^\prime}_2 + {\delta}   \\
    \leq& C \norm{\bm{\epsilon}^\prime}_2 + \delta 
\end{align*}
This concludes the proof of \eqref{eqn:lemma-7-3-1-4}.

\textbf{Step 2: Show Lemma \ref{lemma:conditional-mean-variance}.2.}

The expectations in Lemma \ref{lemma:conditional-mean-variance}.2 condition on $\estsigmasq_{\fcs,t}(N)$, $\estxisq_{\fcs,t}(N)$ and $\estsigmasq_{\ad,1,t}(N)$, for any $2 \leq t \leq T$. Since both $\estsigmasq_{\fcs,t}(N)$ and $\estxisq_{\fcs,t}(N)$ are estimated using NTU only, the randomness of both $\estsigmasq_{\fcs,t}(N)$ and $\estxisq_{\fcs,t}(N)$ come from $\varepsilon_{is}$ for $i $ in NTU ($i \in \mathcal{S}_\fcs$) and $1 \leq s\leq t$. Similarly, the randomness of $\estsigmasq_{\ad,1,t}(N)$ comes from $\varepsilon_{is}$ for $i \in \mathcal{S}_{\ad,1}$ and $1 \leq s\leq t$.  Furthermore, since $\varepsilon_{it}$ is i.i.d.,  $\varepsilon_{is}$ is independent of $\varepsilon_{ju}$ for $i \in \mathcal{S}_\fcs \cup \mathcal{S}_{\ad,1}$ and $j \not \in \mathcal{S}_\fcs \cup \mathcal{S}_{\ad,1}$, and for any $s$ and $u$. Therefore if $j \not \in \mathcal{S}_{\fcs}  \cup \mathcal{S}_{\ad,1}$ and for $u \in [T]$, the asymptotic expectations conditional on $\estsigmasq_{\fcs,t}(N)$, $\estxisq_{\fcs,t}(N)$ and $\estsigmasq_{\ad,1,t}(N)$ equal to the corresponding unconditional expectations, that is, for $i \not \in \mathcal{S}_{\fcs} \cup \mathcal{S}_{\ad,1}$,
\begin{align*}
&\+E\left[\varepsilon_{iu}  \mid \tilde{N} = N, \tilde{T} = T, \bm{\upvarphi}_T(N) - \bar{\bm{\upvarphi}} = {\bm{\epsilon}}, Z \right] = \+E\left[\varepsilon_{iu}  \mid \tilde{N} = N, \tilde{T} = T, \bm{\upvarphi}^\prime_T(N) - \bar{\bm{\upvarphi}}^\prime = {\bm{\epsilon}}^\prime \right] = \+E\left[\varepsilon_{iu}  \right] = 0  \\
&\+E\left[\varepsilon_{iu}^2  \mid \tilde{N} = N, \tilde{T} = T, \bm{\upvarphi}_T(N) - \bar{\bm{\upvarphi}} = {\bm{\epsilon}}, Z \right] =    \+E\left[\varepsilon_{iu}^2  \mid \tilde{N} = N, \tilde{T} = T, \bm{\upvarphi}^\prime_T(N) - \bar{\bm{\upvarphi}}^\prime = {\bm{\epsilon}}^\prime \right] = \+E\left[\varepsilon_{iu}^2  \right] = \sigma_\varepsilon^2 \\
& \+E\left[\varepsilon_{iu} \varepsilon_{js}  \mid \tilde{N} = N, \tilde{T} = T, \bm{\upvarphi}_T(N) - \bar{\bm{\upvarphi}} = {\bm{\epsilon}}, Z \right] =    \+E\left[\varepsilon_{iu} \varepsilon_{js}  \mid \tilde{N} = N, \tilde{T} = T, \bm{\upvarphi}^\prime_T(N) - \bar{\bm{\upvarphi}}^\prime = {\bm{\epsilon}}^\prime \right] = \+E\left[\varepsilon_{iu} \varepsilon_{js}  \right] = 0 
\end{align*}
This concludes the proof of \eqref{eqn:lemma-7-3-2-1}, \eqref{eqn:lemma-7-3-2-2}, and \eqref{eqn:lemma-7-3-2-3}. \halmos

\end{proof}

\bigskip
Now we are ready to present the proof of Theorem \ref{theorem:asymptotic-page}.

\begin{proof}{Proof of Theorem \ref{theorem:asymptotic-page}.}

\textbf{Step 1: Show the asymptotic normal distribution of $\hat{\tau}_{\all,T}$.}

Based on Lemma \ref{lemma:asymptotic-tau-sigma-general},
the estimation error of $\hat{\tau}_{\all,\tilde{T}}$ can be written as 
\[\hat{\tau}_{\all,\tilde{T}} - \tau = \frac{1}{N \tilde{T}} \sum_{i,s} \underbrace{\left(\frac{1}{N \tilde{T}} \sum_{i,s} \dot{z}^2_{is}  \right)^\I}_{\funfrac(\bm{\omega}_{\all,1:\tilde{T}}(N),\tilde{T})^\I} \underbrace{   \left(z_{is} - \sum_i \bar{z}_{i,\cdot} - \sum_s \bar{z}_{\cdot,s} + \bar{z}\right) }_{\zeta_{is}}
 \varepsilon_{is}. \]

  Let us consider the mean of the estimation error (multiplied by $N$ and $\tilde{T}\funfrac(\bm{\omega}_{\all,1:\tilde{T}}(N),\tilde{T})/\sigma_\varepsilon^2$) 
 \begin{flalign*}
 & \+E\left[ N \tilde{T}\funfrac(\bm{\omega}_{\all,1:\tilde{T}}(N),\tilde{T})/\sigma_\varepsilon^2 \cdot (\hat{\tau}_{\all,\tilde{T}} - \tau) \mid \tilde{N} = N \right]\\
 =& \sum_{T \leq T_{\rm max}} \+E\left[ N \tilde{T}\funfrac(\bm{\omega}_{\all,1:T}(N),{T})/\sigma_\varepsilon^2 \cdot (\hat{\tau}_{\all,{T}} - \tau) \mid \tilde{N} = N, \tilde{T} = T \right] P(\tilde{T} = T \mid \tilde{N} = N)\\
 =&   \sum_{T \leq T_{\rm max}} \sum_{\bm{\epsilon}  } (\sigma_\varepsilon^2)^{-1} \bigg[  \sum_{i,s \leq T}  \underbrace{\+E\left[\zeta_{is} \cdot \varepsilon_{is} \mid \tilde{N} = N, \tilde{T} = T, \bm{\upvarphi}_T(N) = \bar{\bm{\upvarphi}} + \bm\epsilon \right]}_{\coloneqq a_{is}} \bigg] \\ & \cdot P(\tilde{T} = T, \bm{\upvarphi}_T(N) = \bar{\bm{\upvarphi}} + \bm\epsilon \mid \tilde{N} = N) \\
 =& 0 
 \end{flalign*}
 where the last line uses that for any $i$ and $s$
\begin{align*}
    a_{is} =& \+E\Bigg[ \zeta_{is} \cdot \underbrace{\+E\left[\varepsilon_{is} \mid  \tilde{N} = N, \tilde{T} = T, \bm{\upvarphi}_T(N) = \bar{\bm{\upvarphi}} + \bm\epsilon, Z\right]}_{= 0 \text{ from \eqref{eqn:lemma-7-3-1-1} and \eqref{eqn:lemma-7-3-2-1} in Lemma \ref{lemma:conditional-mean-variance}}}\mid \tilde{N} = N, \tilde{T} = T, \bm{\upvarphi}_T(N) = \bar{\bm{\upvarphi}} + \bm\epsilon \Bigg] \\
    =& 0.
\end{align*}
 
Let us consider the second moment of the estimation error (multiplied by $N$ and $\tilde{T}\funfrac(\bm{\omega}_{\all,1:\tilde{T}}(N),\tilde{T})/\sigma_\varepsilon^2$)
\begin{align*}
     & \+E[N \tilde{T}\funfrac(\bm{\omega}_{\all,1:\tilde{T}}(N),\tilde{T})/\sigma_\varepsilon^2 \cdot (\hat{\tau}_{\all,\tilde{T}} - \tau)^2 - 1\mid \tilde{N}= N] \\
     =& \sum_{T \leq T_{\rm max}} \+E\left[ N \tilde{T}\funfrac(\bm{\omega}_{\all,1:T}(N),\tilde{T})/\sigma_\varepsilon^2 \cdot (\hat{\tau}_{\all,\tilde{T}} - \tau)^2  - 1\mid \tilde{N} = N, \tilde{T} = T \right] P(\tilde{T} = T \mid \tilde{N} = N)\\
     =& (\sigma_\varepsilon^2)^{-1} \sum_{T \leq T_{\rm max}} \underbrace{\bigg[ \frac{1}{NT}  \sum_{i,s \leq T} \+E\left[ \funfrac(\bm{\omega}_{\all,1:T}(N),{T})^\I \cdot \zeta_{is}^2 (\varepsilon_{is}^2 - \sigma_\varepsilon^2) \mid \tilde{N} = N, \tilde{T} = T \right] \bigg] }_{a_{NT}} P(\tilde{T} = T \mid \tilde{N} = N)\\
     & + (\sigma_\varepsilon^2)^{-1} \sum_{T \leq T_{\rm max}} \underbrace{\bigg[ \frac{1}{NT} \sum_{i,s,u\leq T} \+E\left[ \funfrac(\bm{\omega}_{\all,1:T}(N),{T})^\I \cdot \zeta_{is} \zeta_{iu}  \varepsilon_{is} \varepsilon_{iu} \mid \tilde{N} = N, \tilde{T} = T \right] \bigg] }_{b_{NT}} P(\tilde{T} = T \mid \tilde{N} = N)\\
     & + (\sigma_\varepsilon^2)^{-1} \sum_{T \leq T_{\rm max}} \underbrace{\bigg[ \frac{1}{NT} \sum_{i,j: i \neq j,s,u\leq T} \+E\left[ \funfrac(\bm{\omega}_{\all,1:T}(N),{T})^\I \cdot \zeta_{is} \zeta_{ju} \varepsilon_{is} \varepsilon_{ju} \mid \tilde{N} = N, \tilde{T} = T \right] \bigg] }_{c_{NT}}  P(\tilde{T} = T \mid \tilde{N} = N)
\end{align*}

As each entry $\bm{\upvarphi}_T(N)$ in converge (in probability) to $\bar{\bm{\upvarphi}}$, have for any generic sequence of bounded random variables $X_{NT}$, for any $(\bm\epsilon, \delta)$, there exist $N_0$ such that for any $N > N_0$, we have 
\begin{align}
   & \nonumber \left|\+E\left[X_{NT} \mid \tilde{N} = N, \tilde{T} = T\right] \right| \\
  \nonumber =& |\+E\left[X_{NT} \mid  \tilde{N} = N, \tilde{T} = T, \bm{\upvarphi}_T(N) - \bar{\bm{\upvarphi}} \leq  \bm\epsilon \mid \right] \cdot P(|\bm{\upvarphi}_T(N) - \bar{\bm{\upvarphi}}| \leq  \bm\epsilon \mid \tilde{N} = N, \tilde{T} = T) \\ 
   \nonumber & + \+E\left[X_{NT} \mid  \tilde{N} = N, \tilde{T} = T, |\bm{\upvarphi}_T(N) - \bar{\bm{\upvarphi}}| >  \bm\epsilon \right] \cdot \underbrace{P(|\bm{\upvarphi}_T(N) - \bar{\bm{\upvarphi}}| >  \bm\epsilon \mid \tilde{N} = N, \tilde{T} = T) }_{\substack{< P(|\bm{\upvarphi}_{\fcs,T}(N) - \bar{\bm{\upvarphi}}_{\fcs}| >  \bm\epsilon_{\fcs} \mid \tilde{N} = N, \tilde{T} = T) \\ < P(|\bm{\upvarphi}_{\fcs,T}(N) - \bar{\bm{\upvarphi}}_{\fcs}| >  \bm\epsilon_{\fcs} \mid \tilde{N} = N) < \delta }} \tag{$\bm{\upvarphi}_{\fcs,T}(N)$ is independent of $\tilde T$} \\
    <& \left|\+E\left[X_{NT}  \mid \tilde{N} = N, \tilde{T} = T, |\bm{\upvarphi}_T(N) - \bar{\bm{\upvarphi}}| \leq  \bm\epsilon \right] \right|\cdot P(|\bm{\upvarphi}_T(N) - \bar{\bm{\upvarphi}}| \leq  \bm\epsilon \mid \tilde{N} = N, \tilde{T} = T) + C\delta  \\
     <& \left|\+E\left[X_{NT}  \mid \tilde{N} = N, \tilde{T} = T, |\bm{\upvarphi}_T(N) - \bar{\bm{\upvarphi}}| \leq  \bm\epsilon \right] \right| + C\delta \label{eqn:abs-prob-bound}
\end{align}
for some constant $C$, where 
\[\bm{\upvarphi}_{\fcs,T}(N) = \left(\estsigmasq_{\fcs,2}(N), \estxisq_{\fcs,2}(N), \cdots, \estsigmasq_{\fcs,T}(N), \estxisq_{\fcs,T}(N) \right) \]
and $\bar{\bm{\upvarphi}}_{\fcs}$ and $\bm{\epsilon}$ are defined analogously. Note that $\bm{\upvarphi}_{\fcs,T}(N)$ is independent of $\tilde{T}$ because $\bm{\upvarphi}_{\fcs,T}(N)$ is a function of $\varepsilon_{is}$ for $i \in \mathcal{S}_{\fcs}$, $\tilde{T}$ is a function of $\estsigmasq_{\ad,1,2}(N), \cdots, \estsigmasq_{\ad,1,T}(N)$ and therefore a function of $\varepsilon_{js}$ for $j \in \mathcal{S}_{\ad,1}$, and $\varepsilon_{is}$ is i.i.d. for all $i$ and $s$. 

Using \eqref{eqn:abs-prob-bound}, we have for any $(\bm\epsilon, \delta_1, \delta_2)$, there exist $N_0$ such that for any $N > N_0$,
\begin{align*}
    |a_{NT}| <& \Bigg|\+E \left[ \frac{1}{NT} \sum_{i,s} \funfrac(\bm{\omega}_{\all,1:T}(N),T)^\I \cdot   \zeta_{is}^2  \cdot (\varepsilon_{is}^2 - \sigma_\varepsilon^2) \mid \tilde{N} = N, \tilde{T} = T, |\bm{\upvarphi}_T(N) - \bar{\bm{\upvarphi}}|  \leq  \bm\epsilon \right]\Bigg|  + C_1 \delta_1 \\
    <& C_2 \sup_{i,s} \Bigg|\+E \left[ \varepsilon_{is}^2 - \sigma_\varepsilon^2 \mid \tilde{N} = N, \tilde{T} = T, |\bm{\upvarphi}_T(N) - \bar{\bm{\upvarphi}}|  \leq  \bm\epsilon, Z \right]\Bigg|  + C_1 \delta_1 \tag{$g_\tau(\cdot)$ and $\zeta_{is}^2$ are bounded positives} \\
    <& C_2 \left(C_3 \norm{{\bm{\epsilon}}^\prime}_2  + \delta_2\right) + C_1 \delta_1. \tag{from \eqref{eqn:lemma-7-3-1-2} and \eqref{eqn:lemma-7-3-2-2}}
\end{align*}
for some constant $C_1, C_2$ and $C_3$, and ${\bm{\epsilon}}^\prime$ is equal to ${\bm{\epsilon}}$ excluding the first coordinate. 

Similarly, for any $(\bm\epsilon, \delta_1, \delta_2)$, there exist $N_0$ such that for any $N > N_0$,
\begin{align*}
    |b_{NT}| <& \Bigg|\+E \left[ \frac{1}{NT} \sum_{i,s,u\leq T} \funfrac(\bm{\omega}_{\all,1:T}(N),T)^\I \cdot   \zeta_{is} \zeta_{iu} \varepsilon_{is} \varepsilon_{iu} \mid \tilde{N} = N, \tilde{T} = T, |\bm{\upvarphi}_T(N) - \bar{\bm{\upvarphi}}|  \leq  \bm\epsilon \right]\Bigg|  + C_1 \delta_1 \\
    <& \left( \sup_{\bm{\omega}_{\all,1:T}(N)}  \funfrac(\bm{\omega}_{\all,1:T}(N),T)^\I \right) \cdot \left(\frac{1}{NT} \sum_{i,s,u\leq T}  \zeta^2_{is} \zeta^2_{iu} \right)^{1/2} \\
    & \cdot \sup_{i,s,u\leq T} \Bigg|\+E \left[ \varepsilon_{is} \varepsilon_{iu} \mid \tilde{N} = N, \tilde{T} = T, |\bm{\upvarphi}_T(N) - \bar{\bm{\upvarphi}}|  \leq  \bm\epsilon, Z \right]\Bigg|  + C_1 \delta_1 \tag{by Cauchy-Schwarz inequality} \\
    <& C_2 \sup_{i,s,u\leq T} \Bigg|\+E \left[ \varepsilon_{is} \varepsilon_{iu} \mid \tilde{N} = N, \tilde{T} = T, |\bm{\upvarphi}_T(N) - \bar{\bm{\upvarphi}}|  \leq  \bm\epsilon, Z \right]\Bigg|  + C_1 \delta_1 \tag{$g_\tau(\cdot)$ is bounded positive and $\zeta_{is} \in [-1,1]$} \\
    <& C_2 \left(C_3 \norm{{\bm{\epsilon}}^\prime}_2  + \delta_2\right) + C_1 \delta_1. \tag{from \eqref{eqn:lemma-7-3-1-3} and \eqref{eqn:lemma-7-3-2-3}}
\end{align*}
for some constants $C_1$, $C_2$ and $C_3$, and ${\bm{\epsilon}}^\prime$ is equal to ${\bm{\epsilon}}$ excluding the first coordinate. 
Moreover, for any $(\bm\epsilon, \delta_1, \delta_2)$, there exist $N_0$ such that for any $N > N_0$,
\begin{align*}
    |c_{NT}| <& \Bigg|\+E \left[ \frac{1}{NT} \sum_{i,j:i\neq j,s,u\leq T} \funfrac(\bm{\omega}_{\all,1:T}(N),T) \cdot   \zeta_{is} \zeta_{ju} \varepsilon_{is} \varepsilon_{ju} \mid \tilde{N} = N, \tilde{T} = T, |\bm{\upvarphi}_T(N) - \bar{\bm{\upvarphi}}|  \leq  \bm\epsilon \right]\Bigg|  + C_1 \delta_1 \\
    <& \left( \sup_{\bm{\omega}_{\all,1:T}(N)}  \funfrac(\bm{\omega}_{\all,1:T}(N),T) \right) \cdot \left(\frac{1}{N^2 T} \sum_{i,j:i\neq j,s,u\leq T}  \zeta^2_{is} \zeta^2_{ju} \right)^{1/2} \\
    & \cdot \sup_{i,j:i\neq j,s,u\leq T} \Bigg|\+E \left[ N\varepsilon_{is} \varepsilon_{ju} \mid \tilde{N} = N, \tilde{T} = T, |\bm{\upvarphi}_T(N) - \bar{\bm{\upvarphi}}|  \leq  \bm\epsilon, Z \right]\Bigg|  + C_1 \delta_1 \tag{by Cauchy-Schwarz inequality} \\
    <& C_2 \sup_{i,s,u\leq T} \Bigg|\+E \left[ N\varepsilon_{is} \varepsilon_{iu} \mid \tilde{N} = N, \tilde{T} = T, |\bm{\upvarphi}_T(N) - \bar{\bm{\upvarphi}}|  \leq  \bm\epsilon, Z \right]\Bigg|  + C_1 \delta_1 \tag{$g_\tau(\cdot)$ and $\zeta_{is}$ are bounded} \\
    <& C_2 \left(C_3 \norm{{\bm{\epsilon}}^\prime}_2  + \delta_2\right) + C_1 \delta_1. \tag{from \eqref{eqn:lemma-7-3-1-4} and \eqref{eqn:lemma-7-3-2-3}}
\end{align*}
for some constants $C_1$, $C_2$ and $C_3$, and ${\bm{\epsilon}}^\prime$ is equal to ${\bm{\epsilon}}$ excluding the first coordinate. 
Combining $a_{NT}$, $b_{NT}$ and $c_{NT}$ together, and since there are only finitely many choices of  $T$ is bounded above, for any $(\bm\epsilon, \delta_1, \delta_2)$, there exist $N_0$ such that for any $N > N_0$,
\begin{align*}
    \left|\+E[N \tilde{T}\funfrac(\bm{\omega}_{\all,1:\tilde{T}}(N),\tilde{T})/\sigma_\varepsilon^2 \cdot (\hat{\tau}_{\all,\tilde{T}} - \tau)^2 - 1\mid \tilde{N}= N]\right| < C_2 \left(C_3 \norm{{\bm{\epsilon}}^\prime}_2  + \delta_2\right) + C_1 \delta_1
\end{align*}
for some constants $C_1$, $C_2$ and $C_3$. This implies that 
\begin{align}
    \lim_{N \rightarrow \infty} \+E\left[N \cdot \left[(\tilde{T}\funfrac(\bm{\omega}_{\all,1:\tilde{T}}(N),\tilde{T}))^{1/2}/\sigma_\varepsilon \cdot (\hat{\tau}_{\all,\tilde{T}} - \tau) \right]^2 \mid \tilde{N}= N\right] = 1 \label{eqn:conditional-mean-square}
\end{align}
Therefore $\hat{\tau}_{\all,\tilde{T}}$ converges in mean square. This implies $\hat{\tau}_{\all,\tilde{T}}$ converges in probability ({\it i.e.}, consistency). Following the expression of \eqref{eqn:conditional-mean-square}, the convergence rate of $\hat{\tau}_{\all,\tilde{T}}$ is $\sqrt{N}$. Moreover, both the mean and asymptotic variance of $(\tilde{T}\funfrac(\bm{\omega}_{\all,1:\tilde{T}}(N),\tilde{T}))^{1/2}/\sigma_\varepsilon \cdot (\hat{\tau}_{\all,\tilde{T}} - \tau)$ do not depend on $\tilde{T}$. 

The mean and asymptotic second moment of $\hat{\tau}_{\all,\tilde{T}} - \tau$ is identical to that of $ \hat{\tau} - \tau$ in Lemma \ref{lemma:asymptotic-tau-sigma-general}, where $\hat{\tau}$ is estimated from the non-adaptive experimental data with the same number of units and time periods and with the same $\bm{\omega}$. This implies that the correlations between $\varepsilon_{it}$ and $\varepsilon_{js}$ conditional on $\tilde{T}$ are sufficiently ``weak'', so that the correlations do not change the second moment. We can further verify the conditions in martingale CLT holds and apply the martingale CLT to $\hat{\tau}_{\all,\tilde{T}}$, yielding
\[\sqrt{N} \big(\tilde{T}\funfrac(\bm{\omega}_{\all,1:\tilde{T}},\tilde{T})/\sigma_\varepsilon^2\big)^{1/2} (\hat{\tau}_{\all,\tilde{T}} - \tau) \xrightarrow{d} \mathcal{N} \left(0,1 \right), \]
conditional on $\tilde{T} = T$. As this asymptotic distribution does not depend on the value of $T$, and 
$\tilde{T}$ can only take finitely many values, the asymptotic distribution of $\hat{\tau}_{\all,\tilde{T}}$ unconditional on $\tilde{T}$ stays the same. 
This concludes the proof of the asymptotic distribution of $\hat{\tau}_{\all,\tilde{T}}$. 

\textbf{Step 2: Show the asymptotic normal distribution of $\estsigmasq_{\ad,2,\tilde{T}}$.}

Note that $Z_{\ad,2,\tilde{T}}$ is chosen based on $\estsigmasq_{\fcs,2}, \estxisq_{\fcs,2}, \cdots, \estsigmasq_{\fcs,\tilde{T}}, \estxisq_{\fcs,\tilde{T}}$ that are estimated using $\varepsilon_{ju}$ for $j$ in $\mathcal{S}_{\fcs}$. Moreover, the length of $Z_{\ad,2,\tilde{T}}$ is $\tilde{T}$, where $\tilde{T}$ is determinated by $\estsigmasq_{\ad,1,2}, \cdots, \estsigmasq_{\ad,1,\tilde{T}}$ that are estimated using $\varepsilon_{ju}$ using $j$ in $\mathcal{S}_{\ad,1}$.  Therefore, for any $i \in \mathcal{S}_{\ad,2}$ and any $t$, $\varepsilon_{it}$ is independent of $Z_{\ad,2,\tilde{T}}$ conditional on $\tilde{T}$, following that $\varepsilon_{it}$ i.i.d. in $i$ and $t$ for $i$ in both NTU and ATU. Therefore, conditional on $\tilde{T}$, we can directly apply Lemma \ref{lemma:asymptotic-tau-sigma-general}, use Slutsky's theorem, and obtain 
\[\sqrt{N}  \big(\tilde{T} p_{\ad,2}/\xi^{\dagger 2}_{\varepsilon,\tilde{T}} \big)^{1/2} \cdot \big(\estsigmasq_{\ad,2,\tilde{T}} - \sigma_\varepsilon^2 \big) \xrightarrow{d} N\left(0, 1  \right) \, \]
where 
$\xi^{\dagger 2}_{\varepsilon,\tilde{T}}  = \xi_\varepsilon^2 + 2/(\tilde{T}-1) \cdot \sigma_\varepsilon^4 $. As this asymptotic distribution does not depend on the value of $\tilde{T}$ and $\tilde{T}$ can only take finitely many values, this asymptotic distribution also holds unconditional on $\tilde{T}$. 
This concludes the proof of the asymptotic distribution of $\estsigmasq_{\ad,2}$.

\textbf{Step 3: Show the joint asymptotic distribution of $\tau_{\all,\tilde{T}}$ and $\estsigmasq_{\ad,2,\tilde{T}}$}

To show the joint asymptotic distribution in Theorem \eqref{theorem:asymptotic-page}, we need to show the asymptotic covariance between $\tau_{\all,\tilde{T}} - \tau$ and $\estsigmasq_{\ad,2,\tilde{T}} - \sigma_\varepsilon^2$ is 0. As an intermediate step, we first show their asymptotic covariance conditional on $\tilde{T} = T$ is 0. 
\begin{align*}
   & \estsigmasq_{\ad,2,\tilde{T}} - \sigma_\varepsilon^2 = \underbrace{\frac{1}{|\mathcal{S}_{\ad,2}| \tilde{T}} \sum_{i \in \mathcal{S}_{\ad,2},t \leq \tilde{T}} (\varepsilon_{it}^2 - \sigma_\varepsilon^2 ) }_{e_1}
  + \underbrace{\frac{1}{|\mathcal{S}_{\ad,2}| \tilde{T} (\tilde{T}-1)} \sum_{i\in \mathcal{S}_{\ad,2}, t,s\leq \tilde{T}: t\neq s} \varepsilon_{it} \varepsilon_{is}}_{e_2}   +  O_p\left( \frac{1}{N} \right)
\end{align*}

Note that the non-leading terms of $\sqrt{N} \left(\estsigmasq_{\ad,2,\tilde{T}} - \sigma^2_\varepsilon\right)$ is at the order of $O_p\left(1/\sqrt{N}\right)$, and the order of $\sqrt{N}(\hat{\tau}_{\all,\tilde{T}} - \tau)$ is $O_p(1)$. Therefore, their product is at the order of $O_p\left(1/\sqrt{N}\right) = o_p(1)$. Equivalently, the asymptotic covariance between non-leading terms of $\estsigmasq_{\ad,1,\tilde{T}} - \sigma^2_\varepsilon$ and $\hat{\tau}_{\all,\tilde{T}} - \tau$ is $0$ conditional on $\tilde{T} = T$. For notation simplicity, let 
\[\tilde{\zeta}_{is} = \funfrac(\bm{\omega}_{\all,1:\tilde{T}}(N),\tilde{T})^\I \zeta_{is}  \]
and then we can write the estimation error of $\hat{\tau}_{\all,\tilde{T}}$ as 
\[\hat{\tau}_{\all,\tilde{T}} - \tau = \frac{1}{N\tilde{T}} \sum_{i, s\leq \tilde{T}} \tilde{\zeta}_{is} \varepsilon_{is}. \]

Next we show the covariance between $\hat{\tau}_{\all,T} - \tau$ and $e_1$ conditional on $\tilde{T} = T$. Note that
\begin{align*}
    & \+E\left[ N \tilde{T} \cdot \left( \frac{1}{N\tilde{T}} \sum_{i,t} \tilde\zeta_{it} \varepsilon_{it} \right) \left(  \frac{1}{|\mathcal{S}_{\ad,2}|\tilde{T}} \sum_{i\in\mathcal{S}_{\ad,2},t\leq\tilde{T}} (\varepsilon_{it}^2 - \sigma_\varepsilon^2 ) \right)\mid \tilde{N} = N,  \tilde{T} = T \right] \\
    =& \underbrace{\frac{1}{ |\mathcal{S}_{\ad,2}| T} \sum_{i \in \mathcal{S}_{\ad,2},t \leq T} \+E\left[ \tilde\zeta_{it} \varepsilon_{it} (\varepsilon_{it}^2 - \sigma_\varepsilon^2)\mid \tilde{N} = N,  \tilde{T} = T\right]}_{A} \\ 
    &+ \underbrace{\frac{1}{ |\mathcal{S}_{\ad,2}| T} \sum_{(i,t) \neq (j,s), j \in \mathcal{S}_{\ad,2}} \+E\left[ \tilde\zeta_{it} \varepsilon_{it} (\varepsilon_{js}^2 - \sigma_\varepsilon^2) \mid \tilde{N} = N,  \tilde{T} = T\right]}_{B}.
\end{align*}

For term $A$,  for any $i \in \mathcal{S}_{\ad,2}$, and for any $N$, 
\begin{flalign*}
    & \+E\left[ \tilde\zeta_{it} \varepsilon_{it} (\varepsilon_{it}^2 - \sigma_\varepsilon^2)\mid \tilde{N} = N, \tilde{T} = T \right] \\
    =& \+E\left[ \tilde\zeta_{it} \left( \+E\left[  \varepsilon_{it}^3  \mid \tilde{N} = N, \tilde{T} = T, Z\right] - \+E[\varepsilon_{it} \mid \tilde{N} = N, \tilde{T} = T, Z] \cdot \sigma_\varepsilon^2  \right)\mid \tilde{N} = N, \tilde{T} = T \right]   \\
    =& \+E\left[ \zeta_{it} \left( \+E\left[  \varepsilon_{it}^3 \right] - \+E[\varepsilon_{it} ] \cdot \sigma_\varepsilon^2  \right) \mid \tilde{N} = N, \tilde{T} = T\right]  & \tag{$\varepsilon_{it}$ is independent of $\tilde{T}$ and $Z$ for $i \in \mathcal{S}_{\ad,2}$ } \\
    =& 0
\end{flalign*}
and therefore $A = 0$. 

For term $B$, for any $j \in \mathcal{S}_{\ad,2}$ and any $i$ (with $(i,t) \neq (j,s)$), and for any $N$
\begin{flalign*}
    & \+E\left[ \tilde\zeta_{it} \varepsilon_{it} (\varepsilon_{js}^2 - \sigma_\varepsilon^2)\mid \tilde{N} = N, \tilde{T} = T \right] \\
    =&  \+E\left[\tilde\zeta_{it} \cdot \+E\left[\varepsilon_{it} (\varepsilon_{js}^2 - \sigma_\varepsilon^2)\mid \tilde{N} = N, \tilde{T} = T, Z \right] \mid \tilde{N} = N, \tilde{T} = T\right] \\
    =& \+E\left[\tilde\zeta_{it} \cdot \+E\left[ \varepsilon_{it} \cdot \+E\left[  \varepsilon_{js}^2 - \sigma_\varepsilon^2 \mid \tilde{N} = N, \tilde{T} = T, Z, \varepsilon_{it} \right]\mid \tilde{N} = N, \tilde{T} = T, Z \right]\mid \tilde{N} = N, \tilde{T} = T\right]  \\
    =&  \+E\left[\tilde\zeta_{it} \cdot \+E\left[ \varepsilon_{it} \cdot \+E\left[  \varepsilon_{js}^2 - \sigma_\varepsilon^2 \right]\mid \tilde{N} = N, \tilde{T} = T, Z \right]\mid \tilde{N} = N, \tilde{T} = T\right] & \tag{$\varepsilon_{js}$ is independent of $\tilde{N}$, $\tilde{T}$, $Z$ and $\varepsilon_{it}$}\\
    =& \tilde\zeta_{it} \cdot \+E\left[ \varepsilon_{it} \cdot 0\mid \tilde{N} = N,  \tilde{T} = T, Z \right] = 0
\end{flalign*}
and therefore $B = 0$. As both $A = 0$ and $B = 0$, the covariance between $\hat{\tau}_{\all,T} - \tau$ and $a_1$ is 0 for any $N$ and $T$.

For the covariance between $\hat{\tau}_{\all,\tilde{T}} - \tau$ and $e_2$ conditional on $\tilde{T} = T$, 
\begin{align*}
        & \+E\left[NT \cdot \left( \frac{1}{N \tilde{T}} \sum_{i,t \leq \tilde{T}} \tilde\zeta_{it} \varepsilon_{it} \right) \left( \frac{1}{|\mathcal{S}_{\ad,1}| \tilde{T} (\tilde{T}-1)} \sum_{i \in \mathcal{S}_{\ad,1}, t\neq s \leq \tilde{T}} \varepsilon_{it} \varepsilon_{is}  \right) \mid \tilde{N} = N, \tilde{T} = T \right]\\
    =&  \frac{1}{|\mathcal{S}_{\ad,1}|  T (T-1)} \sum_{i \in [N],j \in \mathcal{S}_{\ad,1}, t, s, u: u \neq s} \+E\bigg[ \tilde\zeta_{it} \cdot \underbrace{\+E \left[ \varepsilon_{it} \varepsilon_{ju} \varepsilon_{js} \mid \tilde{N} = N,  \tilde{T} = T, Z\right]}_{0} \mid \tilde{N} = N,  \tilde{T} = T\bigg] = 0
\end{align*}
where $\+E \left[ \varepsilon_{it} \varepsilon_{ju} \varepsilon_{js} \mid \tilde{N} = N, \tilde{T} = T, Z\right] = 0$ follows that at least one of $(j,u)$ and $(j,s)$ does not equal to $(i,t)$, given that $u \neq s$. Suppose $(j,s) \neq (i,t)$. Then 
\begin{flalign*}
    & \+E \left[ \varepsilon_{it} \varepsilon_{ju} \varepsilon_{js} \mid \tilde{N} = N, \tilde{T} = T, Z\right] \\ 
    =&\+E \left[ \varepsilon_{it} \varepsilon_{ju} \+E\left[\varepsilon_{js} \mid \tilde{N} = N,  \tilde{T} = T, Z, \varepsilon_{it}, \varepsilon_{ju}\right]  \mid \tilde{N} = N, \tilde{T} = T, Z\right] \\
    =&\+E \left[ \varepsilon_{it} \varepsilon_{ju} \+E\left[\varepsilon_{js}\right]  \mid \tilde{N} = N, \tilde{T} = T, Z\right] & \tag{$\varepsilon_{js}$ is independent of $\tilde{N}$, $\tilde{T}$, $Z$, $\varepsilon_{it}$ and $\varepsilon_{ju}$}\\
    =& \+E \left[ \varepsilon_{it} \varepsilon_{ju} \cdot 0 \mid \tilde{N} = N, \tilde{T} = T, Z\right] = 0
\end{flalign*}

Therefore, the covariance between $\hat{\tau}_{\all,\tilde{T}} - \tau$ and $e_2$ is 0 conditional on $\tilde{T}= T$. Together with the zero covariance between $\hat{\tau}_{\all,\tilde T} - \tau$ and $e_1$ conditional on $\tilde{T} = T$, and zero asymptotic covariance between $\hat{\tau}_{\all,\tilde T} - \tau$ and non-leading terms of $\estsigmasq_{\ad,2,\tilde T} - \sigma_\varepsilon^2 $ conditional on $\tilde{T} = T$, we have finished showing that the asymptotic covariance between $\hat{\tau}_{\all,\tilde T} - \tau$ and $\estsigmasq_{\ad,2,\tilde T} - \sigma_\varepsilon^2 $ is zero conditional on $\tilde T= T$. As this holds on any value of $\tilde T$ and $\tilde T$ can only take finitely many values, the unconditional asymptotic covariance between $\hat{\tau}_{\all,\tilde T} - \tau$ and $\estsigmasq_{\ad,2,\tilde T} - \sigma_\varepsilon^2 $ is also zero. 

Next we apply multivariate martingale CLT to $\hat{\tau}_{\all,\tilde T}$ and $\estsigmasq_{\ad,2}$. The conditions for the CLT can be verified similarly to the previous steps. Then we have  
\begin{align}
        \sqrt{N} \cdot \begin{bmatrix}
    \big(\tilde{T}\funfrac(\bm{\omega}_{\all,1:\tilde{T}},\tilde{T})/\sigma_\varepsilon^2\big)^{1/2} \cdot \left( \hat{\tau}_{\all,\tilde{T}} - \tau\right) \vspace{0.3cm}  \\ \big(\tilde{T} p_{\ad,2}/\xi^{\dagger 2}_{\varepsilon,\tilde{T}} \big)^{1/2} \cdot \big(\estsigmasq_{\ad,2,\tilde{T}} - \sigma_\varepsilon^2 \big)
        \end{bmatrix}
        \stackrel{d}{\longrightarrow } \mathcal{N} \left(\bm{0}, I_2 \right)
    \end{align}
This concludes the proof of Theorem \ref{theorem:asymptotic-page}. \halmos

\end{proof}

%% file: appendix_E.tex
\section{A Machine Learning Estimator for Treatment Effects}\label{sec:algorithm} 
	
	One could directly estimate $L$ with $\bm{\tau}$ using the following objective function,
	\begin{equation}\label{eqn:obj-mf-fe}
	    \begin{aligned}
	         \hat{\bm{\tau}}, \hat{\bm{\alpha}}, \hat{\bm{\beta}}_{(\ell+1):T}, \hat L =& \argmin_{\bm{\tau}, \bm{\alpha}, \bm{\beta}_{(\ell+1):T}, L } \frac{1}{N(T-\ell)}  \\ &  \norm{Y_{:,(\ell+1):T} - \bm{\alpha} \bm{1}_{T-\ell}^\T - \bm{1}_N \bm{\beta}_{(\ell+1):T}^\T -  L - \tau_0 Z_{:,(\ell+1):T} - \tau_1 Z_{:,\ell:(T-1)} - \cdots - \tau_\ell Z_{:,1:(T-\ell)}}_F^2 \\ & \qquad+ \mu \norm{L}_\ast\,,
	    \end{aligned}
	\end{equation}
	where we refer to this objective as low-rank matrix estimation with fixed effects (LRME). Here, $\norm{L}_\ast$ is the nuclear norm (or trace norm) of matrix $L$, which is equal to the sum of its singular values. Also, $\norm{\cdot}_F$ refers to the Frobenius norm of a matrix. The rank of $\hat L$ tends to decrease with the regularization parameter $\mu$. Note that the bias tends to increase with $\mu$, but the variance tends to decrease with $\mu$. With a properly chosen $\mu$, we can reduce the root mean squared error (RMSE) of $\hat L$, and that of  $\hat{\tau}_0, \hat{\tau}_1, \cdots, \hat{\tau}_\ell$.

	The objective function \eqref{eqn:obj-mf-fe} is convex in $\tau, \alpha, \beta$ and $L$, which has $N(T-\ell) + N + T+1$ variables in total. Finding the global optimal solution of convex program \eqref{eqn:obj-mf-fe} can be slow with off-the-shelf software for convex optimization problems such as {\tt cvxpy}. Alternatively, we propose to use the {\it iterative singular value thresholding and ordinary least squares (iterative SVT and OLS)} algorithm to efficiently solve convex program \eqref{eqn:obj-mf-fe}. The details of this algorithm are described in Algorithm \ref{algo-est-svt-ls}. We can justify SVT using   Theorem 1 in  \cite{hastie2015matrix} that shows the optimal solution of
	\[\hat L =  \argmin_{rank(\ell) \leq k_0 } \frac{1}{2} \norm{Y -  L}_F^2 + \mu \norm{L}_\ast, \]
	is  $\hat L = U_{k_0} S_{\mu} (D_{k_0}) V_{k_0}^\T$,
	where the rank-$k_0$ SVD of $Y$ is $U_{k_0} D_{k_0} V_{k_0}^\T$ and $S_{\mu}(D_{k_0}) $ is a diagonal $k_0$ by $k_0$ matrix with its diagonal entries to be $(\sigma_1 - \mu)_+, \cdots, (\sigma_{k_0} - \mu)_+$.
	When we have historical control data, we can use cross-validation to find the optimal $\mu$ by the grid search algorithm.

	\begin{algorithm}
		\SetAlgoLined
		\SetKwInOut{Input}{Inputs}
		\SetKwInOut{Output}{Outputs}
		\Input{$Y, Z, k_0, \mu_{NT}$, $\Delta_{\tau}$, and $t_{\max}$}
		$\hat \tau^{(-1)} \leftarrow 0$ \;
		At $t = 0$, $\hat{\bm{\tau}}^{(0)}, \hat{\bm{\alpha}}^{(0)}, \hat{\bm{\beta}}_{(\ell+1):T}^{(0)} \leftarrow \argmin_{\tau, \alpha, \beta} \frac{1}{2} \norm{Y_{:,(\ell+1):T} - \bm{\alpha} \bm{1}_{T-\ell}^\T - \bm{1}_N \bm{\beta}_{(\ell+1):T}^\T - \tau_0 Z_{:,(\ell+1):T} - \tau_1 Z_{:,\ell:(T-1)} - \cdots - \tau_\ell Z_{:,1:(T-\ell)}}_F^2 $\;
		$\hat Y_e^{(0)} \leftarrow Y_{:,(\ell+1):T} - \hat{\bm{\alpha}}^{(0)} \bm{1}_{T-\ell}^\T - \bm{1}_N (\hat{\bm{\beta}}^{(0)}_{(\ell+1):T})^\T - \hat{\tau}_0^{(0)} Z_{:,(\ell+1):T} - \hat{\tau}_1^{(0)} Z_{:,\ell:(T-1)} - \cdots - \hat{\tau}_\ell^{(0)} Z_{:,1:(T-\ell)}$ \;
		
		\While{ $\max_j |\hat{\tau}_j^{(t)} - \hat{\tau}_j^{(t-1)} | > \Delta_{\tau}$ and $t < t_{\max}$ }{
			The rank-$k_0$ SVD of $\hat Y_e^{(t)}$ is $U_{k_0}^{(t)} D_{k_0}^{(t)}  (V_{k_0}^{(t)})^\T$, where $D_{k_0}^{(t)}  = \diag(d^{(t)}_1, \cdots, d^{(t)}_{k_0})$ \;
			$S_{\mu_{NT}} (D_{k_0}^{(t)})  \leftarrow \diag((d^{(t)}_1 - \mu_{NT})_+, \cdots, (d^{(t)}_{k_0} - \mu_{NT})_+)$ \;
			$\hat L^{(t+1)} = U_{k_0}^{(t)} S_{\mu_{NT}} (D_{k_0}^{(t)}) (V_{k_0}^{(t)})^\T$ \;
			$\hat{\bm{\tau}}^{(t+1)}, \hat{\bm{\alpha}}^{(t+1)}, \hat{\bm{\beta}}_{(\ell+1):T}^{(t+1)} = \argmin_{\bm{\tau}, \bm{\alpha}, \bm{\beta}_{(\ell+1):T}} \frac{1}{2} \norm{Y_{:,(\ell+1):T} - \bm{\alpha} \bm{1}_{T-\ell}^\T - \bm{1}_N \bm{\beta}_{(\ell+1):T}^\T - \tau_0 Z_{:,(\ell+1):T}  - \cdots - \tau_\ell Z_{:,1:(T-\ell)} - \hat L^{(t+1)} }_F^2 $\;
			$\hat Y_e^{(t+1)} = Y_{:,(\ell+1):T} -\hat{\bm{\alpha}}^{(t+1)} \bm{1}^\T - \bm{1}(\hat{\bm{\beta}}_{(\ell+1):T}^{(t+1)})^\T - \hat{\tau}_0^{(t+1)} Z_{:,(\ell+1):T} - \cdots - \hat{\tau}_\ell^{(t+1)} Z_{:,1:(T-\ell)}$ \;
			$t \leftarrow t+1$ \;
		}
		\Output{$\hat{\bm{\tau}}^{(t-1)}, \hat{\bm{\alpha}}^{(t-1)}, \hat{\bm{\beta}}_{(\ell+1):T}^{(t-1)}, \hat L^{(t-1)}$}
		\caption{Iterative SVT and OLS}
		\label{algo-est-svt-ls}
	\end{algorithm}

%% file: appendix_F.tex
\section{Supplementary Empirical Results for Non-Adaptive and Adaptive Experiments}\label{sec:supplementary-empirical}

\subsection{Data Description}\label{subsec:description-data-sets}

We first provide more details about the flu data. We use ICD-9 diagnosis codes to select the inpatient and outpatient records whose primary diagnosis of the patient is influenza.\footnote{These databases only have claim records with ICD-9 diagnosis codes.} The ICD-9 diagnosis codes for influenza are 488, 487.0, 487.1, 487.8, 488.0, 488.1, 488.01, 488.02, 488.09, 488.11, 488.12, 488.19, 488.81, 488.82, and 488.89.
Since these are claims data, unlike electronic medical records that may be restricted to only a few healthcare providers, we can see all clinical visits of every patient for the duration of their enrollment. Empowered by this unbiased coverage of patient visits in our data, we observe that patients are not typically admitted to hospitals for influenza, as there are 21,277 inpatient admissions versus 9,678,572 outpatient visits with primary diagnosis influenza. We denote all of these as influenza visits. 

Next, we provide further details about the three data sets, home medical visits, grocery store transactions, and Lending Club data, initially introduced in Section \ref{subsec:data-description}.
\begin{itemize}
    \item Home medical visits data set has 40,079 records of home medical visits from Jan 2016 to Dec 2018 in the metropolitan area of Barcelona Spain.\footnote{This data set is publicly available on Kaggle and can be downloaded at \url{https://www.kaggle.com/ckroxigor/home-medical-visits-eda/}.} This data set has been used to study how environmental factors adversely affect vulnerable people to environmental agents (climate, pollution, etc). We aggregate this data at the city level, as many environmental policies are carried out at an aggregate level. Given the high noise in the number of visits, we consider the 16-week moving average of medical visits. Then we obtain a panel of $61$ cities across $144$ weeks.

\item Grocery store transactions data set contains 17,880,248 transactions from a large grocery store between May 2005 and May 2007.\footnote{This data set is available to researchers at Stanford and Berkeley by application. Papers that use this data set are available at \url{https://are.berkeley.edu/SGDC/publications.html}.} We aggregate the transactions by household and week. Our analysis focuses on ``frequent'' households, defined as those who had expenditures in at least half of the weeks in the data set. These households tend to pay more attention to changes in the loyalty program. We then obtain a panel of $7,130$ frequent households over $97$ weeks.

\item Lending Club loan data contains 2,260,668 loans issued from June 2007 to December 2018 on Lending Club.\footnote{This data set can be downloaded at \url{https://www.kaggle.com/wordsforthewise/lending-club}}   This data set contains information, such as the current loan status (Current, Late, Fully Paid, etc.), latest payment information, first three digits of zip codes and issued month. We aggregate the number of loans issued by month and by the first three digits of zip codes. We get a panel of $956$ units over $139$ months.

\end{itemize}

\subsection{Supplementary Results for Non-Adaptive Experiments}

\subsubsection{Robustness to Additional Data Sets}\label{subsubsec:robustness-additional-data-set}
\begin{figure}[H]
	\centering
	\begin{subfigure}{1\textwidth}
		\centering
		\includegraphics[width=0.8\linewidth]{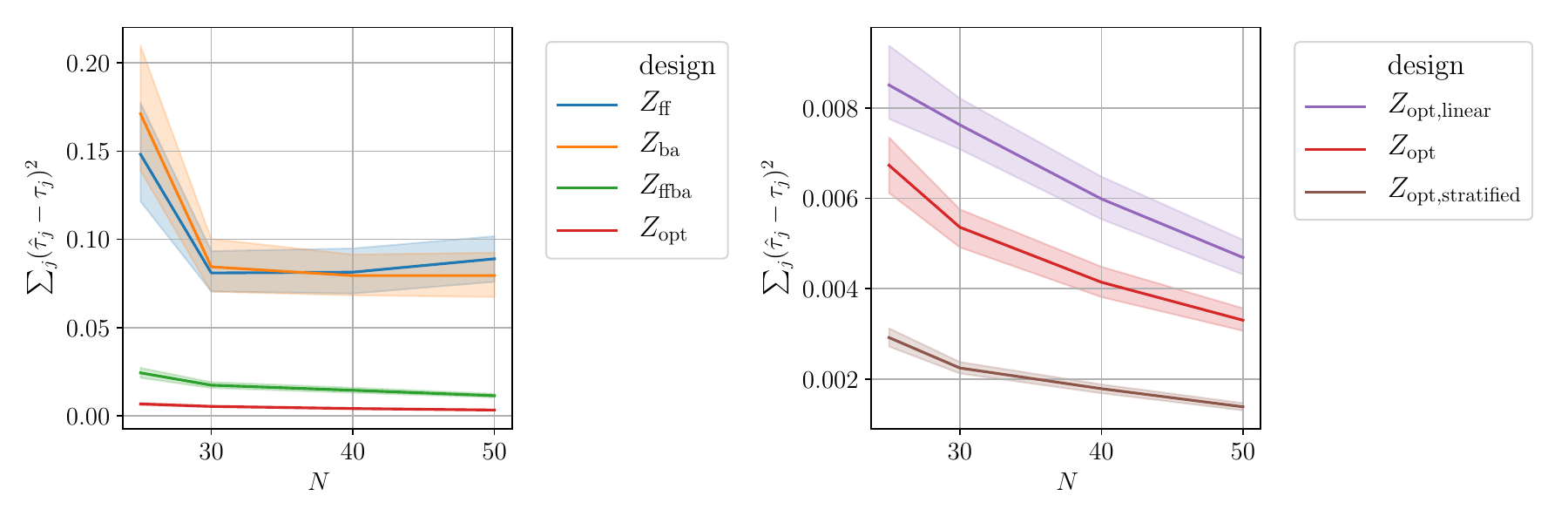}
		\caption{Home medical visit data}
	\end{subfigure}
	\begin{subfigure}{1\textwidth}
		\centering
		\includegraphics[width=0.8\linewidth]{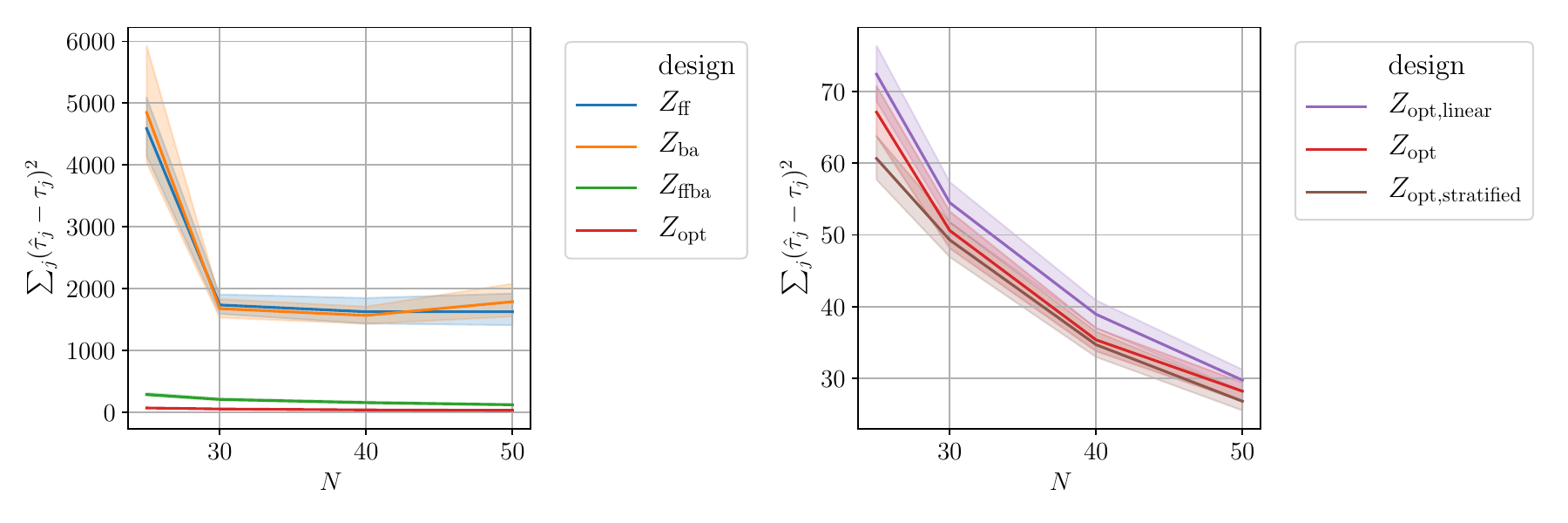}
		\caption{Grocery data}
	\end{subfigure}
	\begin{subfigure}{1\textwidth}
		\centering
		\includegraphics[width=0.8\linewidth]{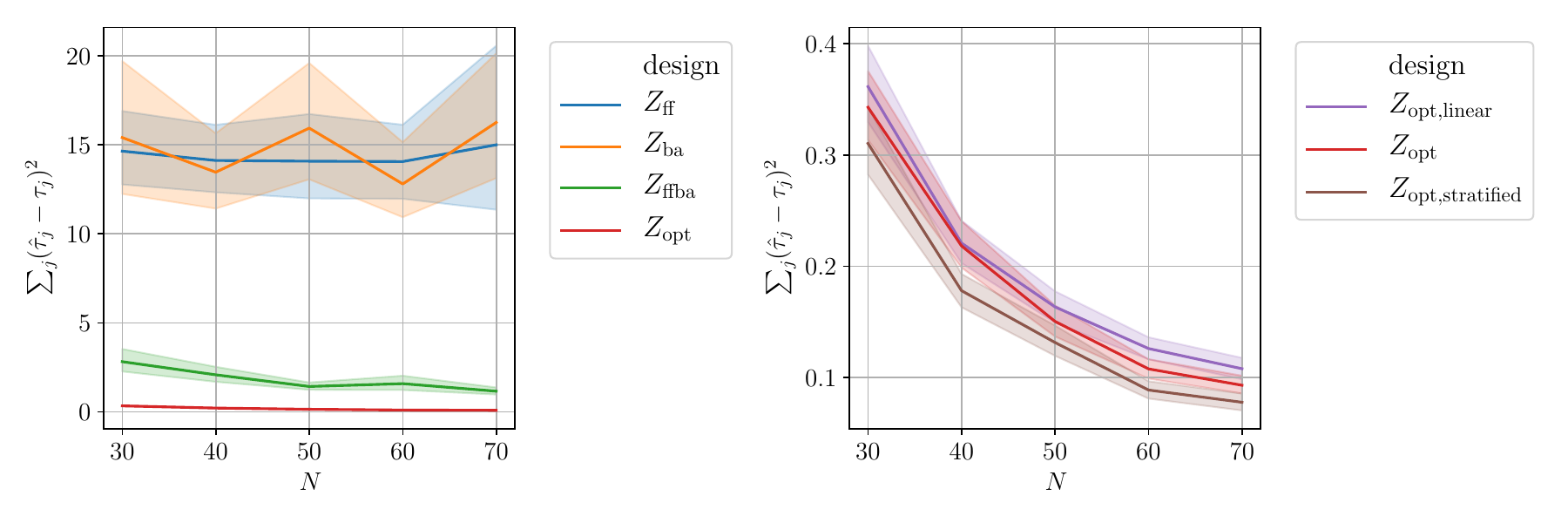}
		\caption{Loan data}
	\end{subfigure}
	\caption{\textbf{Varying $N$ (additional data sets).} These figures show the mean and 95\% confidence band of $\sum_{j}(\hat{\tau}_j - \tau_j)^2$ for various designs, based on 2,000 synthetic non-adaptive experiments with $\ell = 2$ and varying $N$. For the medical data, $T$ is 10, and $\sum_{j= 0}^\ell \tau_j$ is $-10$\% of the average monthly visit rate. For the grocery data, $T$ is 20, and $\sum_{j= 0}^\ell \tau_j$ is $10$\% of the average weekly expenditure. For the loan data, $T$ is 20, and $\sum_{j= 0}^\ell \tau_j$ is $10$\% of the average monthly number of loans issued. The size of $\sum_{j= 0}^\ell \tau_j$ is for illustrative purpose, and the estimation error does not vary with the value of $\tau_j$ for all $j$.}
	\label{fig:additional-varying-N}
\end{figure}

\begin{figure}[H]
	\centering
	\begin{subfigure}{1\textwidth}
		\centering
		\includegraphics[width=0.75\linewidth]{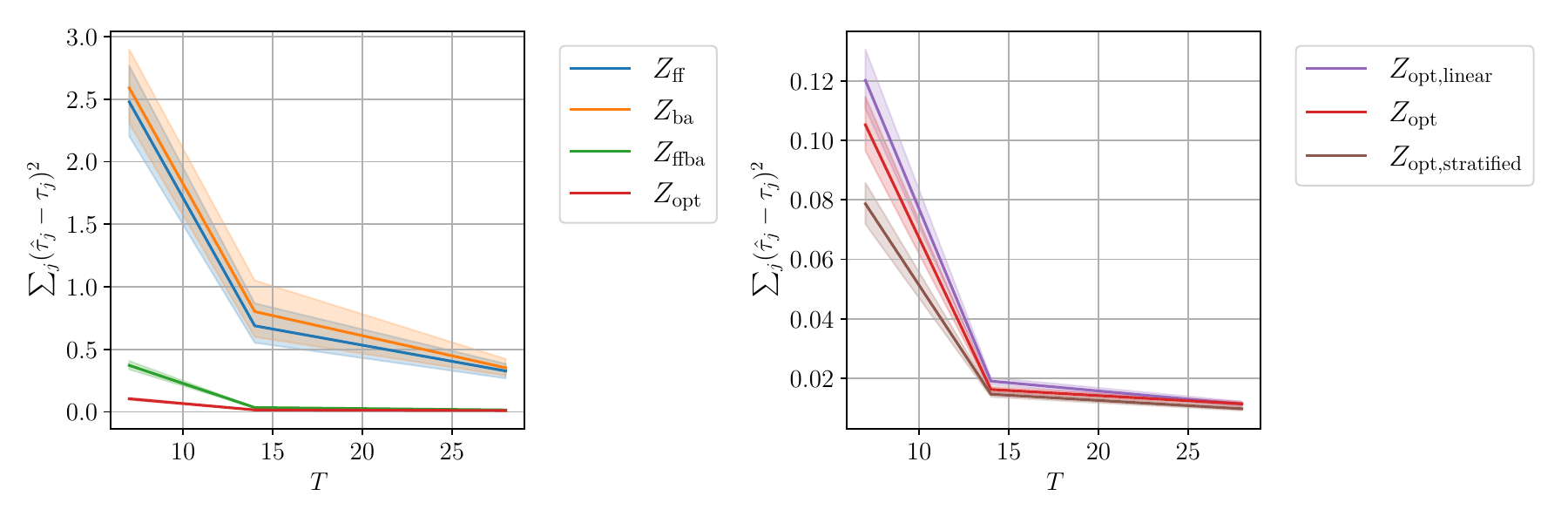}
		\caption{Flu data}
	\end{subfigure}
	\begin{subfigure}{1\textwidth}
		\centering
		\includegraphics[width=0.75\linewidth]{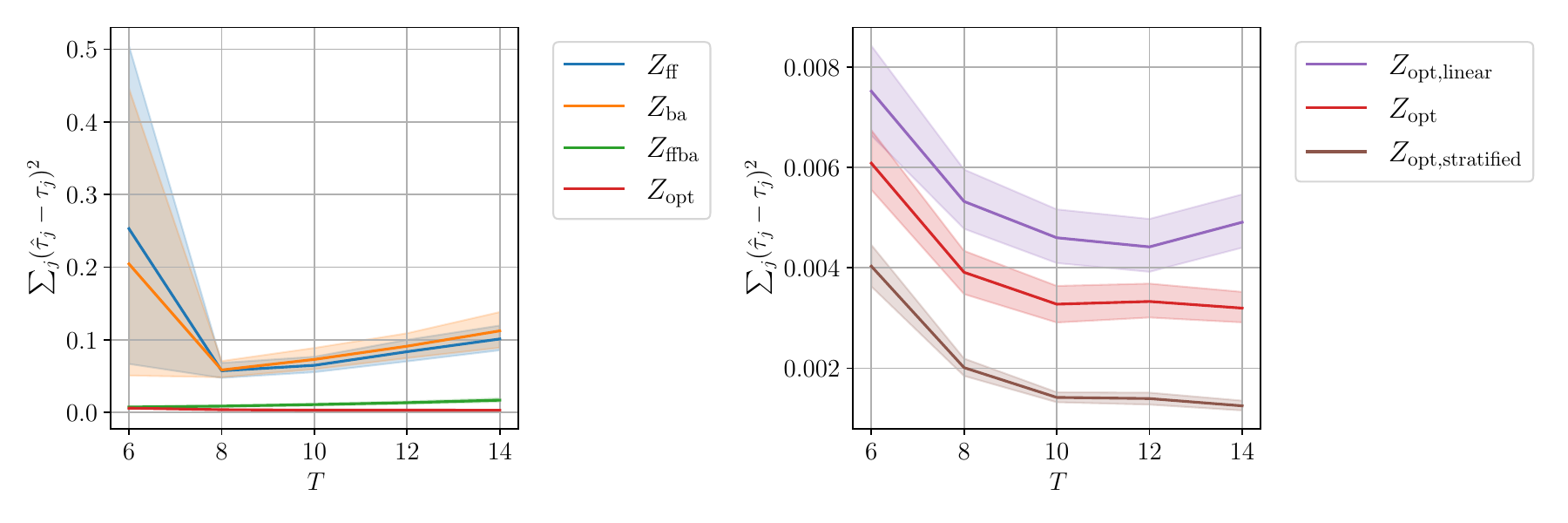}
		\caption{Home medical visit data}
	\end{subfigure}
	\begin{subfigure}{1\textwidth}
		\centering
		\includegraphics[width=0.75\linewidth]{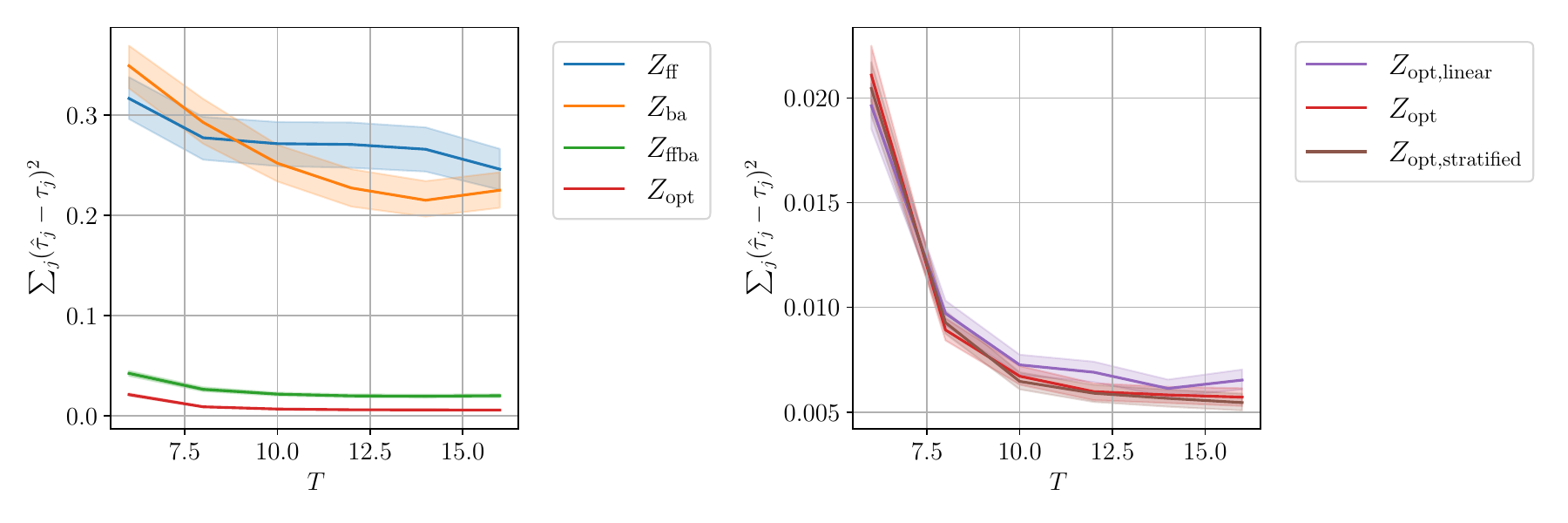}
		\caption{Grocery data}
	\end{subfigure}
	\begin{subfigure}{1\textwidth}
		\centering
		\includegraphics[width=0.75\linewidth]{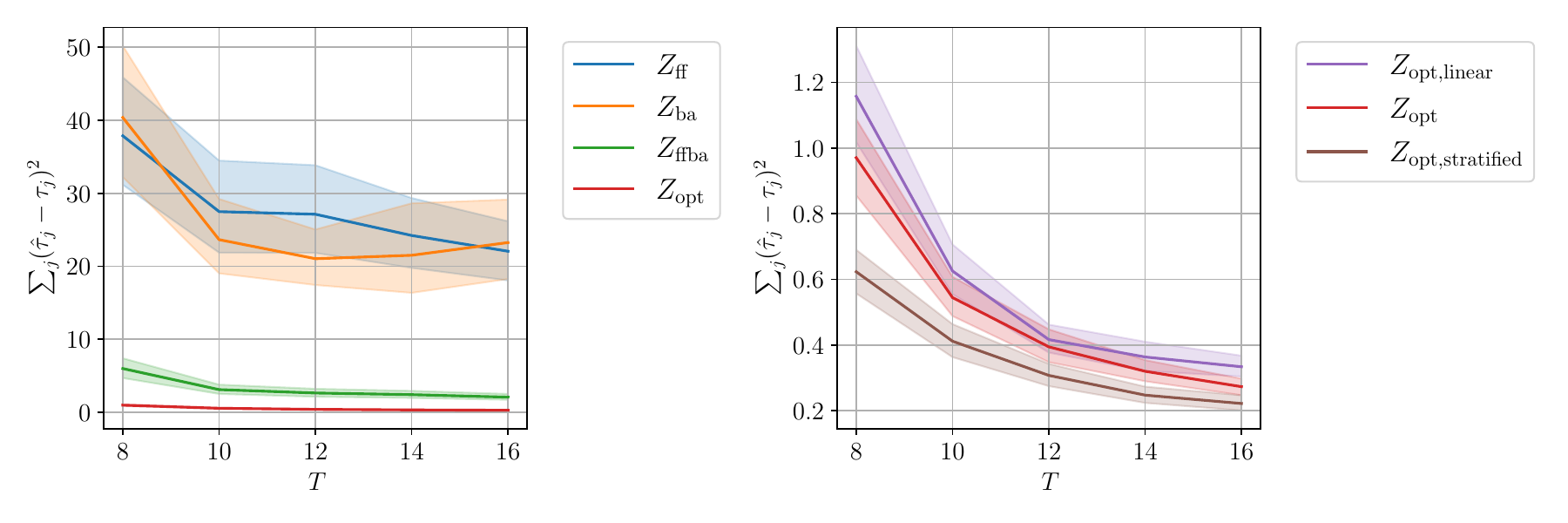}
		\caption{Loan data}
	\end{subfigure}
	\caption{\textbf{Varying $T$.} These figures show the mean and 95\% confidence band of $\sum_{j}(\hat{\tau}_j - \tau_j)^2$ for various designs, based on 2,000 synthetic non-adaptive experiments with $\ell = 2$, $N = 50$, and varying $T$. The data generating process is identical to that in Figure \ref{fig:additional-varying-N}.}
	\label{fig:additional-varying-T}
\end{figure}

Figure \ref{fig:additional-varying-N} shows that the three findings in Section \ref{subsec:fixed-sample-empirical-results} continue to hold on the other three data sets, as $N$ is varied. Figure \ref{fig:additional-varying-T} shows the three findings in Section \ref{subsec:fixed-sample-empirical-results} continue to hold on all four data sets, as $T$ is varied.

\clearpage

\subsubsection{Robustness to Specification of Estimator}\label{subsubsec:robustness-to-specification}

We compare the performance of various treatment designs, when the specification of the estimator varies:
\begin{itemize}
    \item ``no fe'': Least squares estimator of $\bm{\tau}$ using the specification $Y_{it} = c + \tau_0 z_{it} + \cdots + \tau_{\ell} z_{i,t-\ell} + \varepsilon_{it}$, which is the same as the difference-in-means estimator. 
    \item ``unit fe only'': Least squares estimator of $\bm{\tau}$ using the specification $Y_{it} = \alpha_i + \tau_0 z_{it} + \cdots + \tau_{\ell} z_{i,t-\ell} + \varepsilon_{it}$. 
    \item ``time fe only'': Least squares estimator of $\bm{\tau}$ using the specification $Y_{it} = \beta_t + \tau_0 z_{it} + \cdots + \tau_{\ell} z_{i,t-\ell} + \varepsilon_{it}$. 
    \item ``two-way fe'': Least squares estimator of  $\bm{\tau}$ using the specification $Y_{it} = \alpha_i + \beta_t +  \tau_0 z_{it} + \cdots + \tau_{\ell} z_{i,t-\ell} + \varepsilon_{it}$. (equivalent to the \within estimator of $\bm{\tau}$).
    \item ``two-way fe$+$covar'': Weighted least squares estimator of $\bm{\tau}$ using the specification \eqref{eqn:model-setup}, with $\*W \propto \big(\hat{\*U} \hat{\*\Sigma}_v \hat{\*U}^\T + \hat\sigma_\varepsilon^2 \*I_N \big)^\I$.
\end{itemize}

\begin{figure}[H]
	\centering
	\begin{subfigure}{.95\textwidth}
		\centering
		\includegraphics[width=1\linewidth]{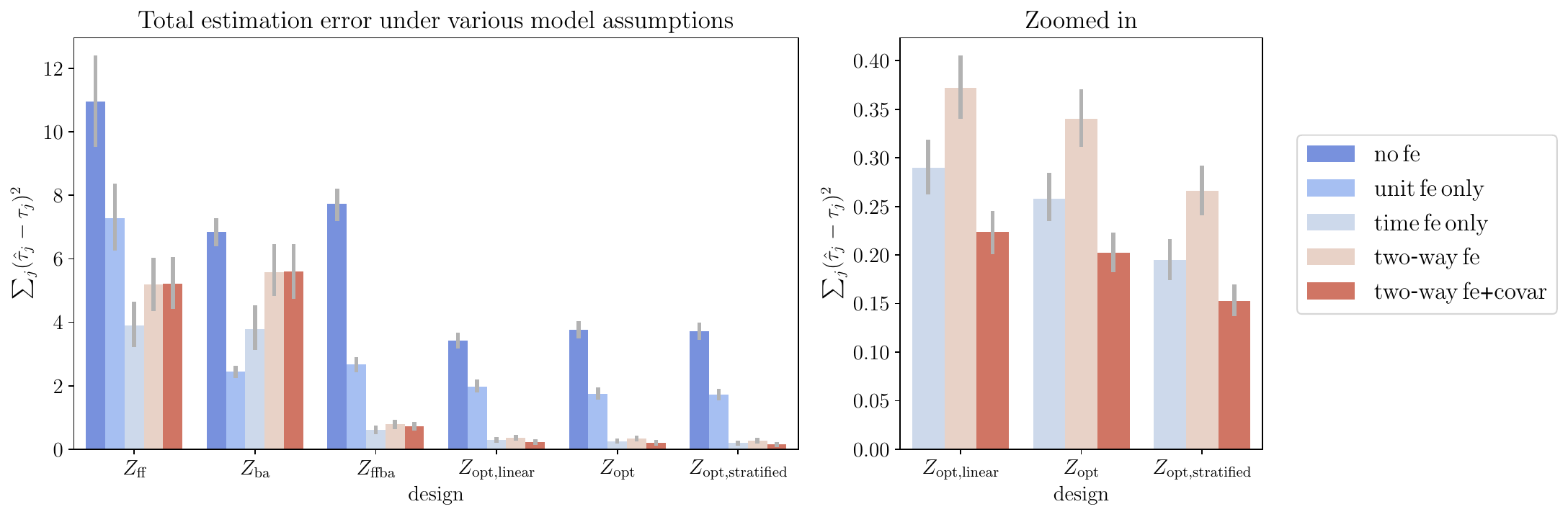}
	\end{subfigure}
	\caption{\textbf{Estimation error under various specifications.} Instantaneous and lagged effects are estimated under various specifications from 1,000 synthetic experiments of dimension $25\times 7$ with $\ell = 2$ on the flu data. }
	\label{fig:various-estimation-method}
\end{figure}

\begin{figure}[H]
	\centering
	\begin{subfigure}{.9\textwidth}
		\centering
		\includegraphics[width=1\linewidth]{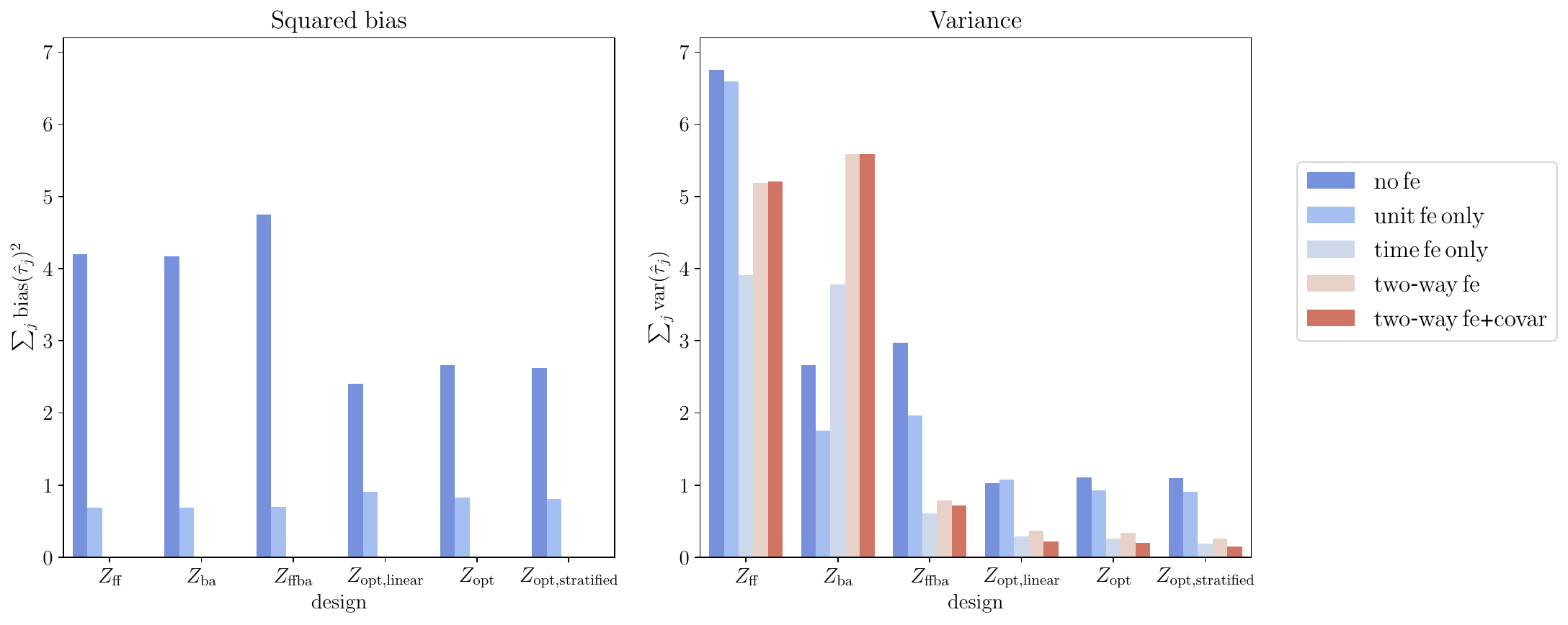}
	\end{subfigure}
	\caption{\textbf{Bias and variance decomposition.} Instantaneous and lagged effects are estimated under various specifications from 1,000 synthetic experiments of dimension $25\times 7$ with $\ell = 2$ on the flu data. }
	\label{fig:bias-variance}
\end{figure}

Figure \ref{fig:various-optimal-design} provides examples of T-optimal designs that maximize \eqref{eqn:obj} under each of the above five specifications. Figure \ref{fig:various-estimation-method} compares the total mean-squared error (MSE) of $\bm{\tau}$ of various designs under the above specifications of the estimator. Figure \ref{fig:bias-variance} decomposes the total MSE into bias squared and variance of various designs. There are three findings from Figure \ref{fig:various-estimation-method}:
\begin{enumerate}
    \item Allowing for time-fixed effects in the specification can significantly reduce MSE. The reduction mainly comes from bias reduction, and also comes from variance reduction for most designs. As the flu occurrence rate fluctuates by month, allowing for time-fixed effects can control this seasonality effect in the estimation of $\bm{\tau}$. This is particularly useful for estimation error reduction when $T$ is small (which is the case in this experiment).
    \item Allowing for covariates in the specification ({\it i.e.}, ``two-way fe$+$covar'') can further reduce MSE. The reduction mainly comes from variance reduction. 
    \item 
    Among all the designs, $Z_{\mathrm{opt,linear}}$ and $Z_{\mathrm{opt,stratified}}$ have the smallest MSE and variance under various specifications. This implies that our designs are robust to specification of the the estimator (especially when misspecification is a concern).
\end{enumerate}
In summary, to maximally reduce the estimation error, both the treatment decisions (design) and the specification of estimator play major roles.

\subsubsection{Robustness to Alternative Metrics}\label{subsubsec:robustness-to-alternative-metrics}
We further evaluate various designs by alternative metrics. Specifically, we consider the squared estimation error of cumulative effect, that is, $\big(\sum_{j = 0}^{\ell} (\hat \tau_{j} - \tau_j) \big)^2$, and metrics related to hypothesis testing. When we conduct hypothesis testing, we are interested in true positives (TP), true negatives (TN), false positives (FP), and false negatives (FN) in the confusion matrix. For each $j \in \{0\} \cup [\ell]$, the positive class of $\tau_j$ is defined as $\{|\tau_j|=a\}$, and the negative class is defined as $\{|\tau_j|=0\}$. {\blue  We use the same $a$ for different $j$, so that the results of $\tau_j$ for different $j$ are comparable.\footnote{
In contrast to the estimation error metrics in which $\tau_j$'s are different, as $j$ varies, in the hypothesis testing metrics, they are all equal. However, we empirically confirm that varying $a$ does not change the ranking of the performance of various designs, with respect to the estimation error.
} }
On each randomly selected block from the original control data, we then run a pair of synthetic experiments such that $\tau_j$ is set to $a$ (positive class) in one experiment, or to $0$ (negative class) in the other experiment. We then calculate the ${\sf t}$-statistic of $\hat{\tau}_j$ for all of these $2m$ experiments: If the absolute ${\sf t}$-statistic is above some threshold $\iota$, the estimated class of $\tau_j$ is positive; otherwise, the estimated class is negative. We can then compare the true class with the estimated class on all $2m$ experiments, and count TP, TN, FP and FN for each $j$. If we vary the threshold $\iota$, in the same spirit as when the receiver operating characteristic (ROC) curve for a binary classification problem is generated, then the estimated class may change, and TP, TN, FP and FN may change as well. In fact, FP and FN are type I and type II errors of the test.
Therefore, we can vary $\iota$ and study, for various treatment designs, how the TP rate, or equivalently, power, varies with the FP rate, or equivalently, significance level. Similarly, we can study how the ``precision", $\mathrm{TP}/(\mathrm{TP}+\mathrm{FP})$, varies with the recall, $\mathrm{TP}/(\mathrm{TP}+\mathrm{FN})$.\footnote{We put ``precision" in quotes here to distinguish between this notion of ``precision" that is standard in statistical learning literature and our main precision metric defined in Section \ref{sec:model}.} Overall, for each $j$, $0\le j\le \ell$, we obtain one ROC curve and one ``precision"-recall curve.

\begin{figure}[h!]
\captionsetup{singlelinecheck = false, format= hang,justification=raggedright}
	\begin{subfigure}{0.286\textwidth}
		\centering
		\includegraphics[width=1\linewidth]{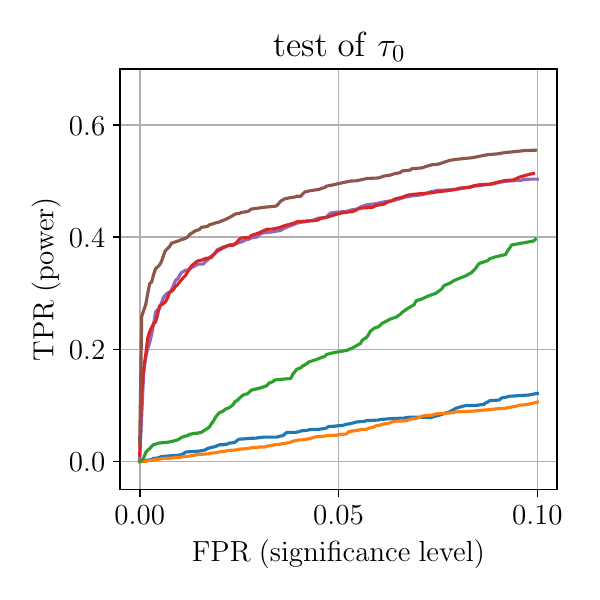}
	\end{subfigure}%
	\begin{subfigure}{0.286\textwidth}
		\centering
		\includegraphics[width=1\linewidth]{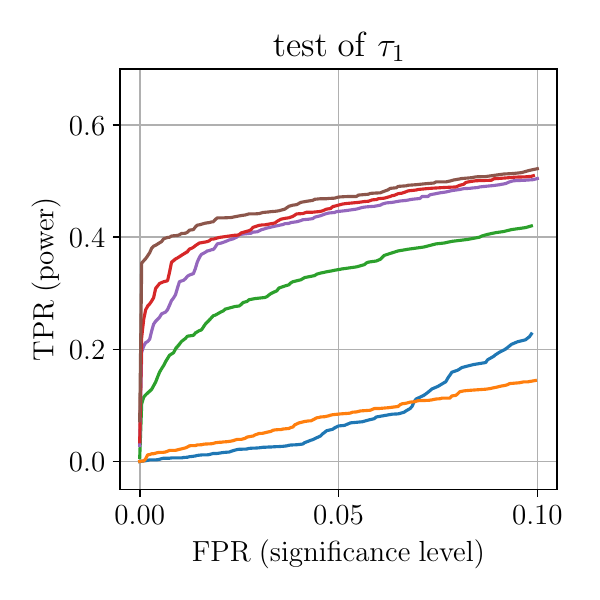}
	\end{subfigure}%
	\begin{subfigure}{0.429\textwidth}
		\centering
		\includegraphics[width=1\linewidth]{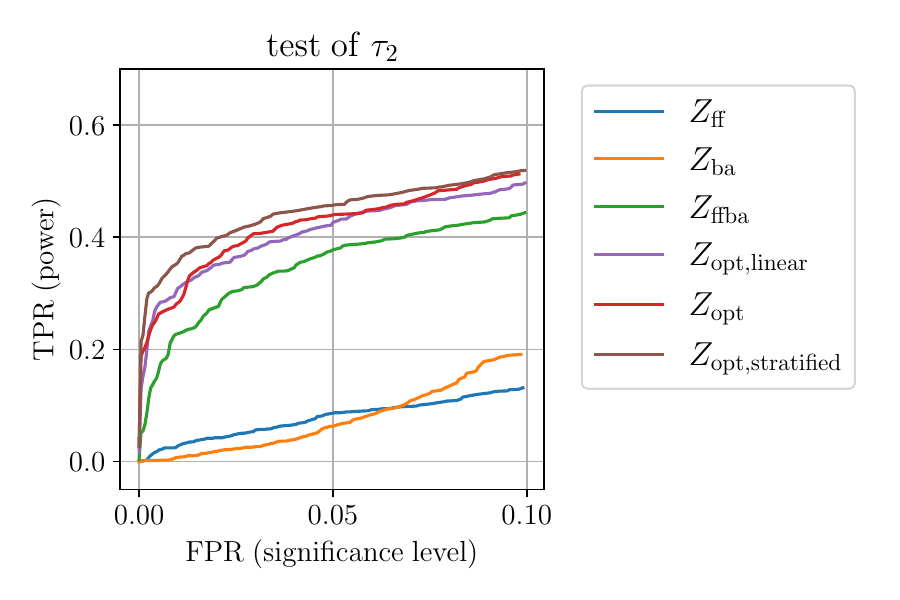}
	\end{subfigure}
	\caption{\textbf{ROC curve.} The TP and FP rates are calculated from 2,000 pairs of synthetic experiments with dimension $50\times 7$ and $\ell = 2$ on the flu data. The true positive class of $\tau_j$ is defined as $\{|\tau_j| = 0.1(NT)^{-1} \sum_{i,t} Y_{it}(-\bm{1}_{\ell+1})\}$. }
	\label{fig:roc-equal-tau}
\end{figure}

\begin{figure}[H]
	\centering
	\begin{subfigure}{1\textwidth}
		\centering
		\includegraphics[width=0.65\linewidth]{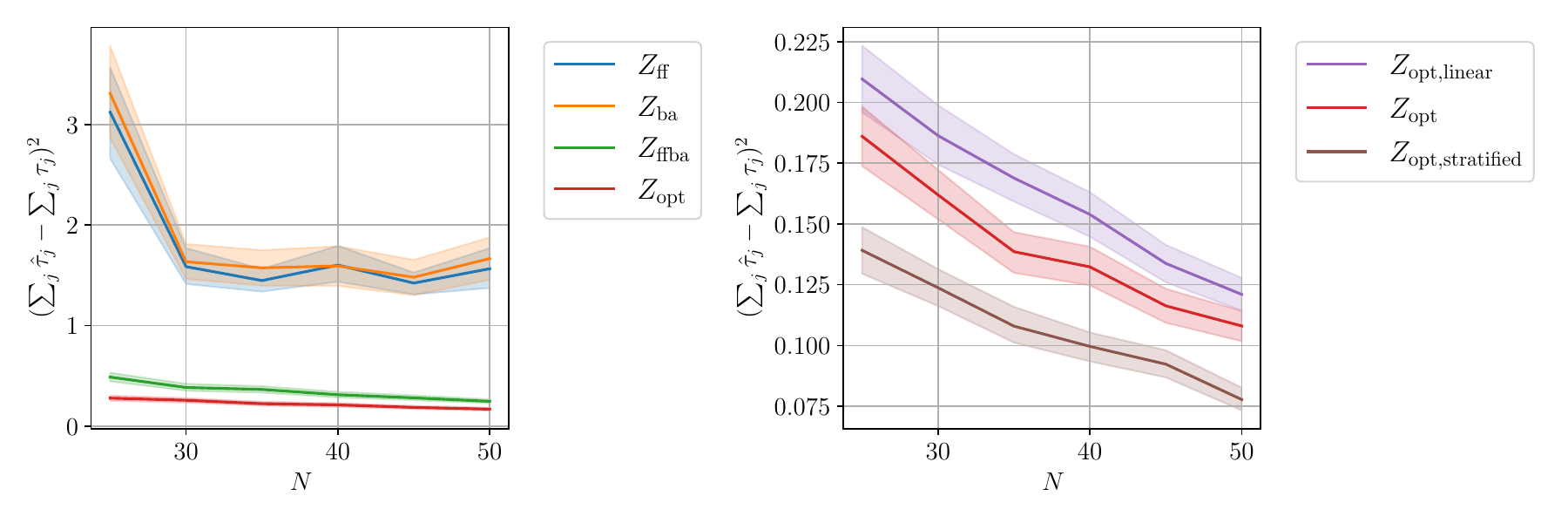}
		\caption{Flu data}
	\end{subfigure}
	\begin{subfigure}{1\textwidth}
		\centering
		\includegraphics[width=0.65\linewidth]{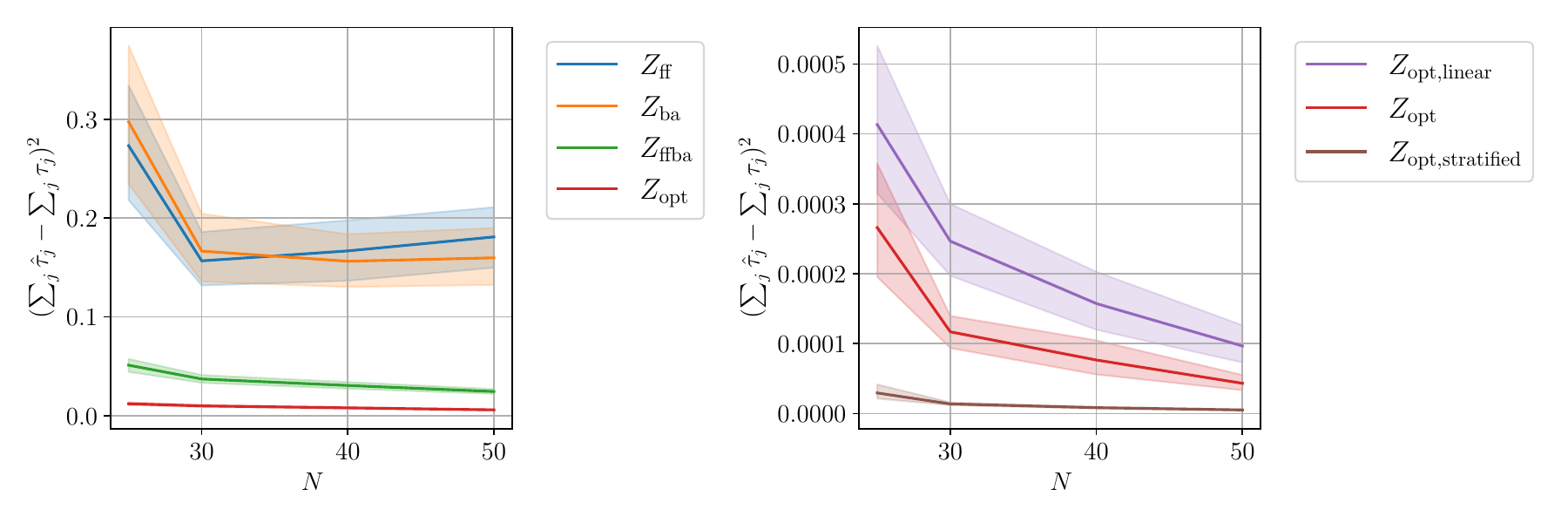}
		\caption{Home medical visit data}
	\end{subfigure}
	\begin{subfigure}{1\textwidth}
		\centering
		\includegraphics[width=0.65\linewidth]{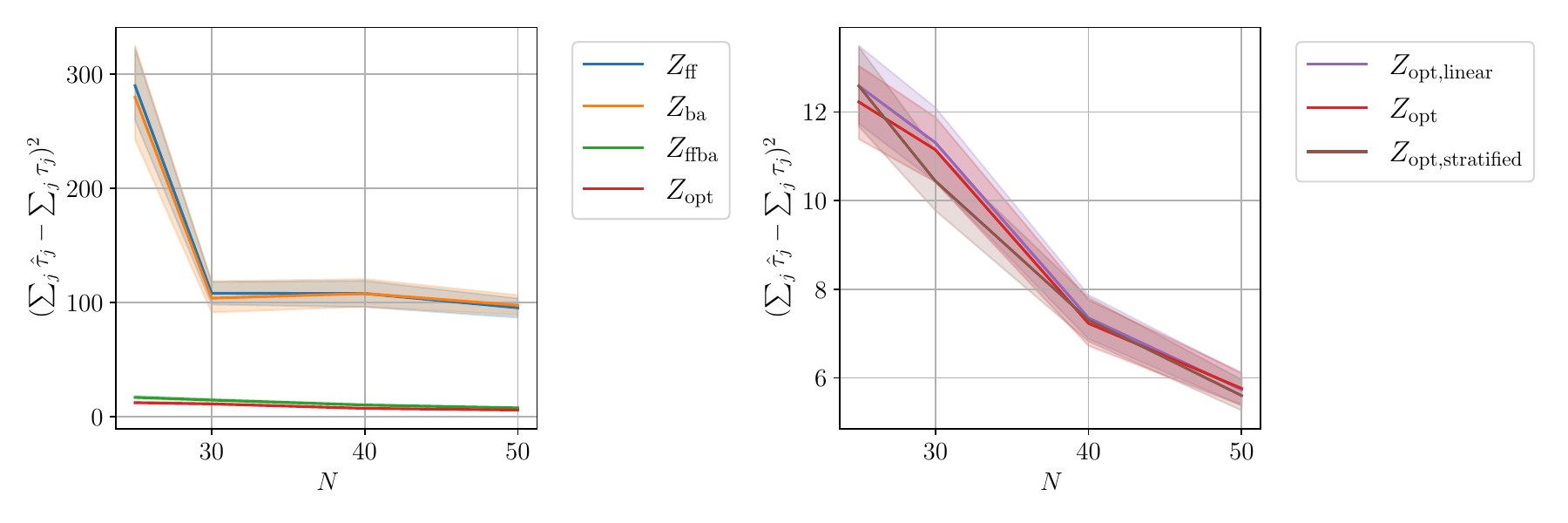}
		\caption{Grocery data}
	\end{subfigure}
	\begin{subfigure}{1\textwidth}
		\centering
		\includegraphics[width=0.65\linewidth]{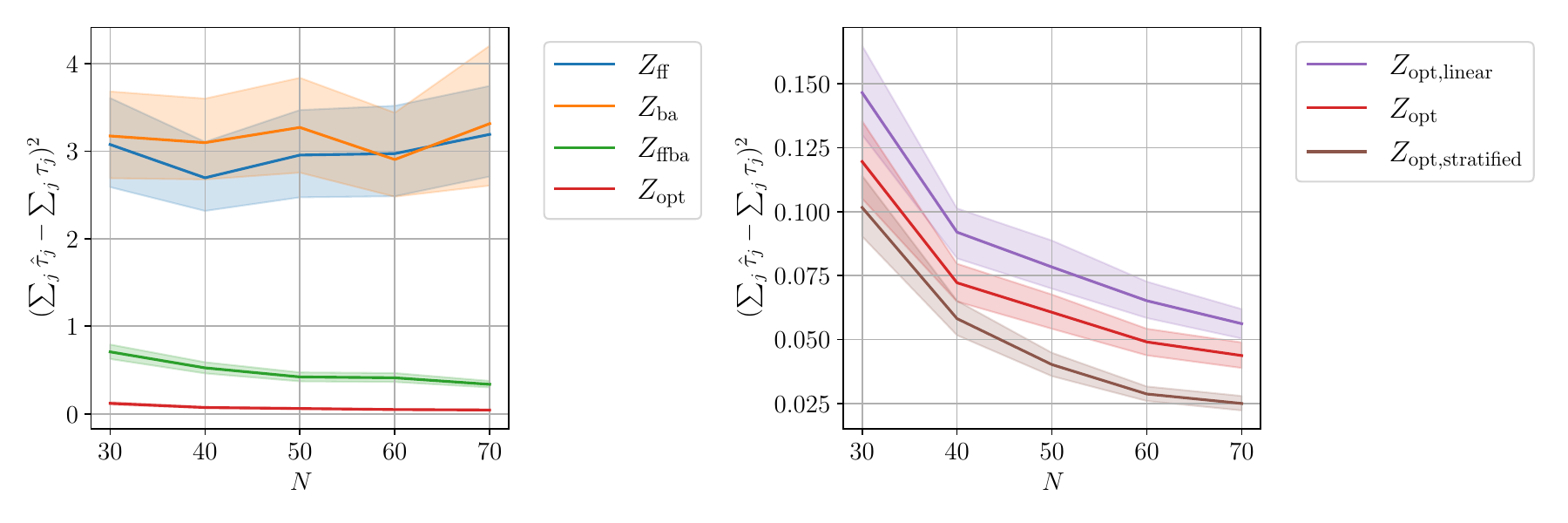}
		\caption{Loan data}
	\end{subfigure}
	\vspace{0.05in}
	\caption{\textbf{Alternative metrics: squared estimation error of cumulative effect.} These figures show the mean and 95\% confidence band of the squared estimation error of the cumulative effect of various designs, based on 2,000 randomly sampled blocks with $\ell = 2$. The data generating process is identical to that in Figure \ref{fig:additional-varying-N}.}
	\label{fig:varying-N-other-metrics}
\end{figure}

The findings in Section \ref{subsec:fixed-sample-empirical-results} are robust to alternative evaluation metrics. Figure \ref{fig:varying-N-other-metrics} shows the squared estimation error of cumulative effect of various treatment designs on all four data sets. The findings from Figure \ref{fig:varying-N-other-metrics} are aligned with those from Figure \ref{fig:varying-N-flu}.
Figure \ref{fig:roc-equal-tau} shows the ROC curve of various designs (i.e., power vs. significance level) on the flu data for testing each of $\tau_0$, $\tau_1$, and $\tau_2$, for when significance level is up to 10\%. $Z_{\OPT,\mathrm{stratified}}$ has consistently higher power than all other designs, at all significance levels. $Z_{\OPT}$ also outperforms or nearly ties with $Z_{\OPT,\mathrm{linear}}$. These three of these designs dominate benchmarks ($Z_{\FF}$, $Z_{\BA}$, and $Z_{\FFBA}$). 
Their corresponding area under the curve (AUC) values (for the full ROC curves) are shown in 
Table \ref{tab:auc}. The AUC of $Z_{\FF}$ and $Z_{\BA}$ are consistently and significantly lower than the AUC of other treatment designs in the test of $\tau_0$, $\tau_1$ and $\tau_2$. The AUC of $Z_{\OPT}$ is consistently higher than that of $Z_{\FFBA}$ and the improvement is more noticeable for $\tau_1$. Consistent with the previous metric, $Z_{\OPT,\mathrm{stratified}}$ further improves upon $Z_{\OPT}$. 

\begin{table}[h!]
    \centering
    {
    \begin{tabular}{l|p{1.5cm}p{1.5cm}p{1.5cm}p{1.5cm}p{1.5cm}p{1.8cm}}
\toprule
 &  $Z_{\FF}$ &  $Z_{\BA}$ &  $Z_{\FFBA}$ &  $Z_{\OPT}$ &  $Z_{\OPT,\mathrm{linear}}$ &  $Z_{\OPT,\mathrm{stratified}}$  \\
\midrule
$\tau_0$   &              0.614 &              0.633 &                0.744 &               0.750 &                      0.740 &                          0.761 \\
$\tau_1$   &              0.598 &              0.563 &                0.702 &               0.745 &                      0.740 &                          0.756 \\
$\tau_2$   &              0.590 &              0.629 &                0.738 &               0.748 &                      0.736 &                          0.762 \\
\bottomrule
\end{tabular}
    }
    \caption{\textbf{AUC.} AUC of various treatment designs and hypothesis tests in Figure \ref{fig:roc-equal-tau}.}
    \label{tab:auc}
\end{table}

\clearpage

\clearpage

\subsubsection{Varying Treatment Effects.}\label{subsubsec:magnitude-treatment-effect}

\begin{figure}[H]
	\centering
	\begin{subfigure}{1\textwidth}
		\centering
		\includegraphics[width=0.75\linewidth]{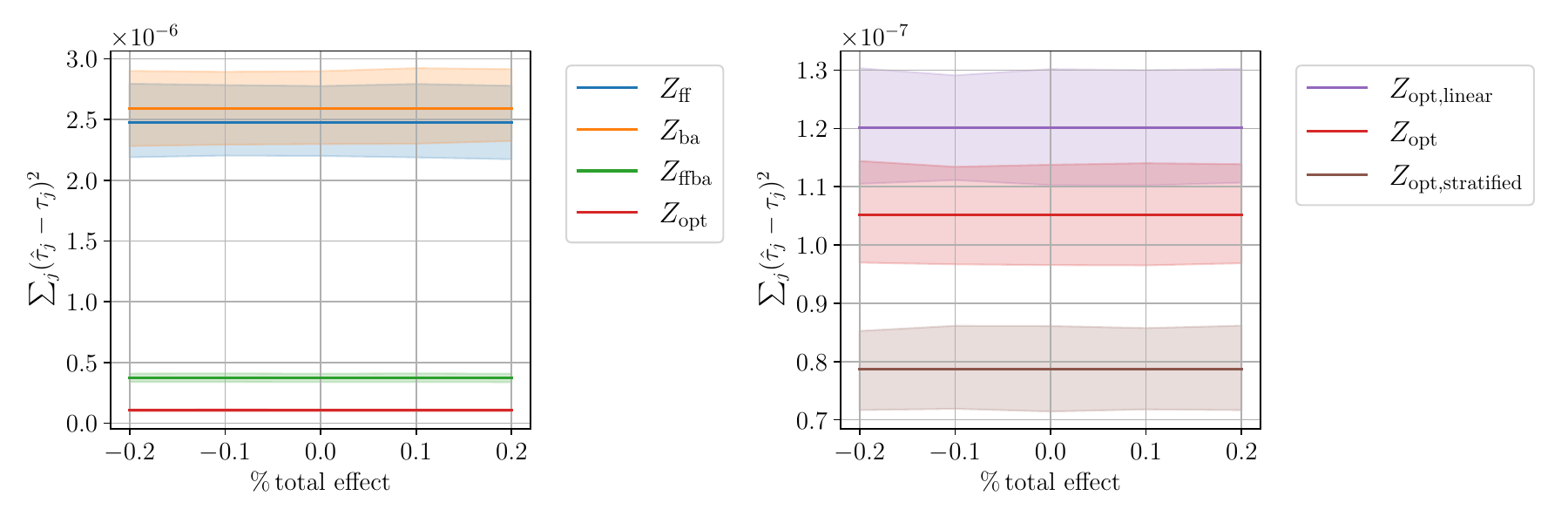}
	\end{subfigure}
	\vspace{0.01in}
	\caption{\textbf{Varying magnitude of treatment effects.} These figures show the mean and 95\% confidence band of $\sum_{j}(\hat{\tau}_j - \tau_j)^2$ for various designs, based on 1,000 synthetic non-adaptive experiments with $\ell = 2$, $T = 7$ and $N = 50$ on the flu data. The red curve in two figures are identical. The right figure zooms in the left one. The total cumulative effect varies from $-20$\% to $20$\% of the average monthly flu occurrence rate. The total estimation error stays constant with varying total cumulative effect.}
	\label{fig:varying-effect}
\end{figure}

The total estimation error does not vary with the magnitude of treatment effects, as shown in Figure \ref{fig:varying-effect}. This is because the (weighted) least squares estimator is linear $Y_{it}$, and $Y_{it}$ is specified as linear in $\bm{\tau}$. Therefore, the estimation error of the (weighted) least squares estimator can be written as a linear function of residuals $\varepsilon_{it}$ that does not depend on $\bm{\tau}$, implying that the estimation error does not depend on the magnitude of treatment effects. 

\begin{figure}[H]
	\centering
	\begin{subfigure}{1\textwidth}
		\centering
		\includegraphics[width=0.75\linewidth]{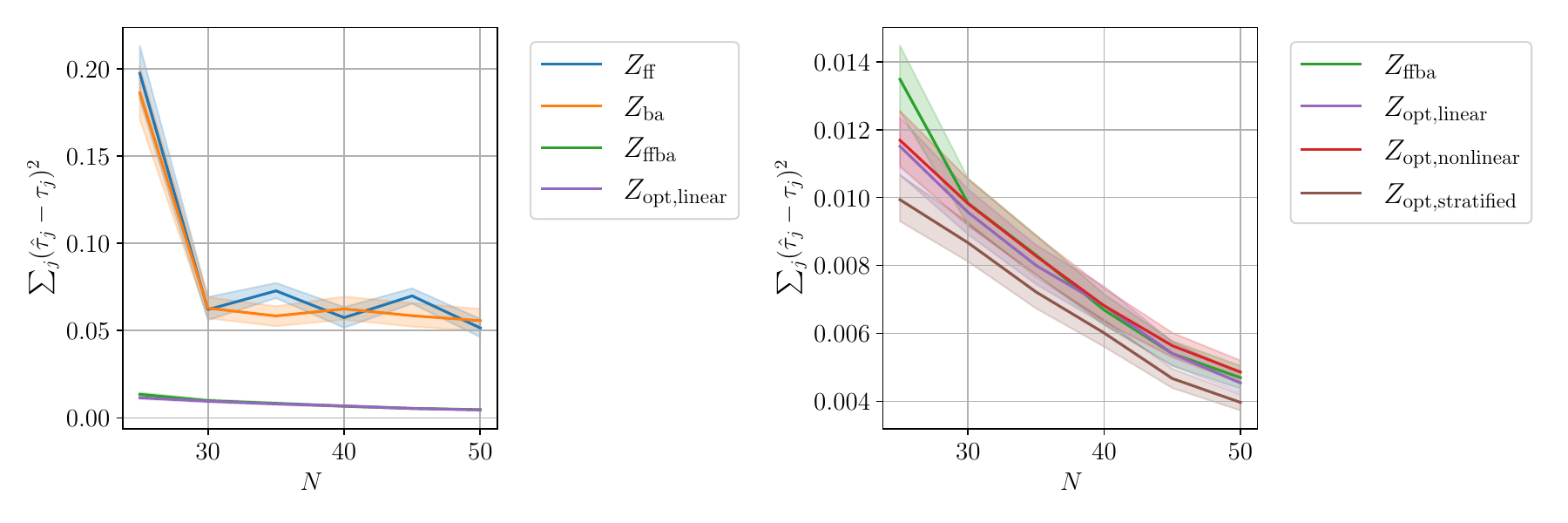}
	\end{subfigure}
	\caption{\textbf{Instantaneous effect only.} 
 These figures show the mean and 95\% confidence band of $(\hat{\tau}_0 - \tau_0)^2$ for various designs, based on 1,000 synthetic non-adaptive experiments with $\ell = 0$, $T = 14$ and varying $N$. The red curve in two figures are identical. The right figure zooms in the left one. The instantaneous effect equals $-10$\% of the average monthly flu occurrence rate. }
	\label{fig:instantaneous-effect}
\end{figure}

Furthermore, we compare various designs when the treatment only has the instantaneous effect in Figure \ref{fig:instantaneous-effect}. In this case, the treated fractions in $Z_{\OPT,\mathrm{linear}}$ and $Z_{\OPT,\mathrm{stratified}}$ are optimal, but the treated fractions in $Z_{\OPT,\mathrm{nonlinear}}$ are sub-optimal. $Z_{\OPT,\mathrm{stratified}}$, as the stratified version of $Z_{\OPT,\mathrm{linear}}$, has lowest error among all designs.

\subsubsection{Varying duration of carryover effects $\ell$}\label{ecsub:varying-ell}

\subsubsection{Specification of $\ell$}\label{ecsec:spec-ell}

\begin{figure}[H]
	\centering
	\begin{subfigure}{1.0\textwidth}
		\centering
		\includegraphics[width=1\linewidth]{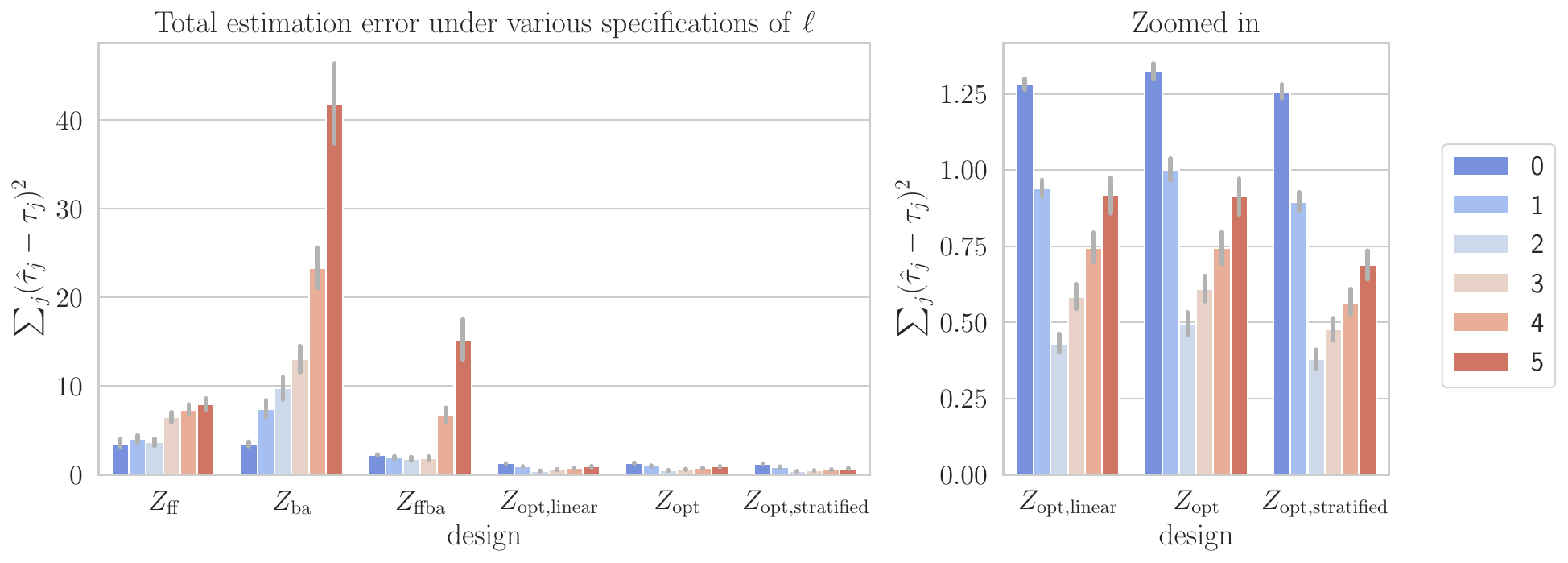}
	\end{subfigure}
	\caption{\textbf{Varying $\ell$.} Instantaneous and lagged effects are estimated under various specifications of $\ell$ from 1,000 synthetic experiments of dimension $50\times 7$ with the true number of lags as $2$ on the flu data. The model assumptions are identical to those in Figure \ref{fig:various-optimal-design}.}
	\label{fig:varying-ell}
\end{figure}

Figure \ref{fig:varying-ell} shows the estimation error under various specifications of $\ell$. The bias of $\hat{\bm{\tau}}$ generally decreases with $\ell$ when $\ell$ is smaller than the true $\ell$. However, the variance of $\hat{\bm{\tau}}$ generally increases with $\ell$. When $\ell$ is correctly specified, the estimation error of $\hat{\bm{\tau}}$ is the lowest, as compared to other specifications of $\ell$, for all treatment designs. 

\subsection{Supplementary Results for Adaptive Experiments}

\begin{figure}[h!]
    \centering
    \begin{subfigure}{1\textwidth}
		\centering
		\includegraphics[width=1\linewidth]{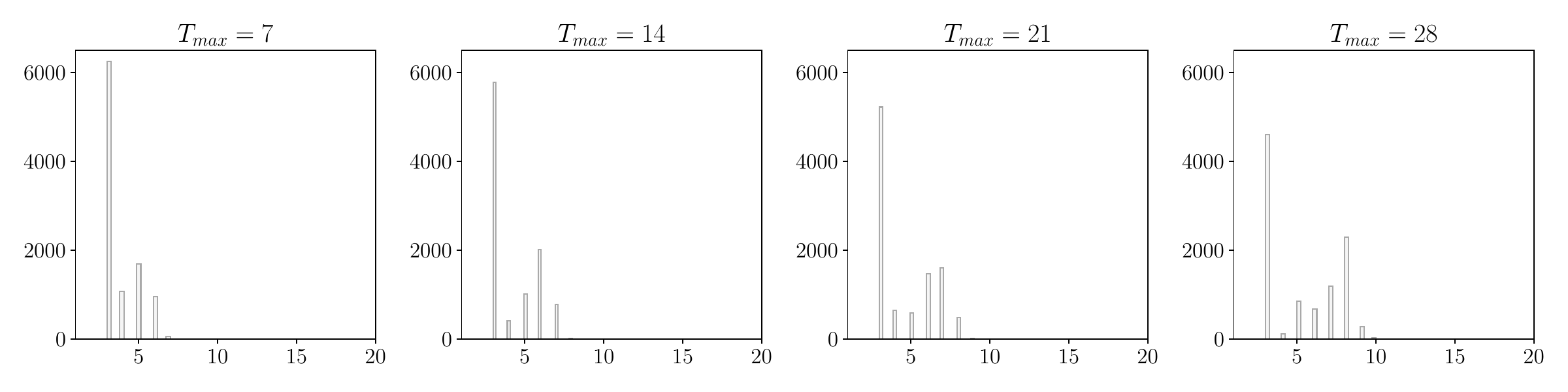}
		\caption{small threshold $c = 0.01 \cdot N/\tau_0^2$}
	\end{subfigure}
	\begin{subfigure}{1\textwidth}
		\centering
		\includegraphics[width=1\linewidth]{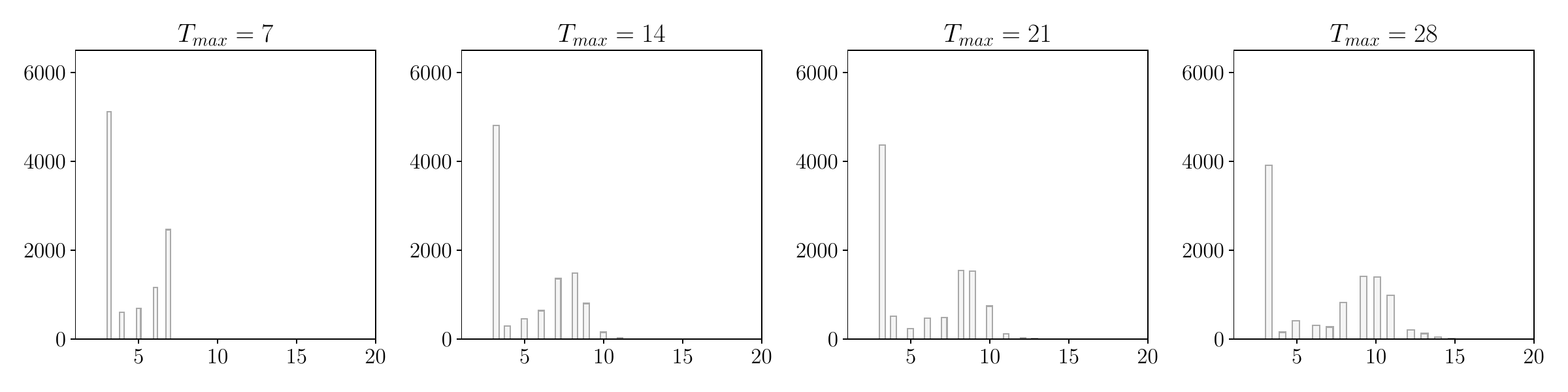}
		\caption{large threshold $c = 0.02 \cdot N/\tau_0^2$}
	\end{subfigure}
	\smallskip
	\caption{\textbf{Empirical distribution of termination time $\tilde{T}$ with alternative thresholds} This figure shows the histogram of the termination time $\tilde{T}$ for $T_{\max} \in \{7,14,21,28\}$ based on 10,000 sequential experiments. $N$ is chosen at $50$. In PGAE, the set of NTU has $N p_{\fcs} = 50 \times 0.2 = 10$ units, and both first and second sets of ATU has $N p_{\ad,1} =N p_{\ad,2} =  50 \times 0.4 = 20$ units.  $\tau_0$ is chosen at $/{-0.1}/{(NT_{\max})} \sum_{i,t} Y_{it}(-1)$ (i.e., 10\% of the average monthly flu occurrence rate). The earliest termination time $t_0$ is set as $3$. 
	}
	\label{fig:experiment-termination-time-supp}
\end{figure}

\begin{figure}[h!]
    \centering
    \begin{subfigure}{0.5\textwidth}
		\centering
		\includegraphics[width=1\linewidth]{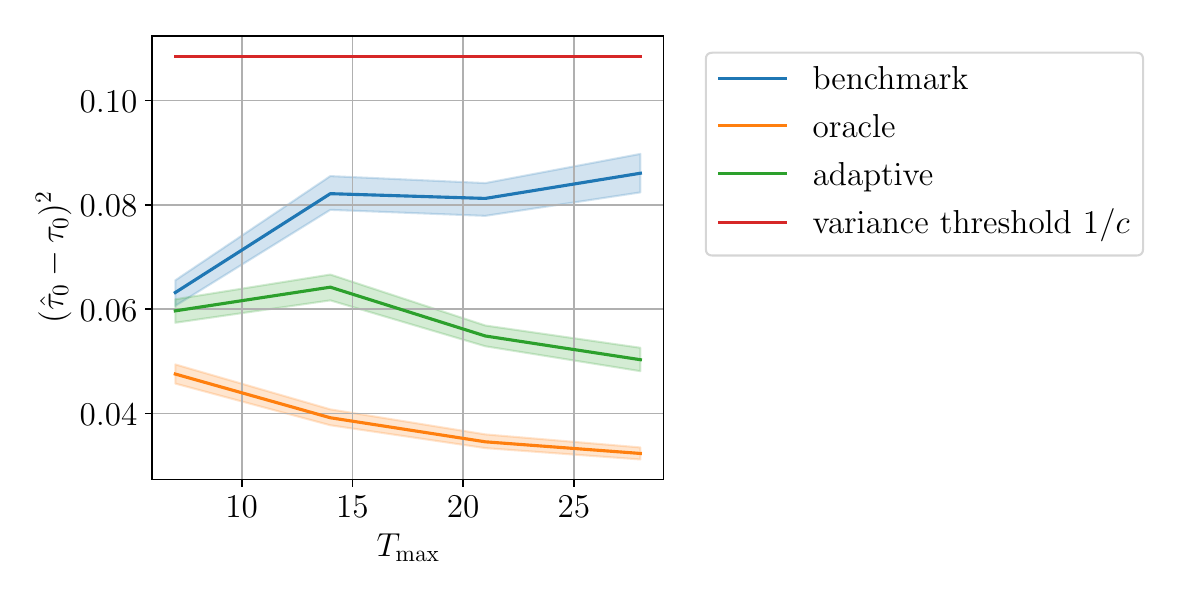}
		\caption{small threshold $c = 0.01 \cdot N/\tau_0^2$}
	\end{subfigure}%
	\begin{subfigure}{0.5\textwidth}
		\centering
		\includegraphics[width=1\linewidth]{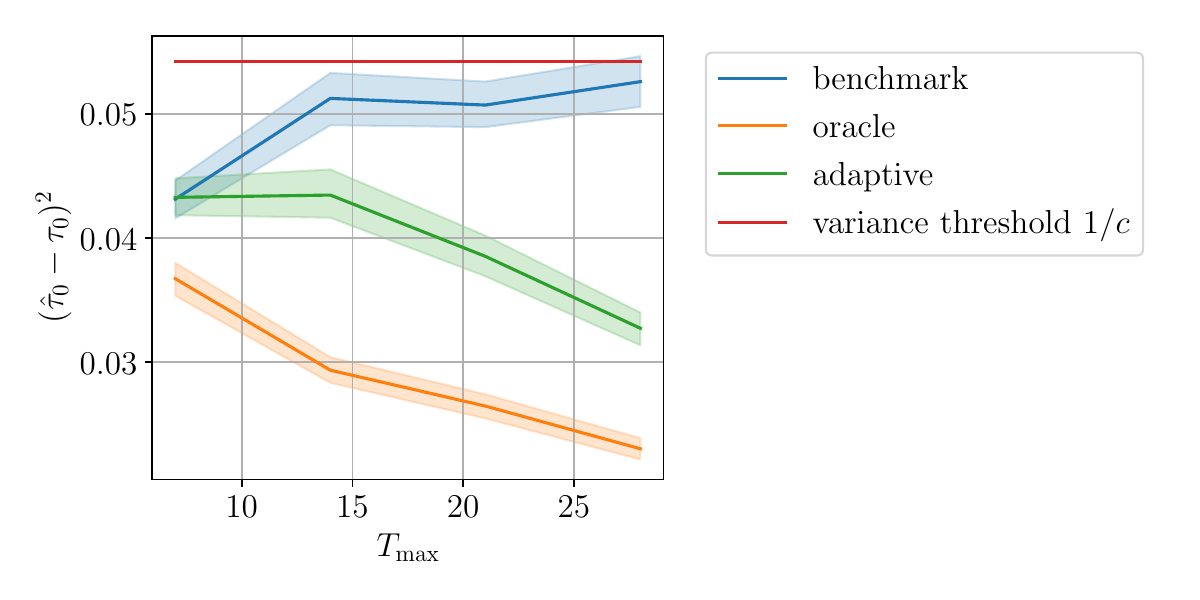}
		\caption{large threshold $c = 0.02 \cdot N/\tau_0^2$}
	\end{subfigure}
	\smallskip
	\caption{\textbf{Comparison of various designs in adaptive experiments with alternative thresholds} This figure shows the mean and 95\% confidence band of $(\hat{\tau}_0 - \tau_0)^2$ for adaptive, benchmark, and oracle designs, based on 10,000 synthetic adaptive experiments. 
	}
	\label{fig:various-opt-design-supp}
\end{figure}

The results in Section \ref{subsec:sequential-result} are robust to the choice of experiment termination threshold $c$, as shown by the histograms of experiment termination times in Figure \ref{fig:experiment-termination-time-supp} and the estimation errors of various designs in Figure \ref{fig:various-opt-design-supp}. In addition, by comparing Figure \ref{fig:experiment-termination-time} with Figure \ref{fig:experiment-termination-time-supp}, the experiment termination times tend to increase with the threshold $c$, which is as expected. Moreover, by comparing Figure \ref{fig:various-opt-design} and Figure \ref{fig:various-opt-design-supp}, the estimation errors tend to increase with the threshold $c$ for all designs, which is also as expected.

%% file: appendix_G.tex
\section{Supplementary Simulation Results for Adaptive Experiments}\label{sec:additional-simulation-sequential}

\subsection{Finite Sample Properties of Lemma \ref{lemma:asymptotic-tau-sigma}}\label{subsec:finite-sample-lemma}

We verify the finite sample properties of Lemma \ref{lemma:asymptotic-tau-sigma} using simulated data. The simulated data with $N$ units and $T$ time periods is generated as follows 
\[Y_{is} = \alpha_i + \beta_s + \tau_0 z_{is} + \varepsilon_{is},\]
where 
\[\alpha_i \stackrel{\iid}{\sim} \mathcal{N}(0,1), \qquad \beta_s \stackrel{\iid}{\sim} \mathcal{N}(0,1), \qquad \varepsilon_{is} \stackrel{\iid}{\sim} \mathcal{N}(0,\sigma^2_\varepsilon),  \]
and $Z = [z_{is}]_{i\in[N],s\in [T]}$ is randomly sampled from the treatment designs that satisfy $N^{-1} \sum_{i} z_{is} = {(2s - 1 - T)}/{T}$ ({\it i.e.}, the optimal design with instantaneous effect only). In this case, all $N$ units are NTU. 

Let $\hat{\tau}$ be the \within estimator of $\hat{\tau}_0$ and let $\estsigmasq$ be the estimated $\sigma_\varepsilon^2$ using the formula \eqref{eqn:second-moment-sigma-estimator}. We report the standardized $\hat{\tau}$, that is, 
\[ \hat{\tau}_{\mathsf{ss}} =  \sqrt{N} \cdot \frac{\hat{\tau} - \tau_0}{[\estsigmasq /(T \cdot \funfrac(\bm{\omega}_{\fcs, 1:T}, T))]^{1/2}} \]
and the standardized $\estsigmasq$, that is, \[\estsigmasq_{\mathsf{ss}} =  \sqrt{N} \cdot \frac{\estsigmasq - \sigma^2}{[\estxidaggersq/T]^{1/2}} \]
where $\estxidaggersq = \estxisq + {2}/{(T-1)} \cdot (\estsigmasq)^2$ and $\estxisq$ is estimated using the formula \eqref{eqn:fourth-moment-sigma-estimator}. 

We also report the standardized $\estsigmasq$ by a naive plug-in estimator of the variance of $\estsigmasq$, that is, 
\[\estsigmasq_{\mathsf{ss},\mathrm{naive}} =  \sqrt{N} \cdot \frac{\estsigmasq - \sigma^2}{[\estxisq_{\mathrm{naive}}/T]^{1/2}}, \]
where 
\[\estxisq_{\mathrm{naive}} = \frac{1}{NT} \sum_{i,s} \left( (\dot{y}_{is} - \hat{\tau} \cdot \dot{z}_{is})^2 - \estsigmasq \right)^2. \]

\begin{figure}[h!]
	\centering
	\begin{subfigure}{0.3\textwidth}
		\centering
		\includegraphics[width=1\linewidth]{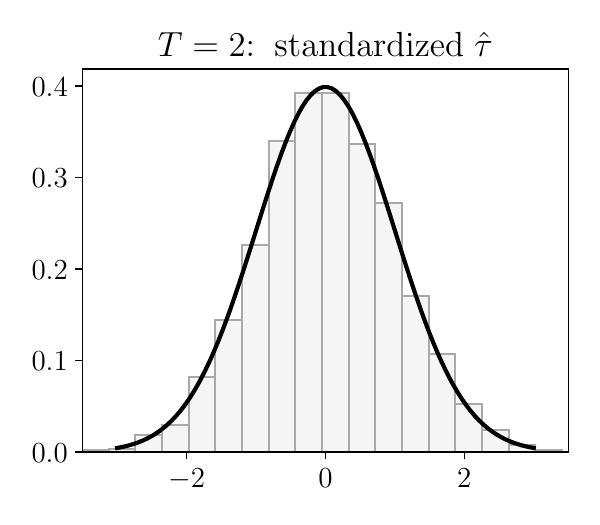}
	\end{subfigure}
	\begin{subfigure}{0.3\textwidth}
		\centering
		\includegraphics[width=1\linewidth]{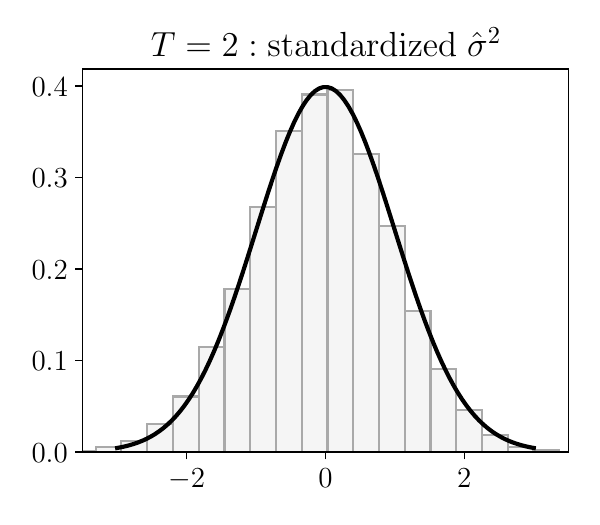}
	\end{subfigure}
	\begin{subfigure}{0.3\textwidth}
		\centering
		\includegraphics[width=1\linewidth]{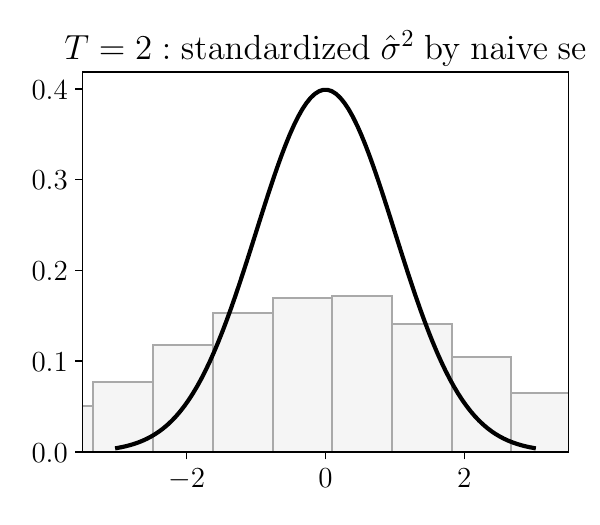}
	\end{subfigure}
	\begin{subfigure}{0.3\textwidth}
		\centering
		\includegraphics[width=1\linewidth]{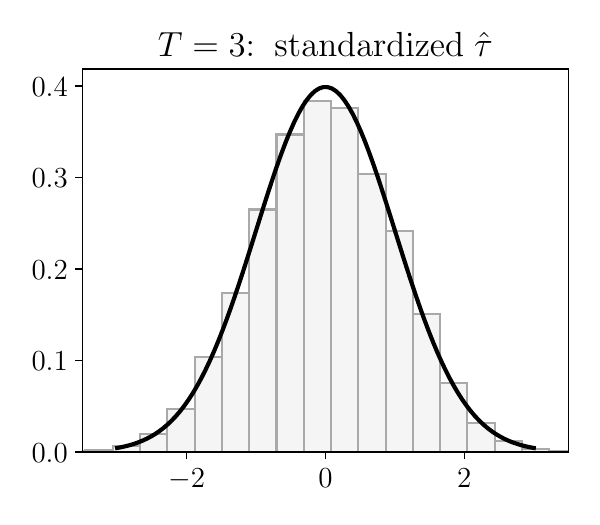}
	\end{subfigure}
	\begin{subfigure}{0.3\textwidth}
		\centering
		\includegraphics[width=1\linewidth]{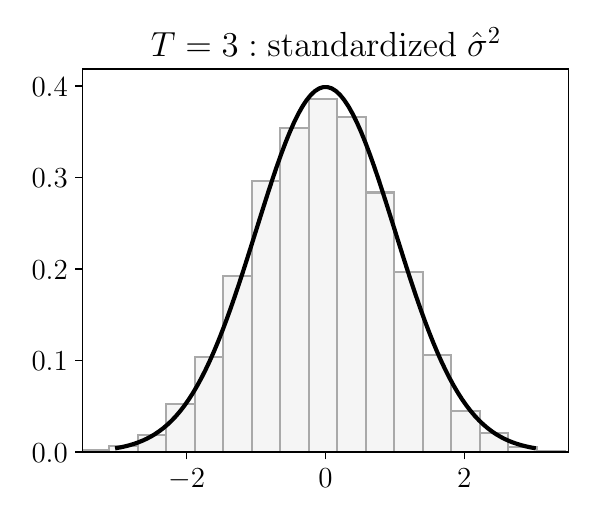}
	\end{subfigure}
	\begin{subfigure}{0.3\textwidth}
		\centering
		\includegraphics[width=1\linewidth]{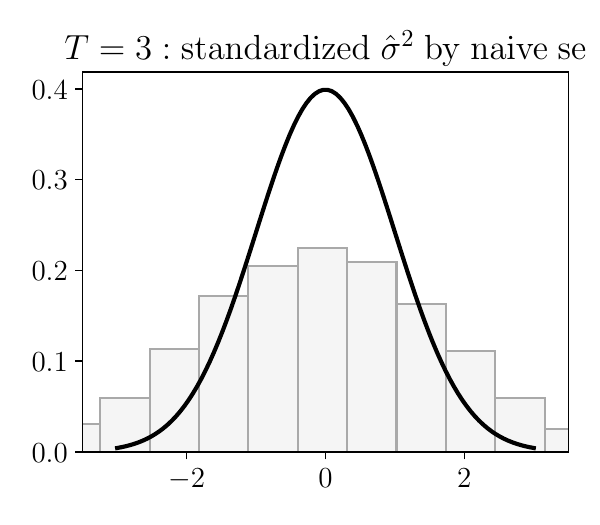}
	\end{subfigure}
	\begin{subfigure}{0.3\textwidth}
		\centering
		\includegraphics[width=1\linewidth]{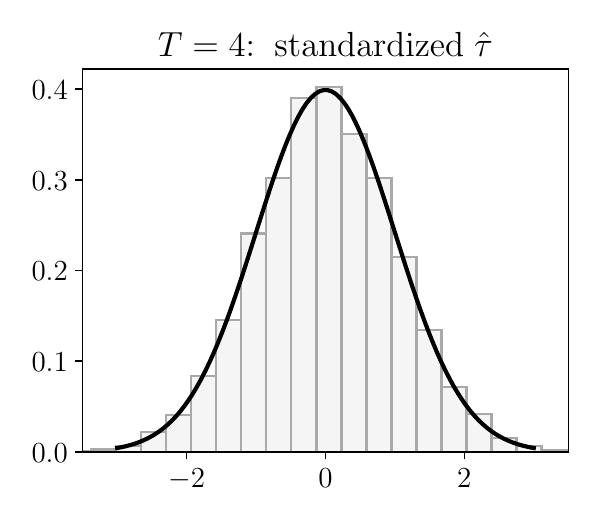}
	\end{subfigure}
	\begin{subfigure}{0.3\textwidth}
		\centering
		\includegraphics[width=1\linewidth]{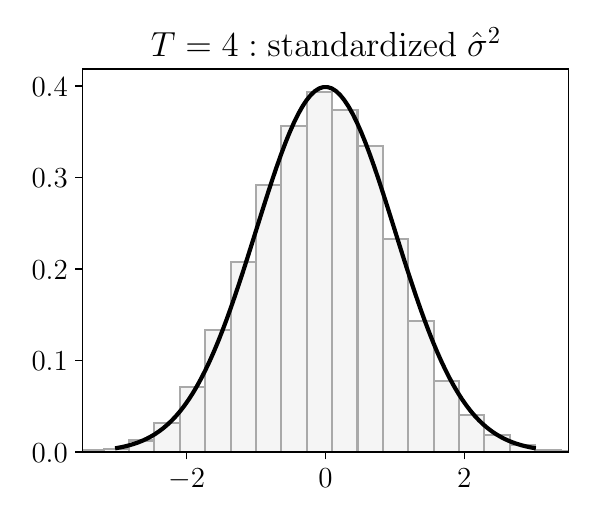}
	\end{subfigure}
	\begin{subfigure}{0.3\textwidth}
		\centering
		\includegraphics[width=1\linewidth]{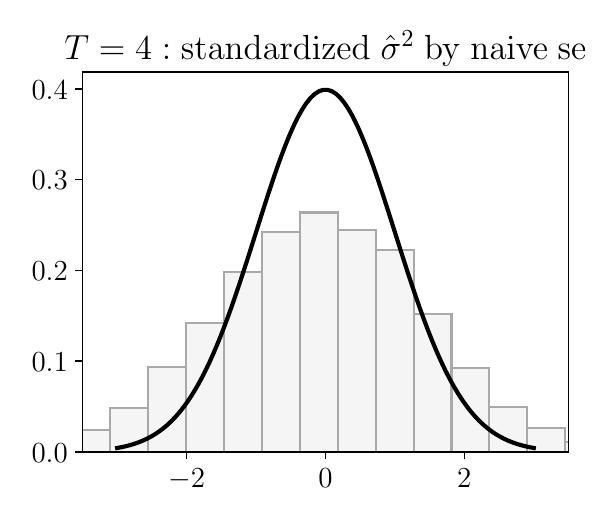}
	\end{subfigure}
	\caption{Finite sample properties of Lemma \ref{lemma:asymptotic-tau-sigma}: Histograms of $\hat{\tau}_{\mathsf{ss}}$, $\estsigmasq_{\mathsf{ss}}$, and $\estsigmasq_{{\mathsf{ss}},\mathrm{naive}} $.  The standard normal density function is superimposed on the histograms. $N = 10,000$, $\tau_0 = 3$ and $\sigma=1$.}
	\label{fig:test-asymptotics}
\end{figure}

We repeat the above data generating and estimation procedure for 1,000 times and report the histograms of $\hat{\tau}_{\mathsf{ss}}$, $\estsigmasq_{\mathsf{ss}}$, and $\estsigmasq_{\mathsf{ss},\mathrm{naive}}$ in Figure \ref{fig:test-asymptotics} for various $T$.  As shown in Figure \ref{fig:test-asymptotics}, the histograms of $\hat{\tau}_{\mathsf{ss}}$ and  $\estsigmasq_{\mathsf{ss}}$ are close to the standard normal density function, showing the good finite sample properties of $\hat{\tau}_{\mathsf{ss}}$ and  $\estsigmasq_{\mathsf{ss}}$. However, the histograms of $\estsigmasq_{\mathsf{ss},\mathrm{naive}} $ deviate from the standard normal density function and the deviation is larger for a smaller $T$, implying that the naive plug-in estimator $\estxisq_{\mathrm{naive}}$ underestimates the variance of $\estsigmasq$. 

Furthermore, we report the mean of $\hat{\tau}_{\mathsf{ss}}^2 $, $(\estsigmasq_{\mathsf{ss}})^2$ and $\hat{\tau}_{\mathsf{ss}} \cdot \estsigmasq_{\mathsf{ss}}$, which serve as the estimates of $\var(\hat{\tau}_{\mathsf{ss}})$, $\var(\estsigmasq_{\mathsf{ss}})$, and $\mathrm{Cov}(\hat{\tau}_{\mathsf{ss}}, \estsigmasq_{\mathsf{ss}})$, in Table \ref{tab:test-asymptotics}. As shown in Table \ref{tab:test-asymptotics}, the estimates of  $\var(\hat{\tau}_{\mathsf{ss}})$ and $\var(\estsigmasq_{\mathsf{ss}})$ are close to $1$ for various $T$, serving as additional supports of the good finite sample properties of $\hat{\tau}_{\mathsf{ss}}$ and $\estsigmasq_{\mathsf{ss}}$. Moreover, the estimate of $\mathrm{Cov}(\hat{\tau}_{\mathsf{ss}}, \estsigmasq_{\mathsf{ss}})$ is close to $0$, verifying the asymptotic independence between $\hat{\tau}_{\mathsf{ss}}$ and $ \estsigmasq_{\mathsf{ss}}$.

\begin{table}[h!]
    \centering
\begin{tabular}{l|p{2cm}p{2cm}p{2cm}}
\toprule
  $T$ &  $\widehat{\var}(\hat{\tau}_{\mathsf{ss}})$ &  $\widehat{\var}(\estsigmasq_{\mathsf{ss}})$ &  $\widehat{\mathrm{Cov}}(\hat{\tau}_{\mathsf{ss}}, \estsigmasq_{\mathsf{ss}})$   \\
\midrule
2 & 0.994 & 1.012 & $-0.008$ \\
3 & 1.022 & 1.002 & $-0.008$ \\
4 & 1.005 & 0.994 & 0.001 \\
\bottomrule
\end{tabular}
    \caption{Finite sample properties of Lemma \ref{lemma:asymptotic-tau-sigma}: Estimates of $\var(\hat{\tau}_{\mathsf{ss}})$, $\var(\estsigmasq_{\mathsf{ss}})$, and $\mathrm{Cov}(\hat{\tau}_{\mathsf{ss}}, \estsigmasq_{\mathsf{ss}})$ with $N = 10,000$, $\tau_0 = 3$  and $\sigma=1$. }
    \label{tab:test-asymptotics}
\end{table}

\clearpage

\subsection{Finite Sample Properties of Theorem \ref{theorem:asymptotic-page}}\label{subsec:finite-sample-pgae}

We verify the finite sample properties of Theorem \ref{theorem:asymptotic-page} using simulated data. The data generating process is the same as that in Section \ref{subsec:finite-sample-lemma}. We run adaptive experiments using PGAE that allow for the early termination of the experiment. Let $\tilde{T}$ be the duration of the adaptive experiment.

We report the standardized $\hat{\tau}_\all$, that is, 
\[ \hat{\tau}_{\all,{\mathsf{ss}}} =  \sqrt{N} \cdot \frac{\hat{\tau} - \tau_0}{[\estsigmasq_{\ad,2} /(\tilde{T} \cdot \funfrac(\bm{\omega}_{\fcs, 1:\tilde{T}}, \tilde{T}))]^{1/2}}, \]
Moreover, we report the standardized $\estsigmasq_{\ad,2}$, that is, \[\estsigmasq_{\ad,2,{\mathsf{ss}}} =  \sqrt{N} \cdot \frac{\estsigmasq_{\ad,2} - \sigma^2}{[\estxidaggersq_{\ad,2}/\tilde{T}]^{1/2} }, \]
where $\estxidaggersq_{\ad,2} = \estxisq_{\ad,2} + {2}/{(\tilde{T}-1)} \cdot (\estsigmasq_{\ad,2})^2$ and $\estxisq_{\ad,2}$ is estimated using the formula \eqref{eqn:fourth-moment-sigma-estimator} on $\mathcal{S}_{\ad,2}$. 

\begin{figure}[h!]
	\centering
	\begin{subfigure}{0.25\textwidth}
		\centering
		\includegraphics[width=1\linewidth]{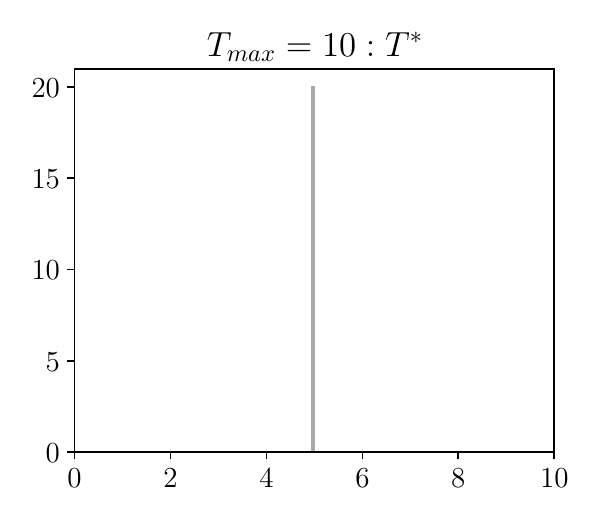}
	\end{subfigure}%
	\begin{subfigure}{0.25\textwidth}
		\centering
		\includegraphics[width=1\linewidth]{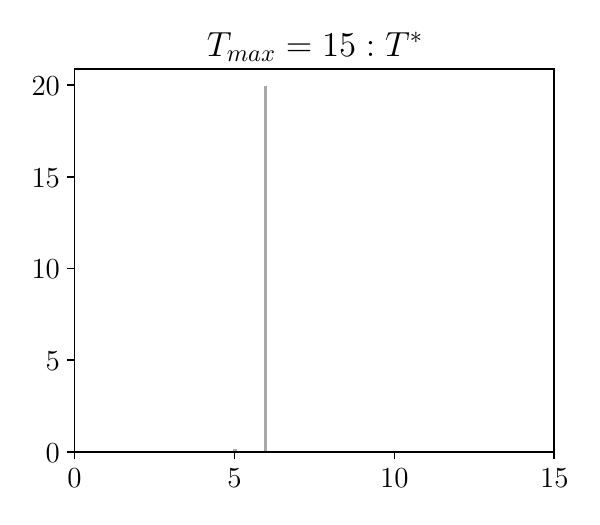}
	\end{subfigure}%
	\begin{subfigure}{0.25\textwidth}
		\centering
		\includegraphics[width=1\linewidth]{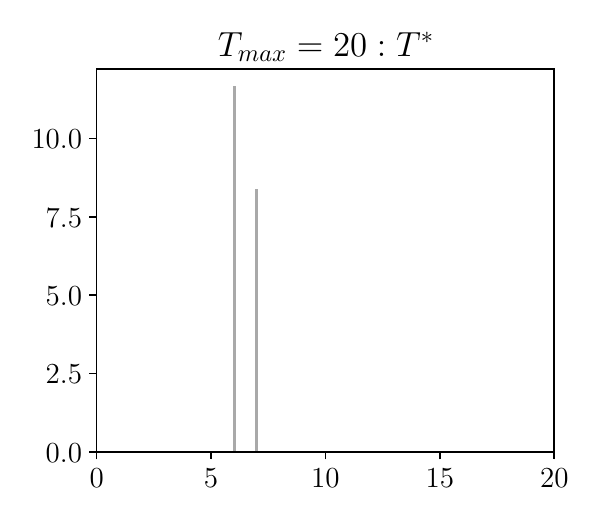}
	\end{subfigure}%
	\begin{subfigure}{0.25\textwidth}
		\centering
		\includegraphics[width=1\linewidth]{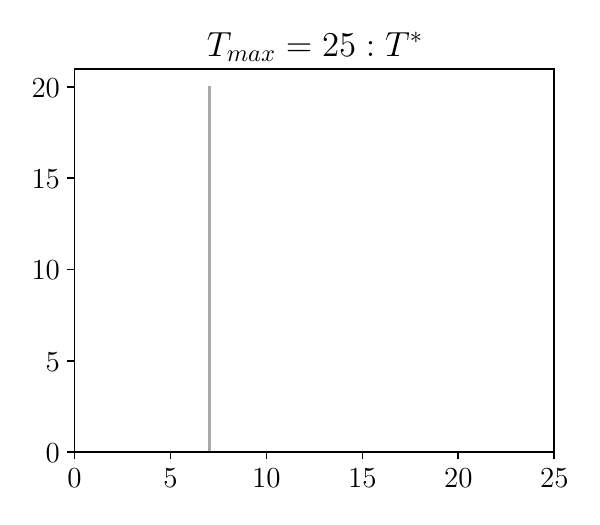}
	\end{subfigure}
	\caption{Histogram of the experiment termination times from PGAE for various $T_{\max}$ with $N = 500$, $\sigma_\varepsilon = 3$ and the threshold for precision $c = N/\sigma_\varepsilon^2 $.}
	\label{fig:experiment-termination-page}
\end{figure}

We generate the simulated data and run adaptive experiments for 1,000 times. Note that the experiment termination time varies across the 1,000 iterations, as shown in Figure \ref{fig:experiment-termination-page}.

\begin{figure}[h!]
	\centering
	\begin{subfigure}{0.3\textwidth}
		\centering
		\includegraphics[width=1\linewidth]{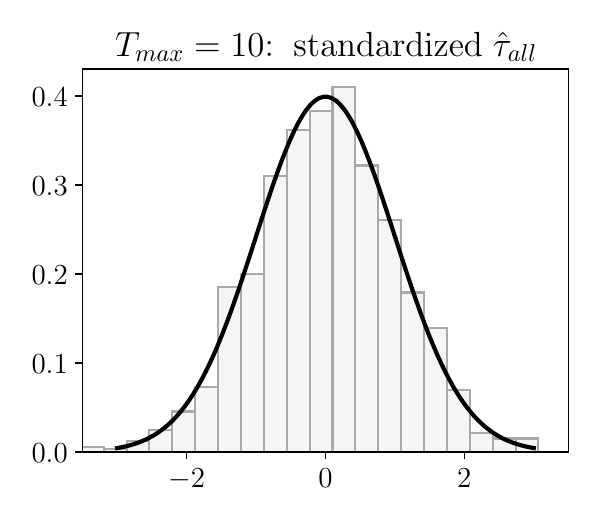}
	\end{subfigure}%
	\begin{subfigure}{0.3\textwidth}
		\centering
		\includegraphics[width=1\linewidth]{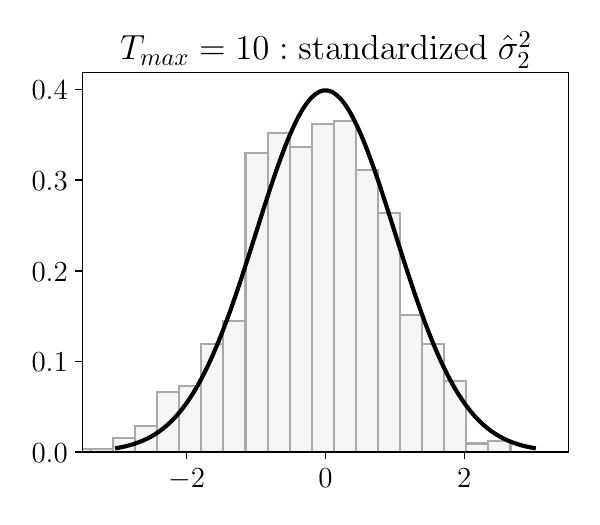}
	\end{subfigure}
	\begin{subfigure}{0.3\textwidth}
		\centering
		\includegraphics[width=1\linewidth]{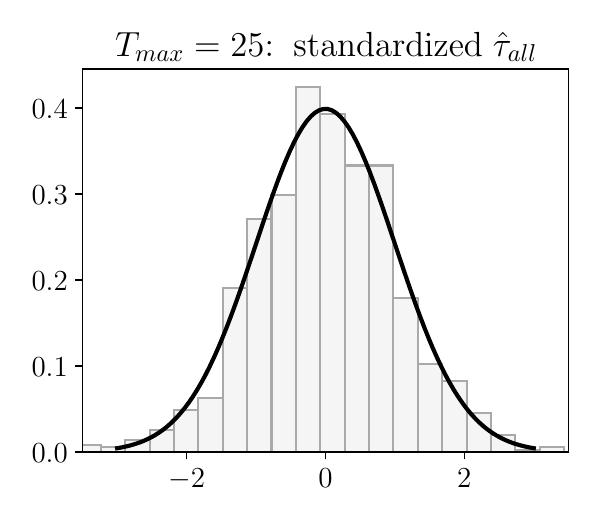}
	\end{subfigure}%
	\begin{subfigure}{0.3\textwidth}
		\centering
		\includegraphics[width=1\linewidth]{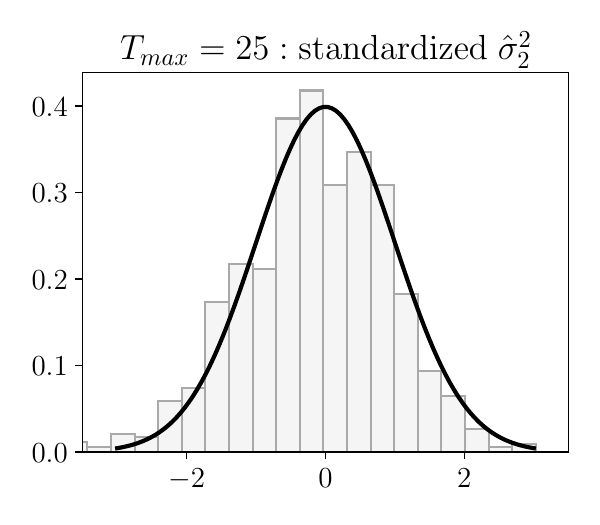}
	\end{subfigure}
	\caption{Finite sample properties of Theorem  \ref{theorem:asymptotic-page}: Histograms of $\hat{\tau}_{\all,\mathsf{ss}}$ and $\estsigmasq_{\ad,2,\mathsf{ss}}$.  The standard normal density function is superimposed on the histograms. $N = 500$, $\tau_0 = 1$, and $\sigma_\varepsilon=1$}
	\label{fig:test-asymptotics-page}
\end{figure}

We report the histograms of $\hat{\tau}_{\all,s}$ and $\estsigmasq_{\ad,2,{\mathsf{ss}}}$ in Figure \ref{fig:test-asymptotics-page}. The histograms of $\hat{\tau}_{\mathsf{ss}}$ and  $\estsigmasq_{\mathsf{ss}}$ are close to the standard normal density function, showing the good finite sample properties of $\hat{\tau}_{\all,{\mathsf{ss}}}$ and $\estsigmasq_{\ad,2,{\mathsf{ss}}}$ (even though the experiment termination times vary across iterations). 

Furthermore, we report the mean of $\hat{\tau}_{\all,{\mathsf{ss}}}$ and $\estsigmasq_{\ad,2,{\mathsf{ss}}}$, respectively, which serve as the estimates of $\var(\hat{\tau}_{\all,{\mathsf{ss}}})$ and $\var(\estsigmasq_{\ad,2,{\mathsf{ss}}})$ in Table \ref{tab:test-asymptotics}.
They are close to $1$ for various $T_{\max}$, that additionally show the good finite sample properties and verify the asymptotic distribution of $\hat{\tau}_{\all,{\mathsf{ss}}}$ and $\estsigmasq_{\ad,2,{\mathsf{ss}}}$. Moreover, we report the mean of $\hat{\tau}_{\all,{\mathsf{ss}}} \cdot \estsigmasq_{\ad,2,{\mathsf{ss}}}$, which serve as the estimates of $\mathrm{Cov}(\hat{\tau}_{\all,{\mathsf{ss}}}, \estsigmasq_{\ad,2,{\mathsf{ss}}})$, verifying the mutual asymptotic independence between $\hat{\tau}_{\all,{\mathsf{ss}}}$ and $\estsigmasq_{\ad,2,{\mathsf{ss}}}$.

\begin{table}[h!]
    \centering
{\footnotesize
\begin{tabular}{l|cccccc}
\toprule
  $T_{\max}$ &  $\widehat{\var}(\hat{\tau}_{\all,{\mathsf{ss}}})$ & $\widehat{\var}(\estsigmasq_{\ad,2,{\mathsf{ss}}})$ & $\widehat{\mathrm{Cov}}(\hat{\tau}_{\all,{\mathsf{ss}}}, \estsigmasq_{\ad,2,{\mathsf{ss}}})$   \\
\midrule
5 &  1.062 &  1.113 &  0.001 \\
10 &  1.013 &  1.076 &  0.006 \\
15 &  1.033 &  1.038 & -0.050 \\
20 &  1.007 &  1.078 & -0.001 \\
25 &  1.035 &  1.113 & -0.016 \\
30 &  1.072 &  1.099 &  0.028 \\
35 &  0.931 &  1.026 & -0.014 \\
\bottomrule
\end{tabular}
}
    \caption{Finite sample properties of Theorem  \ref{theorem:asymptotic-page}: Estimates of $\var(\hat{\tau}_{\all,{\mathsf{ss}}})$, $\var(\estsigmasq_{\ad,2,{\mathsf{ss}}})$, and $\Cov(\hat{\tau}_{\all,{\mathsf{ss}}}, \estsigmasq_{\ad,2,{\mathsf{ss}}})$ with $N = 500$, $\tau_0 = 1$, $\sigma_\varepsilon=1$.}
    \label{tab:theorem-asymptotic-covariance}
\end{table}

%% file: main.bbl
\begin{thebibliography}{75}
\providecommand{\natexlab}[1]{#1}
\providecommand{\url}[1]{\texttt{#1}}
\providecommand{\urlprefix}{URL }

\bibitem[{Abadie et~al.(2010)Abadie, Diamond, \protect\BIBand{}
  Hainmueller}]{abadie2010synthetic}
Abadie A, Diamond A, Hainmueller J (2010) Synthetic control methods for
  comparative case studies: Estimating the effect of california’s tobacco
  control program. \emph{Journal of the American statistical Association}
  105(490):493--505.

\bibitem[{Abadie \protect\BIBand{} Zhao(2021)}]{abadie2021synthetic}
Abadie A, Zhao J (2021) Synthetic controls for experimental design. \emph{arXiv
  preprint arXiv:2108.02196} .

\bibitem[{Abaluck et~al.(2021)Abaluck, Kwong, Styczynski, Haque, Kabir,
  Bates-Jefferys, Crawford, Benjamin-Chung, Raihan, Rahman, Benhachmi, Bintee,
  Winch, Hossain, Reza, Jaber, Momen, Rahman, Banti, Huq, Luby,
  \protect\BIBand{} Mobarak}]{abaluck2021impact}
Abaluck J, Kwong LH, Styczynski A, Haque A, Kabir MA, Bates-Jefferys E,
  Crawford E, Benjamin-Chung J, Raihan S, Rahman S, Benhachmi S, Bintee NZ,
  Winch PJ, Hossain M, Reza HM, Jaber AA, Momen SG, Rahman A, Banti FL, Huq TS,
  Luby SP, Mobarak AM (2021) Impact of community masking on covid-19: A
  cluster-randomized trial in bangladesh. \emph{Science} .

\bibitem[{Angrist et~al.(1999)Angrist, Imbens, \protect\BIBand{}
  Krueger}]{angrist1999jackknife}
Angrist JD, Imbens GW, Krueger AB (1999) Jackknife instrumental variables
  estimation. \emph{Journal of Applied Econometrics} 14(1):57--67.

\bibitem[{Angrist \protect\BIBand{} Krueger(1995)}]{angrist1995split}
Angrist JD, Krueger AB (1995) Split-sample instrumental variables estimates of
  the return to schooling. \emph{Journal of Business \& Economic Statistics}
  13(2):225--235.

\bibitem[{Angrist \protect\BIBand{} Pischke(2008)}]{angrist2008mostly}
Angrist JD, Pischke JS (2008) \emph{Mostly harmless econometrics: An
  empiricist's companion} (Princeton university press).

\bibitem[{Athey et~al.(2021)Athey, Bayati, Doudchenko, Imbens,
  \protect\BIBand{} Khosravi}]{athey2018matrix}
Athey S, Bayati M, Doudchenko N, Imbens G, Khosravi K (2021) Matrix completion
  methods for causal panel data models. \emph{Journal of the American
  Statistical Association} 116(536):1716--1730.

\bibitem[{Athey \protect\BIBand{} Imbens(2016)}]{athey2016recursive}
Athey S, Imbens G (2016) Recursive partitioning for heterogeneous causal
  effects. \emph{Proceedings of the National Academy of Sciences}
  113(27):7353--7360.

\bibitem[{Atkinson et~al.(2007)Atkinson, Donev, \protect\BIBand{}
  Tobias}]{atkinson2007optimum}
Atkinson A, Donev A, Tobias R (2007) \emph{Optimum experimental designs, with
  SAS}, volume~34 (Oxford University Press).

\bibitem[{Atkinson \protect\BIBand{}
  Fedorov(1975{\natexlab{a}})}]{atkinson1975design}
Atkinson A, Fedorov V (1975{\natexlab{a}}) The design of experiments for
  discriminating between two rival models. \emph{Biometrika} 62(1):57--70.

\bibitem[{Atkinson \protect\BIBand{}
  Fedorov(1975{\natexlab{b}})}]{atkinson1975optimal}
Atkinson AC, Fedorov VV (1975{\natexlab{b}}) Optimal design: Experiments for
  discriminating between several models. \emph{Biometrika} 62(2):289--303.

\bibitem[{Auer(2003)}]{auer2003using}
Auer P (2003) Using confidence bounds for exploitation-exploration trade-offs
  3(null):397–422, ISSN 1532-4435.

\bibitem[{Bai(2003)}]{bai2003inferential}
Bai J (2003) Inferential theory for factor models of large dimensions.
  \emph{Econometrica} 71(1):135--171.

\bibitem[{Bai \protect\BIBand{} Ng(2002)}]{bai2002determining}
Bai J, Ng S (2002) Determining the number of factors in approximate factor
  models. \emph{Econometrica} 70(1):191--221.

\bibitem[{Bajari et~al.(2023)Bajari, Burdick, Imbens, Masoero, McQueen,
  Richardson, \protect\BIBand{} Rosen}]{bajari2021multiple}
Bajari P, Burdick B, Imbens GW, Masoero L, McQueen J, Richardson TS, Rosen IM
  (2023) Experimental design in marketplaces. \emph{Statistical Science}
  1(1):1--19.

\bibitem[{Basse et~al.(2023)Basse, Ding, \protect\BIBand{}
  Toulis}]{basse2019minimax}
Basse GW, Ding Y, Toulis P (2023) Minimax designs for causal effects in
  temporal experiments with treatment habituation. \emph{Biometrika}
  110(1):155--168.

\bibitem[{Bastani \protect\BIBand{} Bayati(2020)}]{bastani2020online}
Bastani H, Bayati M (2020) Online decision making with high-dimensional
  covariates. \emph{Operations Research} 68(1):276--294.

\bibitem[{Bertsekas(2012)}]{bertsekas2012dynamic}
Bertsekas D (2012) \emph{Dynamic programming and optimal control: Volume I},
  volume~1 (Athena scientific).

\bibitem[{Bertsimas et~al.(2015)Bertsimas, Johnson, \protect\BIBand{}
  Kallus}]{bertsimas2015power}
Bertsimas D, Johnson M, Kallus N (2015) The power of optimization over
  randomization in designing experiments involving small samples.
  \emph{Operations Research} 63(4):868--876.

\bibitem[{Bertsimas et~al.(2019)Bertsimas, Korolko, \protect\BIBand{}
  Weinstein}]{bertsimas2019covariate}
Bertsimas D, Korolko N, Weinstein AM (2019) Covariate-adaptive optimization in
  online clinical trials. \emph{Operations Research} .

\bibitem[{Bhat et~al.(2019)Bhat, Farias, Moallemi, \protect\BIBand{}
  Sinha}]{bhat2019near}
Bhat N, Farias VF, Moallemi CC, Sinha D (2019) Near optimal ab testing.
  \emph{Management Science} .

\bibitem[{Bojinov et~al.(2023)Bojinov, Simchi-Levi, \protect\BIBand{}
  Zhao}]{bojinov2020design}
Bojinov I, Simchi-Levi D, Zhao J (2023) Design and analysis of switchback
  experiments. \emph{Management Science} 69(7):3759--3777.

\bibitem[{Brown \protect\BIBand{} Lilford(2006)}]{brown2006stepped}
Brown CA, Lilford RJ (2006) The stepped wedge trial design: a systematic
  review. \emph{BMC medical research methodology} 6(1):54.

\bibitem[{Bubeck et~al.(2012)Bubeck, Cesa-Bianchi et~al.}]{bubeck2012regret}
Bubeck S, Cesa-Bianchi N, et~al. (2012) Regret analysis of stochastic and
  nonstochastic multi-armed bandit problems. \emph{Foundations and
  Trends{\textregistered} in Machine Learning} 5(1):1--122.

\bibitem[{Cachon et~al.(2019)Cachon, Gallino, \protect\BIBand{}
  Olivares}]{cachon2019does}
Cachon GP, Gallino S, Olivares M (2019) Does adding inventory increase sales?
  evidence of a scarcity effect in us automobile dealerships. \emph{Management
  Science} 65(4):1469--1485.

\bibitem[{Card \protect\BIBand{} Krueger(1994)}]{card1994minimum}
Card D, Krueger AB (1994) Minimum wages and employment: A case study of the
  fast-food industry in new jersey and pennsylvania. \emph{American Economic
  Review} 84(4):772--93.

\bibitem[{Chernozhukov et~al.(2018)Chernozhukov, Chetverikov, Demirer, Duflo,
  Hansen, Newey, \protect\BIBand{} Robins}]{chernozhukov2018double}
Chernozhukov V, Chetverikov D, Demirer M, Duflo E, Hansen C, Newey W, Robins J
  (2018) Double/debiased machine learning for treatment and structural
  parameters: Double/debiased machine learning.

\bibitem[{Chow \protect\BIBand{} Robbins(1965)}]{chow1965asymptotic}
Chow YS, Robbins H (1965) On the asymptotic theory of fixed-width sequential
  confidence intervals for the mean. \emph{The Annals of Mathematical
  Statistics} 36(2):457--462.

\bibitem[{Cui et~al.(2019)Cui, Zhang, \protect\BIBand{}
  Bassamboo}]{cui2019learning}
Cui R, Zhang DJ, Bassamboo A (2019) Learning from inventory availability
  information: Evidence from field experiments on amazon. \emph{Management
  Science} 65(3):1216--1235.

\bibitem[{De~Stavola \protect\BIBand{} Cox(2008)}]{de2008consequences}
De~Stavola B, Cox D (2008) On the consequences of overstratification.
  \emph{Biometrika} 95(4):992--996.

\bibitem[{Deshpande et~al.(2021)Deshpande, Javanmard, \protect\BIBand{}
  Mehrabi}]{deshpande2021online}
Deshpande Y, Javanmard A, Mehrabi M (2021) Online debiasing for adaptively
  collected high-dimensional data with applications to time series analysis.
  \emph{Journal of the American Statistical Association} 0(0):1--14,
  \urlprefix\url{http://dx.doi.org/10.1080/01621459.2021.1979011}.

\bibitem[{Deshpande et~al.(2018)Deshpande, Mackey, Syrgkanis, \protect\BIBand{}
  Taddy}]{deshpande2017accurate}
Deshpande Y, Mackey L, Syrgkanis V, Taddy M (2018) Accurate inference for
  adaptive linear models 1194--1203.

\bibitem[{Dette et~al.(2015)Dette, Melas, \protect\BIBand{}
  Guchenko}]{dette2015bayesian}
Dette H, Melas VB, Guchenko R (2015) Bayesian t-optimal discriminating designs.
  \emph{Annals of statistics} 43(5):1959.

\bibitem[{Dette et~al.(2012)Dette, Melas, Shpilev et~al.}]{dette2012t}
Dette H, Melas VB, Shpilev P, et~al. (2012) T-optimal designs for
  discrimination between two polynomial models. \emph{The Annals of Statistics}
  40(1):188--205.

\bibitem[{Dette et~al.(2013)Dette, Melas, Shpilev et~al.}]{dette2013robust}
Dette H, Melas VB, Shpilev P, et~al. (2013) Robust $ t $-optimal discriminating
  designs. \emph{Annals of Statistics} 41(4):1693--1715.

\bibitem[{Doudchenko et~al.(2019)Doudchenko, Gilinson, Taylor,
  \protect\BIBand{} Wernerfelt}]{doudchenkodesigning2021}
Doudchenko N, Gilinson D, Taylor S, Wernerfelt N (2019) Designing experiments
  with synthetic controls .

\bibitem[{Doudchenko et~al.(2021)Doudchenko, Khosravi, Pouget-Abadie, Lahaie,
  Lubin, Mirrokni, Spiess et~al.}]{doudchenko2021synthetic}
Doudchenko N, Khosravi K, Pouget-Abadie J, Lahaie S, Lubin M, Mirrokni V,
  Spiess J, et~al. (2021) Synthetic design: An optimization approach to
  experimental design with synthetic controls. \emph{Advances in Neural
  Information Processing Systems} 34:8691--8701.

\bibitem[{Efron(1971)}]{efron1971forcing}
Efron B (1971) Forcing a sequential experiment to be balanced.
  \emph{Biometrika} 58(3):403--417.

\bibitem[{Fox(2000)}]{fox2000separability}
Fox BL (2000) Separability in optimal allocation. \emph{Operations Research}
  48(1):173--176.

\bibitem[{Girling \protect\BIBand{} Hemming(2016)}]{girling2016statistical}
Girling AJ, Hemming K (2016) Statistical efficiency and optimal design for
  stepped cluster studies under linear mixed effects models. \emph{Statistics
  in medicine} 35(13):2149--2166.

\bibitem[{Glynn et~al.(2020)Glynn, Johari, \protect\BIBand{}
  Rasouli}]{glynn2020adaptive}
Glynn PW, Johari R, Rasouli M (2020) Adaptive experimental design with temporal
  interference: A maximum likelihood approach. \emph{Advances in Neural
  Information Processing Systems} 33:15054--15064.

\bibitem[{Glynn \protect\BIBand{}
  Whitt(1992{\natexlab{a}})}]{glynn1992asymptotic}
Glynn PW, Whitt W (1992{\natexlab{a}}) The asymptotic efficiency of simulation
  estimators. \emph{Operations research} 40(3):505--520.

\bibitem[{Glynn \protect\BIBand{}
  Whitt(1992{\natexlab{b}})}]{glynn1992asymptoticstopping}
Glynn PW, Whitt W (1992{\natexlab{b}}) The asymptotic validity of sequential
  stopping rules for stochastic simulations. \emph{The Annals of Applied
  Probability} 2(1):180--198.

\bibitem[{Goldenshluger \protect\BIBand{}
  Zeevi(2013)}]{goldenshluger2013linear}
Goldenshluger A, Zeevi A (2013) A linear response bandit problem.
  \emph{Stochastic Systems} 3(1):230--261.

\bibitem[{Gupta et~al.(2019)Gupta, Kohavi, Tang, Xu, Andersen, Bakshy, Cardin,
  Chandran, Chen, Coey, Curtis, Deng, Duan, Forbes, Frasca, Guy, Imbens,
  Saint~Jacques, Kantawala, Katsev, Katzwer, Konutgan, Kunakova, Lee, Lee, Liu,
  McQueen, Najmi, Smith, Trehan, Vermeer, Walker, Wong, \protect\BIBand{}
  Yashkov}]{gupta2019top}
Gupta S, Kohavi R, Tang D, Xu Y, Andersen R, Bakshy E, Cardin N, Chandran S,
  Chen N, Coey D, Curtis M, Deng A, Duan W, Forbes P, Frasca B, Guy T, Imbens
  GW, Saint~Jacques G, Kantawala P, Katsev I, Katzwer M, Konutgan M, Kunakova
  E, Lee M, Lee M, Liu J, McQueen J, Najmi A, Smith B, Trehan V, Vermeer L,
  Walker T, Wong J, Yashkov I (2019) Top challenges from the first practical
  online controlled experiments summit. \emph{SIGKDD Explor. Newsl.}
  21(1):20–35, ISSN 1931-0145.

\bibitem[{Hamidi et~al.(2019)Hamidi, Bayati, \protect\BIBand{}
  Gupta}]{hamidi2019personalizing}
Hamidi N, Bayati M, Gupta K (2019) Personalizing many decisions with
  high-dimensional covariates. volume~32.

\bibitem[{Hastie et~al.(2015)Hastie, Mazumder, Lee, \protect\BIBand{}
  Zadeh}]{hastie2015matrix}
Hastie T, Mazumder R, Lee JD, Zadeh R (2015) Matrix completion and low-rank svd
  via fast alternating least squares. \emph{The Journal of Machine Learning
  Research} 16(1):3367--3402.

\bibitem[{Hayes(2002)}]{hayes2002computing}
Hayes B (2002) Computing science: The easiest hard problem. \emph{American
  Scientist} 90(2):113--117.

\bibitem[{Hayes \protect\BIBand{} Moulton(2017)}]{hayes2017cluster}
Hayes RJ, Moulton LH (2017) \emph{Cluster randomised trials} (Chapman and
  Hall/CRC).

\bibitem[{Hemming et~al.(2015)Hemming, Haines, Chilton, Girling,
  \protect\BIBand{} Lilford}]{hemming2015stepped}
Hemming K, Haines TP, Chilton PJ, Girling AJ, Lilford RJ (2015) The stepped
  wedge cluster randomised trial: rationale, design, analysis, and reporting.
  \emph{Bmj} 350:h391.

\bibitem[{Hussey \protect\BIBand{} Hughes(2007)}]{hussey2007design}
Hussey MA, Hughes JP (2007) Design and analysis of stepped wedge cluster
  randomized trials. \emph{Contemporary Clinical Trials} 28(2):182--191.

\bibitem[{Imbens \protect\BIBand{} Rubin(2015)}]{imbens2015causal}
Imbens GW, Rubin DB (2015) \emph{Causal inference in statistics, social, and
  biomedical sciences} (Cambridge University Press).

\bibitem[{Johari et~al.(2017)Johari, Koomen, Pekelis, \protect\BIBand{}
  Walsh}]{johari2017peeking}
Johari R, Koomen P, Pekelis L, Walsh D (2017) Peeking at a/b tests: Why it
  matters, and what to do about it. \emph{Proceedings of the 23rd ACM SIGKDD
  International Conference on Knowledge Discovery and Data Mining}, 1517--1525.

\bibitem[{Johari et~al.(2022)Johari, Li, Liskovich, \protect\BIBand{}
  Weintraub}]{johari2020experimental}
Johari R, Li H, Liskovich I, Weintraub G (2022) Experimental design in
  two-sided platforms: An analysis of bias. \emph{Management Science} .

\bibitem[{Ju et~al.(2019)Ju, Hu, Henderson, \protect\BIBand{}
  Hong}]{ju2019sequential}
Ju N, Hu D, Henderson A, Hong L (2019) A sequential test for selecting the
  better variant: Online a/b testing, adaptive allocation, and continuous
  monitoring. \emph{Proceedings of the Twelfth ACM International Conference on
  Web Search and Data Mining}, 492--500.

\bibitem[{Kernan et~al.(1999)Kernan, Viscoli, Makuch, Brass, \protect\BIBand{}
  Horwitz}]{kernan1999stratified}
Kernan WN, Viscoli CM, Makuch RW, Brass LM, Horwitz RI (1999) Stratified
  randomization for clinical trials. \emph{Journal of clinical epidemiology}
  52(1):19--26.

\bibitem[{Lai \protect\BIBand{} Wei(1982)}]{lai1982least}
Lai TL, Wei CZ (1982) {Least Squares Estimates in Stochastic Regression Models
  with Applications to Identification and Control of Dynamic Systems}.
  \emph{The Annals of Statistics} 10(1):154 -- 166,
  \urlprefix\url{http://dx.doi.org/10.1214/aos/1176345697}.

\bibitem[{Lattimore \protect\BIBand{}
  Szepesv{\'a}ri(2020)}]{lattimore2018bandit}
Lattimore T, Szepesv{\'a}ri C (2020) Bandit algorithms .

\bibitem[{Lawrie et~al.(2015)Lawrie, Carlin, \protect\BIBand{}
  Forbes}]{lawrie2015optimal}
Lawrie J, Carlin JB, Forbes AB (2015) Optimal stepped wedge designs.
  \emph{Statistics \& Probability Letters} 99:210--214.

\bibitem[{Li et~al.(2018)Li, Turner, \protect\BIBand{}
  Preisser}]{li2018optimal}
Li F, Turner EL, Preisser JS (2018) Optimal allocation of clusters in cohort
  stepped wedge designs. \emph{Statistics \& Probability Letters} 137:257--263.

\bibitem[{Mertens(2006)}]{mertens2006easiest}
Mertens S (2006) The easiest hard problem: Number partitioning.
  \emph{Computational Complexity and Statistical Physics} 125(2):125--139.

\bibitem[{Mulvey(1983)}]{mulvey1983multivariate}
Mulvey JM (1983) Multivariate stratified sampling by optimization.
  \emph{Management Science} 29(6):715--724.

\bibitem[{Ng et~al.(2001)Ng, Jordan, \protect\BIBand{} Weiss}]{ng2001spectral}
Ng A, Jordan M, Weiss Y (2001) On spectral clustering: Analysis and an
  algorithm. \emph{Advances in neural information processing systems} 14.

\bibitem[{Nikolaev et~al.(2013)Nikolaev, Jacobson, Cho, Sauppe,
  \protect\BIBand{} Sewell}]{nikolaev2013balance}
Nikolaev AG, Jacobson SH, Cho WKT, Sauppe JJ, Sewell EC (2013) Balance
  optimization subset selection (boss): An alternative approach for causal
  inference with observational data. \emph{Operations Research} 61(2):398--412.

\bibitem[{Robinson(1988)}]{robinson1988root}
Robinson PM (1988) Root-n-consistent semiparametric regression.
  \emph{Econometrica: Journal of the Econometric Society} 931--954.

\bibitem[{Siegmund(1985)}]{siegmund1985sequential}
Siegmund D (1985) \emph{Sequential analysis: tests and confidence intervals}
  (Springer Science \& Business Media).

\bibitem[{Singham \protect\BIBand{} Schruben(2012)}]{singham2012finite}
Singham DI, Schruben LW (2012) Finite-sample performance of absolute precision
  stopping rules. \emph{INFORMS Journal on Computing} 24(4):624--635.

\bibitem[{Uci{\'n}ski \protect\BIBand{} Bogacka(2005)}]{ucinski2005t}
Uci{\'n}ski D, Bogacka B (2005) T-optimum designs for discrimination between
  two multiresponse dynamic models. \emph{Journal of the Royal Statistical
  Society: Series B (Statistical Methodology)} 67(1):3--18.

\bibitem[{Wager \protect\BIBand{} Athey(2018)}]{wager2018estimation}
Wager S, Athey S (2018) Estimation and inference of heterogeneous treatment
  effects using random forests. \emph{Journal of the American Statistical
  Association} 113(523):1228--1242.

\bibitem[{Wager \protect\BIBand{} Xu(2021)}]{wager2021experimenting}
Wager S, Xu K (2021) Experimenting in equilibrium. \emph{Management Science}
  67(11):6694--6715.

\bibitem[{Wald(2004)}]{wald2004sequential}
Wald A (2004) \emph{Sequential analysis} (Courier Corporation).

\bibitem[{Wallace \protect\BIBand{} Hussain(1969)}]{wallace1969use}
Wallace TD, Hussain A (1969) The use of error components models in combining
  cross section with time series data. \emph{Econometrica: Journal of the
  Econometric Society} 55--72.

\bibitem[{Wiens(2009)}]{wiens2009robust}
Wiens DP (2009) Robust discrimination designs. \emph{Journal of the Royal
  Statistical Society: Series B (Statistical Methodology)} 71(4):805--829.

\bibitem[{Woertman et~al.(2013)Woertman, de~Hoop, Moerbeek, Zuidema, Gerritsen,
  \protect\BIBand{} Teerenstra}]{woertman2013stepped}
Woertman W, de~Hoop E, Moerbeek M, Zuidema SU, Gerritsen DL, Teerenstra S
  (2013) Stepped wedge designs could reduce the required sample size in cluster
  randomized trials. \emph{Journal of Clinical Epidemiology} 66(7):752--758.

\bibitem[{Xiong et~al.(2023)Xiong, Chin, \protect\BIBand{}
  Taylor}]{xiong2023bias}
Xiong R, Chin A, Taylor S (2023) Bias-variance tradeoffs for designing
  simultaneous temporal experiments. \emph{The KDD'23 Workshop on Causal
  Discovery, Prediction and Decision}, 115--131 (PMLR).

\end{thebibliography}
